\documentclass[a4paper,11pt]{article}
\usepackage{jheppub} 
\usepackage{lineno}
\usepackage{bm}
\usepackage{mathrsfs}
\usepackage{booktabs}

\newcommand{\ii}{\mathrm{i}}
\newcommand{\dd}{\mathrm{d}}
\newcommand{\cC}{\mathcal C}
\newcommand{\cD}{\mathcal D}

\newcommand{\tlam}{\widetilde\lambda}
\newcommand{\tnu}{\widetilde\nu}
\newcommand{\mR}{\widetilde M}
\newcommand{\TO}{\mathrm{TO}}

\newcommand{\tF}[4]{%
{}_{2}\mathbf F_{1}\!\left[\begin{matrix}#1,#2\\#3\end{matrix};#4\right]}

\title{\boldmath Cosmological Collider Signals from Fermion Bubble Loops with a Chemical Potential}

\author[a,b]{Chengcheng Han,}
\author[a]{ Zhanhong Lei,}
\author[c]{ Zhehan Qin,}
\author[a]{ Wuzhou Yin}
\affiliation[a]{School of Physics, Sun Yat-Sen University, Guangzhou 510275, P. R. China}
\affiliation[b]{Asia Pacific Center for Theoretical Physics, Pohang 37673, Korea}
\affiliation[c]{Department of Applied Mathematics and Theoretical Physics, University of Cambridge, Wilberforce Road, Cambridge, CB3 0WA, UK}

\emailAdd{hanchch@mail.sysu.edu.cn}
\emailAdd{leizhh3@alumni.sysu.edu.cn}
\emailAdd{zq242@cam.ac.uk}
\emailAdd{yinwzh3@mail2.sysu.edu.cn}

\abstract{
We present the first systematic analytical calculation of the local and nonlocal cosmological collider (CC) signals generated by massive fermion bubble loops in the presence of an effective chemical potential during inflation. Working in the de Sitter approximation, we investigate the four-point correlator of inflaton perturbations in the single-squeezed limit. By combining CC cutting rules, late-time expansions, and partial Mellin–Barnes (MB) techniques, we derive analytical expressions for both local and nonlocal signals. In particular, we identify two distinct contributions to the local signal, arising from spectrum and integer poles and from ultraviolet poles of the loop-momentum integral. We further investigate the three-point limit, where the separately extracted local and nonlocal contributions develop apparent folded singularities. By reorganizing the MB representation, we demonstrate that these singularities cancel in the complete signal and obtain a convergent double-MB integral suitable for numerical evaluation. Independent calculations using complementary MB representations provide nontrivial consistency checks of our results. We find that the chemical potential can significantly enhance the oscillatory signal while increasing its logarithmic oscillation frequency. Our analytical results provide a foundation for further studies of chemical-potential-enhanced fermionic CC signals, including fermion triangle loops.
}

\begin{document}
\maketitle
\flushbottom

\section{Introduction}
\label{sec:introduction}
Observations of the cosmic microwave background (CMB)~\cite{Planck:2018jri,Planck:2019kim} and the large-scale structure (LSS)~\cite{Cabass:2022wjy,DAmico:2022gki} of the Universe probe the statistics of the primordial curvature perturbation, providing a window into inflationary physics~\cite{Starobinsky:1980te,Guth:1980zm,Linde:1981mu,Linde:1983gd,Albrecht:1982wi}. While its two-point correlation function defines the primordial power spectrum, its connected higher-point functions encode primordial non-Gaussianities and reveal additional information about inflationary interactions and particle content~\cite{Maldacena:2002vr,Bartolo:2004if,Seery:2005wm,Lyth:2005fi,Chen:2006nt,Cheung:2007st,Wands:2010af,Chen:2010xka,Wang:2013zva,Renaux-Petel:2015bja,Meerburg:2019qqi,Achucarro:2022qrl,Wang:2022eop,Salcedo:2024smn,Pinol:2023oux}.
Searches for primordial non-Gaussianities in CMB and LSS data have
already placed important constraints on their amplitudes~\cite{Planck:2019kim,DAmico:2022gki}.  Future LSS surveys and $21\,\mathrm{cm}$ tomography could further tighten constraints on primordial non-Gaussianities~\cite{Meerburg:2019qqi,Achucarro:2022qrl,Munoz:2015eqa,Meerburg:2016zdz}. Of particular interest are the signatures of additional particles with masses of order the inflationary Hubble scale, $m\sim H$, which can be produced during inflation through gravitational effects or couplings to the rolling inflaton background~\cite{Parker:1968mv,Gibbons:1977mu,Adshead:2015kza,Flauger:2016idt,Adshead:2018oaa,Sou:2021juh,Racco:2024aac,Kolb:2023ydq,Feiteira:2025phi,Lust:2025auk,Deng:2026fcx}. Through their subsequent propagation and interactions with the inflaton, these particles can leave characteristic imprints on primordial non-Gaussianities, allowing their properties to be probed at energies potentially far beyond the reach of terrestrial experiments. This connection between massive particles and primordial non-Gaussianities, explored in early studies of quasi-single-field inflation, was subsequently developed into the broader framework of cosmological collider (CC) physics~\cite{Chen:2009we,Chen:2009zp,Baumann:2011nk,Assassi:2012zq,Sefusatti:2012ye,Chen:2012ge,Pi:2012gf,Noumi:2012vr,Gong:2013sma,Arkani-Hamed:2015bza,Chen:2015lza,Lee:2016vti,Chen:2016uwp,Chen:2016hrz,Chen:2018sce,Chen:2018xck,Kumar:2019ebj,Wang:2019gbi}.
\par 
A characteristic prediction of cosmological collider (CC) physics is the appearance of oscillatory momentum dependence in primordial higher-point correlation functions. In soft momentum limits, massive particles can generate nonanalytic contributions with complex scaling exponents, whose oscillation frequencies and angular dependence carry information about their masses and spins. Extracting these signals requires evaluating inflationary correlators, conventionally formulated through the in-in formalism or its equivalent Schwinger--Keldysh (SK) path-integral representation~\cite{Weinberg:2005vy,Adshead:2009cb,Musso:2006pt,Chen:2017ryl}. At loop level, the calculation becomes particularly involved because the time evolution of massive fields must be integrated together with the internal loop momenta, while nonanalytic signals must be separated from contributions analytic in the relevant momentum variables. Significant progress has been made using symmetry-based bootstrap techniques~\cite{Arkani-Hamed:2018kmz,Baumann:2019oyu,Baumann:2020dch,Qin:2022fbv,Wang:2022eop}, Mellin-space representations~\cite{Sleight:2019mgd,Sleight:2019hfp,Sleight:2020obc}, and spectral and dispersive approaches~\cite{Xianyu:2022jwk,Loparco:2023rug,Werth:2024mjg,Liu:2024xyi, Aoki:2026smw}. For the explicit calculation of loop-induced CC signals, two particularly useful developments are the partial Mellin--Barnes (MB) representation, which exposes the relevant nonanalyticities through the pole structure of the integrand~\cite{Qin:2022lva,Qin:2023bjk,Qin:2023nhv,Xianyu:2023ytd,Aoki:2026vbc}, and cosmological cutting rules, which isolate the signal contributions through the factorization properties of inflationary correlators~\cite{Melville:2021lst,Goodhew:2021oqg,Tong:2021wai,Qin:2023bjk,Ema:2024hkj,Colipi-Marchant:2025oin,Qin:2026yuo}. These developments provide the analytical tools needed to investigate loop effects involving massive particles with nontrivial spin and time-dependent interactions.
\par 
Fermions are an essential part of the CC program. Beyond the fermionic sector of the Standard Model, heavy gauge-singlet fermions, such as right-handed Majorana neutrinos, are well motivated by the seesaw mechanism for neutrino masses~\cite{Minkowski:1977sc,Yanagida:1979as,Yanagida:1980xy,Gell-Mann:1979vob,Mohapatra:1979ia,Mohapatra:1980yp,Schechter:1980gr,Schechter:1981cv} and by leptogenesis as a possible origin of the cosmic matter asymmetry~\cite{Fukugita:1986hr,Luty:1992un,Plumacher:1996kc,Barbieri:1999ma}. Studies of the Standard Model spectrum during
inflation~\cite{Chen:2016uwp,Chen:2016hrz} and neutrino signatures in primordial non-Gaussianities~\cite{Chen:2018xck,You:2024hit,Han:2024qbw,You:2026xoq}, together with proposals to probe leptogenesis through CC observables~\cite{Cui:2021iie}, further motivate the investigation of fermionic CC signals. Since fermions enter inflaton interaction vertices through bilinears, the internal fermion lines in correlators with only inflaton external legs must form closed loops. Fermionic contributions therefore first arise at loop level. The resulting CC signals are therefore suppressed by loop factors. For fermions with masses $m\gtrsim H$, gravitational production in minimal scenarios introduces an additional Boltzmann suppression, characterized by $e^{-2\pi m/H}$. A promising mechanism involves derivative inflaton--fermion couplings compatible with the approximate shift symmetry of the inflaton. On a rolling inflaton background, these couplings can induce an effective chemical potential, leading to helicity-dependent fermion production. The production of one helicity state can be exponentially enhanced by this chemical potential, thereby alleviating the heavy-mass suppression~\cite{Adshead:2015kza,Adshead:2018oaa,Deng:2026fcx}. This enhanced fermion production can therefore amplify the oscillatory CC signals generated by fermion loops~\cite{Chen:2018xck,Wang:2019gbi,Hook:2019zxa,Tong:2023krn,You:2026xoq}.
\par 
Quantifying this enhancement requires including the chemical potential explicitly in the internal fermion propagators. The chemical potential breaks de Sitter boost symmetry and modifies the mode functions in a helicity-dependent manner. The resulting calculation involves time integrals over products of Whittaker functions with mass- and chemical potential-dependent parameters, coupled to the loop momentum integration. Spinor contractions and sums over helicity contributions further complicate the analytical evaluation. Exact bubble-loop signals for fermions without a chemical potential have recently been obtained using spectral
decomposition and the partial Mellin--Barnes
representation~\cite{Aoki:2026vbc}. The nonzero chemical potential
case is substantially more challenging, however, requiring
a dedicated treatment of the helicity-dependent propagators
and the resulting Mellin pole structure. Extracting both
local and nonlocal signals and evaluating their combined
three-point limit therefore require further analytical
developments.
\par 
A further motivation for computing fermion bubble loops with a chemical potential is to provide building blocks for calculating CC signals from fermion triangle loops in soft limits. Ref.~\cite{Qin:2026yuo} demonstrated that, for scalar triangle loops, a directional cutting rule allows the leading signal to factorize into a tree-level single-exchange seed associated with the hard subgraph and a remaining bubble-type contribution. Equivalently, the leading triangle signal can be reproduced by a bubble diagram with an effective pinched coupling, including both local and nonlocal signals. The same factorization mechanism is expected to apply to fermion triangle loops, although the spinor contractions and helicity structures require additional care. The fermion bubble signals derived in this work therefore provide important building blocks for calculating local and nonlocal CC signals from fermion triangle loops with a chemical potential.
\par 
In this work, we compute the local and nonlocal CC signals from fermion bubble-loop contributions to the four-point correlator of inflaton perturbations in de Sitter spacetime with an effective chemical potential. Here, local and nonlocal signals are distinguished by their analytic structure in the internal momentum: the local signal is analytic in this momentum but oscillatory in ratios of external energy sums, whereas the nonlocal signal is nonanalytic in the internal momentum. We consider inflaton interactions involving the scalar fermion bilinear \(\psi\psi+\psi^\dagger\psi^\dagger\), which are distinct from the derivative coupling responsible for generating the effective chemical potential. While the chemical potential is incorporated into the fermion propagators, the bubble-loop vertices considered here arise from independent effective interactions. We encode their vertex-time dependence in a family of seed integrals. 
We extract the CC signals in the single-squeezed limit ($k_s=|\mathbf{k}_1+\mathbf{k}_2|\ll k_{12}$), where the $s$-channel naturally describes the exchange of soft internal momentum between the two pairs of external inflaton modes. The corresponding $t$- and $u$-channel contributions follow by permutations of the external momenta.
We expand the fermion modes at late times at one vertex while retaining their exact form at the other. Combining CC cutting rules with the partial Mellin--Barnes representation, we perform the time and loop-momentum integrations and identify the Mellin pole families responsible for the different signals. We derive residue representations for the nonlocal signal and for two distinct contributions to the local signal: one from spectrum poles combined with integer poles, and the other from UV poles generated by the loop-momentum integral. Both local contributions are included in the total oscillatory signal. We also derive the leading-pole expressions in the double-squeezed limit and examine how subleading poles modify the momentum dependence away from this regime. We independently derive the signals using two complementary MB representations and verify the agreement of the resulting expressions.
\par 
To obtain the three-point limit, we develop a complementary representation that keeps the different signal contributions combined. Although the separately extracted nonlocal and local signals can exhibit singular behavior at the folded endpoint, we demonstrate that the potentially singular terms cancel in the complete sum. By reorganizing the MB integration variables and analytically performing two contour integrations, we obtain a double-MB integral representation suitable for numerical evaluation, including at the folded endpoint. We specify the integration contours, account for the additional residues required by contour deformation, and establish absolute convergence for the contour and parameter domain considered. We numerically compare the results obtained from this integral representation with the sum of the separately computed nonlocal and local signals, providing a consistency check of the two methods. We then compute the CC signal in the three-point correlator and analyze how its oscillation amplitude and frequency depend on the fermion mass and chemical potential.
\par 
The paper is organized as follows. In Section~\ref{sec2}, we
introduce the fermion setup with an effective chemical potential
in de Sitter spacetime, derive the mode functions and SK
propagators, and define the bubble seed integrals.
In Section~\ref{sec3}, we present the extraction of CC signals
using the cutting rule, the late-time expansion, and the MB
representation, and analyze the relevant pole structure.
In Section~\ref{4}, we calculate the nonlocal and local signals
separately. In Section~\ref{sec5}, we develop a reorganized MB
representation for evaluating the three-point limit and present
the corresponding numerical results. In Section~\ref{sec6},
we summarize our conclusions. Supporting derivations and the
convergence analysis are collected in the appendices.
\label{sec:intro}
\paragraph{Notations and conventions.}
We approximate the slow-roll inflationary background spacetime by de Sitter spacetime, with the metric given by
\begin{equation}
\dd s^2 =  a^2(\tau)\left(-\dd\tau^2 + \dd \mathbf{x}^2\right)=\frac{1}{H^2\tau^2}\left(-\dd\tau^2 + \dd \mathbf{x}^2\right).
\end{equation}
Here $\tau\in(-\infty,0)$ denotes conformal time, which is related to cosmic time $t$ through $\dd\tau=\dd t/a$, while $\mathbf{x}$ denotes the comoving spatial coordinates. The scale factor $a(\tau)$ takes the form $a(\tau)=-1/(H\tau)$ in the de Sitter approximation. The Hubble parameter $H=\dot{a}/a=a'/a^2$ is approximately constant during slow-roll inflation. Here and throughout this paper, we adopt the notation $\dot{f}\equiv\dd f/\dd t=\partial_t f$ and $f'\equiv\dd f/\dd\tau=\partial_\tau f$ to denote derivatives of a function $f$ with respect to cosmic time and conformal time, respectively.
\par 
Throughout this paper, we describe the fermion field using a two-component Weyl spinor $\psi$ and its Hermitian conjugate $\psi^\dagger$ and adopt the conventions of Refs.~\cite{Martin:1997ns,Dreiner:2008tw}. The components of the two-component spinors $\psi$ and $\psi^\dagger$ are labeled by the spinor indices $\alpha$ and $\dot{\alpha}$, respectively. The contraction of two Weyl spinors takes the form
\begin{equation}
    \psi_1\psi_2
    =\psi_1{}^\alpha\psi_{2\alpha}
    =\epsilon^{\alpha\beta}\psi_{1\beta}\psi_{2\alpha},\quad 
    \psi_1^\dagger\psi_2^\dagger
    =\psi_{1\dot{\alpha}}^\dagger\psi_{2}^{\dagger\dot{\alpha}}
    =\epsilon_{\dot{\alpha}\dot{\beta}}\psi_{1}^{\dagger\dot{\beta}}\psi_{2}^{\dagger\dot{\alpha}},
\end{equation}
where $\psi_{\dot{\alpha}}^\dagger=(\psi_\alpha)^\dagger$, $(\psi^{\dagger\dot{\alpha}})^\dagger=\psi^\alpha$, and $\epsilon^{\dot{\alpha}\dot{\beta}}=\epsilon_{\dot{\beta}\dot{\alpha}}=\epsilon^{\alpha\beta}=\epsilon_{\beta\alpha}=-\ii\sigma^2$ is the unit antisymmetric tensor, where $\sigma^2$ is the second Pauli matrix among $\sigma^i\, (i=1,2,3)$. We can also define the sigma matrices as $\sigma^\mu=(1,\sigma^i)^\mu$ and $\bar{\sigma}^\mu=(1,-\sigma^i)^\mu$. We will also encounter spinor contractions involving sigma matrices, such as $\psi_{1}^\dagger\bar{\sigma}^\mu\psi_2 = \psi_{1\dot{\alpha}}^\dagger\bar{\sigma}^{\mu \dot{\alpha}\alpha}\psi_{2\alpha}$.
\par 
Throughout our calculations, we frequently encounter special functions, including the Gamma function $\Gamma[x]$ and the generalized hypergeometric function ${}_p\text F_q$ (most commonly ${}_{p+1}\text F_p$). For products and ratios of Gamma functions, we adopt the shorthand notation
\begin{equation}
    \frac{\Gamma[x_1]\cdots \Gamma[x_n]}{\Gamma[y_1]\cdots \Gamma[y_n]} \equiv\Gamma
    \!\left[
    \begin{matrix}
        x_1,\cdots,x_n\\
        y_1,\cdots,y_n
    \end{matrix}
    \right].
\end{equation}
The generalized hypergeometric function ${}_{p+1}\text F_p(z)$ is defined by the following series within its domain of convergence (typically $|z|<1$, while convergence on the boundary $|z|=1$ requires additional discussion, see Ref.~\cite{NIST:DLMF} for details)
\begin{equation}
\label{pFqseries}
    {}_{p+1}\text F_p\!
\left[
\begin{matrix}
a_1,\cdots,a_{p+1}\\
b_1,\cdots,b_{p}
\end{matrix}
;z
\right]
=
\sum_{n=0}^\infty \frac{(a_1)_n\cdots (a_{p+1})_n}{(b_1)_n\cdots(b_p)_n}\frac{z^n}{n!},
\end{equation}
where $(a)_n=\Gamma[a+n]/\Gamma[a]$ denotes the Pochhammer symbol. For a nonterminating series, the above representation is no longer convergent for $|z|>1$, and the generalized hypergeometric function must be understood through analytic continuation. An important case is the convergence of the series at $z=1$, which is governed by the parametric excess
\begin{equation}
    \mathfrak s=\sum_{i=1}^{p}b_i-\sum_{j=1}^{p+1}a_j.
\end{equation}
For a nonterminating series, the series representation does not converge when $\text{Re}\, \mathfrak s\leq0$. We define the regularized generalized hypergeometric function as
\begin{equation}
\label{reghyperfun}
    {}_{p}\mathbf F_q\!
\left[
\begin{matrix}
a_1,\cdots,a_{p}\\
b_1,\cdots,b_{q}
\end{matrix}
;z
\right]
\equiv 
\frac{1}{\Gamma[b_1,\cdots,b_{q}]}\, 
{}_{p}\text F_q
\!\left[
\begin{matrix}
a_1,\cdots,a_{p}\\
b_1,\cdots,b_{q}
\end{matrix}
;z
\right].
\end{equation}
\par 
Unless otherwise stated, we consider connected correlators of inflaton
perturbations evaluated at the end of inflation. In the late-time limit,
we write
\begin{equation}
\begin{aligned}
    \lim_{\tau\to 0^-}
    \langle\delta\phi(\tau,\mathbf{k}_1)\cdots
    \delta\phi(\tau,\mathbf{k}_n)\rangle
    &\equiv
    \langle\delta\phi_{\mathbf{k}_1}\cdots
    \delta\phi_{\mathbf{k}_n}\rangle
    =
    (2\pi)^3\delta^{(3)}
    \left(\mathbf{k}_1+\cdots+\mathbf{k}_n\right)
    \langle\delta\phi_{\mathbf{k}_1}\cdots
    \delta\phi_{\mathbf{k}_n}\rangle'.
\end{aligned}
\end{equation}
Here, a prime indicates that the overall momentum-conserving factor
$(2\pi)^3\delta^{(3)}(\mathbf{k}_1+\cdots+\mathbf{k}_n)$
has been removed. For the four-point correlator, we denote the external three-momenta carried by the inflaton perturbations as $\mathbf{k}_1$, $\mathbf{k}_2$, $\mathbf{k}_3$, and $\mathbf{k}_4$. We use $k=|\mathbf{k}|$ to denote the magnitude of a three-momentum. Throughout our calculations, we frequently use the following notation for combinations of momenta
\begin{equation}
    \mathbf{k}_s\equiv \mathbf{k}_1+\mathbf{k}_2,\quad 
    k_{ij}=k_i+k_j,\quad r_1\equiv k_s/k_{12},\quad r_2\equiv k_s/k_{34}.
\end{equation}

\section{Fermions with a Chemical Potential in de Sitter Spacetime\label{sec2}}
\label{sec:setup}

In this section, we establish the ingredients needed to compute cosmological collider signals from a fermion bubble loop in the presence of a chemical potential. We approximate the spacetime of slow-roll inflation by de Sitter spacetime and consider a Weyl fermion whose derivative coupling to the rolling inflaton induces an effective chemical potential. We first quantize the fermion field and derive its mode functions. We then construct the corresponding propagators, which are needed to compute the correlators using the Schwinger--Keldysh (SK) path-integral formalism. Finally, we introduce the fermion bubble function and the associated seed integrals, which form the basis of the analysis in the following sections. For related discussions of fermions in inflationary backgrounds,
see Refs.~\cite{Aoki:2026vbc,You:2026xoq,Tong:2023krn,Chen:2018xck,Sou:2021juh}.

\subsection{Inflationary Setup and Fermion Mode Functions}
\label{subsec:fermion_modes}
In de Sitter spacetime, we consider a two-component Weyl fermion $\psi$ with a Majorana mass $M$, coupled to the inflaton field
$\phi$ through a derivative interaction. The fermionic part of the action is
\begin{equation}
    S_\psi
    =
    \int \dd^4x\, \sqrt{-g}
    \left[
        \ii\psi^\dagger\bar\varsigma^\mu \text{D}_\mu\psi
        -\frac{M}{2}
         \bigl(\psi\psi+\psi^\dagger\psi^\dagger\bigr)
        -\frac{\partial_\mu\phi}{\Lambda}
         \psi^\dagger\bar\varsigma^\mu\psi
    \right],
    \label{eq:setup_action}
\end{equation}
where $\bar\varsigma^\mu=e^\mu{}_a\bar\sigma^a$ denotes the Pauli matrices in curved spacetime, with $e^\mu{}_a=a^{-1}(\tau )\delta^\mu_a$ the vierbein of de Sitter spacetime, $\text{D}_\mu$ is the covariant derivative acting on the Weyl spinor, whose explicit form is given in Appendix~\ref{app:fermions_dS}, and $\Lambda$ denotes the UV scale suppressing the higher-dimensional derivative interaction. In the above action, we work in conformal coordinates of de Sitter spacetime, for which $\partial_0\phi=\partial_\tau\phi(\tau)\equiv\phi'(\tau)$ and the metric determinant factor is $\sqrt{-g}=a^4(\tau)$.
\par 
During inflation, the inflaton field $\phi(\tau,\mathbf{x})$ can be decomposed into a spatially homogeneous slow-roll background $\phi_0(\tau)$ and an inhomogeneous perturbation $\delta\phi(\tau,\mathbf{x})$ as
\begin{equation}
    \phi(\tau,\mathbf{x})=\phi_0(\tau)+\delta\phi(\tau,\mathbf{x}).
\end{equation}
For a slowly rolling background, we impose the slow-roll condition
\begin{equation}
    \left|\delta\right|
    \equiv
    \left|
    \frac{\ddot{\phi}_0}{H\dot{\phi}_0}
    \right|
    \ll 1,
\end{equation}
such that the rolling velocity of the inflaton,
$\dot{\phi}_0\equiv\partial_t\phi_0(t)$, can be approximately treated as a constant. Due to the derivative coupling between the fermion and the inflaton, the rolling inflaton background generates an additional contribution to the quadratic fermion action, leading to a nontrivial modification of the fermion equation of motion. 
\par 
To simplify the fermion action, we rescale the Weyl spinor $\psi$ as
\begin{equation}
    \widetilde\psi\equiv a^{3/2} \psi.
\end{equation}
After this field rescaling, the quadratic part of the fermion action can be written as
\begin{equation}
\label{Spsi2}
    S_\psi^{(2)}
    =
    \int \dd^4 x\, 
    \left[
        \ii\widetilde\psi^\dagger
        \bar\sigma^\mu\partial_\mu\widetilde\psi
        -a\lambda\widetilde\psi^\dagger
        \bar\sigma^0\widetilde\psi
        -\frac{aM}{2}
         \bigl(
             \widetilde\psi\widetilde\psi
             +\widetilde\psi^\dagger\widetilde\psi^\dagger
         \bigr)
    \right],
\end{equation}
where we have combined the inflaton rolling velocity $\dot{\phi}_0$ with the UV scale $\Lambda$ to define
\begin{equation}
    \lambda \equiv \frac{\dot{\phi}_0}{\Lambda},
\end{equation}
which plays the role of an effective chemical potential for the fermion. Varying the quadratic fermion action $S_\psi^{(2)}$, we directly obtain the equation of motion for the Weyl fermion field $\widetilde{\psi}$
\begin{equation}
\label{psiEOM}
\begin{aligned}
i\bar{\sigma}^\mu \partial_\mu \widetilde{\psi} &= a\lambda\,\bar{\sigma}^0 \widetilde{\psi} + a M \widetilde{\psi}^\dagger,\\
i\sigma^\mu\partial_\mu\widetilde{\psi}^\dagger&=-a\lambda\sigma^0\widetilde{\psi}^\dagger+aM\widetilde{\psi}.
\end{aligned}
\end{equation}

To solve this set of equations of motion explicitly, we first canonically quantize the fermion field $\widetilde{\psi}$. We expand $\widetilde{\psi}$ and $\widetilde{\psi}^\dagger$ in plane-wave modes as
\begin{equation}
\label{planewave}
\begin{aligned}
\widetilde{\psi}_{\alpha}\left(\tau,\mathbf{x}\right)&=\int\frac{\dd^{3}\mathbf{k}}{(2\pi)^{3}}\sum_{s=\pm}\left[h_{\alpha}^{s}(\hat{\mathbf{k}})u_s(k,\tau)b_{\mathbf{k}}^{s}e^{\ii\mathbf{k}\cdot\mathbf{x}}+h_{\dot{\gamma}}^{s\dagger}(\hat{\mathbf{k}})\bar{\sigma}^{0\dot{\gamma}\beta}\epsilon_{\alpha\beta}v_{s}^{*}(k,\tau)b_{\mathbf{k}}^{s\dagger}e^{-\ii\mathbf{k}\cdot\mathbf{x}}\right],\\
\widetilde{\psi}^{\dagger\dot{\alpha}}\left(\tau,\mathbf{x}\right)&=\int\frac{\dd^{3}\mathbf{k}}{(2\pi)^{3}}\sum_{s=\pm}\left[\epsilon^{\dot{\alpha}\dot{\beta}}h_{\dot{\beta}}^{s\dagger}(\hat{\mathbf{k}})u^*_s(k,\tau)b_{\mathbf{k}}^{s\dagger}e^{-\ii\mathbf{k}\cdot\mathbf{x}}+\bar{\sigma}^{0\dot{\alpha}\gamma}h_{\gamma}^{s}(\hat{\mathbf{k}})v_{s}(k,\tau)b_{\mathbf{k}}^{s}e^{\ii\mathbf{k}\cdot\mathbf{x}}\right].
\end{aligned}
\end{equation}
Here, $s=\pm$ labels the two helicity states of the spinor, while $b_{\mathbf{k}}^{s\dagger}$ and $b_{\mathbf{k}}^{s}$ denote the creation and annihilation operators, respectively, for a fermionic mode with momentum $\mathbf{k}$ and helicity $s$. They satisfy the anticommutation relations
\begin{equation}
\{b_{\mathbf{k}}^{s},b_{\mathbf{k}'}^{s^\prime\dagger}\}=(2\pi)^3\delta_{ss'}\delta^3(\mathbf{k}-\mathbf{k}').    
\end{equation}
The helicity eigenspinors $h^s_\alpha(\hat{\mathbf{k}})$ satisfy the eigenvalue equations
\begin{equation}
-\hat{\mathbf{k}}^{i}\sigma_{\alpha\dot{\beta}}^{0}\bar{\sigma}^{i\dot{\beta}\gamma}h_{\gamma}^{s}(\hat{\mathbf{k}})=s h_{\alpha}^{s}(\hat{\mathbf{k}}),\quad h_{\dot{\gamma}}^{s\dagger}(\hat{\mathbf{k}})\bar{\sigma}^{0\dot{\gamma}\beta}\sigma_{\beta\dot{\alpha}}^{i}\hat{\mathbf{k}}^{i}=s h_{\dot{\alpha}}^{s\dagger}(\hat{\mathbf{k}}),
\end{equation}
and the orthonormality and completeness relations
\begin{equation}
h_{\dot{\alpha}}^{s\dagger}\bar{\sigma}^{0\dot{\alpha}\beta}h_{\beta}^{s^{\prime}}=\delta^{ss^{\prime}},\quad\sum_{s=\pm}h_{\alpha}^{s}h_{\dot{\beta}}^{s\dagger}\bar{\sigma}^{0\dot{\beta}\beta}=\delta_{\alpha}^{\beta}.  
\end{equation}
For a unit momentum vector $\hat{\mathbf{k}}=(\sin\theta\cos\varphi,\sin\theta\sin\varphi,\cos\theta)$, the helicity eigenspinors can be chosen as
\begin{equation}
    h_\alpha^+(\theta,\varphi)=
    \begin{pmatrix}
        e^{-\ii\frac{\varphi}{2}}\cos\frac{\theta}{2}\\
        e^{\ii\frac{\varphi}{2}}\sin\frac{\theta}{2}
    \end{pmatrix}_\alpha,
    \qquad
    h_\alpha^-(\theta,\varphi)=
    \begin{pmatrix}
        -e^{-\ii\frac{\varphi}{2}}\sin\frac{\theta}{2}\\
        e^{\ii\frac{\varphi}{2}}\cos\frac{\theta}{2}
    \end{pmatrix}_\alpha.
\end{equation}
In the following derivations, we will make use of a useful identity satisfied by the helicity eigenspinors
\begin{equation}
\label{eigenspinoridentity}
h_\alpha^s(\hat{\mathbf{k}})h_{\dot{\beta}}^{s\dagger}(\hat{\mathbf{k}})=\frac{1}{2}\left(\sigma^0+s\hat{\mathbf{k}}\cdot\vec{\sigma}\right)_{\alpha\dot{\beta}}.
\end{equation}
\par 
Substituting the plane-wave expansion of the Weyl spinor in Eq.~\eqref{planewave} into the equation of motion \eqref{psiEOM}, we obtain the equations of motion for the fermion mode functions
\begin{equation}
\begin{aligned}
\ii u^\prime_s+sku_s&=a\lambda u_s+aMv_s,\\
\ii v^\prime_s-skv_s&=-a\lambda v_s+aMu_s.
\end{aligned}
\end{equation}
We see that the mode functions $u$ and $v$ are coupled in this system of equations, with the coupling vanishing in the massless limit $M\to 0$. To decouple the mode functions and facilitate the solution, we can convert the coupled first-order differential equations into a set of decoupled second-order differential equations
\begin{equation}
\begin{aligned}
u_{s}^{\prime\prime}-aHu_{s}^{\prime}+\left[(s k-a\lambda)^{2}+a^{2}M^{2}+\mathrm{i}saHk\right]u_{s}=0,\\v_{s}^{\prime\prime}-aHv_{s}^{\prime}+\left[(s k-a\lambda)^{2}+a^{2}M^{2}-\mathrm{i}saHk\right]v_{s}=0.
\end{aligned}
\end{equation}
Using the scale factor $a(\tau)=-1/(H\tau)$ during inflation and appropriately rescaling the mode functions, we can recast these second-order differential equations into the standard Whittaker form. By imposing the Bunch--Davies vacuum initial conditions in the limit $\tau\to-\infty$, together with the canonical normalization conditions on the mode functions, we obtain their explicit expressions as
\begin{align}
  u_+(k,\tau)
  &= \frac{e^{+\pi\tlam/2}}{\sqrt{-2k\tau}}\,
  \mR\,W_{-1/2-\ii\tlam,\,\ii\tnu}( 2\ii k\tau),
  \\
  u_-(k,\tau)
  &= \ii \frac{e^{-\pi\tlam/2}}{\sqrt{-2k\tau}}
  W_{1/2+\ii\tlam,\,\ii\tnu}(2\ii k\tau),
  \\
  v_+(k,\tau)
  &= \ii \frac{e^{+\pi\tlam/2}}{\sqrt{-2k\tau}}
  W_{1/2-\ii\tlam,\,\ii\tnu}(2\ii k\tau),
  \\
  v_-(k,\tau)
  &= \frac{e^{-\pi\tlam/2}}{\sqrt{-2k\tau}}\,
  \mR\,W_{-1/2+\ii\tlam,\,\ii\tnu}(2\ii k\tau).
\end{align}
The detailed derivation can be found in Appendix~\ref{app:fermions_dS}. In these solutions, $W_{\kappa,\mu}(z)$ denotes the Whittaker function. We also introduce the following dimensionless parameters normalized by the Hubble scale during inflation
\begin{equation}
    \mR = \frac{M}{H},\quad \tlam = \frac{\lambda}{H},\quad \tnu = \sqrt{\mR^2+\tlam^2}.
\end{equation}
Here, the parameter $\tnu$ appearing in the Whittaker functions sets the characteristic oscillation frequency of the CC signal on a logarithmic scale. In the vanishing chemical potential limit $\tlam\to 0$, $\tnu$ reduces to the fermion mass $\mR$.
\par 
In the following analysis, we will extract the CC signals from correlators. It is therefore useful to examine the late-time behavior of the fermion mode functions. Using the asymptotic behavior of the Whittaker function in the limit $|k\tau|\to 0$
\begin{equation}
  W_{\kappa,\ii\tnu}(2\ii k\tau)
  \sim
  \frac{\Gamma[-2\ii\tnu]}{\Gamma[1/2-\ii\tnu-\kappa]}(2\ii k\tau)^{1/2+\ii\tnu}
  +
  \frac{\Gamma[2\ii\tnu]}{\Gamma[1/2+\ii\tnu-\kappa]}(2\ii k\tau)^{1/2-\ii\tnu},
\end{equation}
we obtain the late-time asymptotic expressions for the fermion mode functions as
\begin{align}
\label{latetimeexp1}
\lim_{|k\tau|\to 0}u_{+}(k,\tau)
&\simeq
e^{-\ii\pi/4+\pi(\tilde{\lambda}+\tnu )/2}
\widetilde{M}
\frac{
\Gamma[-2\ii\tilde{\nu}]}
{\Gamma[1+\ii\tilde{\lambda}-\ii\tilde{\nu}]}
(-2k\tau)^{\ii\tilde{\nu}}
+(\tilde{\nu}\to-\tilde{\nu}),
\\
\lim_{|k\tau|\to 0}u_{-}(k,\tau)
&\simeq
\ii e^{-\ii\pi/4-\pi(\tilde{\lambda}-\tnu)/2}
\frac{
\Gamma[-2\ii\tilde{\nu}]}
{\Gamma[-\ii\tilde{\lambda}-\ii\tilde{\nu}]}
(-2k\tau)^{\ii\tilde{\nu}}
+(\tilde{\nu}\to-\tilde{\nu}),
\\
\lim_{|k\tau|\to 0}v_{+}(k,\tau)
&\simeq
\ii e^{-\ii\pi/4+\pi(\tilde{\lambda}+\tnu)/2}
\frac{\Gamma[-2\ii\tilde{\nu}]}
{\Gamma[\ii\tilde{\lambda}-\ii\tilde{\nu}]}
(-2k\tau)^{\ii\tilde{\nu}}
+(\tilde{\nu}\to-\tilde{\nu}),
\\
\label{latetimeexp4}
\lim_{|k\tau|\to 0}v_{-}(k,\tau)
&\simeq
e^{-\ii\pi/4-\pi(\tilde{\lambda}-\tnu)/2}\widetilde{M}
\frac{\Gamma[-2\ii\tilde{\nu}]}
{\Gamma[1-\ii\tilde{\lambda}-\ii\tilde{\nu}]}
(-2k\tau)^{\ii\tilde{\nu}}
+(\tilde{\nu}\to-\tilde{\nu}),
\end{align}
where $(\tilde{\nu}\to-\tilde{\nu})$ denotes the term obtained by replacing $\tnu$ with $-\tnu$ in the first term. To simplify our notation, we define the late-time limit of a generic mode function $f(k,\tau)$ as
\begin{equation}
\label{latetimeexp}
  \lim_{|k\tau|\rightarrow 0}f(k,\tau) = \sum_{\rho = \pm}C_f^{(\rho)}(-2k\tau)^{\rho\ii\tnu},
\end{equation}
where $C_f^{(\rho)}$ denotes the late-time expansion coefficient of the mode function $f$, which is completely determined by Eqs.~\eqref{latetimeexp1}--\eqref{latetimeexp4}. For example, for $u_+(k,\tau)$, we have
\begin{equation}
    C_{u_+}^{(+)}=e^{-\ii\pi/4+\pi(\tilde{\lambda}+\tnu )/2}
\widetilde{M}
\frac{
\Gamma[-2\ii\tilde{\nu}]}
{\Gamma[1+\ii\tilde{\lambda}-\ii\tilde{\nu}]},\ \ \  
C_{u_+}^{(-)}=e^{-\ii\pi/4+\pi(\tilde{\lambda}-\tnu )/2}
\widetilde{M}
\frac{
\Gamma[2\ii\tilde{\nu}]}
{\Gamma[1+\ii\tilde{\lambda}+\ii\tilde{\nu}]}.
\end{equation}

\subsection{Propagators in the SK Formalism during Inflation}
\label{subsec:fermion_bubble}
We are interested in the contribution of the fermion bubble loop to the correlators of inflaton perturbations. To compute these correlators, we employ the Schwinger--Keldysh (SK) path-integral formalism, for which we need the SK propagators associated with the external inflaton perturbations and the internal fermion lines. For detailed discussions of the SK path-integral formalism and its application to fermionic fields, see Refs.~\cite{Chen:2017ryl,Chen:2018xck}. In this subsection, we only present the final expressions for these propagators in terms of the corresponding mode functions, while the detailed derivations are provided in Appendix~\ref{appendixB}.
\par 
We first consider the fermion propagators. Depending on the contractions of the Weyl fermion fields, there are four types of propagators: $\langle\widetilde{\psi}\widetilde{\psi}\rangle$, $\langle\widetilde{\psi}\widetilde{\psi}^\dagger\rangle$, $\langle\widetilde{\psi}^\dagger\widetilde{\psi}\rangle$, and $\langle\widetilde{\psi}^\dagger\widetilde{\psi}^\dagger\rangle$. In the SK path-integral formalism, the fermion field $\widetilde{\psi}$ is doubled into $\widetilde{\psi}_+$ and $\widetilde{\psi}_-$, corresponding to the forward time-ordered $(+)$ and backward anti-time-ordered $(-)$ branches of the time contour, respectively. These two branches are labeled by the SK indices $\texttt{a},\texttt{b}=\pm$. Consequently, each type of fermion propagator has four distinct SK components. For example, the propagator of the form $\langle\widetilde{\psi}\widetilde{\psi}\rangle$ is denoted by $D_{\texttt{ab},\alpha\beta}$, where the SK index pair $(\texttt{a},\texttt{b})$ takes the four possible values $\{(++),(+-),(-+),(--)\}$. We summarize all types of fermion propagators below.
\begin{itemize}
\item \textbf{Propagator of type $\langle\widetilde{\psi}\widetilde{\psi}\rangle$:} 
\begin{equation}
\label{psipropagator1}
    \begin{aligned}
D_{-+\alpha\beta}(\mathbf{k};\tau_{1},\tau_{2})&=\sum_{s=\pm}u_s(k,\tau_1)v_s^*(k,\tau_2)h^s_\alpha(\hat{\mathbf{k}})h^{s\dagger}_{\dot{\gamma}}(\hat{\mathbf{k}})\bar{\sigma}^{0\dot{\gamma}\delta}\epsilon_{\beta\delta},\\
D_{+-\alpha\beta}(\mathbf{k};\tau_{1},\tau_{2})&=
-\sum_{s=\pm}u_s(k,\tau_2)v_s^*(k,\tau_1)h^s_\beta(-\hat{\mathbf{k}})h^{s\dagger}_{\dot{\gamma}}(-\hat{\mathbf{k}})\bar{\sigma}^{0\dot{\gamma}\delta}\epsilon_{\alpha\delta},\\
D_{++\alpha\beta}(\mathbf{k};\tau_{1},\tau_{2})&=D_{-+\alpha\beta}(\mathbf{k};\tau_{1},\tau_{2})\theta(\tau_1-\tau_2)+D_{+-\alpha\beta}(\mathbf{k};\tau_{1},\tau_{2})\theta(\tau_2-\tau_1),\\
D_{--\alpha\beta}(\mathbf{k};\tau_{1},\tau_{2})&=D_{+-\alpha\beta}(\mathbf{k};\tau_{1},\tau_{2})\theta(\tau_1-\tau_2)+D_{-+\alpha\beta}(\mathbf{k};\tau_{1},\tau_{2})\theta(\tau_2-\tau_1).
\end{aligned}
\end{equation}
\item \textbf{Propagator of type $\langle\widetilde{\psi}\widetilde{\psi}^\dagger\rangle$:}
\begin{equation}
    \begin{aligned}
D_{-+\alpha}{}^{\dot{\beta}}(\mathbf{k};\tau_{1},\tau_{2})&=\sum_{s=\pm}u_s(k,\tau_1)u_s^*(k,\tau_2)h^s_\alpha(\hat{\mathbf{k}})\epsilon^{\dot{\beta}\dot{\gamma}}h^{s\dagger}_{\dot{\gamma}}(\hat{\mathbf{k}}),\\
D_{+-\alpha}{}^{\dot{\beta}}(\mathbf{k};\tau_{1},\tau_{2})&=-\sum_{s=\pm}v_{s}(k,\tau_{2})v_{s}^{*}(k,\tau_{1})\bar{\sigma}^{0\dot{\beta}\gamma}h_{\gamma}^{s}(-\hat{\mathbf{k}})h_{\dot{\gamma}}^{s\dagger}(-\hat{\mathbf{k}})\bar{\sigma}^{0\dot{\gamma}\delta}\epsilon_{\alpha\delta},\\
D_{++\alpha}{}^{\dot{\beta}}(\mathbf{k};\tau_{1},\tau_{2})&=D_{-+\alpha}{}^{\dot{\beta}}(\mathbf{k};\tau_{1},\tau_{2})\theta(\tau_1-\tau_2)+D_{+-\alpha}{}^{\dot{\beta}}(\mathbf{k};\tau_{1},\tau_{2})\theta(\tau_2-\tau_1),\\
D_{--\alpha}{}^{\dot{\beta}}(\mathbf{k};\tau_{1},\tau_{2})&=D_{+-\alpha}{}^{\dot{\beta}}(\mathbf{k};\tau_{1},\tau_{2})\theta(\tau_1-\tau_2)+D_{-+\alpha}{}^{\dot{\beta}}(\mathbf{k};\tau_{1},\tau_{2})\theta(\tau_2-\tau_1).
\end{aligned}
\end{equation}
\item \textbf{Propagator of type $\langle\widetilde{\psi}^\dagger\widetilde{\psi}\rangle$:} 
\begin{equation}
    \begin{aligned}
D_{-+}{}^{\dot{\alpha}}{}_{\beta}(\mathbf{k};\tau_{1},\tau_{2})&=\sum_{s=\pm}v_s(k,\tau_1)v_s^*(k,\tau_2)\bar{\sigma}^{0\dot{\alpha}\gamma}h^s_\gamma(\hat{\mathbf{k}})h^{s\dagger}_{\dot{\gamma}}(\hat{\mathbf{k}})\bar{\sigma}^{0\dot{\gamma}\delta}\epsilon_{\beta\delta},\\
D_{+-}{}^{\dot{\alpha}}{}_{\beta}(\mathbf{k};\tau_{1},\tau_{2})&=-\sum_{s=\pm}u_{s}(k,\tau_{2})u_{s}^{*}(k,\tau_{1})h_{\beta}^{s}(-\hat{\mathbf{k}})\epsilon^{\dot{\alpha}\dot{\gamma}}h_{\dot{\gamma}}^{s\dagger}(-\hat{\mathbf{k}}),\\
D_{++}{}^{\dot{\alpha}}{}_{\beta}(\mathbf{k};\tau_{1},\tau_{2})&=D_{-+}{}^{\dot{\alpha}}{}_{\beta}(\mathbf{k};\tau_{1},\tau_{2})\theta(\tau_1-\tau_2)+D_{+-}{}^{\dot{\alpha}}{}_{\beta}(\mathbf{k};\tau_{1},\tau_{2})\theta(\tau_2-\tau_1),\\
D_{--}{}^{\dot{\alpha}}{}_{\beta}(\mathbf{k};\tau_{1},\tau_{2})&=D_{+-}{}^{\dot{\alpha}}{}_{\beta}(\mathbf{k};\tau_{1},\tau_{2})\theta(\tau_1-\tau_2)+D_{-+}{}^{\dot{\alpha}}{}_{\beta}(\mathbf{k};\tau_{1},\tau_{2})\theta(\tau_2-\tau_1).
\end{aligned}
\end{equation}
\item \textbf{Propagator of type $\langle\widetilde{\psi}^\dagger\widetilde{\psi}^\dagger\rangle$:} 
\begin{equation}
\label{psipropagator4}
    \begin{aligned}
D_{-+}{}^{\dot{\alpha}\dot{\beta}}(\mathbf{k};\tau_{1},\tau_{2})&=\sum_{s=\pm}v_s(k,\tau_1)u_s^*(k,\tau_2)\bar{\sigma}^{0\dot{\alpha}\delta}h^s_\delta(\hat{\mathbf{k}})\epsilon^{\dot{\beta}\dot{\gamma}}h^{s\dagger}_{\dot{\gamma}}(\hat{\mathbf{k}}),\\
D_{+-}{}^{\dot{\alpha}\dot{\beta}}(\mathbf{k};\tau_{1},\tau_{2})&=
-\sum_{s=\pm}v_s(k,\tau_2)u_s^*(k,\tau_1)\bar{\sigma}^{0\dot{\beta}\delta}h^s_\delta(-\hat{\mathbf{k}})\epsilon^{\dot{\alpha}\dot{\gamma}}h^{s\dagger}_{\dot{\gamma}}(-\hat{\mathbf{k}})
,\\
D_{++}{}^{\dot{\alpha}\dot{\beta}}(\mathbf{k};\tau_{1},\tau_{2})&=D_{-+}{}^{\dot{\alpha}\dot{\beta}}(\mathbf{k};\tau_{1},\tau_{2})\theta(\tau_1-\tau_2)+D_{+-}{}^{\dot{\alpha}\dot{\beta}}(\mathbf{k};\tau_{1},\tau_{2})\theta(\tau_2-\tau_1),\\
D_{--}{}^{\dot{\alpha}\dot{\beta}}(\mathbf{k};\tau_{1},\tau_{2})&=D_{+-}{}^{\dot{\alpha}\dot{\beta}}(\mathbf{k};\tau_{1},\tau_{2})\theta(\tau_1-\tau_2)+D_{-+}{}^{\dot{\alpha}\dot{\beta}}(\mathbf{k};\tau_{1},\tau_{2})\theta(\tau_2-\tau_1).
\end{aligned}
\end{equation}
\end{itemize}
Here, $\theta(\tau)$ denotes the Heaviside step function. We distinguish the different types of propagators by the spinor indices carried by $D_{\texttt{ab}}$. For propagators of the same type, we can raise and lower the spinor indices using the antisymmetric tensor $\epsilon$. For example, for the propagator of the form $\langle\widetilde{\psi}\widetilde{\psi}^\dagger\rangle$, we can write
\begin{equation}
\begin{aligned}
    D_{-+}{}^{\alpha}{}_{\dot{\beta}}(\mathbf{k};\tau_{1},\tau_{2})
    &=\epsilon^{\alpha\gamma}D_{-+}{}_{\gamma\dot{\beta}}(\mathbf{k};\tau_{1},\tau_{2})
    =\epsilon^{\alpha\gamma}\epsilon_{\dot{\beta}\dot{\delta}}D_{-+}{}_{\gamma}{}^{\dot{\delta}}(\mathbf{k};\tau_{1},\tau_{2})
    \\&=
    \sum_s u_s(k,\tau_1)u_s^*(k,\tau_2)\epsilon^{\alpha\gamma}h_\gamma^s(\hat{\mathbf{k}})h_{\dot{\beta}}^{s\dagger}(\hat{\mathbf{k}}).
\end{aligned}
\end{equation}
This will be useful in the subsequent calculations. 
\par 
We now discuss some additional properties of the fermion propagators that will be useful in the subsequent calculations. Under the exchange of SK indices, spinor indices, and spacetime coordinates, the fermion propagators satisfy the following relations
\begin{equation}
\label{fermionpropagatorrelation1}
\begin{aligned}
D_{\texttt{ab}\alpha\beta}(\mathbf{k};\tau_{1},\tau_{2})&=-D_{\texttt{ba}\beta\alpha}(-\mathbf{k};\tau_2,\tau_1),\\
D_{\texttt{ab}\alpha}{}^{\dot{\beta}}(\mathbf{k};\tau_{1},\tau_{2})&=-D_{\texttt{ba}}{}^{\dot{\beta}}{}_{\alpha}(-\mathbf{k};\tau_2,\tau_1),\\
D_{\texttt{ab}}{}^{\dot{\alpha}}{}_{\beta}(\mathbf{k};\tau_{1},\tau_{2})&=-D_{\texttt{ba}\beta}{}^{\dot{\alpha}}(-\mathbf{k};\tau_2,\tau_1),\\
D_{\texttt{ab}}{}^{\dot{\alpha}\dot{\beta}}(\mathbf{k};\tau_{1},\tau_{2})&=-D_{\texttt{ba}}{}^{\dot{\beta}\dot{\alpha}}(-\mathbf{k};\tau_2,\tau_1).
\end{aligned}
\end{equation}
In addition, the propagators satisfy the following conjugation relations
\begin{equation}
\label{fermionpropagatorrelation2}
\begin{aligned}
\left[D_{\texttt{ab}\alpha\beta}(\mathbf{k};\tau_1,\tau_2)\right]^\ast&=-D_{-\texttt{a}-\texttt{b}\dot{\alpha}\dot{\beta}}(-\mathbf{k};\tau_1,\tau_2),\\
\left[D_{\texttt{ab}\alpha}{}^{\dot{\beta}}(\mathbf{k};\tau_1,\tau_2)\right]^\ast&=-D_{-\texttt{a}-\texttt{b}\dot{\alpha}}{}^\beta(-\mathbf{k};\tau_1,\tau_2).
\end{aligned}
\end{equation}
These relations also show that conjugation interchanges the propagator types $\langle\widetilde{\psi}\widetilde{\psi}\rangle\leftrightarrow\langle\widetilde{\psi}^\dagger\widetilde{\psi}^\dagger\rangle$ and $\langle\widetilde{\psi}\widetilde{\psi}^\dagger\rangle\leftrightarrow\langle\widetilde{\psi}^\dagger\widetilde{\psi}\rangle$.
\par 
Next, we turn to the inflaton propagators. In the slow-roll approximation, we treat the inflaton perturbation as a massless minimally coupled scalar in de Sitter spacetime. Imposing the Bunch--Davies initial condition, we obtain the following bulk-to-bulk propagators
\begin{equation}
\label{bulktobulk}
    \begin{aligned}
        G_{-+}(k,\tau_1,\tau_2)&=\frac{H^2}{2k^3}(1+\ii k\tau_1)(1-\ii k\tau_2)e^{-\ii k (\tau_1-\tau_2)},
        \\
        G_{+-}(k,\tau_1,\tau_2)&=G_{-+}^\ast (k,\tau_1,\tau_2),
        \\
        G_{++}(k,\tau_1,\tau_2)&=G_{-+}(k,\tau_1,\tau_2)\theta(\tau_1-\tau_2)+G_{+-}(k,\tau_1,\tau_2)\theta(\tau_2-\tau_1),
        \\
        G_{--}(k,\tau_1,\tau_2)&=G_{++}^\ast (k,\tau_1,\tau_2).
    \end{aligned}
\end{equation}
Since the inflaton propagators serve as external legs in our calculation, we also need the bulk-to-boundary propagators, which can be obtained by taking the boundary limit $\tau_2=\tau_f\to 0^-$ of the bulk-to-bulk propagators. We define the bulk-to-boundary propagators as
\begin{equation}
    G_{\texttt{a}}(k,\tau)\equiv G_{\texttt{a}+}(k,\tau,0^-).
\end{equation}
Using Eq.~\eqref{bulktobulk}, we obtain the bulk-to-boundary propagators for different SK indices as
\begin{equation}
\label{bulk-to-boundary}
    G_{\texttt{a}}(k,\tau) = \frac{H^2}{2k^3}(1-\ii \texttt{a}k\tau)e^{\ii \texttt{a}k\tau}.
\end{equation}

\subsection{Fermion Bubble Loop and Seed Integrals}
\begin{figure}[htbp]
\centering
\includegraphics[width=.5\textwidth]{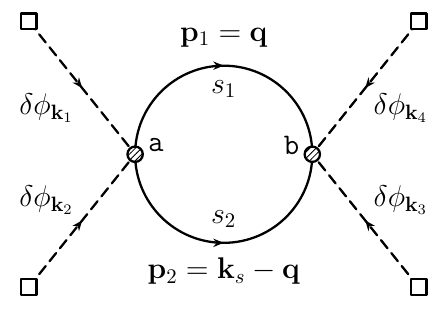}
\qquad
\caption{An $s$-channel SK diagram for the four-point correlator with a
fermion bubble loop. The left and right vertices carry the SK
indices $\mathtt{a}$ and $\mathtt{b}$, respectively, with
$\mathtt{a},\mathtt{b}=\pm$. The dashed lines represent the
bulk-to-boundary propagators of the inflaton, carrying the external
momenta $\mathbf{k}_1$, $\mathbf{k}_2$, $\mathbf{k}_3$, and
$\mathbf{k}_4$, while the solid lines represent the internal
fermion propagators with momenta $\mathbf{p}_1=\mathbf{q}$ and
$\mathbf{p}_2=\mathbf{k}_s-\mathbf{q}$ and helicity labels $s_1$
and $s_2$, respectively, with $s_1,s_2=\pm$.
Here, $\mathbf{q}$ denotes the loop momentum and
$\mathbf{k}_s=\mathbf{k}_1+\mathbf{k}_2$ denotes the $s$-channel
momentum. We show the general SK diagram configuration without
specifying the types of internal fermion propagators.\label{fig:1}}
\end{figure}

Having defined the various propagators, we are now ready to consider the three-point and four-point correlators of inflaton perturbations induced by the fermion bubble loop. In the SK formalism, the four-point correlator with a fermion bubble loop can be represented diagrammatically. Figure~\ref{fig:1} shows the SK diagram for its $s$-channel contribution. The four dashed external lines represent the bulk-to-boundary propagators of the inflaton, carrying external momenta $\mathbf{k}_1$, $\mathbf{k}_2$, $\mathbf{k}_3$, and $\mathbf{k}_4$, which satisfy momentum conservation $\mathbf{k}_1+\mathbf{k}_2+\mathbf{k}_3+\mathbf{k}_4=0$. The solid lines represent the internal fermion propagators, carrying momenta $\mathbf{p}_1=\mathbf{q}$ and $\mathbf{p}_2=\mathbf{k}_s-\mathbf{q}$, respectively, where $\mathbf{q}$ is the loop momentum and $\mathbf{k}_s=\mathbf{k}_1+\mathbf{k}_2=-(\mathbf{k}_3+\mathbf{k}_4)$ is the $s$-channel momentum. Each interaction vertex can carry either of the two SK indices, $+$ or $-$. In the single-folded limit with $\mathbf{k}_4=0$, the four-point correlator can be reduced to the corresponding three-point correlator by appropriately modifying the interaction vertices.
\par 
In the SK formalism, the full four-point correlator involves a sum over the SK indices ($\texttt{a,b}=\pm$) of the vertices, as well as a sum over all possible SK diagrams with different propagator types and momentum channels. For the SK diagram configuration shown in Fig.~\ref{fig:1}, the $s$-channel four-point correlator takes the following general form in the absence of derivative couplings
\begin{equation}
\begin{aligned}
\langle\delta\phi_{\mathbf{k}_1}\delta\phi_{\mathbf{k}_2}\delta\phi_{\mathbf{k}_3}\delta\phi_{\mathbf{k}_4}\rangle_s'
\sim\sum_{\texttt{a},\texttt{b}=\pm}-\texttt{ab}\int_{-\infty}^0\dd\tau_1\dd\tau_2&(-\tau_1)^{c_1}(-\tau_2)^{c_2}G_\texttt{a}(k_1,\tau_1)G_\texttt{a}(k_2,\tau_1)
\\&
G_\texttt{b}(k_3,\tau_2)
G_\texttt{b}(k_4,\tau_2)\mathcal{Q}_{\texttt{ab}}(k_s;\tau_1,\tau_2),
\end{aligned}
\end{equation}
where $\tau_1$ and $\tau_2$ denote the time variables associated with the left and right vertices, respectively. The integers $c_1$ and $c_2$ are determined by the time factors appearing in the inflaton-fermion interactions at the left and right vertices, respectively. We refer to $\mathcal{Q}_{\mathtt a\mathtt b}(k_s;\tau_1,\tau_2)$ as the fermion bubble function, which includes the loop-momentum integral and the sum over connected fermionic Wick contractions between the two vertices at fixed SK indices $\mathtt a,\mathtt b$. Different types\footnote{
Here, the interaction type refers to the spinor-index structure of
the fermion bilinear in the inflaton--fermion interaction.
For example, $\psi^\dagger\bar{\sigma}^{\mu}\psi$ and
$\psi\psi+\psi^\dagger\psi^\dagger$ correspond to different
interaction types.} of interactions lead to different fermion bubble functions. In the following analysis, we focus on the interaction term
\begin{equation}
    \propto (\psi\psi+\psi^\dagger\psi^\dagger).
\end{equation}
The fermion bubble function can then be written as
\begin{equation}
\label{bubblefunction}
\begin{aligned}
\mathcal{Q}_{\texttt{ab}}(k_s;\tau_1,\tau_2)=-2\int\frac{\dd^3\mathbf{p}_1}{(2\pi)^3}&\left\{D_{\texttt{ab}}{}^{\alpha\beta}(\mathbf{p}_1;\tau_1,\tau_2)D_{\texttt{ab}\alpha\beta}(\mathbf{p}_2;\tau_1,\tau_2)\right.\\&+D_{\texttt{ab}}{}^{\alpha}{}_{\dot{\beta}}(\mathbf{p}_1;\tau_1,\tau_2)D_{\texttt{ab}\alpha}{}^{\dot{\beta}}(\mathbf{p}_2;\tau_1,\tau_2)\\&+D_{\texttt{ab}}{}_{\dot{\alpha}}{}^{\beta}(\mathbf{p}_1;\tau_1,\tau_2)D_{\texttt{ab}}{}^{\dot{\alpha}}{}_{\beta}(\mathbf{p}_2;\tau_1,\tau_2)\\&\left.+D_{\texttt{ab}\dot{\alpha}\dot{\beta}}(\mathbf{p}_1;\tau_1,\tau_2)D_{\texttt{ab}}{}^{\dot{\alpha}\dot{\beta}}(\mathbf{p}_2;\tau_1,\tau_2)\right\}.    
\end{aligned}
\end{equation}
The detailed derivation can be found in Appendix~\ref{appendixB}. Note that the SK indices $\texttt{a},\texttt{b}$ are not summed over, while the spinor indices $\alpha,\beta,\dot{\alpha},\dot{\beta}$ are summed. The minus sign arises from the exchange of order when contracting fermion fields. The factor of $2$ originates from the two possible Wick contractions between the fermion fields at the two vertices, which give identical contributions after the antisymmetric spinor indices are contracted. We see that the bubble function incorporates contributions from all four types of fermion propagators, corresponding to different SK diagrams.
\par 
Next, we discuss the properties of the fermion bubble function and present its explicit expression in terms of the mode functions. Using the properties of the fermion propagators in Eqs.~\eqref{fermionpropagatorrelation1} and \eqref{fermionpropagatorrelation2}, we directly obtain
\begin{equation}
\label{Qrelation1}
\mathcal{Q}_{+-}(k_s,\tau_1,\tau_2)=\begin{bmatrix}\mathcal{Q}_{-+}(k_s,\tau_1,\tau_2)\end{bmatrix}^*=\mathcal{Q}_{-+}(k_s,\tau_2,\tau_1).
\end{equation}
Following the decomposition of $D_{++}$ and $D_{--}$ in terms of $D_{-+}$ and $D_{+-}$, we can similarly decompose $\mathcal{Q}_{++}$ and $\mathcal{Q}_{--}$ as
\begin{equation}
\label{Qrelation2}
\begin{aligned}
\mathcal{Q}_{++}(k_s,\tau_{1},\tau_{2})=\mathcal{Q}_{-+}(k_s,\tau_{1},\tau_{2})\theta(\tau_{1}-\tau_{2})+\mathcal{Q}_{+-}(k_s,\tau_{1},\tau_{2})\theta(\tau_{2}-\tau_{1}),\\\mathcal{Q}_{--}(k_s,\tau_{1},\tau_{2})=\mathcal{Q}_{+-}(k_s,\tau_{1},\tau_{2})\theta(\tau_{1}-\tau_{2})+\mathcal{Q}_{-+}(k_s,\tau_{1},\tau_{2})\theta(\tau_{2}-\tau_{1}).
\end{aligned}
\end{equation}
Therefore, using Eqs.~\eqref{Qrelation1} and \eqref{Qrelation2}, we can express all types of $\mathcal{Q}_{\texttt{ab}}$ ($\texttt{a,b}=\pm$) entirely in terms of $\mathcal{Q}_{-+}$. It is thus sufficient to focus on $\mathcal{Q}_{-+}$ in the following analysis. Substituting the propagator expressions in Eqs.~\eqref{psipropagator1}--\eqref{psipropagator4} into the bubble function \eqref{bubblefunction} and using the helicity eigenspinor identity \eqref{eigenspinoridentity}, we directly obtain
\begin{equation}
\label{bubblefunctionexpress}
\begin{aligned}
\mathcal{Q}_{-+}(k_s;\tau_1,\tau_2)=2\int\frac{\dd^3\mathbf{p}_1}{(2\pi)^3}\sum_{s_1,s_2=\pm}&F_{s_1s_2}(\hat{\mathbf{p}}_1,\hat{\mathbf{p}}_2)\left[u_{s_1}(p_1,\tau_1)u_{s_2}(p_2,\tau_1)-v_{s_1}(p_1,\tau_1)v_{s_2}(p_2,\tau_1)\right]\\
&\times\left[u^*_{s_1}(p_1,\tau_2)u^*_{s_2}(p_2,\tau_2)-v_{s_1}^*(p_1,\tau_2)v^*_{s_2}(p_2,\tau_2)\right],
\end{aligned}
\end{equation}
where we have defined
\begin{equation}
F_{s_1s_2}(\hat{\mathbf{p}}_1,\hat{\mathbf{p}}_2)=\frac{1}{2}(1-s_1s_2\, \hat{\mathbf{p}}_1\cdot\hat{\mathbf{p}}_2).
\end{equation}
The detailed derivation can be found in Appendix~\ref{appendixB}.
\par 
We can now summarize the general structure of the cosmological correlators arising from the fermion bubble loop as follows
\begin{itemize}
\item Time integrals at the left and right vertices, including the associated SK indices and factors of the imaginary unit: $\ii\texttt{a}\int_{-\infty}^0 \dd\tau_1$ and $\ii\texttt{b}\int_{-\infty}^0 \dd\tau_2$.
\item The exponential factors arising from the inflaton bulk-to-boundary propagators $G_{\texttt{a}}$, as given in Eq.~\eqref{bulk-to-boundary}. Specifically, $G_{\texttt{a}}(k_1,\tau_1)G_{\texttt{a}}(k_2,\tau_1)$ and $G_{\texttt{b}}(k_3,\tau_2)G_{\texttt{b}}(k_4,\tau_2)$ contribute the exponential factors $e^{\ii\texttt{a}k_{12}\tau_1}$ and $e^{\ii\texttt{b}k_{34}\tau_2}$, respectively, where $k_{12}=|\mathbf{k}_1|+|\mathbf{k}_2|$ and $k_{34}=|\mathbf{k}_3|+|\mathbf{k}_4|$.
\item The time factors $(-\tau_1)^{\tilde{q}_1}$ and $(-\tau_2)^{\tilde{q}_2}$. The powers of conformal time arise from three sources: (i) the scale factors $a(\tau)=-1/(H\tau )$ appearing in the interactions between the inflaton perturbation $\delta\phi$ and the fermion $\psi$; (ii) the time dependence of the inflaton bulk-to-boundary propagators $G_{\texttt{a}}$, which contributes additional powers of conformal time. In the presence of a derivative coupling between the inflaton and the fermion, the time derivatives $\partial_\tau G_{\texttt{a}}$ can also give rise to such contributions; and (iii) the rescaling of the fermion field. Since our fermion propagators are constructed using the rescaled field $\widetilde{\psi}$, the relation between the physical and rescaled fields, $\psi=a^{-3/2}\widetilde{\psi}$, introduces additional factors of $a^{-3/2}$ and hence additional powers of conformal time. The integers $\tilde{q}_1$ and $\tilde{q}_2$ are determined by the combined contributions from these three sources.
\item The fermion bubble loop $\mathcal{Q}_{\texttt{ab}}(k_s;\tau_1,\tau_2)$.
\end{itemize}
Combining the above four ingredients, we construct seed integrals that serve as building blocks for the four-point correlators of inflaton perturbations involving a fermion bubble loop. For convenience, we introduce the shifted parameters $q_i\equiv\tilde q_i-4$ for $i=1,2$ and use the time variables $\eta_i=-\tau_i>0$. For a specific choice of interactions, the corresponding correlator can be obtained from appropriate combinations of the seed integrals with the relevant values of $q_1$ and $q_2$. We define the seed integrals as
\begin{equation}
\label{seedintdefine}
	\mathcal{I}_{\mathtt{ab}}^{q_1 q_2}
	=
    -\mathtt{ab}\, \frac{k_s^{7+q_{12}}}{r_1^{5+q_1}r_2^{5+q_2}}
	\int_0^\infty \dd \eta_1 \dd \eta_2\,
	\eta_1^{q_1+4}\eta_2^{q_2+4}
	e^{-\ii \mathtt{a} k_{12}\eta_1-\ii\mathtt{b}k_{34}\eta_2}
	\mathcal{Q}_{\mathtt{ab}}(k_s,\eta_1,\eta_2),
\end{equation}
where $q_{12}=q_1+q_2$ and $\mathcal{Q}_{\mathtt a\mathtt b}(k_s;\eta_1,\eta_2)$ denotes the original bubble function evaluated at $\tau_i=-\eta_i$. The normalization factor $k_s^{7+q_{12}}/(r_1^{5+q_1}r_2^{5+q_2})$ is chosen so that the seed integrals are scale invariant, with $r_1=k_s/k_{12}$ and $r_2=k_s/k_{34}$.
\par 
To illustrate how correlation functions can be expressed in terms of seed integrals, we introduce explicit interaction vertices. Specifically, we consider dimension-5 three-point and dimension-7 four-point interactions, with the interaction Lagrangian given by
\begin{equation}
\label{modelint}
\mathcal{L}_{\mathrm{int},3}
=
\frac{a^3}{\Lambda_1}\,
\delta\phi'\left(
\psi\psi+\psi^\dagger\psi^\dagger
\right),
\qquad
\mathcal{L}_{\mathrm{int},4}
=
\frac{a^2}{\Lambda_2^3}\,
(\delta\phi')^2\left(
\psi\psi+\psi^\dagger\psi^\dagger
\right).
\end{equation}

It is important to distinguish the interactions considered in Eq.~\eqref{modelint} from the derivative inflaton–fermion coupling introduced in Eq.~\eqref{eq:setup_action}, which generates the effective chemical potential. The background part of the latter is incorporated into the fermion mode functions and propagators, while its perturbative part gives rise to additional inflaton–fermion interaction vertices involving the fermion current \(\psi^\dagger\bar{\sigma}^\mu\psi\). In this work, we focus on a different class of interactions involving the scalar fermion bilinear \(\psi\psi+\psi^\dagger\psi^\dagger\), with independent effective coupling scales \(\Lambda_1\) and \(\Lambda_2\). Our calculation therefore isolates the bubble-loop contributions generated by these scalar-bilinear interactions in the presence of a chemical potential, rather than the complete set of diagrams induced by all inflaton–fermion couplings.

The four-point and three-point correlation functions can be expressed in terms of seed integrals as follows
\begin{equation}
\begin{aligned}
\langle\delta\phi_{\mathbf{k}_1}\delta\phi_{\mathbf{k}_2}\delta\phi_{\mathbf{k}_3}\delta\phi_{\mathbf{k}_4}\rangle' = 
\frac{H^{10}}{4\Lambda_2^6k_1k_2k_3k_4}\frac{r_1^4r_2^4}{k_s^5}\sum_{\mathtt{a},\mathtt{b}=\pm}\mathcal{I}_{\mathtt{ab}}^{-1,-1}(r_1,r_2) + t\ \text{and}\ u\ \text{channel},
\end{aligned}
\end{equation}
\begin{equation}
\begin{aligned}
\langle\delta\phi_{\mathbf{k}_1}\delta\phi_{\mathbf{k}_2}\delta\phi_{\mathbf{k}_3}\rangle' = 
\frac{-H^{7}}{4\Lambda_1\Lambda_2^3k_1k_2k_{12}^4}\sum_{\mathtt{a},\mathtt{b}=\pm}\lim_{k_4\to 0}\mathcal{I}_{\mathtt{ab}}^{-1,-3}(r_1,r_2) + t\ \text{and}\ u\ \text{channel}.
\end{aligned}
\end{equation}
For the four-point correlator, the $t$- and $u$-channel contributions are obtained from the first term by the permutations
$\mathbf{k}_2\leftrightarrow\mathbf{k}_3$ and
$\mathbf{k}_2\leftrightarrow\mathbf{k}_4$, respectively, with the
corresponding channel momenta $\mathbf{k}_t=\mathbf{k}_1+\mathbf{k}_3$ and $\mathbf{k}_u=\mathbf{k}_1+\mathbf{k}_4$. For the three-point correlator, the $t$- and $u$-channel contributions are obtained from the first term by the permutations $\mathbf{k}_3\leftrightarrow\mathbf{k}_2$ and $\mathbf{k}_3\leftrightarrow\mathbf{k}_1$, respectively.
We focus on the $s$-channel contribution in the single-squeezed limit $r_1=k_s/k_{12}\ll1$, while retaining the full dependence on $r_2=k_s/k_{34}$ for $0<r_2\leq1$. In this limit, the $s$-channel captures the CC signals associated with the soft internal momentum $k_s$, while the corresponding $t$- and $u$-channel internal momenta remain hard for generic external momentum configurations. The other channels are related by permutations of the external momenta and can be analyzed in their respective squeezed limits.
\par 
The seed integrals $\mathcal{I}_{\mathtt{ab}}^{q_1q_2}$ will be the main focus of the subsequent analysis in this paper. In general, a complete analytic evaluation of these seed integrals is challenging, as they involve both time integrals and loop momentum integrals. In the following sections, we will extract the CC signals from the seed integrals and compute the nonlocal and local signals separately using Mellin--Barnes (MB) representations. We will also investigate the CC signals arising from the seed integrals in the three-point limit $k_4\to 0$, for which $r_2\to1$.
\section{Extraction of Cosmological Collider Signals\label{sec3}}
In this section, we extract the cosmological collider (CC) signals from the seed integrals $\mathcal{I}_{\mathtt{ab}}^{q_1q_2}$ defined in the previous section, distinguishing between the nonlocal and local signals. We first introduce the CC cutting rule, which allows us to focus on the factorized seed integrals while discarding the more complicated nested time integrals that contribute only to the background. We then extract the signals in the single-squeezed limit $k_{12}\gg k_s$, or equivalently $r_1\ll 1$. In this limit, the nonanalytic dependence on $r_1$ becomes more manifest. To facilitate the calculation, we employ the Mellin--Barnes (MB) representation to express the seed integrals as MB integrals. When evaluating these integrals using the residue theorem, the poles of the integrands in the MB variables determine the nonanalytic dependence on the external momenta. We therefore analyze the correspondence between these poles and the nonlocal/local CC signals.

\subsection{Cosmological Collider Cutting Rule}
Based on the different analytic properties with respect to the internal and external momenta, the complete seed integral $\mathcal I^{q_1q_2}\equiv \sum_{\texttt{a},\texttt{b}}\mathcal I_{\texttt{ab}}^{q_1q_2}$ can be decomposed into three parts: the nonlocal signal $\mathcal I^{q_1q_2}_{\mathrm{NS}}$, the local signal $\mathcal I^{q_1q_2}_{\mathrm{LS}}$, and the background $\mathcal I^{q_1q_2}_{\mathrm{BG}}$
\begin{equation}
    \mathcal I^{q_1q_2}(r_1,r_2)
    =
    \mathcal I^{q_1q_2}_{\mathrm{NS}}(r_1,r_2)
    +
    \mathcal I^{q_1q_2}_{\mathrm{LS}}(r_1,r_2)
    +
    \mathcal I^{q_1q_2}_{\mathrm{BG}}(r_1,r_2).
\end{equation}
The nonlocal signal $\mathcal I_{\mathrm{NS}}^{q_1q_2}(r_1,r_2)$ contains a non-analytic dependence on the internal momentum $k_s$, which takes the characteristic form $(r_1r_2)^{\ii\omega}$ or $r_{1,2}^{\ii\omega}$. The local signal $\mathcal I_{\mathrm{LS}}^{q_1q_2}(r_1,r_2)$ contains a non-analytic dependence on the external momenta $k_{12}$ or $k_{34}$, while remaining analytic in $k_s$, and is characterized by the nonanalytic behavior $(r_1/r_2)^{\ii\omega}$. By contrast, the background contribution $\mathcal I_{\mathrm{BG}}^{q_1q_2}(r_1,r_2)$ is meromorphic in $r_1$ and $r_2$ and contains no oscillatory complex powers of the momentum ratios.
\par 
To extract the signal part of primary interest, we can apply the CC cutting rules. For detailed discussions of the CC cutting rule, see Refs.~\cite{Qin:2023bjk,Qin:2022fbv,Tong:2021wai,Ema:2024hkj,Aoki:2026vbc}. In the kinematic regime $r_1<r_2$, or equivalently $k_{12}>k_{34}$, we can decompose the time-ordered and anti-time-ordered fermion bubble functions as follows
\begin{equation}
\label{bubblefundec}
    \mathcal{Q}_{\pm\pm}(k_s,\tau_1,\tau_2)
=
\mathcal{Q}_{\mp\pm}(k_s,\tau_1,\tau_2)
+
\theta(\tau_2-\tau_1)
\left[
\mathcal{Q}_{\pm\mp}(k_s,\tau_1,\tau_2)
-
\mathcal{Q}_{\mp\pm}(k_s,\tau_1,\tau_2)
\right].
\end{equation}
Substituting this decomposition into Eq.~\eqref{seedintdefine}, we obtain the following decomposition of the time-ordered and anti-time-ordered seed integrals
\begin{equation}
    \mathcal{I}_{\pm\pm}^{q_1 q_2}
	=\mathcal{I}_{\pm\pm,\text{F}}^{q_1 q_2} + \mathcal{I}_{\pm\pm,\text{TO}}^{q_1 q_2},
\end{equation}
where
\begin{equation}
\mathcal{I}_{\pm\pm,\mathrm{F}}^{q_1q_2}
\equiv
-\frac{k_s^{7+q_{12}}}{r_1^{5+q_1}r_2^{5+q_2}}
\int_{0}^{\infty}
\mathrm{d}\eta_1\,\mathrm{d}\eta_2\,
\eta_1^{q_1+4}
\eta_2^{q_2+4}
e^{\mp \ii k_{12}\eta_1\mp \ii k_{34}\eta_2}
\mathcal{Q}_{\mp\pm}(k_s,\eta_1,\eta_2),
\end{equation}
\begin{equation}
\label{TOseedint}
\begin{aligned}
    \mathcal{I}_{\pm\pm,\mathrm{TO}}^{q_1q_2}
\equiv
-\frac{k_s^{7+q_{12}}}{r_1^{5+q_1}r_2^{5+q_2}}
\int_{0}^{\infty}
\mathrm{d}\eta_1\,&\mathrm{d}\eta_2\,
\eta_1^{q_1+4}
\eta_2^{q_2+4}
e^{\mp \ii k_{12}\eta_1\mp \ii k_{34}\eta_2}
\theta(\eta_1-\eta_2)
\\&\times 
\left[
\mathcal{Q}_{\pm\mp}(k_s,\eta_1,\eta_2)
-
\mathcal{Q}_{\mp\pm}(k_s,\eta_1,\eta_2)
\right].
\end{aligned}
\end{equation}
Here, we refer to $\mathcal{I}_{\pm\pm,\mathrm{F}}^{q_1q_2}$ as the factorized part, since its integrand factorizes with respect to $\eta_1$ and $\eta_2$, and to $\mathcal{I}_{\pm\pm,\mathrm{TO}}^{q_1q_2}$ as the time-ordered part, since its integrand contains the time-ordering factor $\theta(\eta_1-\eta_2)$. The cutting rules imply that $\mathcal{I}_{\pm\pm,\mathrm{TO}}^{q_1q_2}$ does not contain any CC signal and contributes only to the background. A detailed proof is provided in Appendix~\ref{appendixC}. Combining $\mathcal{I}_{\pm\pm,\mathrm{F}}^{q_1q_2}$ and $\mathcal{I}_{\pm\mp}^{q_1q_2}$, we can define the complete factorized seed integrals as
\begin{equation}
\label{Fseedint}
    \mathcal{I}_{\text{F}}^{q_1 q_2}
	\equiv
    \mathcal{I}_{-+}^{q_1 q_2} + \mathcal{I}_{++,\text{F}}^{q_1 q_2}+\mathcal{I}_{+-}^{q_1 q_2}+\mathcal{I}_{--,\text{F}}^{q_1 q_2}
    =
    \mathcal{I}_{-+}^{q_1 q_2} + \mathcal{I}_{++,\text{F}}^{q_1 q_2}+\text{c.c.}\, 
    .
\end{equation}
All the nonlocal and local CC signals of interest arise from the factorized seed integral $\mathcal{I}_{\mathrm{F}}^{q_1q_2}$, namely
\begin{equation}
    \mathcal I^{q_1q_2}_{\mathrm{NS}}
    +
    \mathcal I^{q_1q_2}_{\mathrm{LS}}
    \subset \mathcal{I}_{\mathrm{F}}^{q_1q_2}.
\end{equation}
\par 
We should also explain why we adopt the decomposition of the fermion bubble function in Eq.~\eqref{bubblefundec} for the kinematic regime $r_1<r_2$. In Appendix~\ref{appendixC}, to establish the cutting rule, we evaluate the time-ordered seed integrals $\mathcal{I}_{\pm\pm,\mathrm{TO}}^{q_1q_2}$ in Eq.~\eqref{TOseedint} using the MB representation, which leads to generalized hypergeometric functions ${}_2\mathrm{F}_1$ with argument $-r_1/r_2$. We expand these hypergeometric functions in series to demonstrate that the time-ordered seed integrals do not contribute to the local CC signal. This requires the arguments of the hypergeometric functions to lie within the convergence domain of their series representations, namely $r_1<r_2$. For the kinematic regime $r_1>r_2$, we can instead decompose the fermion bubble functions as
\begin{equation}
    \mathcal{Q}_{\pm\pm}(k_s,\tau_1,\tau_2)
    =
    \mathcal{Q}_{\pm\mp}(k_s,\tau_1,\tau_2)
    +
    \theta(\tau_1-\tau_2)
    \left[
    \mathcal{Q}_{\mp\pm}(k_s,\tau_1,\tau_2)
    -
    \mathcal{Q}_{\pm\mp}(k_s,\tau_1,\tau_2)
    \right].
\end{equation}
With this decomposition, evaluating the corresponding time-ordered seed integrals $\mathcal{I}_{\pm\pm,\mathrm{TO}}^{q_1q_2}$ yields generalized hypergeometric functions ${}_2\mathrm{F}_1$ with argument $-r_2/r_1$. Their series representations converge for $r_1>r_2$, consistent with the kinematic regime under consideration. Both decompositions are algebraic identities valid for arbitrary external energies. The ordering of the external energies determines which decomposition isolates the CC signals in the factorized part.

\subsection{Late-Time Limit \label{3.2}}
We now turn to a detailed analysis of the factorized seed integral in Eq.~\eqref{Fseedint}, which contains two important contributions, $\mathcal{I}_{-+}^{q_1q_2}$ and $\mathcal{I}_{++,\mathrm{F}}^{q_1q_2}$. Since the analysis of $\mathcal{I}_{++,\mathrm{F}}^{q_1q_2}$ proceeds in complete analogy with that of $\mathcal{I}_{-+}^{q_1q_2}$, we focus on $\mathcal{I}_{-+}^{q_1q_2}$ in the following discussion. Since we are primarily interested in the single-squeezed limit $r_1\to 0$, we first perform a late-time expansion of the fermion mode functions associated with $\tau_1$ in the bubble function given in Eq.~\eqref{bubblefunctionexpress}. The equivalence between the single-squeezed limit and this late-time expansion is demonstrated in Appendix~\ref{appendixD}. Using Eq.~\eqref{latetimeexp}, we can expand the fermion mode functions at the left vertex in the late-time limit as
\begin{equation}
\begin{aligned}
    &\lim_{|p_1\tau_1|,|p_2\tau_1|\to 0}
    \left[
    u_{s_1}(p_1,\tau_1)u_{s_2}(p_2,\tau_1)-
v_{s_1}(p_1,\tau_1)v_{s_2}(p_2,\tau_1)
\right]
    \\&=\sum_{\rho_1,\rho_2=\pm }
  \left(C_{u_{s_1}}^{(\rho_1)}C_{u_{s_2}}^{(\rho_2)}-
  C_{v_{s_1}}^{(\rho_1)}C_{v_{s_2}}^{(\rho_2)}
  \right)
  2^{\ii(\rho_1+\rho_2)\tnu}p_1^{\ii\rho_1\tnu}p_2^{\ii\rho_2\tnu}
  (-\tau_1)^{\ii(\rho_1+\rho_2)\tnu}.
\end{aligned}
\end{equation}
Under this late-time expansion, we can directly obtain the expression for $\mathcal{I}_{-+}^{q_1q_2}$ in the limit $r_1\to 0$ as
\begin{equation}
\label{86}
\begin{aligned}
    \lim_{r_1\to 0}\mathcal{I}_{-+}^{q_1q_2}
	&=
    \frac{2\, k_s^{7+q_{12}}}{r_1^{5+q_1}r_2^{5+q_2}}
    \sum_{\rho_1,\rho_2=\pm}\sum_{s_1,s_2=\pm}
    \left(C_{u_{s_1}}^{(\rho_1)}C_{u_{s_2}}^{(\rho_2)}-
  C_{v_{s_1}}^{(\rho_1)}C_{v_{s_2}}^{(\rho_2)}
  \right)
  2^{\ii(\rho_1+\rho_2)\tnu}
  \\&\times 
  \int \frac{\dd^3 \mathbf{q}}{(2\pi)^3}
	\int_0^\infty \dd \eta_1 \dd \eta_2\,
	\eta_1^{q_1+4+\ii(\rho_1+\rho_2)\tnu}\eta_2^{q_2+4}
    p_1^{\ii\rho_1\tnu}p_2^{\ii\rho_2\tnu}
	e^{\ii k_{12}\eta_1-\ii k_{34}\eta_2}
  \\&\times 
  \left[
    u_{s_1}^{\ast}(p_1,\eta_2)
	u_{s_2}^{\ast}(p_2,\eta_2)
    -
	v_{s_1}^{\ast}(p_1,\eta_2)
	v_{s_2}^{\ast}(p_2,\eta_2)
    \right]
    F_{s_1 s_2}(\hat {\mathbf{p}}_1,\hat {\mathbf{p}}_2).
\end{aligned}
\end{equation}
\par 
The time integral over $\eta_1$ appearing in Eq.~\eqref{86} is completely factorized and can therefore be evaluated analytically
\begin{equation}
\label{lefttimeint}
	\int_0^\infty \dd \eta_1\,
	\eta_1^{q_1+4+\ii(\rho_1+\rho_2)\tnu}
	e^{\ii k_{12}\eta_1}
	=
	e^{\ii\frac{\pi}{2}(5+q_1+\ii(\rho_1+\rho_2)\tnu)}
	\Gamma[5+q_1+\ii(\rho_1+\rho_2)\tnu]
	k_{12}^{-(5+q_1+\ii(\rho_1+\rho_2)\tnu)},
\end{equation}
where the time variable is understood to carry the appropriate $\ii\epsilon$ prescription, namely $\eta_1\to(1+\ii\epsilon)\eta_1$ with $\epsilon\to 0^+$. The nonanalytic dependence on the external momentum $k_{12}$ takes the form $k_{12}^{-\ii(\rho_1+\rho_2)\tnu}$. As we shall see below, after performing the loop-momentum integral, this factor combines with $k_s^{\ii(\rho_1+\rho_2)\tnu}$ to yield $r_1^{\ii(\rho_1+\rho_2)\tnu}$. Thus, the nonanalytic dependence on $r_1$ is entirely determined by the choices of $\rho_1$ and $\rho_2$. For the branches with $\rho_1=-\rho_2$, the nonanalytic dependence on $r_1$ disappears, and these branches do not contribute to the CC signal in the three-point limit ($r_2\to 1$). Therefore, to extract the squeezed three-point CC signal, we retain only the branches
\begin{equation}
    \rho_1=\rho_2=\rho,
    \qquad
    \rho=\pm.
\end{equation}
We denote their contribution to Eq.~\eqref{86} by $\mathcal{J}_{-+}^{q_1q_2}$, defined as
\begin{equation}
\begin{aligned}
    \mathcal{J}_{-+}^{q_1q_2}
	&\equiv
    \frac{2\, k_s^{7+q_{12}}}{r_1^{5+q_1}r_2^{5+q_2}}
    \sum_{\rho=\pm}\sum_{s_1,s_2=\pm}
    \left(C_{u_{s_1}}^{(\rho)}C_{u_{s_2}}^{(\rho)}-
  C_{v_{s_1}}^{(\rho)}C_{v_{s_2}}^{(\rho)}
  \right)
  2^{2\ii\rho\tnu}
  e^{\ii\frac{\pi}{2}(5+q_1+2\ii\rho\tnu)}
  \\&\times 
	\Gamma[5+q_1+2\ii\rho\tnu]\, 
	k_{12}^{-(5+q_1+2\ii\rho\tnu)}
  \int \frac{\dd^3 \mathbf{q}}{(2\pi)^3}
	\int_0^\infty \dd \eta_2\,
	\eta_2^{q_2+4}
    p_1^{\ii\rho\tnu}p_2^{\ii\rho\tnu}
	e^{-\ii k_{34}\eta_2}
  \\&\times 
  \left[
    u_{s_1}^{\ast}(p_1,\eta_2)
	u_{s_2}^{\ast}(p_2,\eta_2)
    -
	v_{s_1}^{\ast}(p_1,\eta_2)
	v_{s_2}^{\ast}(p_2,\eta_2)
    \right]
    F_{s_1 s_2}(\hat {\mathbf{p}}_1,\hat {\mathbf{p}}_2).
\end{aligned}
\end{equation}
We now turn to a detailed investigation of the properties of $\mathcal{J}_{-+}^{q_1q_2}$.
\par 
We see that $\mathcal{J}_{-+}^{q_1q_2}$ contains contributions from both $u_{s_1}^{\ast}
	u_{s_2}^{\ast}$ and $v_{s_1}^{\ast}
	v_{s_2}^{\ast}$. Both contributions must be included in the complete calculation. However, since the mode functions $u$ and $v$ satisfy the relation
\begin{equation}
\label{uvrelation}
    u_s(\tlam)=v_{-s}(-\tlam),
\end{equation}
we may temporarily focus only on the $v_{s_1}^{\ast}
	v_{s_2}^{\ast}$ contribution to the $\mathcal{J}_{-+}^{q_1q_2}$ in the explicit analytic derivation. We define
\begin{equation}
\begin{aligned}
\label{Jbranch}
    \widehat{\mathcal{J}}
_{-+,\rho,\lambda}
^{q_1q_2,s_1s_2}
	&\equiv
    \frac{2\, k_s^{7+q_{12}}}{r_1^{5+q_1}r_2^{5+q_2}}\, 
  2^{2\ii\rho\tnu}\, 
  e^{\ii\frac{\pi}{2}(5+q_1+2\ii\rho\tnu)}\, \Gamma[5+q_1+2\ii\rho\tnu]\, 
	k_{12}^{-(5+q_1+2\ii\rho\tnu)}
  \\&\times 
  \int \frac{\dd^3 \mathbf{q}}{(2\pi)^3}
	\int_0^\infty \dd \eta_2\,
	\eta_2^{q_2+4}
    p_1^{\ii\rho\tnu}p_2^{\ii\rho\tnu}
	e^{-\ii k_{34}\eta_2}
	v_{s_1}^{\ast}(p_1,\eta_2)
	v_{s_2}^{\ast}(p_2,\eta_2)
    F_{s_1 s_2}(\hat {\mathbf{p}}_1,\hat {\mathbf{p}}_2),
\end{aligned}
\end{equation}
where $\lambda$ in the subscript denotes the chemical potential before normalization by the Hubble parameter $H$. In the following derivation, we focus exclusively on the contribution $\widehat{\mathcal{J}}_{-+,\rho,\lambda}^{q_1q_2,s_1s_2}$. Using the relation between the mode functions in Eq.~\eqref{uvrelation}, the complete contribution
$\mathcal{J}_{-+}^{q_1q_2}$ can then be expressed directly as
\begin{equation}
\label{partialJ}
    \mathcal{J}_{-+}^{q_1q_2}=-\sum_\rho\sum_{s_1,s_2}\mathcal{C}_{s_1s_2}^{\rho}
    \left[
    \widehat{\mathcal{J}}_{-+,\rho,\lambda}^{q_1q_2,s_1s_2}-\widehat{\mathcal{J}}_{-+,\rho,-\lambda}^{q_1q_2,-s_1-s_2}
    \right],
\end{equation}
where the second term in the square brackets, $\widehat{\mathcal{J}}_{-+,\rho,-\lambda}^{q_1q_2,-s_1-s_2}$, denotes the result obtained from $\widehat{\mathcal{J}}_{-+,\rho,\lambda}^{q_1q_2,s_1s_2}$ by making the replacements
$s_1\to -s_1,\ s_2\to -s_2,\ \lambda\to -\lambda$. We have introduced the following shorthand notation for the combinations of late-time expansion coefficients
\begin{equation}
    \mathcal{C}_{s_1s_2}^{\rho}\equiv 
    \left(C_{u_{s_1}}^{(\rho)}C_{u_{s_2}}^{(\rho)}-
  C_{v_{s_1}}^{(\rho)}C_{v_{s_2}}^{(\rho)}
  \right).
\end{equation}
\par 
The above discussion applies equally to $\mathcal{I}_{++,\text{F}}^{q_1q_2}$. The only difference is that the time integral over $\eta_1$ in Eq.~\eqref{lefttimeint} should be replaced by
\begin{equation}
\label{lefttimeint2}
	\int_0^\infty \dd \eta_1\,
	\eta_1^{q_1+4+\ii(\rho_1+\rho_2)\tnu}
	e^{-\ii k_{12}\eta_1}
	=
	e^{-\ii\frac{\pi}{2}(5+q_1+\ii(\rho_1+\rho_2)\tnu)}
	\Gamma[5+q_1+\ii(\rho_1+\rho_2)\tnu]
	k_{12}^{-(5+q_1+\ii(\rho_1+\rho_2)\tnu)}.
\end{equation}
Analogously to the definition of $\mathcal{J}_{-+}^{q_1q_2}$, we can define $\mathcal{J}_{++,\mathrm{F}}^{q_1q_2}$ in the same manner. Since the integral results in Eqs.~\eqref{lefttimeint} and \eqref{lefttimeint2} differ by a factor $e^{-\ii\pi(5+q_1+\ii(\rho_1+\rho_2)\tnu)}$, while $\mathcal I_{-+}^{q_1q_2}$ and $\mathcal I_{++,\mathrm F}^{q_1q_2}$ carry opposite SK prefactors, we can use Eq.~\eqref{partialJ} to write $\mathcal{J}_{++,\mathrm{F}}^{q_1q_2}$ as
\begin{equation}
\label{JppF}
    \mathcal{J}_{++,\mathrm{F}}^{q_1q_2}
    =
    -\sum_\rho
    e^{-\ii\pi(4+q_1+2\ii\rho\tnu)}
    \sum_{s_1,s_2}\mathcal{C}_{s_1s_2}^{\rho}
    \left[
    \widehat{\mathcal{J}}_{-+,\rho,\lambda}^{q_1q_2,s_1s_2}-\widehat{\mathcal{J}}_{-+,\rho,-\lambda}^{q_1q_2,-s_1-s_2}
    \right].
\end{equation}
\par 
Analogously to the definition of the factorized seed integral in Eq.~\eqref{Fseedint}, we may also define
\begin{equation}
    \mathcal{J}_{\text{F}}^{q_1 q_2}
	\equiv  
    \mathcal{J}_{-+}^{q_1 q_2} + \mathcal{J}_{++,\text{F}}^{q_1 q_2}+\text{c.c.}\, 
    .
\end{equation}
Using Eqs.~\eqref{partialJ} and \eqref{JppF}, $\mathcal{J}_{\mathrm{F}}^{q_1q_2}$ can also be expressed entirely in terms of $\widehat{\mathcal{J}}_{-+,\rho,\lambda}^{q_1q_2,s_1s_2}$
\begin{equation}
\label{JF}
    \mathcal{J}_{\text{F}}^{q_1q_2}=-\sum_\rho
    \left[
    1+e^{-\ii\pi(4+q_1+2\ii\rho\tnu)}
    \right]
    \sum_{s_1,s_2}\mathcal C_{s_1s_2}^\rho
    \left[
    \widehat{\mathcal{J}}_{-+,\rho,\lambda}^{q_1q_2,s_1s_2}-\widehat{\mathcal{J}}_{-+,\rho,-\lambda}^{q_1q_2,-s_1-s_2}
    \right]+\text{c.c.}\, .
\end{equation}
Since both $\mathcal{J}_{-+}^{q_1q_2}$ and $\mathcal{J}_{++,\mathrm{F}}^{q_1q_2}$ exhibit nonanalytic dependence on the external momentum $k_{12}$, $\mathcal{J}_{\mathrm{F}}^{q_1q_2}$ consists entirely of CC signals, without any background contribution. We define the nonlocal and local parts of $\mathcal{J}_{\text{F}}^{q_1 q_2}$ as $\mathcal{J}_{\text{NS}}^{q_1 q_2}$ and $\mathcal{J}_{\text{LS}}^{q_1 q_2}$, respectively. They should satisfy
\begin{equation}
    \mathcal{J}_{\text{F}}^{q_1 q_2}=\mathcal{J}_{\text{NS}}^{q_1 q_2}+\mathcal{J}_{\text{LS}}^{q_1 q_2}.
\end{equation}
Correctly distinguishing and separately evaluating $\mathcal{J}_{\text{NS}}^{q_1 q_2}$ and $\mathcal{J}_{\text{LS}}^{q_1 q_2}$ constitute the main task of our analysis.

\subsection{Mellin--Barnes Representation}
Next, we need to evaluate the $\eta_2$ time integral and the loop-momentum integral in Eq.~\eqref{Jbranch}, which can be accomplished by applying the Mellin--Barnes (MB) representation to the internal propagators. For more detailed discussions of the applications of (Partial)--MB representations to the computation of cosmological correlators, see Refs.~\cite{Aoki:2026vbc,Qin:2023bjk,Qin:2022fbv,Qin:2022lva,Qin:2024gtr}.
\par 
At the right vertex, we perform a Wick rotation of the time variable
\begin{equation}
	\eta_2 \rightarrow -\ii x,
	\qquad x>0 .
\end{equation}
Under this rotation, the Whittaker argument appearing in the mode function $v^\ast $ becomes positive
\begin{equation}
	v_{s}^\ast(p,-\ii x) = \frac{N_s}{\sqrt{2 px}}W_{s(1/2+\ii \tlam),-\ii \tnu}(2px),
\end{equation}
where
\begin{equation}
    N_s=\frac{e^{\pi s\tlam/2}}{\sqrt{-\ii}}\left[
    -\frac{\ii}{2}(1+s)+\frac{\widetilde{M}}{2}(1-s)
    \right].
\end{equation}
Using the relation between the Whittaker function and the Kummer function~\cite{NIST:DLMF}
\begin{equation}
	W_{\kappa,\mu}(z)
	=
	e^{-z/2}z^{\mu+1/2}
	U\!\left(
		\frac{1}{2}+\mu-\kappa,
		1+2\mu,
		z
	\right),
\end{equation}
we can rewrite the Wick-rotated mode function in terms of the Kummer function $U$. Next, we use the MB integral representation (i.e. the inverse Mellin transform) of the Kummer function $U$~\cite{NIST:DLMF}
\begin{equation}
	U(a,b,z)
	=
	\int_{-\ii\infty}^{+\ii\infty}
	\frac{\dd \xi}{2\pi\ii}\,
\Gamma\!
    \left[
    \begin{matrix}
        -\xi,1-b-\xi,a+\xi\\
        a,a-b+1
    \end{matrix}
    \right]\, z^\xi \qquad|\text{Arg}\, z|<\frac{3}{2}\pi.
\end{equation}
We thus obtain the MB representation of $v_s^\ast$ as
\begin{equation}
	v_{s}^\ast(p,-\ii x) =N_s\int_{-\ii\infty}^{+\ii\infty}\frac{\dd \xi}{2\pi\ii}\,
    \Gamma\!\left[
		\begin{matrix}
			-\xi,-\xi+2\ii\tnu,\frac{1-s}{2}-\ii(s\tlam+\tnu)+\xi \\
			\frac{1-s}{2}-\ii(s\tlam+\tnu),\frac{1-s}{2}-\ii(s\tlam-\tnu)
		\end{matrix}
		\right]\, 
        (2px)^{-\ii\tnu +\xi}e^{-px}.
\end{equation}
The contour of the $\xi$ integration is chosen such that it separates the right pole families arising from $\Gamma[-\xi,-\xi+2\ii\tilde\nu]$ and the left pole family arising from $\Gamma[\frac{1-s}{2}-\ii(s\tilde\lambda+\tilde\nu)+\xi]$.
The MB representation of the $v_{s_1}^{\ast}(p_1,-\ii x)v_{s_2}^{\ast}(p_2,-\ii x)$ is 
\begin{equation}
\begin{aligned}
    v_{s_1}^{\ast}(p_1,-\ii x)v_{s_2}^{\ast}(p_2,-\ii x) &=N_{s_1}N_{s_2}\int_{\mathcal C}\frac{\dd \xi\dd \zeta}{(2\pi\ii)^2}\, 
        (2p_1 x)^{-\ii\tnu +\xi}(2p_2x)^{-\ii\tnu +\zeta}e^{-(p_1+p_2)x}
        \\&\times 
        \Gamma\!\left[
		\begin{array}{c}
			-\xi,-\xi+2\ii\tnu,-\zeta,-\zeta+2\ii\tnu,a_{s_1}+\xi,a_{s_2}+\zeta \\
			a_{s_1},\bar{a}_{s_1} ,a_{s_2},\bar{a}_{s_2}
    		\end{array}
		\right],
\end{aligned}
\end{equation}
where, to simplify the notation, we have defined
\begin{equation}
    a_{s_i}\equiv \frac{1-s_i}{2}-\ii(s_i\tlam+\tnu),\quad \bar{a}_{s_i}\equiv a_{s_i}+2\ii\tnu=\frac{1-s_i}{2}-\ii(s_i\tlam-\tnu).
\end{equation}

After applying the MB representation to $v^\ast$, the right-time integral appearing in Eq.~\eqref{Jbranch} is fully factorized. We can now perform the $x$ integral
\begin{equation}
\begin{aligned}
\label{Righttimesint}
    &\int_0^\infty \dd x\,
    x^{q_2+4-2\ii\tnu +\xi+\zeta} e^{-(k_{34}+p_1+p_2)x}
    =
	\Gamma[q_2+5-2\ii\tilde\nu+\xi+\zeta]
	(k_{34}+p_1+p_2)^{-(q_2+5-2\ii\tilde\nu+\xi+\zeta)}.
\end{aligned}
\end{equation}
We observe that the external momentum $k_{34}$ is coupled to the loop momenta $p_1$ and $p_2$ carried by the internal propagators, which is inconvenient for the subsequent loop momentum integration. To factorize the dependence on $p_1$ and $p_2$, we apply an MB integral representation to the coupled factor~\cite{NIST:DLMF}
\begin{equation}
\begin{aligned}
    (k_{34}+p_1+p_2)^{-(q_2+5-2\ii\tilde\nu+\xi+\zeta)}
	&=
	k_{34}^{-(q_2+5-2\ii\tilde\nu+\xi+\zeta)}
	\int\frac{\dd u}{2\pi\ii}
	\int\frac{\dd v}{2\pi\ii}\,
	\left(\frac{p_1}{k_{34}}\right)^u\left(\frac{p_2}{k_{34}}\right)^v 
    \\&\times
    \Gamma\!\left[
		\begin{matrix}
			-u,-v,q_2+5-2\ii\tilde\nu+\xi+\zeta+u+v \\
			q_2+5-2\ii\tilde\nu+\xi+\zeta
    		\end{matrix}
		\right].
\end{aligned}
\end{equation}
Similarly, the integration contours for the $u$ and $v$ variables are chosen such that the left and right pole families are separated from each other. As a result, the loop momentum integral becomes much more tractable.

So far, we have introduced four MB variables, $\xi$, $\zeta$, $u$, and $v$, and have completely evaluated the right-time integral over $\eta_2$ in $\widehat{\mathcal{J}}_{-+,\rho,\lambda}^{q_1q_2,s_1s_2}$. The result can be written as
\begin{equation}
\begin{aligned}
\label{Jhat2}
    \widehat{\mathcal{J}}_{-+,\rho,\lambda}^{q_1q_2,s_1s_2}
	&=
    \frac{2\, k_s^{7+q_{12}}}{r_1^{5+q_1}r_2^{5+q_2}}
  2^{2\ii(\rho-1)\tnu}
	N_{s_1}N_{s_2}e^{\ii\frac{\pi}{2}(q_1-q_2+2\ii\rho\tnu)}
	\Gamma[5+q_1+2\ii\rho\tnu]
    k_{12}^{-(5+q_1+2\ii\rho\tnu)}
    \\&
    \times
    k_{34}^{-(5+q_2-2\ii\tilde\nu)}
    \int \frac{\dd \xi\dd \zeta \dd u\dd v}{(2\pi\ii)^4}
    2^{\xi+\zeta}k_{34}^{-(\xi+\zeta+u+v)}\Gamma[q_2+5-2\ii\tilde\nu+\xi+\zeta+u+v]
    \\&
    \times\Gamma\!\left[
		\begin{array}{c}
			-u,-v,-\xi,-\xi+2\ii\tnu,-\zeta,-\zeta+2\ii\tnu,a_{s_1}+\xi,a_{s_2}+\zeta \\
			a_{s_1},\bar{a}_{s_1} ,a_{s_2},\bar{a}_{s_2}
    		\end{array}
		\right]
    \\&
    \times\int \frac{\dd^3 \mathbf{q}}{(2\pi)^3}\, 
    p_1^{\ii(\rho-1)\tnu+\xi+u}p_2^{\ii(\rho-1)\tnu+\zeta+v}F_{s_1 s_2}(\hat {\mathbf{p}}_1,\hat {\mathbf{p}}_2).
\end{aligned}
\end{equation}
Next, we proceed to evaluate the loop-momentum integral, where the momenta carried by the two internal propagators are defined as $\mathbf{p}_1=\mathbf{q}$ and $\mathbf{p}_2=\mathbf{k}_s-\mathbf{q}$, respectively. First, we can expand $F_{s_1 s_2}(\hat{\mathbf{p}}_1,\hat{\mathbf{p}}_2)$ as
\begin{equation}
  F_{s_1s_2}(\hat {\mathbf{p}}_1,\hat {\mathbf{p}}_2)=
  \frac{1}{2}-\frac{s_1s_2}{2}\hat {\mathbf{p}}_1\cdot\hat {\mathbf{p}}_2=
  \frac{1}{2}-\frac{s_1s_2}{4}
  \left(
  \frac{k_s^2}{p_1p_2}-\frac{p_1}{p_2}-\frac{p_2}{p_1}
  \right).
\end{equation}
Under this expansion, the loop momentum integral can be recast as
\begin{equation}
\label{loopintexpansion}
\begin{aligned}
    &\int \frac{\dd^3 \mathbf{q}}{(2\pi)^3}\, 
    p_1^{\ii(\rho-1)\tnu+\xi+u}p_2^{\ii(\rho-1)\tnu+\zeta+v}F_{s_1 s_2}(\hat {\mathbf{p}}_1,\hat {\mathbf{p}}_2)
    \\&=
    \sum_{\chi=0}^2
    d_\chi
    \int \frac{\dd^3 \mathbf{q}}{(2\pi)^3}\, 
    p_1^{\ii(\rho-1)\tnu+\xi+u+\alpha_\chi}p_2^{\ii(\rho-1)\tnu+\zeta+v+\beta_\chi}
    \\&+k_s^2\, d_3\int \frac{\dd^3 \mathbf{q}}{(2\pi)^3}\, 
    p_1^{\ii(\rho-1)\tnu+\xi+u+\alpha_3}p_2^{\ii(\rho-1)\tnu+\zeta+v+\beta_3},
\end{aligned}
\end{equation}
where the definitions of $d_\chi$, $\alpha_\chi$, and $\beta_\chi$ are summarized in the following table
\begin{equation}
\begin{array}{c|c|c|c|c}
  \chi & d_\chi & \alpha_\chi & \beta_\chi \\ \hline
  0 & 1/2 & 0 & 0 \\
  1 & s_1s_2/4 & 1 & -1 \\
  2 & s_1s_2/4 & -1 & 1 \\
  3 & -s_1s_2/4 & -1 & -1 
\end{array}\ .
\end{equation}
For notational compactness, we define
\begin{equation}
    A_{\rho}^\chi(\xi,u) \equiv  \ii(\rho-1)\tnu +\xi+u+\alpha_\chi ,\quad  B_{\rho}^\chi(\zeta,v)\equiv \ii(\rho-1)\tnu +\zeta+v+\beta_\chi,\quad \delta_\chi\equiv\alpha_\chi+\beta_\chi.
\end{equation}
Within the convergence domain of the parameters, the loop-momentum integral appearing on the right-hand side of Eq.~\eqref{loopintexpansion} can be evaluated by standard methods, such as Feynman parametrization, yielding
\begin{equation}
\label{loopmomentumint}
    \int \frac{\dd^3 \mathbf{q}}{(2\pi)^3}\, 
    p_1^{A_{\rho}^\chi(\xi,u)}p_2^{B_{\rho}^\chi(\zeta,v)}
    =
    \frac{k_s^{A_{\rho}^\chi(\xi,u)+B_{\rho}^\chi(\zeta,v)+3}}{(4\pi)^{3/2}}\, 
    \Gamma\!\left[
		\begin{array}{c}
			-\frac{ A_{\rho}^\chi(\xi,u)+B_{\rho}^\chi(\zeta,v)+3}{2},\frac{A_{\rho}^\chi(\xi,u)+3}{2},\frac{B_{\rho}^\chi(\zeta,v)+3}{2}\\
			\frac{ A_{\rho}^\chi(\xi,u)+B_{\rho}^\chi(\zeta,v)+6}{2},-\frac{A_{\rho}^\chi(\xi,u)}{2},-\frac{B_{\rho}^\chi(\zeta,v)}{2}
    		\end{array}
		\right].
\end{equation}
For parameter values outside the convergence domain, the result of the loop momentum integral should be understood by analytic continuation.
\par 
We have now completed the evaluation of the loop momentum integral in Eq.~\eqref{Jhat2}. After collecting all the factors, the expression for $\widehat{\mathcal{J}}_{-+,\rho,\lambda}^{q_1q_2,s_1s_2}$ can then be written as
\begin{equation}
\begin{aligned}
\label{totalseedint}
   \widehat{\mathcal{J}}_{-+,\rho,\lambda}^{q_1q_2,s_1s_2}
	&=\mathcal{N}_{\rho}^{q_1q_2,s_1s_2}
     r_1^{2\ii\rho\tnu}r_2^{-2\ii\tilde\nu}
     \sum_{\chi=0}^3d_\chi
    \int \frac{\dd \xi\dd \zeta \dd u\dd v}{(2\pi\ii)^4}
    2^{\xi+\zeta}r_2^{\xi+\zeta+u+v}
    \\&\times
    \Gamma\left[-u,-v,q_2+5-2\ii\tilde\nu+\xi+\zeta+u+v\right]\, 
    \widetilde{\Gamma}^{s_1s_2}(\xi,\zeta)
    \, \widehat{\Gamma}_{\rho}^\chi(\xi,\zeta,u,v),
\end{aligned}
\end{equation}
where, for notational compactness, we have extracted all factors independent of the MB variables and the kinematic variables, and define them as
\begin{equation}
    \mathcal{N}_{\rho}^{q_1q_2,s_1s_2}\equiv \frac{N_{s_1}N_{s_2}}{4\pi^{3/2}}
  2^{2\ii(\rho-1)\tnu}
	 e^{\ii\frac{\pi}{2}(q_1-q_2+2\ii\rho\tnu)}
	\Gamma[5+q_1+2\ii\rho\tnu].
\end{equation}
Furthermore, we define the following two combinations of Gamma functions, $\widetilde{\Gamma}^{s_1s_2}(\xi,\zeta)$ and $\widehat{\Gamma}_{\rho}^\chi(\xi,\zeta,u,v)$, respectively, as
\begin{equation}
    \widetilde{\Gamma}^{s_1s_2}(\xi,\zeta)
    \equiv
    \Gamma\!\left[
		\begin{array}{c}
			-\xi,-\xi+2\ii\tnu,-\zeta,-\zeta+2\ii\tnu,a_{s_1}+\xi,a_{s_2}+\zeta \\
			a_{s_1},\bar{a}_{s_1} ,a_{s_2},\bar{a}_{s_2}
    		\end{array}
		\right],
\end{equation}
\begin{equation}
\label{Ghat}
    \widehat{\Gamma}_{\rho}^\chi(\xi,\zeta,u,v)
    \equiv 
    \Gamma\!\left[
		\begin{matrix}
			-\frac{ A_{\rho}^\chi(\xi,u)+B_{\rho}^\chi(\zeta,v)+3}{2},\frac{A_{\rho}^\chi(\xi,u)+3}{2},\frac{B_{\rho}^\chi(\zeta,v)+3}{2}\\
			\frac{ A_{\rho}^\chi(\xi,u)+B_{\rho}^\chi(\zeta,v)+6}{2},-\frac{A_{\rho}^\chi(\xi,u)}{2},-\frac{B_{\rho}^\chi(\zeta,v)}{2}
    		\end{matrix}
		\right].
\end{equation}
Here, $\widetilde{\Gamma}^{s_1s_2}(\xi,\zeta)$ denotes the combination of Gamma functions arising from the MB representations of the Whittaker functions, whereas $\widehat{\Gamma}_{\rho}^\chi(\xi,\zeta,u,v)$ denotes the combination of Gamma functions generated by the loop momentum integration.
\par
We have now evaluated both the time integral at the right vertex and the loop-momentum integral in $\widehat{\mathcal{J}}_{-+,\rho,\lambda}^{q_1q_2,s_1s_2}$, expressing the result as a contour integral over Mellin variables in Eq.~\eqref{totalseedint}. The integrand consists entirely of Gamma functions and power functions. Integrals of this form are commonly known as Mellin--Barnes (MB) integrals~\cite{Barnes1910,Barnes1908,ParisKaminski2001}. The Gamma functions in the integrand generate pole families in the integration variables. For an MB integration variable $z$, factors of the form $\Gamma[z+\cdots]$ generate left pole families, consisting of poles extending indefinitely toward $\operatorname{Re}z\to-\infty$, while factors of the form $\Gamma[-z+\cdots]$ generate right pole families, consisting of poles extending indefinitely toward $\operatorname{Re}z\to+\infty$. When evaluating an MB integral, the initial integration contour must fully separate the left and right pole families. For more detailed discussions, see Ref.~\cite{ParisKaminski2001,Czakon:2005rk,Tausk:1999vh,Smirnov:1999gc}, Appendix of Ref.~\cite{Sleight:2019mgd}, or Ref.~\cite{NIST:DLMF}. To evaluate the integral in Eq.~\eqref{totalseedint}, we therefore need to analyze the poles generated by the Gamma function factors in the integrand. This is the main focus of the next subsection.

\subsection{Pole Analysis \label{poleana}}
At this stage, we have completed the time and loop-momentum integrations in $\widehat{\mathcal{J}}_{-+,\rho,\lambda}^{q_1q_2,s_1s_2}$. However, these integrations are replaced by a four-fold Mellin--Barnes (MB) integral. We can evaluate these MB integrals using the residue theorem, which requires a detailed analysis of the pole structure of the integrand. We observe that the integrand contains the exponential factor $r_2^{\xi+\zeta+u+v}$. In the physical kinematic regime $0<r_2<1$, the convergence of the integral at the boundary of infinite radius requires us to close the integration contours to the right. Therefore, we need to consider all possible right poles\footnote{Here, the term ``right poles'' refers to all poles belonging to the right pole families of the corresponding MB integration variables, namely the poles arising from Gamma functions of the form $\Gamma[-z+\cdots]$, where $z$ denotes a generic MB integration variable (such as $\xi$, $\zeta$, $u$, $v$). It does not specifically refer to poles located in the right half-plane of the complex $z$ plane (i.e. with positive real parts). Even if the leftmost pole of a right pole family lies in the left half-plane, it should still be included in the residue calculation. Conversely, even if the rightmost pole of a left pole family lies in the right half-plane, it should not be considered. This distinction arises because the choice of the vertical integration contours in the MB representation has already separated the left and right pole families.} and identify which poles contribute to the nonlocal signals and which contribute to the local signals.
\par 
The right poles of the MB variables $\xi$ and $\zeta$ arise from the Gamma factors
\begin{equation*}
    \Gamma\!\left[
        -\xi,\,-\xi+2\ii\tnu,\,-\zeta,\,-\zeta+2\ii\tnu
    \right] .
\end{equation*}
We refer to the poles generated by these Gamma factors as the spectrum poles. They are located at
\begin{equation}
    \xi=n_1,
    \qquad
    \xi=2\ii\tnu+n_1,
    \qquad
    n_1=0,1,2,\ldots,
\end{equation}
and
\begin{equation}
    \zeta=n_2,
    \qquad
    \zeta=2\ii\tnu+n_2,
    \qquad
    n_2=0,1,2,\ldots .
\end{equation}
It is useful to introduce the shorthand notation
\begin{equation}
\xi_{\sigma_1}=
\left\{
\begin{aligned}
    2\ii\tnu, & \qquad \sigma_1=+,\\
    0,        & \qquad \sigma_1=-,
\end{aligned}
\right.
\qquad
\zeta_{\sigma_2}=
\left\{
\begin{aligned}
    2\ii\tnu, & \qquad \sigma_2=+,\\
    0,        & \qquad \sigma_2=-.
\end{aligned}
\right.
\end{equation}
Then the spectrum poles can be written compactly as
\begin{equation}
    \xi=\xi_{\sigma_1}+n_1,
    \qquad
    n_1=0,1,2,\ldots,
    \qquad
    \sigma_1=\pm,
\end{equation}
and
\begin{equation}
    \zeta=\zeta_{\sigma_2}+n_2,
    \qquad
    n_2=0,1,2,\ldots,
    \qquad
    \sigma_2=\pm .
\end{equation}
\par 
The MB variables $u$ and $v$ also have independent integer poles, generated by $\Gamma(-u)$ and $\Gamma(-v)$. These poles are located at
\begin{equation}
\label{uvintegerpole}
    u=m_1,
    \qquad
    v=m_2,
    \qquad
    m_1,m_2=0,1,2,\ldots .
\end{equation}
\par 
In addition, the loop momentum integral generates the Gamma factor
\begin{equation*}
    \Gamma\!\left[
        -\frac{
        \ii(2\rho-2)\tnu+\xi+\zeta+u+v+\delta_\chi+3
        }{2}
    \right],
\end{equation*}
which generates a coupled pole family involving all MB integration variables
\begin{equation}
    \xi+\zeta+u+v
    =
    2\ell
    -2\ii(\rho-1)\tnu
    -\delta_\chi-3,
    \qquad
    \ell=0,1,2,\ldots .
    \label{UVpolecondition}
\end{equation}
Since this pole family originates from the UV divergence of the loop-momentum integral in Eq.~\eqref{loopmomentumint}, we refer to these poles as the UV poles.
\par 
The different pole families are associated with different components of the cosmological-collider signal. We first discuss the nonlocal signal. The nonlocal signal is characterized by a non-analytic dependence on the internal momentum $k_s$. In Eq.~\eqref{totalseedint}, the relevant dependence on $k_s$ is encoded in the factor
\begin{equation*}
r_1^{2\ii\rho\tnu}r_2^{-2\ii\tnu+\xi+\zeta+u+v}
    \propto
    k_s^{2\ii\rho\tnu-2\ii\tnu+\xi+\zeta+u+v}.
\end{equation*}
Therefore, the nonlocal contribution requires
\begin{equation}
    \operatorname{Im}(\xi+\zeta+u+v)+2(\rho-1)\tnu\neq 0 .
\end{equation}
The UV pole condition in Eq.~\eqref{UVpolecondition} instead imposes
\begin{equation}
    \operatorname{Im}(\xi+\zeta+u+v)=-2(\rho-1)\tnu .
\end{equation}
Hence, the UV pole family does not contribute to the nonlocal signal.
\par 
By contrast, the spectrum poles are responsible for the non-analytic dependence on the internal momentum $k_s$. Although the integer poles of the MB variables $u$ and $v$ also contribute to the residue expansion, they only shift the powers of $r_2$ by integers and therefore generate analytic descendants within a fixed spectrum-pole branch. In particular, they do not modify the imaginary part of the exponent and hence do not introduce new oscillation frequencies. Consequently, the non-analytic phase of the nonlocal signal is completely determined by the spectrum-pole branches. Among the spectrum-pole branches, we first consider the branch with
\[
\sigma_1=\sigma_2=\rho ,
\]
where $\rho=\pm$ labels the two late-time frequencies of the fermion mode functions. This branch gives a contribution proportional to
\begin{equation*}
    (r_1 r_2)^{2\ii\rho\tilde\nu},
\end{equation*}
which corresponds to a non-analytic dependence on the internal momentum $k_s$ and therefore contributes to the nonlocal signal. The mixed branches
\[
\sigma_1=-\sigma_2=\pm\rho ,
\]
instead generate a dependence of the form
\begin{equation*}
    r_1^{2\ii\rho\tilde\nu}.
\end{equation*}
Although the dependence on $r_2$ is analytic in this case, the remaining non-analytic dependence on $r_1$ implies a non-analytic dependence on $k_s$, and these branches also contribute to the nonlocal signal. Finally, the branch
\[
\sigma_1=\sigma_2=-\rho
\]
produces a purely analytic dependence on the internal momentum $k_s$. Therefore, it does not contribute to the nonlocal signal and should be excluded from $\mathcal J_{\rm NS}^{q_1q_2}$.
\par 
We next consider the local signal. By definition, the local signal is analytic in the internal momentum $k_s$, while it may retain non-analytic dependence on external momentum combinations such as $k_{12}$ and $k_{34}$. Since the factor $r_1^{2\ii\rho\tnu}$ already carries the non-analytic dependence on the external momentum $k_{12}$, extracting the local contribution amounts to imposing analyticity in $k_s$. This gives the condition
\begin{equation}
    \operatorname{Im}(\xi+\zeta+u+v)+2(\rho-1)\tnu=0 .
\end{equation}
This is precisely the imaginary part of the UV pole condition in Eq.~\eqref{UVpolecondition}. Therefore, within the present pole decomposition, the UV pole family provides a local contribution to the signal.
\par 
Besides the UV poles, there is another source of the local signal.
If we select the spectrum-pole branch $\sigma_1=\sigma_2=-\rho$
together with the integer poles of $u$ and $v$, then, since
$\xi_{-\rho}=\zeta_{-\rho}=\ii(1-\rho)\tnu$, this combination
of poles generates the nonanalytic dependence
$\left(r_1/r_2\right)^{2\ii\rho\tnu}$, and therefore also
contributes to the local signal. At these poles, we have
\begin{equation}
    A_{\rho}^{\chi}\bigl(\xi_{-\rho}+n_1,m_1\bigr)
    =
    n_1+m_1+\alpha_{\chi}
    \in \mathbb{Z},
    \qquad n_1,m_1\in\mathbb Z_{\geq 0},
\end{equation}
\begin{equation}
    B_{\rho}^{\chi}\bigl(\zeta_{-\rho}+n_2,m_2\bigr)
    =
    n_2+m_2+\beta_{\chi}
    \in \mathbb{Z},
    \qquad n_2,m_2\in\mathbb Z_{\geq 0}.
\end{equation}
For certain choices of $(n_1,n_2,m_1,m_2)$, these residue
values also satisfy the UV-pole condition,
\begin{equation}
    \frac{A_{\rho}^{\chi}+B_{\rho}^{\chi}+3}{2}
    \in \mathbb{Z}_{\geq 0}.
\end{equation}
At these intersections, the singular Gamma function factors
in the numerator and denominator must be evaluated together
using a specified regularization prescription.
Indeed, because of the Gamma functions in the denominator of
$\widehat{\Gamma}_{\rho}^\chi(\xi,\zeta,u,v)$, when
\begin{equation}
    A_{\rho}^{\chi}\in\mathbb{Z},
    \qquad
    B_{\rho}^{\chi}\in\mathbb{Z},
\end{equation}
a pole of
\begin{equation}
    \Gamma\left[
        -\frac{A_{\rho}^{\chi}+B_{\rho}^{\chi}+3}{2}
    \right]
\end{equation}
is necessarily accompanied by a pole of either
\begin{equation}
    \Gamma\left[-\frac{A_{\rho}^{\chi}}{2}\right]
    \qquad\text{or}\qquad
    \Gamma\left[-\frac{B_{\rho}^{\chi}}{2}\right]
\end{equation}
in the denominator. More explicitly, the condition that
$(A_{\rho}^{\chi}+B_{\rho}^{\chi}+3)/2$ be an integer implies
that $A_{\rho}^{\chi}+B_{\rho}^{\chi}$ is odd, so
exactly one of $A_{\rho}^{\chi}$ and $B_{\rho}^{\chi}$ is even.
Since $\alpha_\chi,\beta_\chi\in\{-1,0,1\}$ and
$n_i,m_i\geq0$, the even integer is nonnegative, and the
corresponding denominator Gamma function has a pole.
The coincident numerator and denominator poles do not, by themselves,
define a unique finite value for their ratio. To preserve symmetry
under interchange of the two internal fermion lines in
Eq.~\eqref{totalseedint}, we introduce a common infinitesimal
shift $\epsilon$ in $\alpha_\chi$ and $\beta_\chi$
\begin{equation}
    \alpha_\chi\rightarrow\alpha_\chi+\epsilon,
    \qquad
    \beta_\chi\rightarrow\beta_\chi+\epsilon.
\end{equation}
We denote by $\mathrm{AC}$ the regularized value obtained by taking
$\epsilon\to0$ along this prescribed deformation. When
$(A_{\rho}^{\chi}+B_{\rho}^{\chi}+3)/2$ is a nonnegative integer
and $A_{\rho}^{\chi}$ is a nonnegative even integer, this gives
\begin{equation}
\label{ACgamma1}
\begin{aligned}
    \mathrm{AC}\left\{
    \frac{
        \Gamma\left[-\frac{A_{\rho}^{\chi}+B_{\rho}^{\chi}+3}{2}\right]
    }{
        \Gamma\left[-\frac{A_{\rho}^{\chi}}{2}\right]
    }
    \right\}
    &\equiv
    \lim_{\epsilon\to0}
    \frac{
        \Gamma\left[
            -\frac{A_{\rho}^{\chi}+B_{\rho}^{\chi}+3+2\epsilon}{2}
        \right]
    }{
        \Gamma\left[-\frac{A_{\rho}^{\chi}+\epsilon}{2}\right]
    }
    =
    \frac{(-1)^{(B_\rho^\chi+3)/2}}{2}
    \frac{
        \Gamma\left[1+\frac{A_{\rho}^{\chi}}{2}\right]
    }{
        \Gamma\left[
            1+\frac{A_{\rho}^{\chi}+B_{\rho}^{\chi}+3}{2}
        \right]
    }.
\end{aligned}
\end{equation}
Similarly, when $(A_{\rho}^{\chi}+B_{\rho}^{\chi}+3)/2$ is a
nonnegative integer and $B_{\rho}^{\chi}$ is a nonnegative even
integer, we obtain
\begin{equation}
\label{ACgamma2}
\begin{aligned}
    \mathrm{AC}\left\{
    \frac{
        \Gamma\left[-\frac{A_{\rho}^{\chi}+B_{\rho}^{\chi}+3}{2}\right]
    }{
        \Gamma\left[-\frac{B_{\rho}^{\chi}}{2}\right]
    }
    \right\}
    &\equiv
    \lim_{\epsilon\to0}
    \frac{
        \Gamma\left[
            -\frac{A_{\rho}^{\chi}+B_{\rho}^{\chi}+3+2\epsilon}{2}
        \right]
    }{
        \Gamma\left[-\frac{B_{\rho}^{\chi}+\epsilon}{2}\right]
    }
    =
    \frac{(-1)^{(A_\rho^\chi+3)/2}}{2}
    \frac{
        \Gamma\left[1+\frac{B_{\rho}^{\chi}}{2}\right]
    }{
        \Gamma\left[
            1+\frac{A_{\rho}^{\chi}+B_{\rho}^{\chi}+3}{2}
        \right]
    }.
\end{aligned}
\end{equation}
Under this prescription, the coincident Gamma-function factors
in the local contribution from the spectrum-pole branch
$\sigma_1=\sigma_2=-\rho$, combined with the integer poles of
$u$ and $v$, have finite limits.

\section{Calculation of Nonlocal and Local Signals\label{4}}
In this section, we use the residue theorem to compute the nonlocal and local signals separately, based on the MB representation of the seed integral derived in the previous section, Eq.~\eqref{totalseedint}, together with the pole analysis of the integrand.

\subsection{Nonlocal Signals\label{6.1}}
We select the three spectral-pole branches in $\xi$ and $\zeta$ that contribute to the nonlocal signal,
$\sigma_1=\sigma_2=\rho$ and $\sigma_1=-\sigma_2=\pm\rho$,
together with the integer poles in $u$ and $v$. Applying the residue theorem to Eq.~\eqref{totalseedint}, we obtain
\begin{equation}
\begin{aligned}
\label{seedintnonlocal}
    \widehat{\mathcal{J}}_{-+,\rho,\lambda,\text{NS}}^{q_1q_2,s_1s_2}
	&=
    \mathcal{N}_{\rho}^{q_1q_2,s_1s_2}
    r_1^{2\ii\rho\tnu}r_2^{-2\ii\tilde\nu}\sum_{(\sigma_1,\sigma_2)^\prime}\sum_{\chi=0}^3d_\chi (2r_2)^{\xi_{\sigma_1}+\zeta_{\sigma_2}}
    \sum_{n_i,m_i=0}^\infty\frac{(-1)^{n_1+n_2+m_1+m_2}}{n_1!n_2!m_1!m_2!}
    \\&\times
    2^{n_1+n_2}r_2^{n_1+n_2+m_1+m_2}\Gamma[q_2+5-2\ii\tilde\nu+\xi_{\sigma_1}+\zeta_{\sigma_2}+n_1+n_2+m_1+m_2]
    \\&\times 
    \widetilde{\Gamma}^{s_1s_2}_{\text{sp},\sigma_1\sigma_2}(n_1,n_2)
    \times \widehat{\Gamma}_{\rho}^\chi(\xi_{\sigma_1}+n_1,\zeta_{\sigma_2}+n_2,m_1,m_2),
\end{aligned}
\end{equation}
where the summation $\sum_{(\sigma_1,\sigma_2)^\prime}$ runs over all spectrum pole branches except for the branch with $\sigma_1=\sigma_2=-\rho$. In the above expression, we have defined the partial residue of $\widetilde{\Gamma}^{s_1s_2}(\xi,\zeta)$ at the spectral poles as $\widetilde{\Gamma}^{s_1s_2}_{\text{sp},\sigma_1\sigma_2}(n_1,n_2)$
\begin{equation}
    \widetilde{\Gamma}^{s_1s_2}_{\text{sp},\sigma_1\sigma_2}(n_1,n_2)\equiv\Gamma\!\left[
		\begin{array}{c}
			-n_1-\sigma_12\ii\tnu,
            a_{s_1}+\xi_{\sigma_1}+n_1,
            -n_2-\sigma_22\ii\tilde\nu,
            a_{s_2}+\zeta_{\sigma_2}+n_2 \\
			a_{s_1},\bar{a}_{s_1} ,a_{s_2},\bar{a}_{s_2} 
		\end{array}
		\right].
\end{equation}
\par 
There is one subtle point that should be noted in this calculation. When applying the residue theorem, we use
\begin{equation}
    \operatorname*{Res}_{\xi = n}\Gamma(-\xi)
    =
    -\frac{(-1)^{n}}{n!},
    \qquad n=0,1,2,\ldots ,
\end{equation}
this gives rise to an extra minus sign. In addition, when applying the residue theorem, we have to close the contour to the right, so that the closed contour is oriented clockwise. However, the residue theorem is conventionally formulated for a counterclockwise contour, which introduces another minus sign. These two minus signs cancel each other, and therefore no extra overall minus sign appears in the final result.
\par 
Next, we would like to simplify the four-fold summation in Eq.~\eqref{seedintnonlocal} and reduce the number of independent sums. To this end, we reorganize the summation variables by defining
\begin{equation}
    n_1+n_2+m_1+m_2\equiv N,\quad n_1+m_1\equiv p,\quad n_2+m_2=N-p
\end{equation}
Under this redefinition, the four summation variables are changed from $(n_1,n_2,m_1,m_2)$ to $(n_1,n_2,p,N)$. The summation can be written as
\begin{equation}
    \sum_{n_1,n_2,m_1,m_2=0}^{\infty}
    =
    \sum_{N=0}^{\infty}
    \sum_{p=0}^{N}
    \sum_{n_1=0}^{p}
    \sum_{n_2=0}^{N-p},
\end{equation}
and then Eq.~\eqref{seedintnonlocal} can be rewritten as
\begin{equation}
\begin{aligned}
\label{JNS1}
    \widehat{\mathcal{J}}_{-+,\rho,\lambda,\text{NS}}^{q_1q_2,s_1s_2}
	&=
    \mathcal{N}_{\rho}^{q_1q_2,s_1s_2}
    r_1^{2\ii\rho\tnu}r_2^{-2\ii\tilde\nu}\sum_{(\sigma_1,\sigma_2)^\prime}\sum_{\chi=0}^3d_\chi (2r_2)^{\xi_{\sigma_1}+\zeta_{\sigma_2}}
    \sum_{N=0}^{\infty}
    \sum_{p=0}^{N}
    \sum_{n_1=0}^{p}
    \sum_{n_2=0}^{N-p}
    \\&\times
    \frac{(-2)^{n_1+n_2}}{n_1!n_2!}
\frac{
(-p)_{n_1}\left(p-N\right)_{n_2}
}{
p!(N-p)!
}\, (-r_2)^N\, 
    \Gamma[q_2+5-2\ii\tilde\nu+\xi_{\sigma_1}+\zeta_{\sigma_2}+N]
    \\&\times 
    \widetilde{\Gamma}^{s_1s_2}_{\text{sp},\sigma_1\sigma_2}(n_1,n_2)
    \times \widehat{\Gamma}_{\rho,\sigma_1\sigma_2}^\chi(N,p),
\end{aligned}
\end{equation}
where we have used the identity
\begin{equation}
    \frac{(-1)^{p-n_1}(-1)^{N-p-n_2}}{(p-n_1)!(N-p-n_2)!}
    =
    (-1)^N
\frac{
(-p)_{n_1}\left(p-N\right)_{n_2}
}{
p!(N-p)!
},
\end{equation}
and defined
\begin{equation}
\label{GammaHat}
    \widehat{\Gamma}_{\rho,\sigma_1\sigma_2}^\chi(N,p)
    \equiv
    \Gamma\!\left[
		\begin{matrix}
			-\frac{ 2\ii(\rho-1)\tnu +\xi_{\sigma_1}+\zeta_{\sigma_2}+N+\delta_\chi+3}{2},\frac{\ii(\rho-1)\tnu+\xi_{\sigma_1}+p+\alpha_\chi+3}{2},\frac{\ii(\rho-1)\tnu+\zeta_{\sigma_2}+N-p+\beta_\chi+3}{2}\\
			\frac{ 2\ii(\rho-1)\tnu +\xi_{\sigma_1}+\zeta_{\sigma_2}+N+\delta_\chi+6}{2},-\frac{\ii(\rho-1)\tnu+\xi_{\sigma_1}+p+\alpha_\chi}{2},-\frac{\ii(\rho-1)\tnu+\zeta_{\sigma_2}+N-p+\beta_\chi}{2}
    		\end{matrix}
		\right].
\end{equation}
It can be seen that $\widehat{\Gamma}_{\rho,\sigma_1\sigma_2}^\chi(N,p)$ has no explicit dependence on the summation variables $n_1$ and $n_2$.
\par 
We can perform the sums over $n_1$ and $n_2$ in Eq.~\eqref{JNS1} analytically. For convenience, we rewrite the $n_1$ and $n_2$ dependent factors in $\widetilde{\Gamma}^{s_1s_2}_{\text{sp},\sigma_1\sigma_2}(n_1,n_2)$ in terms of Pochhammer symbols
\begin{equation}
\begin{aligned}
\label{gammasp}
\widetilde{\Gamma}^{s_1s_2}_{\rm sp,\sigma_1\sigma_2}(n_1,n_2)
&=
(-1)^{n_1+n_2}\,
\Gamma\!\left[
		\begin{matrix}
        -2\ii\sigma_1\tnu,a_{s_1}+\xi_{\sigma_1},-2\ii\sigma_2\tnu,a_{s_2}+\zeta_{\sigma_2}\\
            a_{s_1},\bar{a}_{s_1},a_{s_2},\bar{a}_{s_2}
    		\end{matrix}
		\right]
        \\&\times 
\frac{
\left(
a_{s_1}+\xi_{\sigma_1}
\right)_{n_1}
}{
\left(1+2\ii\sigma_1\tnu\right)_{n_1}
}
\frac{
\left(
a_{s_2}+\zeta_{\sigma_2}
\right)_{n_2}
}{
\left(1+2\ii\sigma_2\tnu\right)_{n_2}
}.
\end{aligned}
\end{equation}
Collecting all factors in Eq.~\eqref{JNS1} that depend on the summation variables $n_1$ and $n_2$, and using the series definition of the hypergeometric function ${}_2\text F_1$ given in Eq.~\eqref{pFqseries}, the sums over $n_1$ and $n_2$ can be expressed as a product of two terminating hypergeometric functions ${}_2\text F_1$.
\begin{equation}
\begin{aligned}
\label{2F1plus2F1NL}
&\sum_{n_1=0}^{p}\sum_{n_2=0}^{N-p}
\frac{2^{n_1+n_2}}{n_1!n_2!}
(-p)_{n_1}\left(p-N\right)_{n_2}
\frac{
\left(
a_{s_1}+\xi_{\sigma_1}
\right)_{n_1}
}{
\left(1+2\ii\sigma_1\tnu\right)_{n_1}
}
\frac{
\left(
a_{s_2}+\zeta_{\sigma_2}
\right)_{n_2}
}{
\left(1+2\ii\sigma_2\tnu\right)_{n_2}
}
\\
&=
{}_2\text F_1\!\left[
		\begin{array}{c}
			-p,a_{s_1}+\xi_{\sigma_1} \\
			1+2\ii\sigma_1\tnu
		\end{array};2
		\right]\, 
{}_2\text F_1\!\left[
		\begin{array}{c}
			-(N-p),a_{s_2}+\zeta_{\sigma_2} \\
			1+2\ii\sigma_2\tnu
		\end{array};2
		\right].
\end{aligned}
\end{equation}
After performing the double summation over $n_1$ and $n_2$, we obtain the following expression for $\widehat{\mathcal{J}}_{-+,\rho,\lambda,\text{NS}}^{q_1q_2,s_1s_2}$
\begin{equation}
\begin{aligned}
\label{JNSfinal}
    \widehat{\mathcal{J}}_{-+,\rho,\lambda,\text{NS}}^{q_1q_2,s_1s_2}
	&=
    \mathcal{N}_{\rho}^{q_1q_2,s_1s_2}
    r_1^{2\ii\rho\tnu}r_2^{-2\ii\tilde\nu}\sum_{(\sigma_1,\sigma_2)^\prime}
    \Gamma\!\left[
		\begin{matrix}
        -2\ii\sigma_1\tnu,a_{s_1}+\xi_{\sigma_1},-2\ii\sigma_2\tnu,a_{s_2}+\zeta_{\sigma_2}\\
            a_{s_1},\bar{a}_{s_1},a_{s_2},\bar{a}_{s_2}
    		\end{matrix}
		\right]
    \\&\times
    (2r_2)^{\xi_{\sigma_1}+\zeta_{\sigma_2}}
    \sum_{\chi=0}^3d_\chi 
    \sum_{N=0}^{\infty}
    \sum_{p=0}^{N}
\frac{
(-r_2)^N
}{
p!(N-p)!
}\, 
    \Gamma[q_2+5-2\ii\tilde\nu+\xi_{\sigma_1}+\zeta_{\sigma_2}+N]
    \\&\times 
    {}_2\text F_1\!\left[
		\begin{array}{c}
			-p,a_{s_1}+\xi_{\sigma_1} \\
			1+2\ii\sigma_1\tnu
		\end{array};2
		\right]\, 
{}_2\text F_1\!\left[
		\begin{array}{c}
			-(N-p),a_{s_2}+\zeta_{\sigma_2} \\
			1+2\ii\sigma_2\tnu
		\end{array};2
		\right]\, 
        \widehat{\Gamma}_{\rho,\sigma_1\sigma_2}^\chi(N,p).
\end{aligned}
\end{equation}
In this expression, $\widehat{\Gamma}_{\rho,\sigma_1\sigma_2}^\chi(N,p)$ is defined as in Eq.~\eqref{GammaHat}. The remaining sums are evaluated numerically below.
\par 
We emphasize that the result for $\widehat{\mathcal{J}}_{-+,\rho,\lambda,\text{NS}}^{q_1q_2,s_1s_2}$ is already in the single-squeezed limit $r_1\ll1$, as discussed in Section~\ref{3.2}. This limit is also understood to apply throughout the subsequent calculations of the other signal contributions. We can further consider the double-squeezed limit $r_1<r_2\ll1$. Taking the limit $r_2\to0$ in Eq.~\eqref{JNSfinal} amounts to retaining only the leading-pole contribution with $N=p=0$, which gives
\begin{equation}
\label{r2to0JNS}
    \begin{aligned}
        \lim_{r_2\to 0}\widehat{\mathcal{J}}_{-+,\rho,\lambda,\text{NS}}^{q_1q_2,s_1s_2}
        =
        \mathcal T^{q_1q_2,s_1s_2}_{\rho,\lambda,\text{NS1}}\, (r_1r_2)^{2\ii\rho\tnu}  + \mathcal T^{q_1q_2,s_1s_2}_{\rho,\lambda,\text{NS2}}\, r_1^{2\ii\rho\tnu},
    \end{aligned}
\end{equation}
where
\begin{equation}
\begin{aligned}
\mathcal T^{q_1q_2,s_1s_2}_{\rho,\lambda,\text{NS1}}
&=
\mathcal{N}_{\rho}^{q_1q_2,s_1s_2}
\left[
s_1s_2
\frac{2(1+\ii\rho\widetilde{\nu})}
{1+2\ii\rho\widetilde{\nu}}
\cosh^2(\pi\widetilde{\nu})
-
\frac{1+2\ii\rho\widetilde{\nu}}
{2(1+\ii\rho\widetilde{\nu})}
\sinh^2(\pi\widetilde{\nu})
\right]
\\
&\times
\frac{\ii\rho\pi\,2^{2\ii(1-\rho)\widetilde{\nu}}}
{4\sinh(4\pi\widetilde{\nu})}\, 
\Gamma\!\left[
\begin{matrix}
-2\ii\rho\widetilde{\nu},\,
q_2+5+2\ii\rho\widetilde{\nu},\,
a_{s_1}+\ii(1+\rho)\widetilde{\nu},\,
a_{s_2}+\ii(1+\rho)\widetilde{\nu}
\\
\frac{5}{2}+2\ii\rho\widetilde{\nu},\,
a_{s_1},\,\bar a_{s_1},\,
a_{s_2},\,\bar a_{s_2}
\end{matrix}
\right]
,
\end{aligned}
\end{equation}
\begin{equation}
\begin{aligned}
\mathcal T^{q_1q_2,s_1s_2}_{\rho,\lambda,\text{NS2}}
&=\mathcal{N}_{\rho}^{q_1q_2,s_1s_2}
\frac{
2^{2\ii\widetilde{\nu}}\sqrt{\pi}\,s_1s_2\,\Gamma[q_2+5]
}{
\widetilde{\nu}\sinh(2\pi\widetilde{\nu})
(1+2\ii\rho\widetilde{\nu})
(3+2\ii\rho\widetilde{\nu})
(2+\ii\rho\widetilde{\nu})
}
\\
&\times
\Bigg\{
\Gamma\!\left[
\begin{matrix}
a_{s_1}+\ii(1+\rho)\widetilde{\nu},\,
a_{s_2}+\ii(1-\rho)\widetilde{\nu}
\\
a_{s_1},\,\bar a_{s_1},\,
a_{s_2},\,\bar a_{s_2}
\end{matrix}
\right]
+
\Gamma\!\left[
\begin{matrix}
a_{s_1}+\ii(1-\rho)\widetilde{\nu},\,
a_{s_2}+\ii(1+\rho)\widetilde{\nu}
\\
a_{s_1},\,\bar a_{s_1},\,
a_{s_2},\,\bar a_{s_2}
\end{matrix}
\right]
\Bigg\}.
\end{aligned}
\end{equation}
Equation~\eqref{r2to0JNS} contains two oscillatory components with kinematic dependences $(r_1r_2)^{2\ii\rho\tnu}$ and $r_1^{2\ii\rho\tnu}$, whose respective amplitudes $\mathcal T^{q_1q_2,s_1s_2}_{\rho,\lambda,\text{NS1}}$ and $\mathcal T^{q_1q_2,s_1s_2}_{\rho,\lambda,\text{NS2}}$ are independent of $r_2$.
\begin{figure}[htbp]
    \centering
    \includegraphics[width=0.5\linewidth]{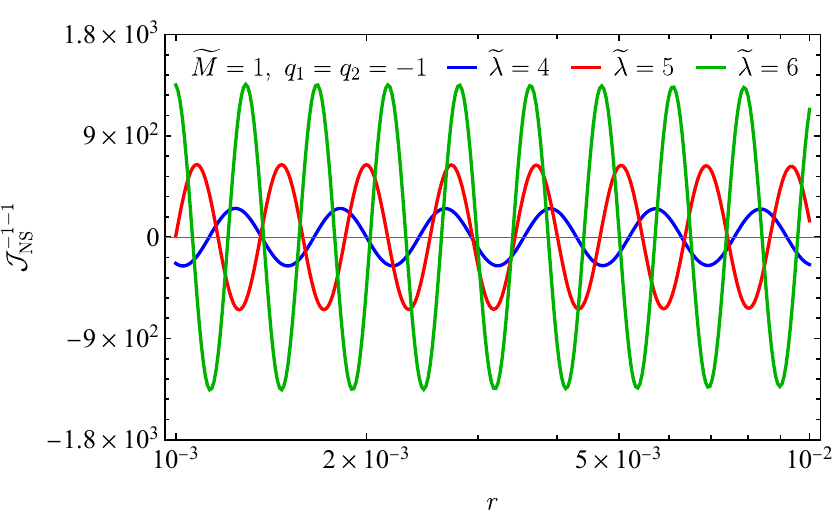}
    \includegraphics[width=0.49\linewidth]{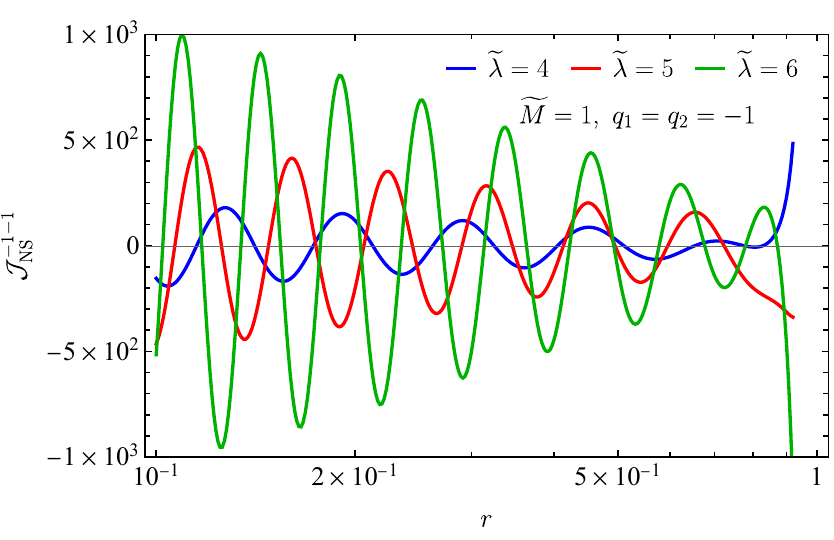}
    \caption{
The oscillatory nonlocal signal $\mathcal{J}_{\mathrm{NS}}^{-1,-1}$ as a function of $r$ for $q_1=q_2=-1$, $\widetilde M=1$, and $10^2\,r_1=r_2=r$. The blue, red, and green curves correspond to the chemical potentials $\widetilde{\lambda}=4$, $5$, and $6$, respectively. The left panel shows the oscillations in the squeezed regime $r_2\ll1$, while the right panel shows those near the folded limit $r_2\to1$.
}
    \label{fig:nonlocal}
\end{figure}
\par 
Using Eq.~\eqref{JF}, the complete nonlocal signal in $\mathcal{J}_{\text{F}}^{q_1q_2}$ can be written as
\begin{equation}
\label{partialJNS}
    \mathcal{J}_{\text{NS}}^{q_1q_2}=-\sum_\rho
    \left[
    1+e^{-\ii\pi(4+q_1+2\ii\rho\tnu)}
    \right]
    \sum_{s_1,s_2}\mathcal C_{s_1s_2}^\rho
    \left[
    \widehat{\mathcal{J}}_{-+,\rho,\lambda,\text{NS}}^{q_1q_2,s_1s_2}-\widehat{\mathcal{J}}_{-+,\rho,-\lambda,\text{NS}}^{q_1q_2,-s_1-s_2}
    \right]+\text{c.c.}\, .
\end{equation}
Using Eq.~\eqref{partialJNS} together with Eq.~\eqref{JNSfinal}, we numerically evaluate the nonlocal oscillatory signal for several values of the chemical potential. The results are shown in Fig.~\ref{fig:nonlocal}, which displays the oscillatory behavior of the nonlocal signal in the squeezed regime $r_2\ll1$ (left panel) and near the folded limit $r_2\to1$ (right panel). The blue, red, and green curves correspond to $\tlam=4$, $5$, and $6$, respectively. We consider the quartic interaction in Eq.~\eqref{modelint}, which corresponds to $q_1=q_2=-1$, and set the fermion mass to $\widetilde M=1$. For the kinematic variables, we take $10^2r_1=r_2=r$, which ensures that $r_1$ always remains in the strongly squeezed regime, $r_1\ll1$.
\par 
As shown in Fig.~\ref{fig:nonlocal}, increasing the chemical potential leads to both a higher oscillation frequency and a larger amplitude. The higher frequency follows from the increase in $\tnu$, which controls the logarithmic oscillations, while the larger amplitude reflects the exponential enhancement of positive-helicity fermion production at larger $\tlam$. In the squeezed regime $r_2\ll1$, the nonlocal signal is dominated by the leading-pole contribution in Eq.~\eqref{r2to0JNS}. For $q_1=q_2=-1$, the leading contribution proportional to $r_1^{2\ii\rho\tilde\nu}$ cancels exactly after summing over helicities and combining the $\rho=\pm$ branches with their complex conjugates in Eq.~\eqref{partialJNS}. The remaining leading contribution is proportional to $(r_1r_2)^{2\ii\rho\tilde\nu}$, with a coefficient independent of the kinematic variables. This accounts for the approximately constant oscillation amplitude in the left panel of Fig.~\ref{fig:nonlocal}. Towards the folded limit $r_2\to1$, subleading poles become important and modify this leading-order behavior. The oscillation amplitude then depends on $r_2$ and initially decreases as $r_2$ increases, before the nonlocal signal develops singular behavior sufficiently close to $r_2=1$, as shown in the right panel.

\subsection{Local Signal}
In this subsection, we compute the complete local signal, including the contribution from the UV poles and that from the spectrum-pole branch $\sigma_1=\sigma_2=-\rho$ together with the integer poles of $u$ and $v$.
\subsubsection{Local Signals from Spectrum and Integer Poles}
In the calculation of the nonlocal signal in Section~\ref{6.1}, we did not include the branch $\sigma_1=\sigma_2=-\rho$ in the sum over $(\sigma_1,\sigma_2)$ in Eq.~\eqref{seedintnonlocal}. As discussed in Section~\ref{poleana}, this spectrum-pole branch contributes to the local signal, with its contribution given by
\begin{equation}
\begin{aligned}
\label{JLS1final}
    \widehat{\mathcal{J}}_{-+,\rho,\lambda,\text{LSSP}}^{q_1q_2,s_1s_2}
	&=
    \mathcal{N}_{\rho}^{q_1q_2,s_1s_2}
    \left(\frac{r_1}{r_2}\right)^{2\ii\rho\tnu}\, 
    \Gamma\!\left[
		\begin{matrix}
        2\ii\rho\tnu,a_{s_1}+\xi_{-\rho},2\ii\rho\tnu,a_{s_2}+\zeta_{-\rho}\\
            a_{s_1},\bar{a}_{s_1},a_{s_2},\bar{a}_{s_2}
    		\end{matrix}
		\right]
    \\&\times
    2^{2\ii(1-\rho)\tnu}
    \sum_{\chi=0}^3d_\chi 
    \sum_{N=0}^{\infty}
    \sum_{p=0}^{N}
\frac{
(-r_2)^N
}{
p!(N-p)!
}\, 
    \Gamma[q_2+5-2\ii\tilde\nu+\xi_{-\rho}+\zeta_{-\rho}+N]
    \\&\times 
    {}_2\text F_1\!\left[
		\begin{matrix}
			-p,a_{s_1}+\xi_{-\rho} \\
			1-2\ii\rho\tnu
		\end{matrix};2
		\right]\, 
{}_2\text F_1\!\left[
		\begin{matrix}
			-(N-p),a_{s_2}+\zeta_{-\rho} \\
			1-2\ii\rho\tnu
		\end{matrix};2
		\right]\, 
        \widehat{\Gamma}_{\rho,-\rho-\rho}^\chi(N,p).
\end{aligned}
\end{equation}
Here, $\widehat{\Gamma}_{\rho,-\rho-\rho}^\chi(N,p)$ is defined in Eq.~\eqref{GammaHat}. In explicit calculations, the ratio of divergent Gamma functions in $\widehat{\Gamma}_{\rho,-\rho-\rho}^\chi(N,p)$ must be evaluated using Eq.~\eqref{ACgamma1} or Eq.~\eqref{ACgamma2}. 
\par 
We can similarly examine the spectrum-pole local signal in the squeezed limit $r_2\ll1$. Retaining only the $N=p=0$ contribution in Eq.~\eqref{JLS1final}, we obtain
\begin{equation}
\lim_{r_2\to 0}
\widehat{\mathcal J}_{-+,\rho,\lambda,\mathrm{LSSP}}^{q_1q_2,s_1s_2}
=
\mathcal T_{\rho,\lambda,\mathrm{LSSP}}^{q_1q_2,s_1s_2}
\left(\frac{r_1}{r_2}\right)^{2i\rho\widetilde{\nu}},
\end{equation}
where
\begin{equation}
\begin{aligned}
\mathcal T_{\rho,\lambda,\mathrm{LSSP}}^{q_1q_2,s_1s_2}
={}&
\mathcal N_{\rho}^{q_1q_2,s_1s_2}
\frac{s_1s_2\,2^{2\ii(1-\rho)\widetilde{\nu}}}
{3\sqrt{\pi}}\, 
\Gamma\!\left[
\begin{matrix}
2\ii\rho\widetilde{\nu},\,
2\ii\rho\widetilde{\nu},\,
q_2+5-2\ii\rho\widetilde{\nu}
\\
a_{s_1}+\ii(1+\rho)\widetilde{\nu},\,
a_{s_2}+\ii(1+\rho)\widetilde{\nu}
\end{matrix}
\right].
\end{aligned}
\end{equation}
\begin{figure}[htbp]
    \centering
    \includegraphics[width=0.8\linewidth]{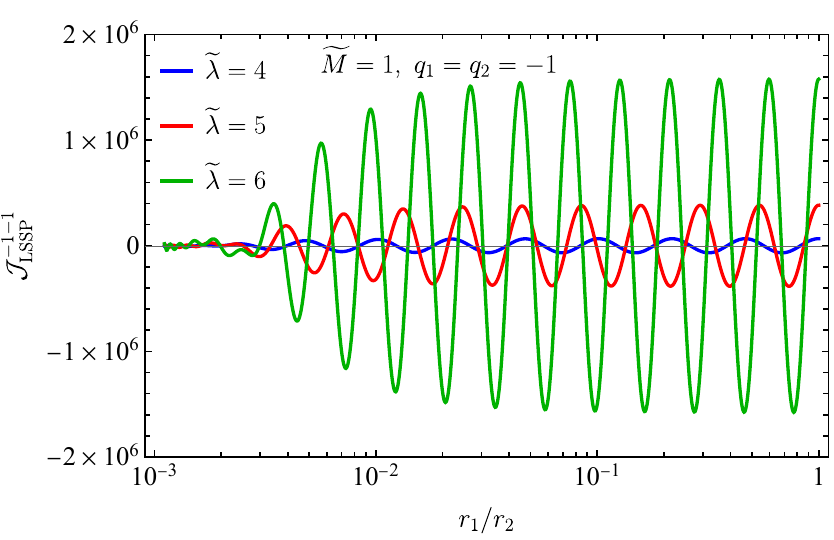}
    \caption{The oscillatory spectrum-pole local signal $\mathcal{J}_{\mathrm{LSSP}}^{-1,-1}$ as a function of $r_1/r_2$ for $q_1=q_2=-1$, $\widetilde M=1$. We fix $r_1=10^{-3}\ll 1$ and vary only $r_2$. The blue, red, and green curves correspond to the chemical potentials
    $\widetilde{\lambda}=4$, $5$, and $6$, respectively.}
    \label{figLSSP}
\end{figure}
\par 
Using Eq.~\eqref{JF}, the complete spectrum-pole local signal in $\mathcal{J}_{\text{F}}^{q_1q_2}$ can be written as
\begin{equation}
\label{partialJLSSP}
    \mathcal{J}_{\text{LSSP}}^{q_1q_2}=-\sum_\rho
    \left[
    1+e^{-\ii\pi(4+q_1+2\ii\rho\tnu)}
    \right]
    \sum_{s_1,s_2}\mathcal C_{s_1s_2}^\rho
    \left[
    \widehat{\mathcal{J}}_{-+,\rho,\lambda,\text{LSSP}}^{q_1q_2,s_1s_2}-\widehat{\mathcal{J}}_{-+,\rho,-\lambda,\text{LSSP}}^{q_1q_2,-s_1-s_2}
    \right]+\text{c.c.}\, .
\end{equation}
Using Eq.~\eqref{partialJLSSP} together with Eq.~\eqref{JLS1final}, we numerically evaluate the spectrum-pole local oscillatory signal for several values of the chemical potential. The results are shown in Fig.~\ref{figLSSP}, which illustrates the oscillatory behavior of the signal as a function of $r_1/r_2$. The blue, red, and green curves correspond to $\tlam=4$, $5$, and $6$, respectively. We choose the parameters $q_1=q_2=-1$ and the fermion mass $\mR=1$. For the kinematic variables, we fix $r_1=10^{-3}\ll 1$ and vary only $r_2$.
\par 
As shown in Fig.~\ref{figLSSP}, increasing the chemical potential raises both the oscillation frequency and the amplitude, as in the nonlocal case. The former follows from the increase in $\tnu$, while the latter reflects enhanced fermion production. For the parameter values used in Fig.~\ref{figLSSP}, the leading-pole contribution dominates over the range $r_2\in[10^{-3},10^{-1}]$, corresponding to $r_1/r_2\in[10^{-2},1]$. The oscillation amplitude is therefore approximately independent of $r_2$ in this regime. As $r_2$ increases towards unity, subleading poles make appreciable contributions, giving rise to an $r_2$-dependent amplitude. Over the range $r_2\in[10^{-1},1)$, corresponding to $r_1/r_2\in(10^{-3},10^{-2}]$, the amplitude decreases as $r_2$ increases.

\subsubsection{Local Signals from UV Poles}
In this subsection, we compute the local signal contributed by the UV poles. Since the UV-pole condition in Eq.~\eqref{UVpolecondition} couples all integration variables, the four-fold MB integral in Eq.~\eqref{totalseedint} has to be evaluated iteratively. We first pick up the UV poles and perform the integral over one of the MB variables, $u$ or $v$. Since performing the $u$ integral first is equivalent to performing the $v$ integral first, we may, without loss of generality, choose to integrate over $v$ first. We then perform the integrals over the remaining MB variables $u$, $\zeta$, and $\xi$. 
\par 
After taking the residue at the UV poles of $v$
\begin{equation}
\label{v_UVpole}
    v
    =
    2\ell
    -2\ii(\rho-1)\tnu
    -\xi-\zeta-u
    -\alpha_\chi-\beta_\chi-3,
\end{equation}
and performing the $v$ integral in Eq.~\eqref{totalseedint}, we obtain
\begin{equation}
\begin{aligned}
\label{seedintLSUV}
    \widehat{\mathcal{J}}_{-+,\rho,\lambda,\text{LSUV}}^{q_1 q_2,s_1 s_2}
	&=\mathcal{N}_{\rho}^{q_1q_2,s_1s_2}
     \left(\frac{r_1}{r_2}\right)^{2\ii\rho\tnu}
    \sum_{\chi}
    d_\chi
    r_2^{-\delta_\chi-3}
    \sum_{\ell=0}^\infty 
    \frac{2}{\ell!}\left(\frac{r_2}{2}\right)^{2\ell}\, 
    \frac{\Gamma[2\ell+q_2+2-2\ii\rho\tnu-\delta_\chi]}{\Gamma\left[\ell+\frac{3}{2}\right]}
    \\&\times 
    \int \frac{\dd \xi\dd \zeta \dd u}{(2\pi\ii)^3}\, 
    2^{\xi+\zeta}\, \Gamma[-u]\, \Gamma\left[2\ii(\rho-1)\tnu +\xi+\zeta+u+\delta_\chi +3-2\ell \right]
    \\&
    \times\widetilde{\Gamma}^{s_1s_2}(\xi,\zeta)\, \left(-1-\ii(\rho-1)\tnu -\xi-u-\alpha_\chi\right)_{2\ell}.
\end{aligned}
\end{equation}
In the calculation, to simplify the residue of $\widehat{\Gamma}_{\rho}^\chi(\xi,\zeta,u,v)$, we use the identity
\begin{equation}
    \Gamma\!\left[
		\begin{array}{c}
			\frac{A_\rho^\chi(\xi,u)+3}{2},\ell-\frac{A_\rho^\chi(\xi,u)}{2}\\
            -\frac{A_\rho^\chi(\xi,u)}{2},\frac{A_\rho^\chi(\xi,u)+3}{2}-\ell
    		\end{array}
		\right]
        =
        (-1)^\ell\, 4^{-\ell}(-1-A_\rho^\chi(\xi,u))_{2\ell}.
\end{equation}

We first aim to evaluate the remaining $u$ integral analytically. We observe that, in the integrand of Eq.~\eqref{seedintLSUV}, the MB variable $u$ appears both in the Pochhammer symbols and in the Gamma functions. To facilitate the evaluation of the integral, we would like to rewrite the integrand such that the $u$ dependence appears only through Gamma functions. For this purpose, we use the following identity
\begin{equation}
(x+a)_m\Gamma[x]
=
\sum_{t=0}^{m}
(-1)^t
\binom{m}{t}
(1-a-t)_t\,
\Gamma[x+m-t],
\end{equation}
and can obtain
\begin{equation}
\begin{aligned}
&(-1-\ii(\rho-1)\tnu-\xi-u-\alpha_{\chi})_{2\ell}
\, 
\Gamma[-u]
\\&=
\sum_{t=0}^{2\ell}
(-1)^t
\binom{2\ell}{t}
(2+\ii(\rho-1)\tnu+\xi+\alpha_{\chi}-t)_t\,
\Gamma(2\ell-t-u).
\end{aligned}
\end{equation}
Define
\begin{equation}
\label{DandDbar}
    D_{\rho\chi}=2\ii(\rho - 1)\tnu +\delta_\chi + 3,\quad \bar{D}_{\rho\chi}=2\ii\rho\tnu +\delta_\chi + 3.
\end{equation}
Direct evaluation of the $u$ integral gives
\begin{equation}
\begin{aligned}
    &\int\frac{\dd u}{2\pi \ii}
    \Gamma(-u)
    \left(-1-\ii(\rho-1)\tnu -\xi-u-\alpha_\chi\right)_{2\ell}\Gamma\left[D_{\rho\chi} +\xi+\zeta+u-2\ell \right]
    \\&=
    \sum_{t=0}^{2\ell}
(-1)^t
\binom{2\ell}{t}
(2+\ii(\rho-1)\tnu +\xi+\alpha_{\chi}-t)_t
\int\frac{\dd u}{2\pi \ii}
\Gamma[2\ell-t-u]\Gamma\left[D_{\rho\chi} +\xi+\zeta+u-2\ell \right]
\\&=
\sum_{t=0}^{2\ell}
(-1)^t
\binom{2\ell}{t}
(2+\ii(\rho-1)\tnu +\xi+\alpha_{\chi}-t)_t\, 2^{t-\xi-\zeta-D_{\rho\chi}}\Gamma[-t+\xi+\zeta+D_{\rho\chi}].
\end{aligned}
\end{equation}
The remaining $\zeta$ integral is a standard Barnes first-lemma integral
\begin{equation}
    \begin{aligned}
        &\int\frac{\dd \zeta}{2\pi \ii}
        \Gamma[-\zeta,-\zeta+2\ii \tnu,a_{s_2}+\zeta,-t+\xi+\zeta+D_{\rho\chi}]
        \\&=
        \Gamma\!\left[
\begin{matrix}
a_{s_2},a_{s_2}+2\ii\tnu,D_{\rho\chi}-t+\xi,\bar{D}_{\rho\chi}-t+\xi\\
a_{s_2}+\bar{D}_{\rho\chi}-t+\xi
\end{matrix}
\right].
    \end{aligned}
\end{equation}
After performing the $u$ and $\zeta$ integrals, only the $\xi$ integral remains. Equation~\eqref{seedintLSUV} can then be written as
\begin{equation}
\label{JLSUVmedi}
\begin{aligned}
    \widehat{\mathcal{J}}_{-+,\rho,\lambda,\text{LSUV}}^{q_1 q_2,s_1 s_2}
	&=\mathcal{N}_{\rho}^{q_1q_2,s_1s_2}
     \left(\frac{r_1}{r_2}\right)^{2\ii\rho\tnu}
    \sum_{\chi}
    d_\chi
    r_2^{-\delta_\chi-3}
    \sum_{\ell=0}^\infty 
    \frac{2}{\ell!}\left(\frac{r_2}{2}\right)^{2\ell}\, 
    \frac{\Gamma[2\ell+q_2+2-2\ii\rho\tnu-\delta_\chi]}{\Gamma\left[\ell+\frac{3}{2}\right]}
    \\&\times 
    \sum_{t=0}^{2\ell}
(-1)^t
\binom{2\ell}{t}\, 
2^{t-D_{\rho\chi}}
    \int \frac{\dd \xi}{2\pi\ii}
    \, 
    (2+\ii(\rho-1)\tnu +\xi+\alpha_{\chi}-t)_t
    \\&
    \times\Gamma\!\left[
\begin{matrix}
-\xi,a_{s_1}+\xi,-\xi+2\ii\tnu,D_{\rho\chi}-t+\xi,\bar{D}_{\rho\chi}-t+\xi\\
a_{s_1},\bar{a}_{s_1},
a_{s_2}+\bar{D}_{\rho\chi}-t+\xi
\end{matrix}
\right].
\end{aligned}
\end{equation}
The details of the $\xi$ integration are provided in Appendix~\ref{appendixE}. Finally, we can express the result of the $\xi$ integration as
\begin{equation}
\label{xiintresult1}
    \begin{aligned}
        &\int \frac{\dd \xi}{2\pi\ii}
    \, 
    (2+\ii(\rho-1)\tnu +\xi+\alpha_{\chi}-t)_t
    \\&\times \Gamma\!\left[
\begin{matrix}
-\xi,a_{s_1}+\xi,-\xi+2\ii\tnu,D_{\rho\chi}-t+\xi,\bar{D}_{\rho\chi}-t+\xi\\
a_{s_1},\bar{a}_{s_1},
a_{s_2}+\bar{D}_{\rho\chi}-t+\xi
\end{matrix}
\right]
\\&=
\sum_{j=0}^{t} \binom{t}{j}\, \mathcal{A}_{tj,\rho \chi}^{s_1 s_2},
    \end{aligned}
\end{equation}
where the definitions of $\mathcal{A}_{tj,\pm\chi}^{\pm s_2}$ are given in Eqs.~\eqref{B++}, \eqref{B+-}, \eqref{B-+}, and \eqref{B--}.
\par 
Substituting Eq.~\eqref{xiintresult1} into Eq.~\eqref{JLSUVmedi}, we obtain the local signal from the UV pole contribution
\begin{equation}
\begin{aligned}
\label{JhatLS2}
    \widehat{\mathcal{J}}_{-+,\rho,\lambda,\text{LSUV}}^{q_1 q_2,s_1 s_2}
	&=\mathcal{N}_{\rho }^{q_1q_2,s_1s_2}
     \left(\frac{r_1}{r_2}\right)^{2\ii\rho \tnu}
    \sum_{\chi}
    d_\chi
    r_2^{-\delta_\chi-3}
    \sum_{\ell=0}^\infty 
    \frac{2}{\ell!}\left(\frac{r_2}{2}\right)^{2\ell}\, 
    \frac{\Gamma[2\ell+q_2+2-2\ii\rho\tnu-\delta_\chi]}{\Gamma\left[\ell+\frac{3}{2}\right]}
    \\&\times 
    \sum_{t=0}^{2\ell}\sum_{j=0}^{t}
(-1)^{t}
\binom{2\ell}{t}\binom{t}{j}\, 
2^{t-\delta_\chi-3-2\ii(\rho-1)\tnu}\times \mathcal{A}_{tj,\rho \chi}^{s_1 s_2}.
\end{aligned}
\end{equation}
In the squeezed limit $r_2\to0$, we have
\begin{equation}
    \lim_{r_2\to 0}\widehat{\mathcal{J}}_{-+,\rho,\lambda,\text{LSUV}}^{q_1 q_2,s_1 s_2}
    \simeq 
    \mathcal T^{q_1q_2,s_1s_2}_{\rho,\lambda,\text{LSUV}}\, r_2^{-3}\left(\frac{r_1}{r_2}\right)^{2\ii\rho\tnu}+\mathcal{O}(r_2^{-1}),
\end{equation}
where
\begin{equation}
\mathcal T_{\rho,\lambda,\text{LSUV}}^{q_1q_2,s_1s_2}
=
\mathcal N_{\rho}^{q_1q_2,s_1s_2}\, 
\frac{(1+s_1s_2)2^{-2\ii(\rho-1)\widetilde{\nu}}
}{4\sqrt{\pi}}\, 
\Gamma[q_2+2-2\ii\rho\widetilde{\nu}]
\,\mathcal A_{00,\rho0}^{s_1s_2}.
\end{equation}
In the squeezed limit $r_2\to0$, the amplitude of $\widehat{\mathcal{J}}_{-+,\rho,\lambda,\text{LSUV}}^{q_1q_2,s_1s_2}$ scales as $r_2^{-3}$ for the same helicity branches, $s_1=s_2$. For opposite helicities, $s_1=-s_2$, the $r_2^{-3}$ contribution cancels, and the amplitude instead scales as $r_2^{-1}$.
\begin{figure}[htbp]
    \centering
    \includegraphics[width=0.49\linewidth]{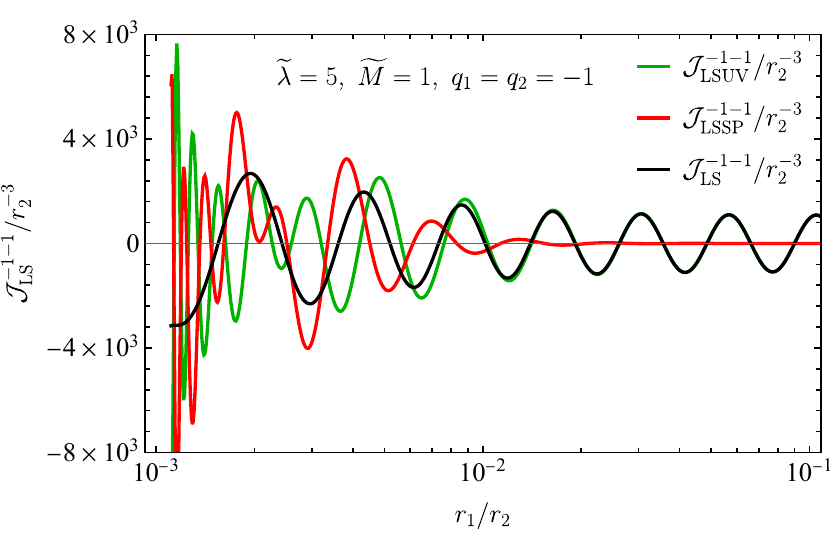}
    \includegraphics[width=0.49\linewidth]{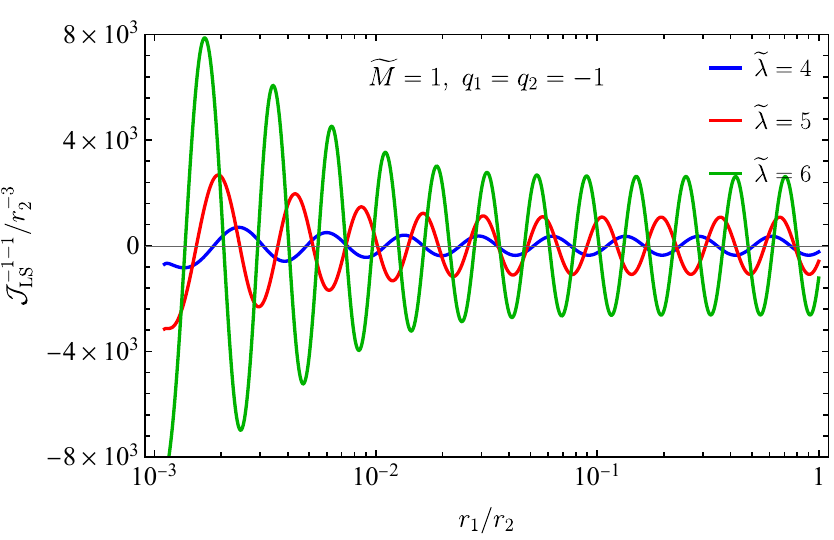}
    \caption{The oscillatory local signal with the overall $r_2^{-3}$ scaling factored out, plotted as a function of $r_1/r_2$ for $q_1=q_2=-1$ and $\widetilde M=1$. We fix $r_1=10^{-3}$ and vary only $r_2$. The left panel shows the oscillatory behavior of the UV-pole local signal $\mathcal{J}_{\text{LSUV}}^{q_1q_2}/r_2^{-3}$, the spectrum-pole local signal $\mathcal{J}_{\text{LSSP}}^{q_1q_2}/r_2^{-3}$, and the total local signal $\mathcal{J}_{\text{LS}}^{q_1q_2}/r_2^{-3}$ as functions of $r_1/r_2$ for $\tilde\lambda=5$, represented by the green, red, and black curves, respectively. The right panel shows the total local signal $\mathcal{J}_{\text{LS}}^{q_1q_2}/r_2^{-3}$ as a function of $r_1/r_2$ for different chemical potentials. The blue, red, and green curves correspond to $\tlam=4$, $5$, and $6$, respectively.}
    \label{fig:local}
\end{figure}
\par 
We write the UV-pole local contribution to $\mathcal{J}_{\text{F}}^{q_1q_2}$ as
\begin{equation}
\label{partialJLSUV}
    \mathcal{J}_{\text{LSUV}}^{q_1q_2}=-\sum_\rho
    \left[
    1+e^{-\ii\pi(4+q_1+2\ii\rho\tnu)}
    \right]
    \sum_{s_1,s_2}\mathcal C_{s_1s_2}^\rho 
    \left[
    \widehat{\mathcal{J}}_{-+,\rho,\lambda,\text{LSUV}}^{q_1q_2,s_1s_2}-\widehat{\mathcal{J}}_{-+,\rho,-\lambda,\text{LSUV}}^{q_1q_2,-s_1-s_2}
    \right]+\text{c.c.}\, .
\end{equation}
The full local signal can then be written as
\begin{equation}
\label{JLStotal}
    \mathcal{J}_{\text{LS}}^{q_1 q_2}= \mathcal{J}_{\text{LSSP}}^{q_1 q_2}+\mathcal{J}_{\text{LSUV}}^{q_1 q_2}.
\end{equation}
Using Eqs.~\eqref{partialJLSSP}, \eqref{partialJLSUV}, and \eqref{JLStotal}, together with Eqs.~\eqref{JLS1final} and \eqref{JhatLS2}, we numerically evaluate the UV-pole local signal $\mathcal{J}_{\text{LSUV}}^{q_1q_2}/r_2^{-3}$, the spectrum-pole local signal $\mathcal{J}_{\text{LSSP}}^{q_1q_2}/r_2^{-3}$, and the total local signal $\mathcal{J}_{\text{LS}}^{q_1q_2}/r_2^{-3}$ as functions of $r_1/r_2$. The results are shown in the left panel of Fig.~\ref{fig:local}. We also evaluate the total local signal for several values of the chemical potential, as shown in the right panel. Since the UV-pole local signal exhibits an $r_2^{-3}$ scaling, we apply a common normalization by dividing all local signals by $r_2^{-3}$ to remove this scaling and better display their oscillatory behavior. Throughout Fig.~\ref{fig:local}, we consider the quartic interaction model in Eq.~\eqref{modelint}, corresponding to $q_1=q_2=-1$, and set the fermion mass to $\widetilde M=1$. For the kinematic variables, we fix $r_1=10^{-3}\ll1$ and vary only $r_2$. In the left panel, we set $\tlam=5$. The green, red, and black curves represent the UV-pole local signal, the spectrum-pole local signal, and their sum, respectively. In the right panel, the blue, red, and green curves represent the total local signal for $\tlam=4$, $5$, and $6$, respectively.
\par 
As shown in the left panel of Fig.~\ref{fig:local}, the normalized total local signal is dominated by the UV-pole contribution in the squeezed regime $r_2\ll1$, where the black and green curves nearly coincide. This behavior follows from the leading-pole expressions above: the UV-pole contribution scales as $r_2^{-3}$, whereas the spectrum-pole contribution has an approximately $r_2$-independent amplitude. Dividing both contributions by $r_2^{-3}$ therefore suppresses the spectrum-pole contribution by a factor of $r_2^3$, while leaving the normalized UV-pole contribution with an approximately constant oscillation amplitude.

As $r_2$ approaches the folded limit, subleading poles become important, and the normalized amplitudes acquire a nontrivial dependence on $r_2$. The spectrum-pole contribution becomes comparable to the UV-pole contribution, and both exhibit increasingly pronounced variations near $r_2=1$. Their oscillations partially cancel, however, so the total local signal has a substantially smoother envelope than either contribution separately. The behavior near the folded limit thus depends on the interplay between the two contributions.

The right panel of Fig.~\ref{fig:local} shows that increasing the chemical potential enhances both the oscillation frequency and the amplitude of the total local signal, as in the nonlocal case. The higher frequency follows from the increase in $\tnu$, which controls the logarithmic oscillations, while the larger amplitude reflects the enhancement of positive-helicity fermion production. For each chemical potential, the normalized oscillation amplitude remains approximately constant in the squeezed regime and develops a dependence on $r_2$ towards the folded limit.

\section{The Three-Point Limit\label{sec5}}
In Section~\ref{4}, we computed the nonlocal and local contributions to the factorized seed integral $\mathcal{J}_{\mathrm{F}}^{q_1q_2}$
separately. In this section, we consider its three-point limit,
obtained by taking $\mathbf{k}_4\to0$, for which $r_2\to1$. For the Bunch--Davies initial state, the complete CC signal is
expected to remain regular in this limit, although its nonlocal
and local contributions can be singular separately. The cancellation of these singularities is not manifest in the representation used in Section~\ref{4}. We therefore reorganize the MB integral for $\widehat{\mathcal{J}}_{-+,\rho,\lambda}^{q_1q_2,s_1s_2}$ to keep these contributions combined and explicitly demonstrate the cancellation of the folded singularities. This representation allows us to evaluate the complete CC signal directly in the three-point limit.
\subsection{Reorganization of the MB Integral}
We introduce the new integration variables
\begin{equation}
    \Lambda=\xi+\zeta+u+v,
    \qquad
    X=\xi+u ,
\end{equation}
while keeping $\xi$ and $\zeta$ as independent variables. 
Thus, the original four independent variables $(\xi,\zeta,u,v)$ are replaced by the new independent variables $(\Lambda,X,\xi,\zeta)$. In terms of the new integration variables, the Eq.~\eqref{totalseedint} can be written as
\begin{equation}
\begin{aligned}
    \widehat{\mathcal{J}}_{-+,\rho,\lambda}^{q_1q_2,s_1s_2}
	&=\mathcal{N}_{\rho}^{q_1q_2,s_1s_2}
     r_1^{2\ii\rho\tnu}r_2^{-2\ii\tilde\nu}
    \int \frac{\dd \Lambda\dd X \dd \xi\dd \zeta}{(2\pi\ii)^4}
    \, 2^{\xi+\zeta}\, r_2^{\Lambda }\, \Gamma[\xi-X]\, \Gamma[X+\zeta-\Lambda]
\\&
    \times
    \Gamma[q_2+5-2\ii\tnu+\Lambda]\, 
    \widetilde{\Gamma}^{s_1s_2}(\xi,\zeta)
    \, \sum_{\chi=0}^3d_\chi\, \widehat{\Gamma}_{\rho}^\chi(\Lambda,X),
\end{aligned}
\end{equation}
where $\widehat{\Gamma}_{\rho}^\chi(\Lambda,X)$ denotes the expression obtained from Eq.~\eqref{Ghat} after replacing the variables $(\xi,\zeta,u,v)$ by $(\Lambda,X)$.
\par 
We want to close the contour to the right and evaluate the $\Lambda$ integral by the residue theorem. In the $\Lambda$ plane, there are two families of right poles, namely
\begin{equation}
\label{LambdapoleSP}
    \Lambda = \ell + X+\zeta,\quad\ell=0,1,2,\cdots, 
\end{equation}
\begin{equation}
\label{LambdapoleUV}
    \Lambda =2\ell - 2\ii(\rho-1)\tnu -\delta_\chi-3\equiv 2\ell +\Lambda_\rho^\chi ,\quad \ell=0,1,2,\cdots.
\end{equation}
After evaluating the residues associated with these two pole families, the integrand of the remaining $X$, $\xi$, and $\zeta$ integrals contains two infinite series. Using the series definition of the generalized hypergeometric function in Eq.~\eqref{pFqseries}, we can reorganize these infinite series into a combination of generalized hypergeometric functions ${}_4\text F_3$. The result can be expressed as
\begin{equation}
\begin{aligned}
\label{seedintLambda}
    \widehat{\mathcal{J}}_{-+,\rho,\lambda}^{q_1q_2,s_1s_2}
	&=\mathcal{N}_{\rho}^{q_1q_2,s_1s_2}
    \int \frac{\dd X \dd \xi\dd \zeta}{(2\pi\ii)^3}\, 
    2^{\xi+\zeta}\, \Gamma[\xi-X]\, 
    \widetilde{\Gamma}^{s_1s_2}(\xi,\zeta)\, 
    \sum_{\chi=0}^3d_\chi
    \\&\times 
    \left[
r_1^{2\ii\rho\tnu}r_2^{-2\ii\tilde\nu+X+\zeta }\widehat{\mathcal{O}}_{1,\rho}^{q_2,\chi}(X,\zeta;r_2)+r_2^{-\delta_\chi-3}
\left(\frac{r_1}{r_2}\right)^{2\ii\rho\tnu}
    \widehat{\mathcal{O}}_{2,\rho}^{q_2,\chi}(X,\zeta;r_2)
    \right].
\end{aligned}
\end{equation}
The factors $\widehat{\mathcal O}_{1,\rho}^{q_2,\chi}$ and $\widehat{\mathcal O}_{2,\rho}^{q_2,\chi}$ appearing in the above expression involve regularized generalized hypergeometric functions ${}_4\mathbf F_3$, which arise from the resummation of the infinite series and are defined in Eq.~\eqref{reghyperfun}. Their expressions are given by

\begin{equation}
\label{O1}
    \begin{aligned}
\widehat{\mathcal O}_{1,\rho}^{q_2,\chi}(X,\zeta;r_2)
&=
\pi^{3/2}
\Gamma\!\left[
\begin{matrix}
R^{q_2}(X+\zeta),\frac{A_\rho^\chi(X)+3}{2},\frac{B_\rho^\chi(\zeta)+3}{2}
\\
-\frac{A_\rho^\chi(X)}{2},1-\frac{B_\rho^\chi(\zeta)+2}{2}
\end{matrix}
\right]
\csc\!\left[
\frac{\pi}{2}
\left(
A_\rho^\chi(X)+B_\rho^\chi(\zeta)+5
\right)
\right]
\\&\times 
{}_4\mathbf F_3\!
\left[
\begin{matrix}
\frac{R^{q_2}(X+\zeta)}{2},
\frac{R^{q_2}(X+\zeta)+1}{2},
\frac{B_\rho^\chi(\zeta)+2}{2},
\frac{B_\rho^\chi(\zeta)+3}{2}
\\
\frac12,
\frac{A_\rho^\chi(X)+B_\rho^\chi(\zeta)+5}{2},
\frac{A_\rho^\chi(X)+B_\rho^\chi(\zeta)+6}{2}
\end{matrix}
;r_2^2
\right]
\\&-r_2\frac{\pi^{3/2}}{2}
\Gamma\!\left[
\begin{matrix}
R^{q_2}(X+\zeta)+1,\frac{A_\rho^\chi(X)+3}{2},\frac{B_\rho^\chi(\zeta)+4}{2}
\\
-\frac{A_\rho^\chi(X)}{2},1-\frac{B_\rho^\chi(\zeta)+3}{2}
\end{matrix}
\right]
\csc\!\left[
\frac{\pi}{2}
\left(
A_\rho^\chi(X)+B_\rho^\chi(\zeta)+6
\right)
\right]
\\&\times 
{}_4\mathbf F_3\!
\left[
\begin{matrix}
\frac{R^{q_2}(X+\zeta)+1}{2},
\frac{R^{q_2}(X+\zeta)+2}{2},
\frac{B_\rho^\chi(\zeta)+3}{2},
\frac{B_\rho^\chi(\zeta)+4}{2}
\\
\frac32,
\frac{A_\rho^\chi(X)+B_\rho^\chi(\zeta)+6}{2},
\frac{A_\rho^\chi(X)+B_\rho^\chi(\zeta)+7}{2}
\end{matrix}
;r_2^2
\right],
\end{aligned}
\end{equation}

\begin{equation}
\label{O2}
\begin{aligned}
\widehat{\mathcal O}_{2,\rho}^{q_2,\chi}(X,\zeta;r_2)
&=
2^{\,1+X+\zeta-\Lambda_\rho^\chi}\, 
\pi^{3/2}\, 
\Gamma\!\left[R^{q_2}(\Lambda_\rho^\chi)\right]\, 
\csc\!\left[
\pi\left(
1-X-\zeta+\Lambda_\rho^\chi
\right)
\right]
\\
&\times
{}_4\mathbf F_3\!
\left[
\begin{matrix}
\frac{R^{q_2}(\Lambda_\rho^\chi)}{2},
\frac{R^{q_2}(\Lambda_\rho^\chi)+1}{2},
-\frac{A_\rho^\chi(X)+1}{2},
-\frac{A_\rho^\chi(X)}{2}
\\
\frac32,
\frac{1-X-\zeta+\Lambda_\rho^\chi}{2},
\frac{2-X-\zeta+\Lambda_\rho^\chi}{2}
\end{matrix}
;r_2^2
\right],
\end{aligned}    
\end{equation}
where, for notational simplicity, we have introduced the following definitions
\begin{equation}
    R^{q_2}(x) = q_2+5-2\ii\tnu+x,
\end{equation}
\begin{equation}
    A_{\rho}^\chi (x) = \ii(\rho - 1)\tnu+\alpha_\chi +x,
\end{equation}
\begin{equation}
    B_{\rho}^\chi (x) = \ii(\rho - 1)\tnu+\beta_\chi+x.
\end{equation}
\par 
It is worth noting that Eqs.~\eqref{O1} and \eqref{O2} contain cosecant factors that introduce apparent singularities. However, these singularities are removable in the complete expression enclosed in the square brackets in Eq.~\eqref{seedintLambda}. Specifically, although $\widehat{\mathcal O}_{1,\rho}^{q_2,\chi}$ and $\widehat{\mathcal O}_{2,\rho}^{q_2,\chi}$ diverge individually at these singular points, their combination within the square brackets in Eq.~\eqref{seedintLambda} remains finite. These removable singularities arise from the coincidence of the two families of $\Lambda$ poles given in Eqs.~\eqref{LambdapoleSP} and \eqref{LambdapoleUV}.
\par 
We can now proceed to analyze the CC signals in Eq.~\eqref{seedintLambda} using the residue theorem. For the first term inside the square brackets in Eq.~\eqref{seedintLambda}, we first evaluate the $X$ integral by taking the residues at the right pole family generated by $\Gamma[\xi-X]$. We then extract the spectrum poles from $\widetilde{\Gamma}^{s_1s_2}(\xi,\zeta)$ and evaluate the remaining $\xi$ and $\zeta$ integrals. This procedure yields both the nonlocal signal and the spectrum-pole local signal. The second term inside the square brackets in Eq.~\eqref{seedintLambda}, on the other hand, already originates from the UV pole family in Eq.~\eqref{LambdapoleUV} and therefore corresponds to the UV-pole local signal. Thus, the nonlocal signal, the spectrum-pole local signal, and the UV-pole local signal, which were calculated separately in Section~\ref{4}, can all be identified in Eq.~\eqref{seedintLambda}.
\par 
We note that the regularized generalized hypergeometric functions ${}_4\mathbf F_3$ appearing in $\widehat{\mathcal O}_{1,\rho}^{q_2,\chi}$ and $\widehat{\mathcal O}_{2,\rho}^{q_2,\chi}$ have argument $r_2^2$. Therefore, the potentially singular behavior of the signal in the limit $r_2\to 1$ is entirely encoded in these ${}_4\mathbf F_3$ functions. For generalized hypergeometric functions ${}_{p+1}\text F_p$, the potentially singular behavior as their argument approaches unity is governed by the parametric excess, defined as the sum of the lower parameters minus the sum of the upper parameters~\cite{NIST:DLMF}. For generic parameter values, these functions do not have a finite limit at unit argument when the parametric excess has a nonpositive real part. In our case, the three regularized generalized hypergeometric functions ${}_4\mathbf F_3$ appearing in $\widehat{\mathcal O}_{1,\rho}^{q_2,\chi}$ and $\widehat{\mathcal O}_{2,\rho}^{q_2,\chi}$ share the same parametric excess, given by
\begin{equation}
\label{sblancepara}
    \mathfrak s=3+A_\rho^\chi(X)-R^{q_2}(X+\zeta) = 
    -2-q_2+\alpha_\chi+\ii(\rho+1)\tnu-\zeta.
\end{equation}
Since the parametric excess $\mathfrak s$ depends on the integration variable $\zeta$, evaluating the $\zeta$ integral by summing over the spectrum-pole residues yields
\begin{equation}
    \text{Re}\, \mathfrak s=-2-q_2+\alpha_\chi-n,
    \qquad n=0,1,2,\ldots.
\end{equation}
As $n$ increases, $\text{Re}\, \mathfrak s$ eventually becomes nonpositive, causing the corresponding regularized generalized hypergeometric functions ${}_4\mathbf F_3$ to develop singularities in the limit $r_2\to 1$.
\par 
Factoring out $r_1^{2\ii\rho\tilde\nu}$ from the square brackets
in Eq.~\eqref{seedintLambda}, we define
\begin{equation}
\mathcal B^{q_2,\chi}_\rho(X,\zeta;r_2)
\equiv
r_2^{-2\ii\widetilde\nu+X+\zeta}
\widehat{\mathcal{O}}_{1,\rho}^{q_2,\chi}(X,\zeta;r_2)
+
r_2^{-\delta_\chi-3-2\ii\rho\widetilde\nu}
\widehat{\mathcal{O}}_{2,\rho}^{q_2,\chi}(X,\zeta;r_2),
\label{eq:complete-bracket}
\end{equation}
Although the ${}_4\mathbf F_3$ functions appearing in $\widehat{\mathcal O}_{1,\rho}^{q_2,\chi}$ and $\widehat{\mathcal O}_{2,\rho}^{q_2,\chi}$ exhibit singular behavior as $r_2\to 1$, these singularities cancel against each other, leaving $\mathcal B^{q_2,\chi}_\rho(X,\zeta;r_2)$ regular in this limit. A detailed proof of this cancellation is provided in Appendix~\ref{appendixF}. Finally, we can express $\mathcal B^{q_2,\chi}_\rho(X,\zeta;r_2)$ as
\begin{equation}
\label{Bregfinalexp}
    \begin{aligned}
       \mathcal B^{q_2,\chi}_\rho(X,\zeta;r_2)
       &=
\frac{2}{\sqrt{\pi}}\frac{
\Gamma\!\left[
q_2+5-2\ii\tnu+X+\zeta
,
\ii(\rho-1)\tnu+\zeta+\beta_\chi+2
,
q_2+1-2\ii\rho\tnu-\delta_\chi
\right]
}
{
\ii(\rho-1)\tnu+X+\alpha_\chi+2
}
\\
&\times
r_2^{-2\ii\tnu+X+\zeta}\, 
{}_2\mathbf F_1\!
\left[
\begin{matrix}
q_2+5-2\ii\tnu+X+\zeta,\,
\ii(\rho-1)\tnu+\zeta+\beta_\chi+2
\\
q_2+3+\zeta-\alpha_\chi-\ii(\rho+1)\tnu
\end{matrix}
;
1-r_2
\right]
\\&+\frac{4}{\sqrt{\pi}}
    \frac{\Gamma\left[q_2+5-2\ii\tnu+X+\zeta,\ii(\rho-1)\tnu+\zeta+\beta_\chi+2,\ii(\rho-1)\tnu+X+\alpha_\chi+2\right]}{\sin\left[\pi(2\ii(\rho-1)\tnu+X+\zeta+\delta_\chi+3)\right]}
    \\&\times 
    \sin\left[\frac{\pi}{2}(\ii(\rho-1)\tnu +X+\alpha_\chi)\right]
    \cos\left[\pi\left(\frac{3}{2}\ii(\rho-1)\tnu +\frac{X}{2}+\zeta+\frac{\alpha_\chi}{2}+\beta_\chi\right)\right]
    \\&\times r_2^{-2\ii\tnu +X+\zeta}
    \tF{q_2+5-2\ii\tnu+X+\zeta}{\ii(\rho-1)\tnu+\zeta+\beta_\chi+2}{2\ii(\rho-1)\tnu+X+\zeta+\delta_\chi+5}{-r_2}
    \\&+
    \frac{
2\sqrt{\pi}\,
\Gamma\!\left[
q_2+1-2\ii\rho\tnu-\delta_\chi
\right]
}{
\ii(\rho-1)\tnu+X+\alpha_\chi+2
}\, 
\csc\left[
\pi(2\ii(\rho-1)\tnu+X+\zeta+\delta_\chi+3)
\right]
\\&\times 
r_2^{-\delta_\chi-4-2\ii\rho\widetilde\nu}\, 
\tF{q_2+1-2\ii\rho\tnu -\delta_\chi}{-\ii(\rho-1)\tnu-X-\alpha_\chi-2}{-2\ii(\rho-1)\tnu-X-\zeta-\delta_\chi-3}{- r_2}.
    \end{aligned}
\end{equation}
As can be seen, the regularized hypergeometric functions ${}_2\mathbf F_1$ appearing in $\mathcal B^{q_2,\chi}_\rho(X,\zeta;r_2)$ remain regular as $r_2\to 1$, with no singularities. We are thus left with evaluating the integral
\begin{equation}
\begin{aligned}
\label{seedintLambdanew}
    \widehat{\mathcal{J}}_{-+,\rho,\lambda}^{q_1q_2,s_1s_2}
	=\mathcal{N}_{\rho}^{q_1q_2,s_1s_2}\, r_1^{2\ii\rho\tnu}
    \int \frac{\dd X \dd \xi\dd \zeta}{(2\pi\ii)^3}\, 
    2^{\xi+\zeta}\, \Gamma(\xi-X)\, 
    \widetilde{\Gamma}^{s_1s_2}(\xi,\zeta)\, 
    \sum_{\chi=0}^3d_\chi\, \mathcal B_\rho^{q_2,\chi}(X,\zeta;r_2).
\end{aligned}
\end{equation}

\subsection{Evaluation of the Remaining Mellin–Barnes Integral\label{5.2}}
So far, we have only demonstrated that the singular part of $\mathcal{B}_{\rho}^{q_2,\chi}(X,\zeta;r_2)$ cancels in the limit $r_2\rightarrow 1$. It remains to prove that the integral in Eq.~\eqref{seedintLambdanew} is convergent throughout the region $0<r_2\leq 1$, and to evaluate the integral explicitly. We provide the convergence proof in Appendix~\ref{appendixG}. In this section, we focus on the method for evaluating this integral.
\par 
Notice that $\mathcal{B}_{\rho}^{q_2,\chi}(X,\zeta;r_2)$ does not depend on the MB variable $\xi$. Therefore, we can first evaluate the $\xi$ integral in Eq.~\eqref{seedintLambdanew} analytically
\begin{equation}
\begin{aligned}
    \int \frac{\dd \xi}{2\pi\ii}\, 
    2^{\xi}\, 
    \Gamma\!\left[
		\begin{array}{c}
			\xi-X,-\xi,a_{s_1}+\xi,-\xi+2\ii\tnu\\
			a_{s_1},\bar{a}_{s_1}
    		\end{array}
		\right]
        =
        2^X\,  \Gamma\!\left[
			-X,2\ii\tnu - X
		\right]\, 
        {}_2\mathbf F_1\!
\left[
\begin{matrix}
-X,2\ii\tnu - X\\
\bar{a}_{s_1} - X
\end{matrix}
;
\frac{1}{2}
\right].
\end{aligned}
\end{equation}
After performing the $\xi$ integral, Eq.~\eqref{seedintLambdanew} becomes
\begin{equation}
\label{totalint}
\begin{aligned}
    \widehat{\mathcal{J}}_{-+,\rho,\lambda}^{q_1q_2,s_1s_2}
	&=
    \mathcal{N}_{\rho}^{q_1q_2,s_1s_2}
    r_1^{2\ii\rho\tnu}
    \int_{-\ii\infty}^{+\ii\infty} \frac{\dd X \dd \zeta}{(2\pi\ii)^2}\, 
    \Gamma\!\left[
		\begin{matrix}
			-X,2\ii\tnu - X,-\zeta,a_{s_2}+\zeta,-\zeta+2\ii\tnu \\
			 a_{s_2},\bar{a}_{s_2}
    		\end{matrix}
		\right]
    \\&\times 2^{X+\zeta}\, 
    {}_2\mathbf F_1\!
\left[
\begin{matrix}
-X,2\ii\tnu - X\\
\bar{a}_{s_1} - X
\end{matrix}
;
\frac{1}{2}
\right]\, 
    \sum_{\chi=0}^3d_\chi\, 
    \mathcal B^{q_2,\chi}_\rho(X,\zeta;r_2).
\end{aligned}
\end{equation}
The complicated dependence of $\mathcal B^{q_2,\chi}_\rho(X,\zeta;r_2)$ on $X$ and $\zeta$ makes further analytic evaluation of the remaining integrals difficult. We therefore focus on the numerical evaluation of this double integral in the following discussion.
\par 
We now specify the integration contours in Eq.~\eqref{totalint} and write the remaining integrals in a form suitable for numerical evaluation. The integral in Eq.~\eqref{totalint} can be regarded as an MB-type integral\footnote{All factors in the integrand of Eq.~\eqref{totalint}, except for the regularized hypergeometric functions ${}_2\mathbf F_1$, can be rewritten in terms of Gamma functions using standard identities. Since these ${}_2\mathbf F_1$ functions are entire in $X$ and $\zeta$, they introduce no poles in either variable. The pole structure of the integrand is therefore determined entirely by the Gamma function factors, and Eq.~\eqref{totalint} can therefore be regarded as an MB-type integral.}, whose integration contours should separate the left and right pole families arising from the Gamma functions. At first sight, identifying these families would require rewriting the non-Gamma-function factors in $\mathcal B^{q_2,\chi}_\rho(X,\zeta;r_2)$ that may introduce poles, namely
\begin{equation}
\label{explicitfactor}
    \frac{1}{\ii(\rho-1)\widetilde\nu+X+\alpha_\chi+2},
    \qquad
    \csc\!\left[
        \pi\bigl(
            2\ii(\rho-1)\widetilde\nu
            +X+\zeta+\delta_\chi+3
        \bigr)
    \right],
\end{equation}
in terms of Gamma functions. However, this step is unnecessary, since all apparent poles introduced by these two factors are removable in the complete kernel $\mathcal B^{q_2,\chi}_\rho(X,\zeta;r_2)$. At
\begin{equation}
    \ii(\rho-1)\widetilde\nu+X+\alpha_\chi+2=0,
\end{equation}
the residues arising from the explicit denominators in the first and third terms cancel. The second term also contains a Gamma function pole at this location, but this pole is canceled by a zero of the accompanying sine factor. Similarly, at
\begin{equation}
    2\ii(\rho-1)\widetilde\nu
    +X+\zeta+\delta_\chi+3=m,
    \qquad m\in\mathbb Z,
\end{equation}
the residues arising from the cosecant factors cancel between the second and third terms. Therefore, neither factor in Eq.~\eqref{explicitfactor} introduces an independent pole family in the complete kernel $\mathcal B_\rho^{q_2,\chi}$, and no additional pole-separation condition is required. For numerical convenience, however, we choose contours that avoid these removable singularities.
\par 
We thus need only consider the poles arising from the explicit Gamma function factors in the integrand of Eq.~\eqref{totalint}. In the following discussion, we adopt a sufficient contour prescription that places all left poles to the left of the contours and all right poles to the right. This prescription does not require checking whether any of these Gamma function poles are canceled in the complete kernel $\mathcal B_\rho^{q_2,\chi}$. First, the Gamma function factors
\begin{equation}
\label{Xpole}
    \Gamma[\ii(\rho-1)\widetilde\nu+X+\alpha_\chi+2],
    \qquad
    \Gamma[-X,2\ii\tnu-X]
\end{equation}
generate the left and right pole families in $X$, respectively. Similarly, the factors
\begin{equation}
\label{zetapole}
    \Gamma\!\left[
        \ii(\rho-1)\tnu+\zeta+\beta_\chi+2,
        \frac{1-s_2}{2}-\ii(s_2\tlam+\tnu)+\zeta
    \right],
    \qquad
    \Gamma[-\zeta,2\ii\tnu-\zeta]
\end{equation}
generate the left and right pole families in $\zeta$, respectively. In addition, the factor
\begin{equation}
    \Gamma\!\left[q_2+5-2\ii\tnu+X+\zeta\right]
\end{equation}
generates a coupled family of left poles involving both $X$ and $\zeta$.
\begin{figure}[htbp]
\centering
\ 
\includegraphics[width=.8\textwidth]{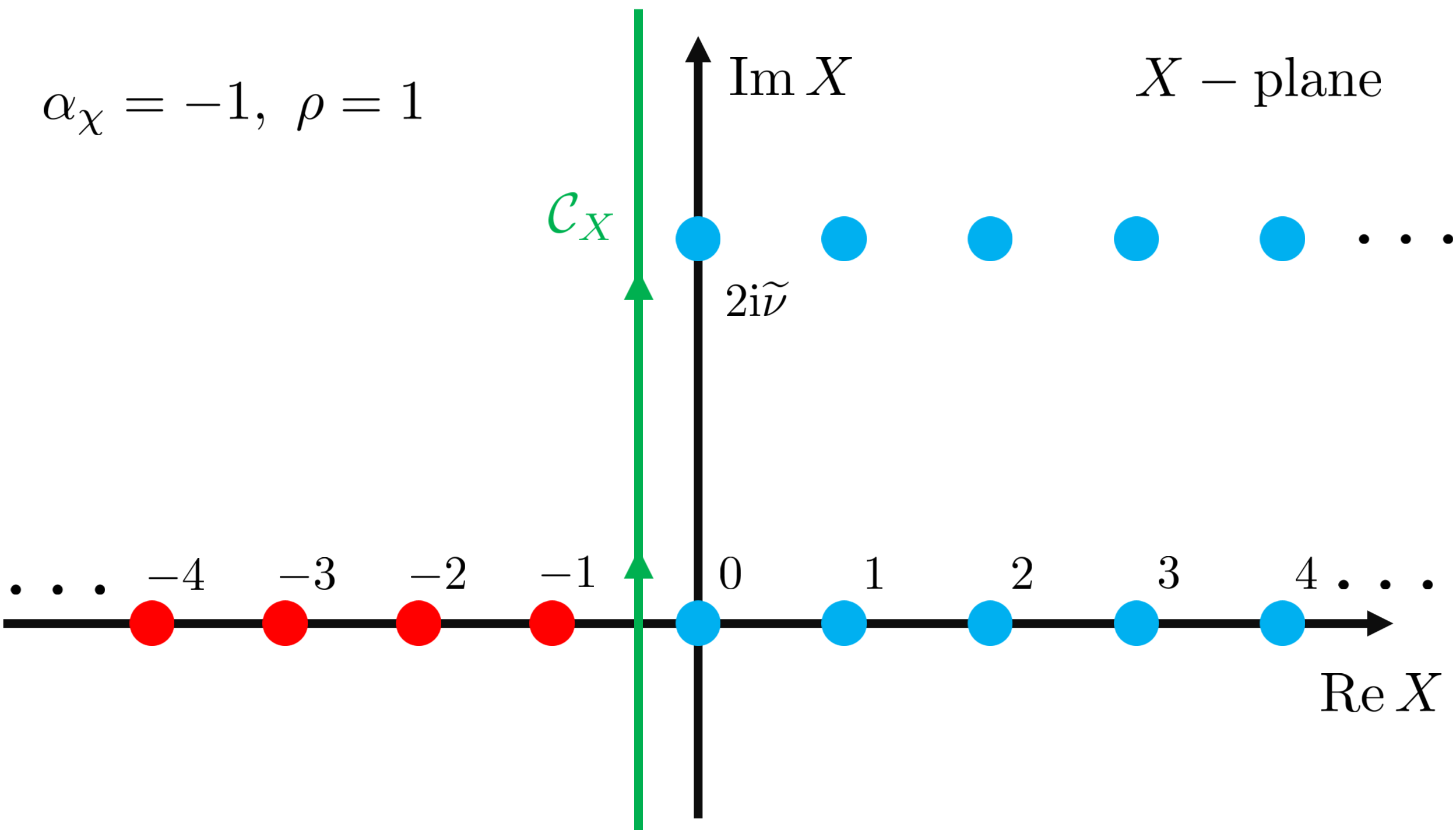}
\\[1cm]
\includegraphics[width=.8\textwidth]{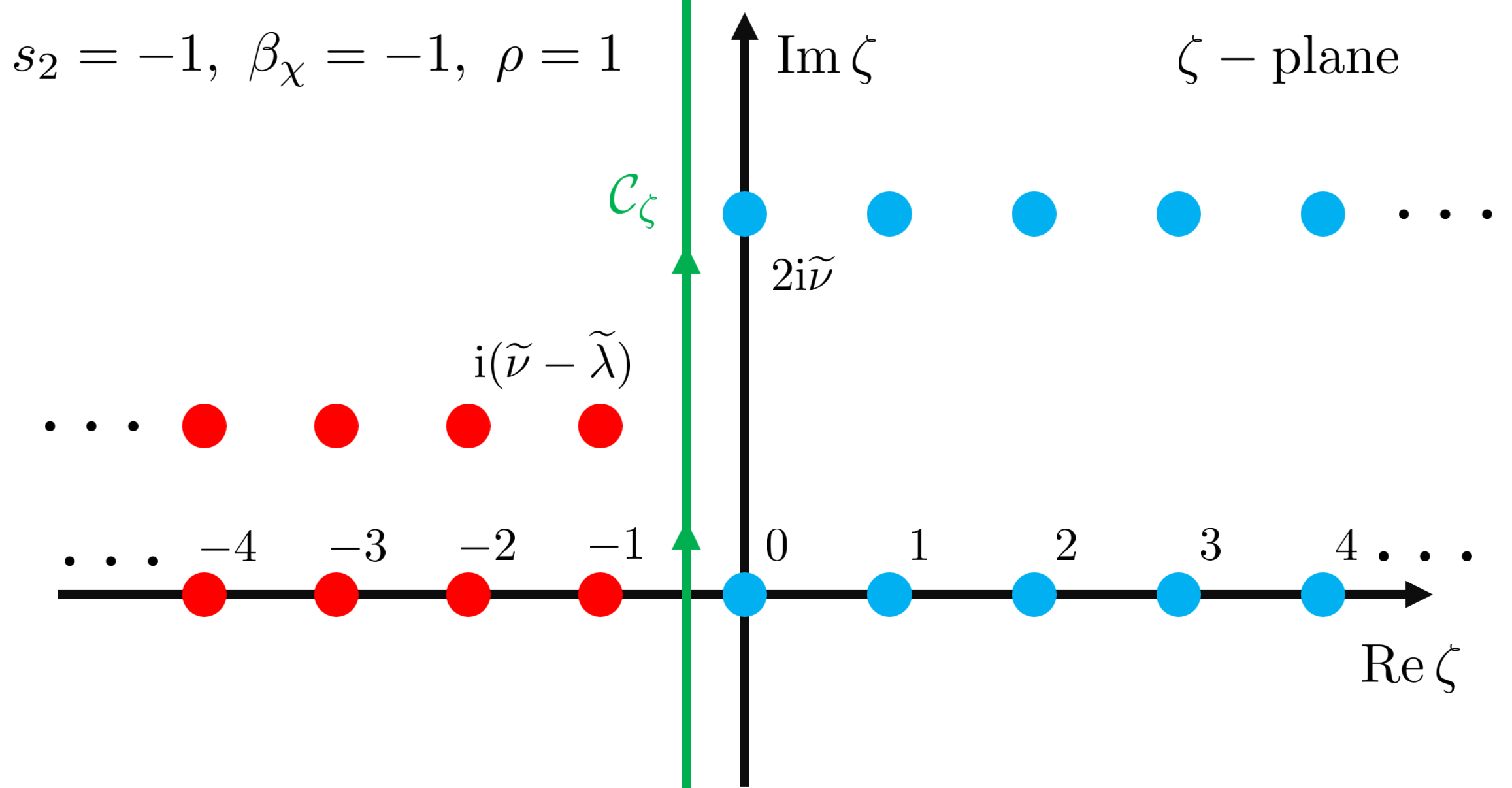}
\qquad
\caption{
Pole structure and contour choices for the integration variables $X$ and $\zeta$.
In the upper panel, blue dots denote the right pole families generated by $\Gamma[-X,2\ii\tnu-X]$, while red dots denote the left pole family generated by $\Gamma[\ii(\rho-1)\tnu+X+\alpha_\chi+2]$. We take $\rho=1$ and $\alpha_\chi=-1$ as an example. The ellipses indicate that the pole families extend indefinitely to the left or right. The integration contour $\mathcal C_X$, shown by the green line with arrows, separates the left and right pole families.
The lower panel shows the pole structure in the $\zeta$ plane for helicity $s_2=-$. Blue dots denote the right pole families generated by $\Gamma[-\zeta,2\ii\tnu-\zeta]$, while red dots denote the left pole families generated by $\Gamma[\ii(\rho-1)\tnu+\zeta+\beta_\chi+2,1-\ii(\tnu-\tlam)+\zeta]$. We take $\rho=1$ and $\beta_\chi=-1$ as an example. The integration contour $\mathcal C_\zeta$, again shown by the green line with arrows, separates the left and right pole families.
The coupled poles involving $X$ and $\zeta$, generated by $\Gamma[q_2+5-2\ii\tnu+X+\zeta]$, are not shown.
}
\label{figcontour1}
\end{figure}
\begin{figure}[htbp]
\centering
\includegraphics[width=1\textwidth]{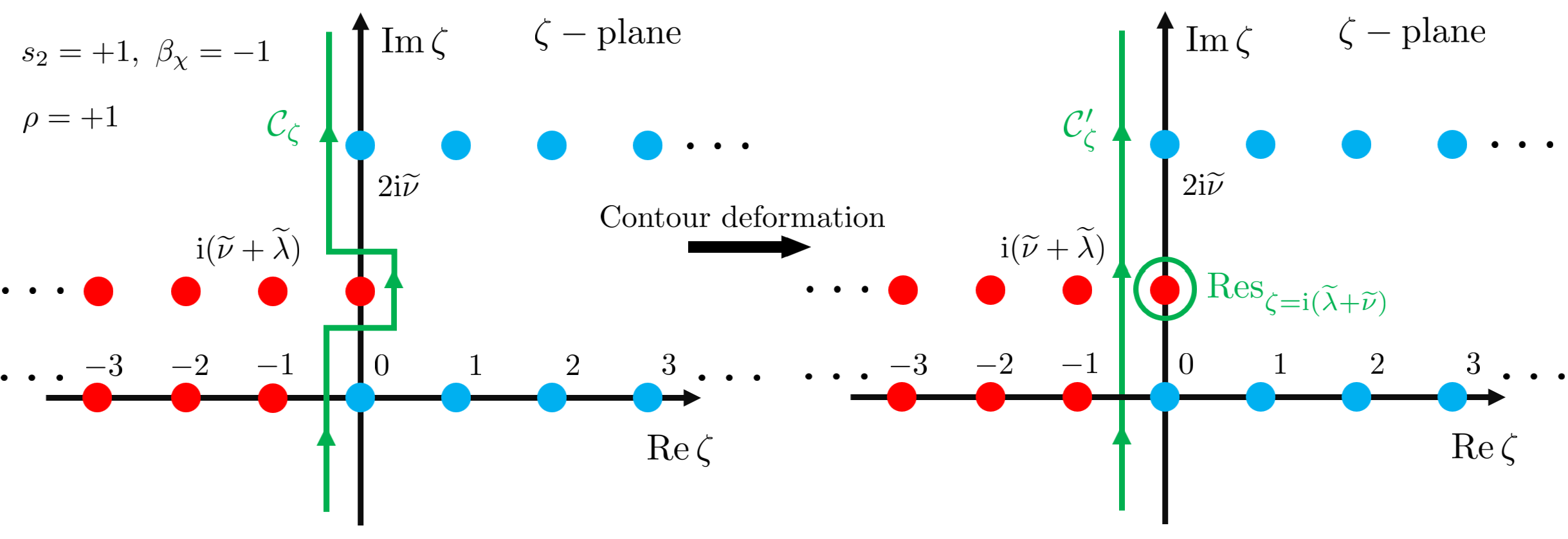}
\caption{
Pole structure and contour choices in the $\zeta$ plane for helicity $s_2=+$, together with the pole crossed during contour deformation.
Blue dots denote the right pole families generated by $\Gamma[-\zeta,2\ii\tnu-\zeta]$, while red dots denote the left pole families generated by $\Gamma[\ii(\rho-1)\tnu+\zeta+\beta_\chi+2,-\ii(\tnu+\tlam)+\zeta]$. We take $\rho=1$ and $\beta_\chi=-1$ as an example. The green lines with arrows indicate the integration contours.
To separate the left and right pole families, the contour $\mathcal C_\zeta$ must bend locally to the right, keeping the pole at $\zeta=\ii(\tnu+\tlam)$ to its left. For numerical convenience, we can deform $\mathcal C_\zeta$ into the vertical contour $\mathcal C_\zeta'$. This deformation crosses the pole at $\zeta=\ii(\tnu+\tlam)$, marked by the green circle. When evaluating the integral along $\mathcal C_\zeta'$, the residue contribution from this pole must therefore be added.
}
\label{figcontour2}
\end{figure}
\par 
The pole structures generated by the Gamma functions in Eqs.~\eqref{Xpole} and \eqref{zetapole} are shown in the $X$ and $\zeta$ planes in Figs.~\ref{figcontour1} and \ref{figcontour2}. The coupled poles involving $X$ and $\zeta$, generated by $\Gamma[q_2+5-2\ii\tnu+X+\zeta]$, are not shown. The upper panel of Fig.~\ref{figcontour1} shows the pole structure in the $X$ plane for $\rho=1$ and $\alpha_\chi=-1$. Blue dots denote the right pole families generated by $\Gamma[-X,2\ii\tnu-X]$, while red dots denote the left pole family generated by $\Gamma[X+1]$. The left and right pole families are naturally separated along the real axis. We can therefore choose a vertical contour $\mathcal C_X$ between them, as indicated by the green line with arrows. Since the rightmost left pole has real part $-1$ and the leftmost right poles have real part $0$, the contour $\mathcal C_X$ should satisfy
\begin{equation}
\label{Xcontcondition1}
    -1<\text{Re}\, X<0,
    \qquad X\in\mathcal C_X.
\end{equation}
For $\rho=-1$, only the imaginary parts of the left poles change, shifting the left pole family upward in the $X$ plane without affecting the contour choice. Since $\alpha_\chi\in\{0,-1,1\}$, the cases $\alpha_\chi=0$ and $1$ shift the left pole family to the left relative to that shown in the upper panel of Fig.~\ref{figcontour1}. The condition in Eq.~\eqref{Xcontcondition1} therefore remains valid in these cases.
\par 
The same analysis applies to the choice of the $\zeta$ integration contour for helicity $s_2=-1$. The lower panel of Fig.~\ref{figcontour1} shows the pole structure and contour choice for $s_2=-1$, $\beta_\chi=-1$, and $\rho=1$. Blue dots denote the right pole families generated by $\Gamma[-\zeta,2\ii\tnu-\zeta]$, while red dots denote the left pole families generated by $\Gamma[\zeta+1,1-\ii(\tnu-\tlam)+\zeta]$. Since the left and right pole families are fully separated along the real axis, we can choose a vertical integration contour $\mathcal C_\zeta$ between them. The only requirement is
\begin{equation}
\label{zetacontcondition1}
    -1<\text{Re}\, \zeta<0,
    \qquad \zeta\in\mathcal C_\zeta.
\end{equation}
This condition also applies to the branches with $\rho=-1$ and to those with $\beta_\chi=0$ or $1$.
\par 
The only case requiring special care is the choice of the $\zeta$ integration contour for helicity $s_2=+1$. Figure~\ref{figcontour2} shows the pole structure and contour choice for $s_2=+1$, $\beta_\chi=-1$, and $\rho=+1$. Blue dots denote the right pole families generated by $\Gamma[-\zeta,2\ii\tnu-\zeta]$, while red dots denote the left pole families generated by $\Gamma[\zeta+1,-\ii(\tnu+\tlam)+\zeta]$. The rightmost left pole and the leftmost right poles all have real part zero. The left and right pole families are therefore not naturally separated along the real axis, and a straight vertical contour cannot separate them. The appropriate contour $\mathcal C_\zeta$ must pass to the right of the pole at $\zeta=\ii(\tnu+\tlam)$, keeping this left pole to its left, as indicated by the green curve with arrows in Fig.~\ref{figcontour2}. For numerical evaluation, it is more convenient to use a vertical $\zeta$ contour, as in the case $s_2=-1$. We therefore deform the curved contour $\mathcal C_\zeta$ into a vertical contour $\mathcal C'_\zeta$ satisfying
\begin{equation}
\label{zetacontcondition2}
    -1<\operatorname{Re}\zeta<0,
    \qquad \zeta\in\mathcal C'_\zeta.
\end{equation}
This deformation crosses the pole at $\zeta=\ii(\tnu+\tlam)$, marked by the green circle in the figure. The corresponding residue contribution must therefore be included when evaluating the integral along $\mathcal C'_\zeta$. The same prescription applies to the other branches of $\rho$ and $\beta_\chi$. Since $\mathcal{B}_{\rho}^{q_2,\chi}$ is regular at $\zeta=\ii(\tnu+\tlam)=-a_+$ along the chosen $X$ contour, the residue of the $\zeta$-dependent part of the integrand in Eq.~\eqref{totalint} at $\zeta=-a_+$ can be evaluated directly
\begin{equation}
\begin{aligned}
    &\operatorname*{Res}_{\zeta=-a_+}
    \left\{
        2^\zeta
        \Gamma\!\left[
            \begin{matrix}
                -\zeta,\;a_++\zeta,\;
                2i\widetilde\nu-\zeta\\
                a_+,\;\bar a_+
            \end{matrix}
        \right]
        \mathcal B^{q_2,\chi}_\rho(X,\zeta;r_2)
    \right\}
    =2^{-a_+}\mathcal B^{q_2,\chi}_\rho(X,-a_+;r_2).
\end{aligned}
\label{eq:zeta_explicit_residue}
\end{equation}
\par 
Finally, we consider the coupled poles in $X$ and $\zeta$ generated by $\Gamma[q_2+5-2\ii\tnu+X+\zeta]$. Since these form a left pole family, the integration contours must satisfy
\begin{equation}
\label{couplepolecondition}
    \text{Re}\, (X+\zeta)>-(q_2+5).
\end{equation}
After deforming the $\zeta$ contour and including the associated
residue, we parameterize the vertical contours used for numerical
integration as
\begin{equation}
    X=c_X+\ii x,\qquad
    \zeta=c_\zeta+\ii y,\qquad
    x,y\in\mathbb R.
\end{equation}
Combining the contour conditions in Eqs.~\eqref{Xcontcondition1}, \eqref{zetacontcondition1}, \eqref{zetacontcondition2}, and \eqref{couplepolecondition}, we obtain the following constraints on $c_X$ and $c_\zeta$
\begin{equation}
    -1<c_X<0,\qquad
    -1<c_\zeta<0,\qquad
    c_X+c_\zeta>-(q_2+5).
\label{eq:Xzeta_sufficient_conditions}
\end{equation}
For $q_2\leq -5$, the above conditions are mutually incompatible, and a separate analysis is required. Here, we restrict our attention to $q_2>-5$.
\par 
We can now express the result entirely in terms of real
integration variables. Since
\begin{equation}
    \frac{\dd X}{2\pi \ii}=\frac{\dd x}{2\pi},
    \qquad
    \frac{\dd\zeta}{2\pi \ii}=\frac{\dd y}{2\pi},
\end{equation}
Eq.~\eqref{totalint}, with the physical contour prescription, becomes
\begin{equation}
\label{seedintNint}
\begin{aligned}
    \widehat{\mathcal{J}}_{-+,\rho,\lambda,\text{NInt}}^{q_1q_2,s_1s_2}
    &=
    \mathcal N_\rho^{q_1q_2,s_1s_2}
    r_1^{2\ii\rho\widetilde\nu}
    \Bigg\{
    \frac{1}{(2\pi)^2}
    \int_{-\infty}^{\infty}\dd x
    \int_{-\infty}^{\infty}\dd y\,
    2^{c_X+c_\zeta+\ii (x+y)}
    \\
    &\times
    \Gamma\!\left[
        \begin{matrix}
        -c_X-\ii x,
            2\ii\widetilde\nu-c_X-\ii x,
            -c_\zeta-\ii y,
            a_{s_2}+c_\zeta+\ii y,
            2\ii\widetilde\nu-c_\zeta-\ii y\\
            a_{s_2},\;\bar a_{s_2}
        \end{matrix}
    \right]
    \\
    &\times
    {}_2\mathbf F_1\!\left[
        \begin{matrix}
            -c_X-\ii x,\;
            2\ii \widetilde\nu-c_X-\ii x\\
            \bar a_{s_1}-c_X-\ii x
        \end{matrix};
        \frac{1}{2}
    \right]
    \sum_{\chi}d_\chi
    \mathcal B_\rho^{q_2,\chi}
        (c_X+\ii x,c_\zeta+\ii y;r_2)
    \\
    &+
    \frac{1+s_2}{2}\,
    \frac{2^{-a_+}}{2\pi}
    \int_{-\infty}^{\infty}\dd x\,
    2^{c_X+\ii x}\, 
    \Gamma\!\left[
        \begin{matrix}
            -c_X-\ii x,\;
            2\ii \widetilde\nu-c_X-\ii x
        \end{matrix}
    \right]
    \\
    &\times
    {}_2\mathbf F_1\!\left[
        \begin{matrix}
            -c_X-\ii x,\;
            2\ii\widetilde\nu-c_X-\ii x\\
            \bar a_{s_1}-c_X-\ii x
        \end{matrix};
        \frac{1}{2}
    \right]
    \sum_{\chi}d_\chi
    \mathcal B_\rho^{q_2,\chi}
        (c_X+\ii x,-a_+;r_2)
    \Bigg\}.
\end{aligned}
\end{equation}
To emphasize that this quantity is evaluated by numerical integration, we attach an additional subscript $\mathrm{NInt}$ to $\widehat{\mathcal{J}}_{-+,\rho,\lambda}^{q_1q_2,s_1s_2}$. For the explicit numerical evaluation, we choose $c_X=c_\zeta=-1/4$. For $s_2=-1$, the factor $(1+s_2)/2$ vanishes, and only the two-dimensional real integral needs to be evaluated. For $s_2=+1$, the deformation of the $\zeta$ integration contour requires an additional residue contribution at $\zeta=-a_+$. This yields the nonvanishing second term, which is a one-dimensional real integral. Using Eq.~\eqref{JF}, we then obtain the full factorized seed integral evaluated by numerical integration
\begin{equation}
\label{JFNINT}
    \mathcal{J}_{\text{F,NInt}}^{q_1q_2}=-\sum_\rho
    \left[
    1+e^{-\ii\pi(4+q_1+2\ii\rho\tnu)}
    \right]
    \sum_{s_1,s_2}\mathcal C_{s_1s_2}^\rho
    \left[
    \widehat{\mathcal{J}}_{-+,\rho,\lambda,\text{NInt}}^{q_1q_2,s_1s_2}-\widehat{\mathcal{J}}_{-+,\rho,-\lambda,\text{NInt}}^{q_1q_2,-s_1-s_2}
    \right]+\text{c.c.}\, .
\end{equation}

\subsection{Numerical Evaluation of the CC Signals in the Three-Point Limit}
Section~\ref{5.2} provides a complete method for numerically evaluating the seed integral in the three-point limit, $\mathcal{J}_{\mathrm{F}}^{q_1q_2}(r_1,r_2=1)$. In fact, Eq.~\eqref{seedintNint} is valid throughout the kinematic range $r_2\in(0,1]$, allowing an independent check of the results obtained in Section~\ref{4}. Before turning to the three-point case, we use this representation to compare the two methods.
\par 
We first summarize the results obtained in Section~\ref{4}, where we separately extracted and evaluated the nonlocal, spectrum-pole local, and UV-pole local contributions to $\mathcal{J}_{\mathrm{F}}^{q_1q_2}$. The results are
\begin{equation}
\label{partialJNSsec5}
    \mathcal{J}_{\text{NS}}^{q_1q_2}=-\sum_\rho
    \left[
    1+e^{-\ii\pi(4+q_1+2\ii\rho\tnu)}
    \right]
    \sum_{s_1,s_2}\mathcal C_{s_1s_2}^\rho
    \left[
    \widehat{\mathcal{J}}_{-+,\rho,\lambda,\text{NS}}^{q_1q_2,s_1s_2}-\widehat{\mathcal{J}}_{-+,\rho,-\lambda,\text{NS}}^{q_1q_2,-s_1-s_2}
    \right]+\text{c.c.}\, ,
\end{equation}
\begin{equation}
\label{partialJNLSSPSsec5}
    \mathcal{J}_{\text{LSSP}}^{q_1q_2}=-\sum_\rho
    \left[
    1+e^{-\ii\pi(4+q_1+2\ii\rho\tnu)}
    \right]
    \sum_{s_1,s_2}\mathcal C_{s_1s_2}^\rho
    \left[
    \widehat{\mathcal{J}}_{-+,\rho,\lambda,\text{LSSP}}^{q_1q_2,s_1s_2}-\widehat{\mathcal{J}}_{-+,\rho,-\lambda,\text{LSSP}}^{q_1q_2,-s_1-s_2}
    \right]+\text{c.c.}\, ,
\end{equation}
\begin{equation}
\label{partialJNLSUVSsec5}
    \mathcal{J}_{\text{LSUV}}^{q_1q_2}=-\sum_\rho
    \left[
    1+e^{-\ii\pi(4+q_1+2\ii\rho\tnu)}
    \right]
    \sum_{s_1,s_2}\mathcal C_{s_1s_2}^\rho
    \left[
    \widehat{\mathcal{J}}_{-+,\rho,\lambda,\text{LSUV}}^{q_1q_2,s_1s_2}-\widehat{\mathcal{J}}_{-+,\rho,-\lambda,\text{LSUV}}^{q_1q_2,-s_1-s_2}
    \right]+\text{c.c.}\, .
\end{equation}
Here, the key building blocks $\widehat{\mathcal{J}}_{-+,\rho,\lambda,\text{NS}}^{q_1q_2,s_1s_2}$, $\widehat{\mathcal{J}}_{-+,\rho,\lambda,\text{LSSP}}^{q_1q_2,s_1s_2}$, and $\widehat{\mathcal{J}}_{-+,\rho,\lambda,\text{LSUV}}^{q_1q_2,s_1s_2}$ are given by Eqs.~\eqref{JNSfinal}, \eqref{JLS1final}, and \eqref{JhatLS2}, respectively. The factorized seed integral $\mathcal{J}_{\mathrm{F}}^{q_1q_2}$ is entirely accounted for by these three contributions
\begin{equation}
    \mathcal{J}_{\mathrm{F}}^{q_1q_2}
    =\mathcal{J}_{\mathrm{NS}}^{q_1q_2}
    +\mathcal{J}_{\mathrm{LSSP}}^{q_1q_2}
    +\mathcal{J}_{\mathrm{LSUV}}^{q_1q_2}.
\end{equation}
The method of Section~\ref{5.2} also allows us to evaluate the factorized seed integral $\mathcal{J}_{\mathrm{F}}^{q_1q_2}$ by numerical integration. We denote the result by $\mathcal{J}_{\mathrm{F,NInt}}^{q_1q_2}$. Consistency between the results of Sections~\ref{4} and \ref{5.2} requires
\begin{equation}
\label{JFcrosscheck}
    \mathcal{J}_{\mathrm{F,NInt}}^{q_1q_2}
    =\mathcal{J}_{\mathrm{NS}}^{q_1q_2}
    +\mathcal{J}_{\mathrm{LSSP}}^{q_1q_2}
    +\mathcal{J}_{\mathrm{LSUV}}^{q_1q_2}.
\end{equation}
\begin{figure}[htbp]
    \centering
    \includegraphics[width=0.48\linewidth]{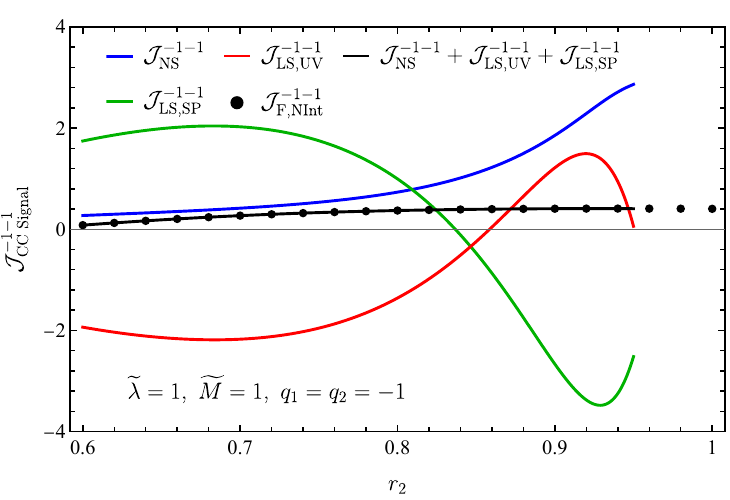}
    \includegraphics[width=0.51\linewidth]{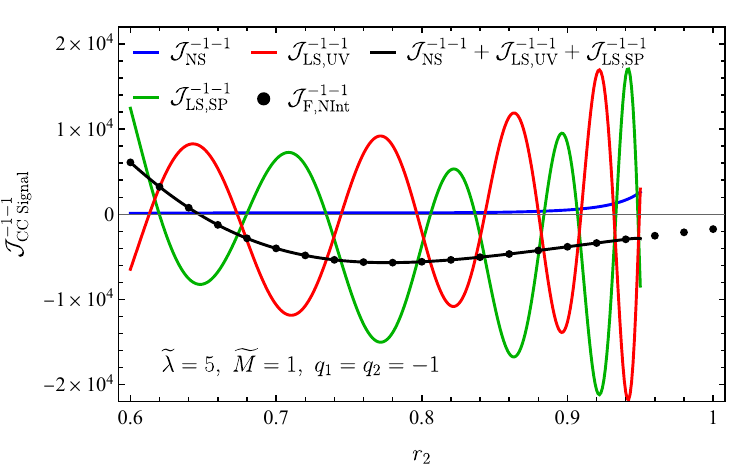}
    \caption{Comparison of the two methods for evaluating the factorized seed integral $\mathcal{J}_{\mathrm{F}}^{q_1q_2}$ as a function of $r_2$ for $q_1=q_2=-1$ and $\widetilde M=1$. We fix $r_1=10^{-3}\ll 1$ and vary only $r_2$. The left and right panels correspond to $\tlam=1$ and $5$, respectively. The blue, green, and red curves represent the nonlocal signal $\mathcal{J}_{\mathrm{NS}}^{-1,-1}$, the spectrum-pole local signal $\mathcal{J}_{\mathrm{LSSP}}^{-1,-1}$, and the UV-pole local signal $\mathcal{J}_{\mathrm{LSUV}}^{-1,-1}$, respectively. The black curve shows their sum, while the black dots show the independent result $\mathcal{J}_{\mathrm{F,NInt}}^{-1,-1}$ obtained by numerical integration.}
    \label{figcrosscheck}
\end{figure}
\par 
We numerically evaluate $\mathcal{J}_{\text{NS}}^{q_1q_2}$, $\mathcal{J}_{\text{LSSP}}^{q_1q_2}$, $\mathcal{J}_{\text{LSUV}}^{q_1q_2}$, and $\mathcal{J}_{\text{F,NInt}}^{q_1q_2}$ as functions of $r_2$ near $r_2\to 1$, using Eqs.~\eqref{partialJNSsec5}, \eqref{partialJNLSSPSsec5}, \eqref{partialJNLSUVSsec5}, and \eqref{JFNINT}, respectively. The results are shown in Fig.~\ref{figcrosscheck}. We consider the quartic interaction model in Eq.~\eqref{modelint}, corresponding to $q_1=q_2=-1$, and set the fermion mass to $\widetilde M=1$. We fix $r_1=10^{-3}\ll1$ and vary $r_2$ towards the folded point. The left and right panels correspond to $\tlam=1$ and $5$, respectively. In each panel, the blue, green, and red curves represent the nonlocal, spectrum-pole local, and UV-pole local signals, respectively. Their sum is shown by the black curve, while the black dots represent the independent result obtained by numerical integration.

Where both methods are evaluated, the black dots lie on the black curve, providing a numerical check of Eq.~\eqref{JFcrosscheck}. Although the individual contributions exhibit pronounced variations towards the folded limit, their sum varies smoothly. For $\tlam=1$, cancellations among the three contributions leave a slowly varying total signal. For $\tlam=5$, the spectrum-pole and UV-pole local contributions exhibit large oscillations that largely cancel, leaving a total signal much smaller than either local contribution separately. The numerical integration further yields a smooth, finite result at $r_2=1$.

These cancellations reflect the fact that the three contributions arise from different pole families of the same seed integral. Separating them makes their distinct oscillatory structures explicit, but also introduces folded singularities that cancel only in the complete sum. The representation used for numerical integration keeps these contributions combined and thus allows the folded limit to be evaluated directly. The agreement between the two methods supports both the pole decomposition in Section~\ref{4} and the integration prescription in Section~\ref{5.2}, while the finite result at $r_2=1$ is consistent with the expected absence of a folded singularity for the Bunch--Davies initial state.
\par 
We now turn to the evaluation of the factorized seed integral in the three-point limit. We denote the three-point limit of the seed integral branch $\widehat{\mathcal{J}}_{-+,\rho,\lambda}^{q_1q_2,s_1s_2}$ by $\widehat{\mathcal{J}}_{-+,\rho,\lambda,\text{3pt}}^{q_1q_2,s_1s_2}$. Using Eq.~\eqref{seedintNint}, we can express it as
\begin{equation}
\label{Jbranch3pt}
    \widehat{\mathcal{J}}_{-+,\rho,\lambda,\text{3pt}}^{q_1q_2,s_1s_2}(r_1)
    =
    \mathcal T^{q_1q_2,s_1s_2}_{\rho,\lambda,\text{3pt}}\, 
    r_1^{2\ii\rho\tnu},
\end{equation}
where $\mathcal T^{q_1q_2,s_1s_2}_{\rho,\lambda,\text{3pt}}$ is given by $\mathcal N_\rho^{q_1q_2,s_1s_2}$ multiplied by the expression inside the curly brackets in Eq.~\eqref{seedintNint}, evaluated at $r_2=1$. In the three-point limit, the nonlocal and local signals become indistinguishable, and the CC signal is entirely captured by the factorized seed integral $\mathcal{J}_{\text{F,3pt}}^{q_1q_2}$. Substituting Eq.~\eqref{Jbranch3pt} into Eq.~\eqref{JF}, we obtain the expression for the factorized seed integral in the three-point limit
\begin{equation}
\label{JF3pt}
    \mathcal{J}_\text{F,3pt}^{q_1q_2}(r_1) = \widetilde{\mathcal T}^{q_1q_2}_{+,\text{3pt}}\,  r_1^{2\ii\tnu} + \widetilde{\mathcal T}^{q_1q_2}_{-,\text{3pt}}\,  r_1^{-2\ii\tnu}
    +
    \text{c.c.},
\end{equation}
where
\begin{equation}
    \widetilde{\mathcal T}^{q_1q_2}_{\rho,\text{3pt}}
    =
    -\left[
    1+e^{-\ii\pi(4+q_1+2\ii\rho\tnu)}
    \right]
    \sum_{s_1,s_2}\mathcal C_{s_1s_2}^\rho
    \left[
    \mathcal T^{q_1q_2,s_1s_2}_{\rho,\lambda,\text{3pt}}-
    \mathcal T^{q_1q_2,-s_1-s_2}_{\rho,-\lambda,\text{3pt}}\right].
\end{equation}
Equation~\eqref{JF3pt} exhibits a simple oscillatory dependence on $r_1$ and can be rewritten as
\begin{equation}
    \mathcal{J}_\text{F,3pt}^{q_1q_2}(r_1)
    =
    \widetilde{\mathcal{T}}_{\text{amp,3pt}}^{q_1q_2}
    \cos(2\tnu\ln r_1 + \varphi),
\end{equation}
where $2\tnu$ determines the oscillation frequency in $\ln r_1$, while the amplitude $\widetilde{\mathcal{T}}_{\text{amp,3pt}}^{q_1q_2}$ and phase $\varphi$ are given by
\begin{equation}
\label{ampandphase}
    \widetilde{\mathcal{T}}_{\text{amp,3pt}}^{q_1q_2}
    =
    2\, \bigg|\widetilde{\mathcal T}^{q_1q_2}_{+,\text{3pt}}+\left(\widetilde{\mathcal T}^{q_1q_2}_{-,\text{3pt}}\right)^\ast\bigg|,\qquad \varphi = \text{Arg}\left[
    \widetilde{\mathcal T}^{q_1q_2}_{+,\text{3pt}}+\left(\widetilde{\mathcal T}^{q_1q_2}_{-,\text{3pt}}\right)^\ast
    \right].
\end{equation}
\begin{figure}[htbp]
    \centering
    \includegraphics[width=0.50\linewidth]{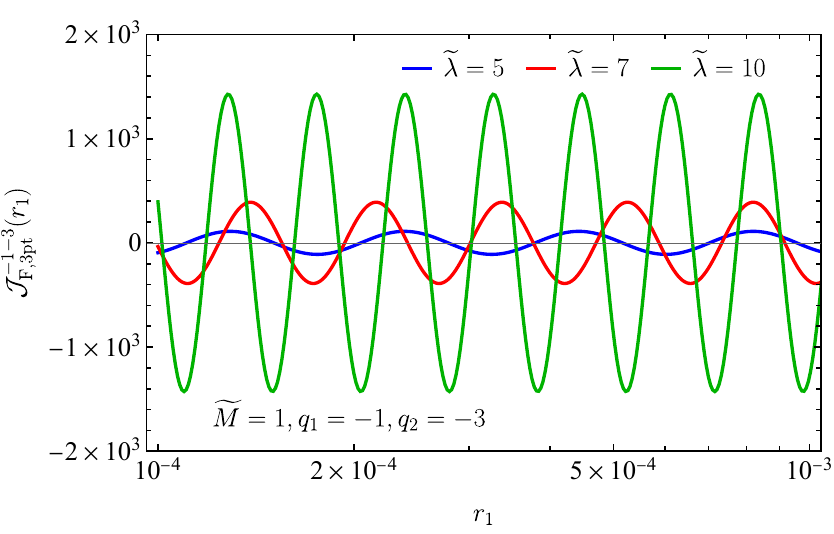}
    \includegraphics[width=0.48\linewidth]{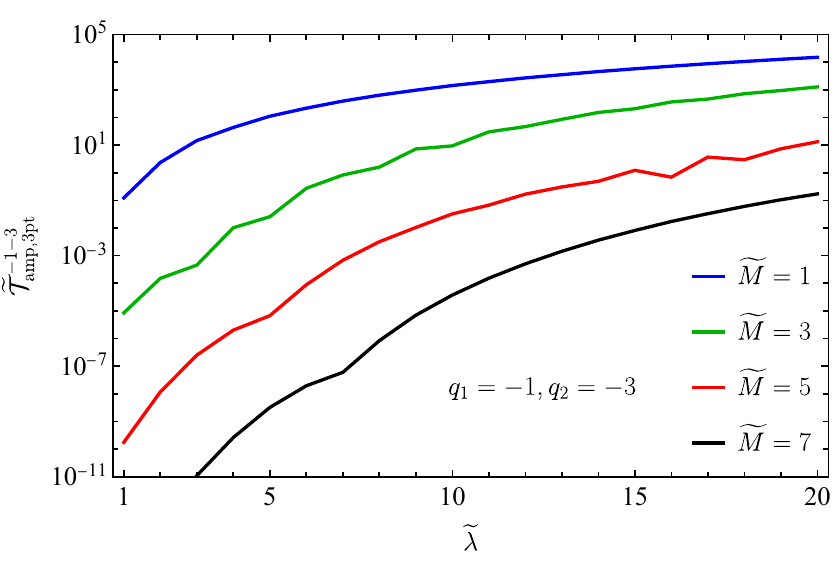}
    \caption{The seed integral in the three-point limit and its oscillation amplitude for $q_1=-1$ and $q_2=-3$. The left panel shows $\mathcal{J}_{\text{F,3pt}}^{-1,-3}(r_1)$ as a function of $r_1$ for $\widetilde M=1$. The blue, red, and green curves correspond to $\tlam=5$, $7$, and $10$, respectively. The right panel shows the oscillation amplitude $\widetilde{\mathcal T}_{\text{amp,3pt}}^{-1,-3}$ as a function of the chemical potential $\tlam$. The blue, green, red and black curves correspond to the fermion masses $\widetilde M=1$, $3$, $5$, and $7$, respectively.}
    \label{fig3pt}
\end{figure}
\par 
We now evaluate $\mathcal{J}_{\text{F,3pt}}^{q_1q_2}$ for a fermion bubble loop with the quartic and cubic interactions in Eq.~\eqref{modelint} at the left and right vertices, respectively, corresponding to $q_1=-1$ and $q_2=-3$. This seed integral provides the key building block for the CC signal in the corresponding cosmological three-point correlator.
Using Eqs.~\eqref{JF3pt} and \eqref{ampandphase}, we numerically evaluate $\mathcal{J}_{\text{F,3pt}}^{-1,-3}(r_1)$ and its oscillation amplitude. The results are shown in Fig.~\ref{fig3pt}. In the left panel, we fix $\widetilde M=1$ and vary $r_1$ over the squeezed range $r_1\in[10^{-4},10^{-3}]$. The blue, red, and green curves correspond to $\tlam=5$, $7$, and $10$, respectively. In the right panel, we plot $\widetilde{\mathcal T}_{\text{amp,3pt}}^{-1,-3}$ as a function of $\tlam$ for $\widetilde M=1$, $3$, $5$, and $7$, represented by the blue, green, red and black curves, respectively.

The left panel shows periodic oscillations in $\ln r_1$ with an
amplitude independent of $r_1$. This follows from Eq.~\eqref{JF3pt}, where the coefficients are independent of $r_1$ and the kinematic dependence enters only through the factors $r_1^{\pm2\ii\tnu}$. Increasing the chemical potential enhances the amplitude and raises the logarithmic oscillation frequency $2\tnu=2\sqrt{\widetilde M^2+\tlam^2}$.

The right panel illustrates how the oscillation amplitude depends on the chemical potential and the fermion mass. The amplitude generally increases with $\tlam$, whereas a larger mass suppresses it at fixed $\tlam$, particularly in the region of small chemical potential. These trends are consistent with the suppression of heavy-fermion production by the mass and its enhancement by the chemical potential. The detailed dependence on these parameters also involves interference between the two $\rho $ branches, since the amplitude is determined by the coherent combination $\widetilde{\mathcal T}^{-1,-3}_{+,\mathrm{3pt}}
+\left(\widetilde{\mathcal T}^{-1,-3}_{-,\mathrm{3pt}}\right)^*$.

\section{Conclusions}
\label{sec6}

In this work, we have computed the local and nonlocal cosmological
collider (CC) signals generated by massive fermion bubble loops
with an effective chemical potential. We approximate the spacetime
of slow-roll inflation by de Sitter spacetime and consider
interactions between inflaton perturbations and the scalar
fermion bilinear $\psi\psi+\psi^\dagger\psi^\dagger$. Our analysis
extracts the leading oscillatory dependence of the four-point
correlator in the single-squeezed limit $r_1\to0$, while retaining
the dependence on $r_2$ throughout $0<r_2\leq1$. The resulting
seed integrals also allow us to obtain the corresponding CC
signal in the squeezed three-point correlator.

The calculation combines the CC cutting rule, a late-time
expansion, and the partial Mellin--Barnes (MB) representation.
The cutting prescription isolates the signal contributions in
factorized time integrals. Expanding the fermion modes at one
vertex while retaining their exact form at the other then makes
the oscillatory dependence on $r_1$ explicit. The MB
representation allows us to perform the remaining time and
loop-momentum integrations and extract the signals from
appropriate families of Mellin poles. We obtain consistent
expressions using two complementary MB representations,
providing a nontrivial check of the calculation.

The pole analysis distinguishes the nonlocal signal from two
separate contributions to the local signal. The nonlocal
oscillations originate from the spectrum poles, whereas the
local signal receives contributions both from a particular
spectrum-pole branch combined with integer poles and from
the ultraviolet poles generated by the loop-momentum integral.
Both local contributions must be retained in the total
oscillatory signal. We derive residue representations for
these contributions and their leading-pole expressions in
the double-squeezed regime. Subleading residues become
important towards the folded endpoint and modify the
momentum dependence of the oscillation amplitudes.

To evaluate the three-point limit, we develop a complementary
organization of the MB integral that keeps the different
signal contributions combined. Although the separately
extracted local and nonlocal signals can exhibit singular
behavior at the folded endpoint, we demonstrate analytically
that the potentially singular terms cancel in the combined
kernel. After performing two contour integrations, we obtain
a double-MB representation suitable for numerical evaluation,
including at the folded endpoint. We specify the integration
contours, include the residues required by contour deformation,
and establish absolute convergence in the contour and
parameter domain considered. Numerical agreement with the
sum of the separately evaluated signals checks both the pole
decomposition and the integration prescription.

In the three-point limit, the local and nonlocal contributions
combine into a single logarithmically oscillating signal with
frequency $2\widetilde{\nu}$, where
$\widetilde{\nu}=\sqrt{\widetilde{M}^{\,2}
+\widetilde{\lambda}^{\,2}}$.
Our numerical results illustrate how the chemical potential
can amplify the signal while increasing its oscillation
frequency. A larger fermion mass suppresses the amplitude
at fixed chemical potential in the examples studied,
particularly when the chemical potential is small.
The detailed amplitude also depends on interference between
the two late-time frequency branches. Quantifying the
enhancement therefore requires the coherent sum of the
signal contributions.

An important direction for future work is to extend the
calculation to fermion triangle loops with a chemical potential.
For scalar triangle loops, a directional cutting rule relates
the leading soft signals to a tree-level single-exchange seed
and a bubble-type contribution~\cite{Qin:2026yuo}.
Establishing an analogous factorization for fermions requires
accounting for the spinor contractions and helicity structure.
The bubble signals obtained
here could serve as building blocks for calculating both local
and nonlocal CC signals from fermion triangle loops in soft
limits.

It would also be useful to extend the calculation beyond
the single-squeezed limit and to include the background
contributions, together with their renormalization.
These extensions would provide more complete correlators
for phenomenological applications. The analytical and
numerical methods developed in this work provide a basis
for pursuing these questions and for studying fermionic
CC signals in the presence of a chemical potential.

\acknowledgments

We acknowledge Zhongzhi Xianyu,  Yuhang Zhu for helpful discussions. This work is supported by the National Key R{\&}D Program of China under grant 2023YFA1606100 and by the National Natural Science Foundation of China under grants Nos. 12435005, 12675144. C.\,H.\ acknowledges support from the Sun Yat-Sen University Science Foundation,  the Fundamental Research Funds for the Central Universities at Sun Yat-sen University under Grant No.\,24qnpy117, and the Key Laboratory of Particle Astrophysics and Cosmology (MOE) of Shanghai Jiao Tong University. Z.Q. is
supported by the China Scholarship Council (CSC).

\appendix
\section{Fermions in dS Spacetime and Their Mode Functions}
\label{app:fermions_dS}

In this appendix, we outline how to quantize Weyl fermions with an effective chemical potential in de Sitter spacetime, and present a detailed derivation from the action to the mode functions.

\subsection{Weyl fermion action in de Sitter spacetime}

\subsubsection*{Vierbein, spin connection and covariant derivative}

Spinors are representations of the double cover $Spin(1,3)$ of the local Lorentz group. To define spinors in curved spacetime, one introduces a local Lorentz frame (vierbein) $e^a{}_\mu$ at each spacetime point, satisfying
\begin{equation}
g_{\mu\nu}=e^a{}_\mu e^b{}_\nu\eta_{ab},\qquad \eta_{ab}=\operatorname{diag}(-1,1,1,1),
\end{equation}
where $a,b$ denote local Lorentz indices and $\mu,\nu$ denote coordinate indices. For the FLRW metric, the vierbein is chosen as
\begin{equation}
    e^a{}_\mu=a(\tau)\delta^a{}_\mu,\qquad e^\mu{}_a=a^{-1}(\tau)\delta^\mu{}_a,
\end{equation}
which satisfies $e^a{}_\mu e^\mu{}_b=\delta^a{}_b,\,\,e^\mu{}_a e^a{}_\nu=\delta^\mu{}_\nu$.

The vierbein defines a local Lorentz frame whose orientation varies
from point to point. Therefore, the parallel transport of the local frame
is characterized by the spin connection $\omega_\mu{}^a{}_b$,
defined through
\begin{equation}
\nabla_\mu e_b
=\omega_\mu{}^a{}_b e_a,
\end{equation}
where $\nabla_\mu$ denotes the Levi-Civita covariant derivative, and $e_b=e^\mu{}_b\partial_\mu$. Equivalently, we have
\begin{equation}
\omega_\mu{}^a{}_b
=e^a{}_\nu
\left(\partial_\mu e^\nu{}_b
+\Gamma^\nu_{\mu\rho}e^\rho{}_b\right).
\end{equation}

Now we consider a Weyl spinor field $\psi$ in de Sitter spacetime. The covariant derivative of $\psi$ can be defined through the spin connection $\omega_{\mu ab}$,
\begin{equation}
\text{D}_\mu\psi=\partial_\mu\psi+\frac{1}{2}\omega_{\mu ab}\Sigma^{ab}\psi,
\end{equation}
where $\Sigma^{ab}$ is the local Lorentz generator of the spinor,
\begin{equation}
    \Sigma^{ab}=-\frac{1}{4}(\sigma^a\bar{\sigma}^b-\sigma^b\bar{\sigma}^a),
\end{equation}
and $\sigma^a=(1,\sigma^i)^a,\,\bar{\sigma}^a=(1,-\sigma^i)^a$. 

For the FLRW metric, the non-zero components of the Levi-Civita connection are
\begin{equation}
\Gamma_{00}^0=\mathcal{H},\qquad\Gamma_{ij}^0=\mathcal{H}\delta_{ij},\qquad\Gamma_{0j}^i=\Gamma_{j0}^i=\mathcal{H}\delta_j^i,
\end{equation}
where $\mathcal{H}=\frac{a^\prime}{a}$. Therefore the non-zero components of the spin-connection are given by
\begin{equation}
\omega_{i0j}=-\omega_{ij0}=-\mathcal{H}\delta_{ij}.
\end{equation}
Since $\Sigma^{0j}=-\Sigma^{j0}=\frac{1}{2}\sigma^j$, we can obtain the covariant derivative of the Weyl spinor in dS spacetime,
\begin{equation}
\label{appeq:covar_deri_weyl_dS}
    \text{D}_0\psi=\partial_0\psi,\qquad \text{D}_i\psi=\partial_i\psi-\frac{\mathcal{H}}{2}\sigma^i\psi.
\end{equation}

\subsubsection*{Weyl kinetic term and conformal rescaling}

The kinetic action for a Weyl spinor in curved
spacetime is
\begin{equation}
    S_{\mathrm{kin}}
    =i\int d^4x\sqrt{-g}\,
    \psi^\dagger\bar\varsigma^\mu D_\mu\psi.
\end{equation}
Using $\sqrt{-g}=a^4$ and the expression
for the covariant derivative derived above,
we obtain
\begin{equation}
    S_{\mathrm{kin}}
    =i\int d^4x\,a^3\psi^\dagger
    \left(
        \bar\sigma^\mu\partial_\mu
        +\frac32\mathcal H
    \right)\psi.
\end{equation}

To simplify the action, we introduce the conformally rescaled field
\begin{equation}
    \widetilde\psi=a^{3/2}\psi.
\end{equation}
Since $a$ depends only on conformal time, we have
\begin{equation}
    \partial_\tau\psi
    =a^{-3/2}\left(
        \partial_\tau\widetilde\psi
        -\frac32\mathcal H\widetilde\psi
    \right),
\end{equation}
therefore the Hubble friction term can be absorbed and the kinetic action reduces to
\begin{equation}
    S_{\mathrm{kin}}
    =i\int d^4x\,
    \widetilde\psi^\dagger
    \bar\sigma^\mu\partial_\mu\widetilde\psi.
\end{equation}
Thus, the massless Weyl kinetic term takes
its flat-spacetime form after conformal
rescaling, as expected from conformal invariance.

\subsubsection*{Effective chemical potential and equations of motion}

In this paper we consider a Weyl fermion $\psi$ with a Majorana
mass $M$ and a derivative coupling to the inflaton
$\phi$. The action is given by
\begin{equation}
    S_\psi
    =\int \dd^4x\,\sqrt{-g}
    \left[
        \ii\psi^\dagger\bar\varsigma^\mu
        \text{D}_\mu\psi
        -\frac{M}{2}
        \left(\psi\psi+\psi^\dagger\psi^\dagger\right)
        -\frac{\partial_\mu\phi}{\Lambda}
        \psi^\dagger\bar\varsigma^\mu\psi
    \right],
\end{equation}
where $\Lambda$ is the UV scale suppressing the
dimension-five operator. Considering only the background evolution of inflaton, we define
\begin{equation}
    \lambda
    \equiv\frac{\dot\phi_0}{\Lambda}
    =\frac{\phi_0'}{a\Lambda},
\end{equation}
then after conformally rescaling the field $\psi$, we can obtain the quadratic
Lagrangian
\begin{equation}
    \mathcal L_\psi^{(2)}
    =\ii\widetilde\psi^\dagger
     \bar\sigma^\mu\partial_\mu\widetilde\psi
    -\frac{aM}{2}
     \left(
        \widetilde\psi\widetilde\psi
        +\widetilde\psi^\dagger
         \widetilde\psi^\dagger
     \right)
    -a\lambda
     \widetilde\psi^\dagger\bar{\sigma}^0\widetilde\psi.
\end{equation}
The derivative coupling thus acts as an effective
chemical potential, while the Majorana mass
breaks the continuous fermion-number symmetry.

Varying the action with respect to
$\widetilde\psi^\dagger_{\dot\alpha}$,
we can obtain the equation of motion
\begin{equation}
    \ii(\bar\sigma^\mu)^{\dot\alpha\alpha}
    \partial_\mu\widetilde\psi_\alpha
    =
    a\lambda(\bar\sigma^0)^{\dot\alpha\alpha}
    \widetilde\psi_\alpha
    +aM\widetilde\psi^{\dagger\dot\alpha}.
    \label{appeq:weyl_eom}
\end{equation}
The conjugate equation reads
\begin{equation}
    \ii(\sigma^\mu)_{\alpha\dot\alpha}
    \partial_\mu\widetilde\psi^{\dagger\dot\alpha}
    =
    -a\lambda(\sigma^0)_{\alpha\dot\alpha}
    \widetilde\psi^{\dagger\dot\alpha}
    +aM\widetilde\psi_\alpha.
    \label{appeq:weyl_eom_conj}
\end{equation}

\subsection{Canonical quantization and mode equations}

\subsubsection*{Canonical quantization and helicity decomposition}

From the quadratic Lagrangian $\mathcal{L}_\psi^{(2)}$, the canonical momentum conjugate to the rescaled Weyl field $\widetilde{\psi}_\alpha$ is
\begin{equation}
    \widetilde{\pi}^\alpha \equiv \frac{\partial\mathcal{L}_\psi^{(2)}}{\partial\widetilde{\psi}'_\alpha} = \ii\widetilde{\psi}^\dagger_{\dot{\alpha}}\bar{\sigma}^{0\dot{\alpha}\alpha}.
\end{equation}
We canonically quantize the theory by imposing the equal-time anticommutation relations
\begin{equation}
    \left\{\widetilde{\psi}_\alpha(\tau,\mathbf{x}),\,\widetilde{\pi}^\beta(\tau,\mathbf{y})\right\} = \ii\delta_\alpha^\beta\,\delta^{(3)}(\mathbf{x}-\mathbf{y}),
\end{equation}
which can equivalently be written as
\begin{equation}
    \left\{\widetilde{\psi}_\alpha(\tau,\mathbf{x}),\,\widetilde{\psi}^\dagger_{\dot{\beta}}(\tau,\mathbf{y})\right\} = \sigma^0_{\alpha\dot{\beta}}\,\delta^{(3)}(\mathbf{x}-\mathbf{y}).
    \label{appeq:equal_time_anticomm}
\end{equation}
Then we can expand $\widetilde{\psi}_\alpha(\tau,\mathbf{x})$ and $\widetilde{\psi}^{\dagger\dot{\alpha}}(\tau,\mathbf{x})$ into plane-wave modes in the helicity basis as in Eq.~\eqref{planewave},
\begin{equation}
\begin{aligned}
\widetilde{\psi}_{\alpha}(\tau,\mathbf{x})&=\int\frac{\dd^{3}\mathbf{k}}{(2\pi)^{3}}\sum_{s=\pm}\left[h_{\alpha}^{s}(\hat{\mathbf{k}})u_s(k,\tau)b_{\mathbf{k}}^{s}e^{\ii\mathbf{k}\cdot\mathbf{x}}+h_{\dot{\gamma}}^{s\dagger}(\hat{\mathbf{k}})\bar{\sigma}^{0\dot{\gamma}\beta}\epsilon_{\alpha\beta}v_{s}^{*}(k,\tau)b_{\mathbf{k}}^{s\dagger}e^{-\ii\mathbf{k}\cdot\mathbf{x}}\right],\\
\widetilde{\psi}^{\dagger\dot{\alpha}}(\tau,\mathbf{x})&=\int\frac{\dd^{3}\mathbf{k}}{(2\pi)^{3}}\sum_{s=\pm}\left[\epsilon^{\dot{\alpha}\dot{\beta}}h_{\dot{\beta}}^{s\dagger}(\hat{\mathbf{k}})u^*_s(k,\tau)b_{\mathbf{k}}^{s\dagger}e^{-\ii\mathbf{k}\cdot\mathbf{x}}+\bar{\sigma}^{0\dot{\alpha}\gamma}h_{\gamma}^{s}(\hat{\mathbf{k}})v_{s}(k,\tau)b_{\mathbf{k}}^{s}e^{\ii\mathbf{k}\cdot\mathbf{x}}\right],
\end{aligned}
\label{appeq:mode_expansion}
\end{equation}
where the creation and annihilation operators satisfy
\begin{equation}
    \left\{b_{\mathbf{k}}^s,\,b_{\mathbf{k}'}^{s'\dagger}\right\} = (2\pi)^3\delta^{ss'}\delta^{(3)}(\mathbf{k}-\mathbf{k}'),
    \qquad
    \left\{b_{\mathbf{k}}^s,\,b_{\mathbf{k}'}^{s'}\right\} = \left\{b_{\mathbf{k}}^{s\dagger},\,b_{\mathbf{k}'}^{s'\dagger}\right\} = 0,
\end{equation}
and the helicity eigenspinors satisfy the helicity eigenvalue equations
\begin{equation}
-\hat{\mathbf{k}}^{i}\sigma_{\alpha\dot{\beta}}^{0}\bar{\sigma}^{i\dot{\beta}\gamma}h_{\gamma}^{s}(\hat{\mathbf{k}})=s h_{\alpha}^{s}(\hat{\mathbf{k}}),\quad h_{\dot{\gamma}}^{s\dagger}(\hat{\mathbf{k}})\bar{\sigma}^{0\dot{\gamma}\beta}\sigma_{\beta\dot{\alpha}}^{i}\hat{\mathbf{k}}^{i}=s h_{\dot{\alpha}}^{s\dagger}(\hat{\mathbf{k}}).
\label{appeq:helicity_eigenfunction}
\end{equation}
The orthonormal condition of the helicity eigenspinors can be naturally chosen as
\begin{equation}
h_{\dot{\alpha}}^{s\dagger}\bar{\sigma}^{0\dot{\alpha}\beta}h_{\beta}^{s^{\prime}}=\delta^{ss^{\prime}},
\end{equation}
then one can obtain the completeness and projection relations
\begin{equation}
    h_\alpha^s(\hat{\mathbf{k}})h_{\dot{\beta}}^{s\dagger}(\hat{\mathbf{k}}) = \frac{1}{2}\left(\sigma^0 + s\hat{\mathbf{k}}\cdot\vec{\sigma}\right)_{\alpha\dot{\beta}},
    \qquad
    \sum_{s=\pm} h_\alpha^s(\hat{\mathbf{k}})h_{\dot{\beta}}^{s\dagger}(\hat{\mathbf{k}})\bar{\sigma}^{0\dot{\beta}\beta} = \delta_\alpha^\beta.
\end{equation}

For a unit momentum vector $\hat{\mathbf{k}}=(\sin\theta\cos\varphi,\sin\theta\sin\varphi,\cos\theta)$, one can obtain the solutions to the helicity eigenvalue equations Eq.~\eqref{appeq:helicity_eigenfunction} up to a phase
\begin{equation}
    h_\alpha^+(\theta,\varphi)=
    \begin{pmatrix}
        e^{-\ii\frac{\varphi}{2}}\cos\frac{\theta}{2}\\
        e^{\ii\frac{\varphi}{2}}\sin\frac{\theta}{2}
    \end{pmatrix}_\alpha,
    \qquad
    h_\alpha^-(\theta,\varphi)=
    \begin{pmatrix}
        -e^{-\ii\frac{\varphi}{2}}\sin\frac{\theta}{2}\\
        e^{\ii\frac{\varphi}{2}}\cos\frac{\theta}{2}
    \end{pmatrix}_\alpha.
\end{equation}

Substituting Eq.~\eqref{appeq:mode_expansion} into Eq.~\eqref{appeq:equal_time_anticomm}, and using the relation $h_{\alpha}^{s}(-\hat{\mathbf{k}})h_{\dot{\beta}}^{s\dagger}(-\hat{\mathbf{k}})=h_{\alpha}^{-s}(\hat{\mathbf{k}})h_{\dot{\beta}}^{-s\dagger}(\hat{\mathbf{k}})$, we obtain the canonical normalization condition for the mode functions of each helicity $s=\pm$
\begin{equation}
    |u_s(k,\tau)|^2 + |v_s(k,\tau)|^2 = 1.
    \label{appeq:uv_norm}
\end{equation}

\subsubsection*{Mode equations}

Substituting the mode expansion \eqref{appeq:mode_expansion} into the equations of motion \eqref{appeq:weyl_eom} and \eqref{appeq:weyl_eom_conj}, we obtain the coupled first-order differential equations for the mode functions $u_s(k,\tau)$ and $v_s(k,\tau)$:
\begin{equation}
\begin{aligned}
    \ii u_s' + sk u_s &= a\lambda u_s + aM v_s,\\
    \ii v_s' - sk v_s &= -a\lambda v_s + aM u_s.
\end{aligned}
\label{appeq:coupled_first_order}
\end{equation}
The Majorana mass $M$ mixes $u_s$ and $v_s$. For $M\neq 0$, we can solve Eq.~\eqref{appeq:coupled_first_order} algebraically for $u_s$ and $v_s$ in terms of each other:
\begin{equation}
\begin{aligned}
    u_s &= \frac{1}{aM}\left(\ii v_s' - sk v_s + a\lambda v_s\right),\\
    v_s &= \frac{1}{aM}\left(\ii u_s' + sk u_s - a\lambda u_s\right).
\end{aligned}
\label{appeq:algebraic_uv}
\end{equation}
Differentiating Eq.~\eqref{appeq:coupled_first_order} with respect to $\tau$, and substituting Eq.~\eqref{appeq:algebraic_uv}, we arrive at the decoupled second-order differential equations
\begin{equation}
\begin{aligned}
    u_s'' - aH u_s' + \left[(sk - a\lambda)^2 + a^2 M^2 + \ii s a H k\right]u_s &= 0,\\
    v_s'' - aH v_s' + \left[(sk - a\lambda)^2 + a^2 M^2 - \ii s a H k\right]v_s &= 0.
\end{aligned}
\label{appeq:decoupled_second_order}
\end{equation}

\subsection{Exact mode functions and late-time asymptotics}

\subsubsection*{Reduction to the Whittaker equation}

In order to simplify the mode equations, in the following we introduce the dimensionless parameters
\begin{equation}
    \mR = \frac{M}{H},\qquad \tlam = \frac{\lambda}{H},\qquad \tnu = \sqrt{\mR^2 + \tlam^2}.
\end{equation}
After substituting the de Sitter scale factor $a(\tau) = -1/(H\tau)$ into Eq.~\eqref{appeq:decoupled_second_order}, we obtain
\begin{equation}
\begin{aligned}
    u_s'' + \frac{1}{\tau}u_s' + \left[k^2 + \frac{sk(2\tlam - \ii)}{\tau} + \frac{\tnu^2}{\tau^2}\right]u_s &= 0,\\
    v_s'' + \frac{1}{\tau}v_s' + \left[k^2 + \frac{sk(2\tlam + \ii)}{\tau} + \frac{\tnu^2}{\tau^2}\right]v_s &= 0.
\end{aligned}
\label{appeq:uv_tau_eq}
\end{equation}
To eliminate the first-derivative terms $u_s'$ and $v_s'$, we rescale the mode functions as $u_s(k,\tau) = (-\tau)^\alpha \tilde{u}_s(k,\tau)$ and $v_s(k,\tau) = (-\tau)^\beta \tilde{v}_s(k,\tau)$. Requiring the coefficients of $\tilde{u}_s'$ and $\tilde{v}_s'$ (proportional to $2\alpha+1$ and $2\beta+1$, respectively) to vanish fixes $\alpha = \beta = -1/2$, which yields
\begin{equation}
\begin{aligned}
    \tilde{u}_s'' + \left[k^2 + \frac{sk(2\tlam - \ii)}{\tau} + \frac{\frac{1}{4}+\tnu^2}{\tau^2}\right]\tilde{u}_s &= 0,\\
    \tilde{v}_s'' + \left[k^2 + \frac{sk(2\tlam + \ii)}{\tau} + \frac{\frac{1}{4}+\tnu^2}{\tau^2}\right]\tilde{v}_s &= 0.
\end{aligned}
\end{equation}
Changing the independent variable to $z = 2\ii k\tau = -2\ii k\eta$, we recast the equations into the standard Whittaker form:
\begin{equation}
\begin{aligned}
    \frac{\dd^2\tilde{u}_s}{\dd z^2} + \left[-\frac{1}{4} - \frac{s\left(\frac{1}{2}+\ii\tlam\right)}{z} + \frac{\frac{1}{4}+\tnu^2}{z^2}\right]\tilde{u}_s &= 0,\\
    \frac{\dd^2\tilde{v}_s}{\dd z^2} + \left[-\frac{1}{4} + \frac{s\left(\frac{1}{2}-\ii\tlam\right)}{z} + \frac{\frac{1}{4}+\tnu^2}{z^2}\right]\tilde{v}_s &= 0.
\end{aligned}
\end{equation}
Comparing with Whittaker's differential equation~\cite{NIST:DLMF},
\begin{equation}
    \frac{\dd^2 W}{\dd z^2} + \left(-\frac{1}{4} + \frac{\kappa}{z} + \frac{\frac{1}{4}-\mu^2}{z^2}\right)W = 0,
\end{equation}
whose two linearly independent solutions can be chosen as $W_{\kappa,\mu}(z)$ and $W_{-\kappa,\mu}(-z)$ due to the invariance of the equation under $(z,\kappa)\to(-z,-\kappa)$, we identify the parameters
\begin{equation}
    \mu = \ii\tnu,\qquad
    \kappa_{u,s} = -s\left(\frac{1}{2}+\ii\tlam\right),\qquad
    \kappa_{v,s} = s\left(\frac{1}{2}-\ii\tlam\right).
    \label{appeq:whittaker_kappa}
\end{equation}

\subsubsection*{Bunch-Davies initial conditions and normalization}

In the early-time limit $\tau \to -\infty$ ($|z|\to\infty$), the two linearly independent Whittaker solutions $W_{\kappa,\mu}(z)$ and $W_{-\kappa,\mu}(-z)$ have the asymptotic behaviors
\begin{equation}
\begin{aligned}
W_{\kappa,\mu}(z) &\sim e^{-z/2}z^\kappa = e^{-ik\tau}(2ik\tau)^\kappa, && |\arg z| < \frac{3\pi}{2},\\
W_{-\kappa,\mu}(-z) &\sim e^{+z/2}(-z)^{-\kappa} = e^{+ik\tau}(-2ik\tau)^{-\kappa}, && |\arg(-z)| < \frac{3\pi}{2}.
\end{aligned}
\end{equation}
Since $z = 2ik\tau$ has $\arg(\pm z) = \mp \pi/2$ for $\tau < 0$, both asymptotic expansions hold strictly in the physical regime. Imposing the Bunch-Davies vacuum condition as $\tau \to -\infty$ requires a pure positive-frequency mode ($\propto e^{-ik\tau}$), so the physical solution is described solely by $W_{\kappa,\mu}(z)$.

Up to normalization constants $\mathcal{N}_u^{(s)}$ and $\mathcal{N}_v^{(s)}$, the mode functions take the form
\begin{equation}
\begin{aligned}
u_s(k,\tau) &= \mathcal{N}_u^{(s)}(-\tau)^{-1/2}W_{\kappa_{u,s},\ii\tnu}(2\ii k\tau),\\    
v_s(k,\tau) &= \mathcal{N}_v^{(s)}(-\tau)^{-1/2}W_{\kappa_{v,s},\ii\tnu}(2\ii k\tau).
\end{aligned}
\end{equation}
In the limit $\tau \to -\infty$, the modes $u_+$ (with $\operatorname{Re}\kappa_{u,+} = -1/2$) and $v_-$ (with $\operatorname{Re}\kappa_{v,-} = -1/2$) decay as $\mathcal{O}((-\tau)^{-1})$, whereas $u_-$ (with $\kappa_{u,-} = 1/2 + \ii\tlam$) and $v_+$ (with $\kappa_{v,+} = 1/2 - \ii\tlam$) remain $\mathcal{O}(1)$. Consequently, the canonical normalization condition \eqref{appeq:uv_norm} reduces at $\tau \to -\infty$ to
\begin{equation}
    \lim_{\tau\to-\infty}|u_-(k,\tau)|^2 = 1,
    \qquad
    \lim_{\tau\to-\infty}|v_+(k,\tau)|^2 = 1.
\end{equation}
Since $\left|(2\ii k\tau)^{\pm\ii\tlam}\right| = e^{\pm\pi\tlam/2}$ for $\arg(2\ii k\tau) = -\pi/2$, choosing the standard Bunch--Davies phase convention yields
\begin{equation}
\begin{aligned}
    u_-(k,\tau) &= \ii\frac{e^{-\pi\tlam/2}}{\sqrt{-2k\tau}}\,W_{\frac{1}{2}+\ii\tlam,\ii\tnu}(2\ii k\tau),\\
    v_+(k,\tau) &= \ii\frac{e^{+\pi\tlam/2}}{\sqrt{-2k\tau}}\,W_{\frac{1}{2}-\ii\tlam,\ii\tnu}(2\ii k\tau).
\end{aligned}
\label{appeq:uminus_vplus}
\end{equation}
The remaining two mode functions $u_+$ and $v_-$ are then uniquely fixed (including their relative phases) by the first-order equations \eqref{appeq:algebraic_uv}:
\begin{equation}
\begin{aligned}
    u_+ &= \frac{-\tau}{\mR}\left(\ii v_+' - k v_+ - \frac{\tlam}{\tau}v_+\right),\\
    v_- &= \frac{-\tau}{\mR}\left(\ii u_-' - k u_- + \frac{\tlam}{\tau}u_-\right).
\end{aligned}
\label{appeq:uplus_vminus_from_eom}
\end{equation}
Using the derivative recurrence relation of the Whittaker function,
\begin{equation}
    W'_{\kappa,\mu}(z) + \left(\frac{1}{2}-\frac{\kappa}{z}\right)W_{\kappa,\mu}(z)
    = -\frac{1}{z}\left[\mu^2 - \left(\kappa-\frac{1}{2}\right)^2\right]W_{\kappa-1,\mu}(z),
\end{equation}
together with $\kappa_{u,-}-1 = \kappa_{v,-}$, $\kappa_{v,+}-1 = \kappa_{u,+}$, and $(\ii\tnu)^2 - (\mp\ii\tlam)^2 = -\tnu^2 + \tlam^2 = -\mR^2$, Eq.~\eqref{appeq:uplus_vminus_from_eom} gives
\begin{equation}
\begin{aligned}
    u_+(k,\tau) &= \frac{e^{+\pi\tlam/2}}{\sqrt{-2k\tau}}\,\mR\,W_{-\frac{1}{2}-\ii\tlam,\ii\tnu}(2\ii k\tau),\\
    v_-(k,\tau) &= \frac{e^{-\pi\tlam/2}}{\sqrt{-2k\tau}}\,\mR\,W_{-\frac{1}{2}+\ii\tlam,\ii\tnu}(2\ii k\tau).
\end{aligned}
\end{equation}
In terms of $\eta = -\tau \in (0,\infty)$, all four mode functions can be written in the compact unified form
\begin{equation}
    f_s(k,\eta) = \frac{\mathfrak{n}_{f,s}}{\sqrt{2k\eta}}\,W_{\kappa_{f,s},\ii\tnu}(-2\ii k\eta),
    \qquad f\in\{u,v\},\quad s=\pm 1,
    \label{appeq:unified_mode_eta}
\end{equation}
where the normalization factors $\mathfrak{n}_{f,s}$ are summarized in Table~\ref{tab:normalization_of_mode_function}. Note also that the mode functions obey the symmetry relation $u_s(\tlam) = v_{-s}(-\tlam)$.

\begin{table}[htbp]
    \centering
    \caption{Normalization prefactors $\mathfrak{n}_{f,h}$ and Whittaker indices $\kappa_{f,h}$ for the fermion mode functions.}
    \vspace{4pt}
    \begin{tabular}{ccccc}
        \toprule
        $f_s$ & $u_+$ & $u_-$ & $v_+$ & $v_-$ \\
        \midrule
        $\kappa_{f,s}$ & $-\frac{1}{2}-\ii\tlam$ & $\frac{1}{2}+\ii\tlam$ & $\frac{1}{2}-\ii\tlam$ & $-\frac{1}{2}+\ii\tlam$ \\[4pt]
        $\mathfrak{n}_{f,s}$ & $e^{+\pi\tlam/2}\mR$ & $\ii e^{-\pi\tlam/2}$ & $\ii e^{+\pi\tlam/2}$ & $e^{-\pi\tlam/2}\mR$ \\
        \bottomrule
    \end{tabular}
    \label{tab:normalization_of_mode_function}
\end{table}

\subsubsection*{Late-time asymptotics}

Using the small-argument expansion of the Whittaker function (DLMF 13.14.18~\cite{NIST:DLMF}) for $\tnu \neq 0$,
\begin{equation}
    W_{\kappa,\ii\tnu}(z)
    = \sum_{\rho=\pm 1}\frac{\Gamma[-2\ii\rho\tnu]}{\Gamma\!\left[\frac{1}{2}-\kappa-\ii\rho\tnu\right]}\,z^{\frac{1}{2}+\ii\rho\tnu} + \mathcal{O}(z^{3/2})
    \qquad (z\to 0),
    \label{appeq:W_small_z}
\end{equation}
and substituting $z = -2\ii k\eta = 2k\eta\,e^{-\ii\pi/2}$ into Eq.~\eqref{appeq:unified_mode_eta}, we obtain the late-time expansion
\begin{equation}
    \lim_{k\eta\to 0^+} f_s(k,\eta) = \sum_{\rho=\pm 1} C_{f_s}^{(\rho)}(2k\eta)^{\ii\rho\tnu},
    \label{appeq:late_time_expansion_general}
\end{equation}
with the unified branch coefficients
\begin{equation}
    C_{f_s}^{(\rho)}
    = \mathfrak{n}_{f,s}\,e^{-\ii\frac{\pi}{4}+\frac{\pi}{2}\rho\tnu}\,
    \frac{\Gamma[-2\ii\rho\tnu]}{\Gamma\!\left[\frac{1}{2}-\kappa_{f,h}-\ii\rho\tnu\right]}.
    \label{appeq:C_fh_rho_general}
\end{equation}
Explicitly, for $u_+$ and $v_+$, we have
\begin{align}
    C_{u_+}^{(+)} &= \mR\,e^{\pi\tlam/2}(-\ii)^{\frac{1}{2}+\ii\tnu}\frac{\Gamma[-2\ii\tnu]}{\Gamma[1-\ii(\tnu-\tlam)]},
    &
    C_{u_+}^{(-)} &= \mR\,e^{\pi\tlam/2}(-\ii)^{\frac{1}{2}-\ii\tnu}\frac{\Gamma[2\ii\tnu]}{\Gamma[1+\ii(\tnu+\tlam)]},
    \\
    C_{v_+}^{(+)} &= \ii e^{\pi\tlam/2}(-\ii)^{\frac{1}{2}+\ii\tnu}\frac{\Gamma[-2\ii\tnu]}{\Gamma[-\ii(\tnu-\tlam)]},
    &
    C_{v_+}^{(-)} &= \ii e^{\pi\tlam/2}(-\ii)^{\frac{1}{2}-\ii\tnu}\frac{\Gamma[2\ii\tnu]}{\Gamma[\ii(\tnu+\tlam)]},
\end{align}
while $C_{u_-}^{(\rho)}$ and $C_{v_-}^{(\rho)}$ follow from Eq.~\eqref{appeq:C_fh_rho_general} (or via $u_s(\tlam)=v_{-s}(-\tlam)$). For the complex-conjugated mode functions $f_s^*(k,\eta)$ appearing at the right vertex, the late-time expansion reads
\begin{equation}
    \lim_{k\eta\to 0^+} f_s^*(k,\eta) = \sum_{\sigma=\pm 1} \overline{C}_{f_s}^{(\sigma)}(2k\eta)^{\ii\sigma\tnu},
    \qquad
    \overline{C}_{f_s}^{(\sigma)} = \left(C_{f_s}^{(-\sigma)}\right)^*.
\end{equation}
Moreover, the ratio between the negative-helicity and positive-helicity late-time coefficients is independent of $f\in\{u,v\}$:
\begin{equation}
    r_\rho \equiv \frac{C_{f_-}^{(\rho)}}{C_{f_+}^{(\rho)}}
    = e^{-\pi\tlam}\,\frac{\rho\tnu-\tlam}{\mR}\,
    \frac{\Gamma[-\ii(\rho\tnu-\tlam)]}{\Gamma[-\ii(\rho\tnu+\tlam)]}.
    \label{appeq:r_rho_def}
\end{equation}

\section{Correlation Functions and Seed Integrals}
\label{appendixB}

In this appendix, we derive the fermion bubble contributions to
late-time inflaton correlation functions in the Schwinger--Keldysh
(SK) path integral formalism. We first construct the inflaton propagators and then derive the fermion propagators from the mode functions obtained in Appendix~\ref{app:fermions_dS}. Then we work out the fermionic Wick contractions that give the bubble function and show how the interaction vertices
and external inflaton propagators determine the normalization
and time weights of the seed integrals. The resulting expressions
are then used to obtain the three-point and four-point functions
presented in Section~\ref{sec2}.

\subsection{Inflaton propagators}
\label{app:sk_contour}

We begin with the external scalar lines. Following the SK path integral
of Ref.~\cite{Chen:2017ryl}, we introduce two copies of the inflaton
perturbation, $\delta\phi_+$ and $\delta\phi_-$, and independent sources
$J_+$ and $J_-$, where $\pm$ is the SK index. The generating functional is
\begin{equation}
\begin{aligned}
 Z_\phi[J_+,J_-]
 ={}&\int\cD\delta\phi_+\cD\delta\phi_-\exp\bigg\{\ii\int_{\tau_i}^{\tau_f}
 \dd\tau\dd^3\mathbf x\,
 \Big[\mathcal L_\phi[\delta\phi_+]
 -\mathcal L_\phi[\delta\phi_-]+J_+\delta\phi_+-J_-\delta\phi_-\Big]\bigg\}.
\end{aligned}
\label{appeq:sk_pathintegral}
\end{equation}
Here $\mathcal L_\phi$ is the Lagrangian density with respect to the
conformal-coordinate measure and includes the scale factors from the
background metric. The two fields live on the forward and backward branches of the closed
time contour $\cC=\cC_+\cup\cC_-$, and they are sewn by the condition $\delta\phi_+(\tau_f)=\delta\phi_-(\tau_f)$. Here $\tau_f=0^-$ corresponds to the conformal future boundary.

In the slow-roll approximation, the inflaton perturbation is treated as a massless scalar in de Sitter spacetime, with the free Lagrangian given by
\begin{equation}
\mathcal L_{\phi,0}[\delta\phi_{\mathtt{a}}]=\frac{1}{2}a^2(\tau)\Big[(\delta\phi_{\mathtt{a}}^\prime)^2-(\nabla\delta\phi_{\mathtt{a}})^2\Big],
\label{appeq:sk_scalar}
\end{equation}
from which we can derive the equation of motion
\begin{equation}
 \delta\phi_{\mathtt{a}}^{\prime\prime}+2\frac{a^\prime}{a}\delta\phi_{\mathtt{a}}^\prime-\nabla^2\delta\phi_{\mathtt{a}}=0.
 \label{appeq:sk_scalareom}
\end{equation}
Substituting the following free-field expansion into Eq.~(\ref{appeq:sk_scalareom}),
\begin{equation}
\begin{aligned}
 \delta\phi(\tau,\mathbf x)
 &=\int\!\frac{\dd^3\mathbf k}{(2\pi)^3}
 \left[a_{\mathbf k}f_k(\tau)e^{\ii\mathbf k\cdot\mathbf x}
 +a^\dagger_{\mathbf k}f_k^*(\tau)e^{-\ii\mathbf k\cdot\mathbf x}\right],\\
 [a_{\mathbf k},a^\dagger_{\mathbf q}]
 &=(2\pi)^3\delta^{(3)}(\mathbf k-\mathbf q),
\end{aligned}
 \label{appeq:sk_scalarexpansion}
\end{equation}
and adopting the Bunch--Davies (BD) boundary condition, one obtains the mode function of the inflaton perturbation,
\begin{equation}
 f_k(\tau)=\frac{H}{\sqrt{2k^3}}(1+\ii k\tau)e^{-\ii k\tau},
 \qquad f_k(\tau_f)=\frac{H}{\sqrt{2k^3}}.
 \label{appeq:sk_scalarmode}
\end{equation}
The normalization follows from the Wronskian condition $a^2(f_kf_k^{*\prime}-f_k^*f_k')=\ii$.

Let $Z_{\phi,0}$ be the Gaussian generating functional defined by
\begin{equation}
\begin{aligned}
 Z_{\phi,0}[J_+,J_-]
 ={}&\int\cD\delta\phi_+\cD\delta\phi_-\exp\bigg\{\ii\int_{\tau_i}^{\tau_f}
 \dd\tau\dd^3\mathbf x\,
 \Big[\mathcal L_{\phi,0}[\delta\phi_+]
 -\mathcal L_{\phi,0}[\delta\phi_-]+J_+\delta\phi_+-J_-\delta\phi_-\Big]\bigg\}.
\end{aligned}
\label{appeq:sk_pathintegral_free}
\end{equation}
The path-integral measure is understood to be normalized such
that $Z_{\phi,0}[0,0]=1$. Then the bulk-to-bulk propagator of $\delta\phi$ is given by
\begin{equation}
\begin{aligned}
G_{\mathtt{ab}}(k;\tau_1,\tau_2)&=\int\mathrm{d}^3\mathbf{x}e^{-\mathrm{i}\mathbf{k}\cdot\mathbf{x}}G_{\mathtt{ab}}(x_1,x_2),\\
G_{\mathtt{ab}}(x_1,x_2)
 &=\left.
 \frac{1}{\ii\mathtt{a}}\frac{\delta}{\delta J_{\mathtt{a}}(x_1)}
 \frac{1}{\ii\mathtt{b}}\frac{\delta}{\delta J_{\mathtt{b}}(x_2)}
 Z_{\phi,0}\right|_{J_\pm=0},
\end{aligned}
 \label{appeq:sk_scalar_propagator}
\end{equation}
where $\mathbf{x}=\mathbf{x}_1-\mathbf{x}_2$. Here and in the following we use $\mathtt{a},\mathtt{b}=\pm1$ to denote the SK index. Using the mode expansion in Eq.~\eqref{appeq:sk_scalarexpansion}
and the SK contour ordering, we obtain the four propagator components
\begin{align}
 G_{-+}(k;\tau_1,\tau_2)
 &=f_k(\tau_1)f_k^*(\tau_2),\nonumber\\
 G_{+-}(k;\tau_1,\tau_2)
 &=f_k^*(\tau_1)f_k(\tau_2),\nonumber\\
 G_{++}(k;\tau_1,\tau_2)&=\theta_{12}G_{-+}+\theta_{21}G_{+-},\nonumber\\
 G_{--}(k;\tau_1,\tau_2)&=\theta_{21}G_{-+}+\theta_{12}G_{+-},
 \label{appeq:sk_scalarsk}
\end{align}
where $\theta_{12}=\theta(\tau_1-\tau_2)$ and
$\theta_{21}=\theta(\tau_2-\tau_1)$. 

For an external line, we can choose $\tau_2=\tau_f\to0^-$, then the bulk-to-boundary propagator is given by
\begin{equation}
\begin{aligned}
 G_{\mathtt a}(k,\tau)
 &=\lim_{\tau_f\to0^-}G_{\mathtt{ab}}(k;\tau,\tau_f)
 =\begin{cases}
 f_k^*(\tau)f_k(0^-),&\mathtt a=+,\\
 f_k(\tau)f_k^*(0^-),&\mathtt a=-,
 \end{cases}\\
 &=\frac{H^2}{2k^3}(1-\ii\mathtt a k\tau)e^{\ii\mathtt a k\tau}.
\end{aligned}
 \label{appeq:sk_bulkboundary}
\end{equation}
A time derivative acting on the inflaton field at an interaction
vertex acts on the corresponding external propagator,
giving
\begin{equation}
 \partial_\tau G_{\mathtt a}(k,\tau)
 =\frac{H^2}{2k}\tau e^{\ii\mathtt a k\tau}.
 \label{appeq:sk_derivative}
\end{equation}

\subsection{Fermion propagators}
\label{app:sk_fermionsk}

For the internal fermion lines we use the canonical Weyl field
$\widetilde\psi=a^{3/2}\psi$. With independent Grassmann-odd sources $I_\pm^\alpha$ and
$I^\dagger_{\pm\dot\alpha}$, the Gaussian generating functional is
\begin{equation}
\begin{aligned}
 Z_{\psi,0}[I_\pm,I^\dagger_\pm]
 ={}&\int\prod_{\mathtt a=\pm}
 \cD\widetilde\psi_{\mathtt a}\cD\widetilde\psi^\dagger_{\mathtt a}\exp\bigg\{\ii S_\psi^{(2)}[\widetilde\psi_+]
 -\ii S_\psi^{(2)}[\widetilde\psi_-]\\
 &\quad+\ii\int\dd\tau\dd^3\mathbf x\,
 \Big[I_+^\alpha\widetilde\psi_{+\alpha}
 +I^\dagger_{+\dot\alpha}\widetilde\psi_+^{\dagger\dot\alpha}
 -I_-^\alpha\widetilde\psi_{-\alpha}
 -I^\dagger_{-\dot\alpha}\widetilde\psi_-^{\dagger\dot\alpha}
 \Big]\bigg\},
\end{aligned}
 \label{appeq:sk_fermion_generating}
\end{equation}
where $I^\alpha$ and $I^\dagger_{\dot\alpha}$ are treated
as independent variables under functional differentiation. The quadratic action is given by
\begin{equation}
\begin{aligned}
 S_\psi^{(2)}=\int\dd^4x\Big[
 &\ii\widetilde\psi^\dagger_{\dot\alpha}
 \bar\sigma^{\mu\dot\alpha\beta}\partial_\mu\widetilde\psi_\beta
 -a\lambda\widetilde\psi^\dagger_{\dot\alpha}
 \bar\sigma^{0\dot\alpha\beta}\widetilde\psi_\beta-\frac{aM}{2}\big(
 \widetilde\psi^\alpha\widetilde\psi_\alpha
 +\widetilde\psi^\dagger_{\dot\alpha}
 \widetilde\psi^{\dagger\dot\alpha}\big)\Big].
\end{aligned}
 \label{appeq:sk_fermion_action}
\end{equation}
Thus the chemical potential is already included in the free theory.

Using left functional derivatives, we obtain the four types of
two-point functions:
\begin{equation}
\begin{aligned}
 D_{\mathtt{ab}\alpha\beta}(x_1,x_2)
 &=\left.\frac{1}{(\ii\mathtt a)(\ii\mathtt b)}
 \frac{\delta}{\delta I_{\mathtt a}^{\alpha}(x_1)}
 \frac{\delta}{\delta I_{\mathtt b}^{\beta}(x_2)}
 Z_{\psi,0}\right|_{I=I^\dagger=0}
 =\langle\widetilde\psi_{\mathtt a\alpha}(x_1)
 \widetilde\psi_{\mathtt b\beta}(x_2)\rangle_0,\\[5pt]
 D_{\mathtt{ab}\alpha}{}^{\dot\beta}(x_1,x_2)
 &=\left.\frac{1}{(\ii\mathtt a)(\ii\mathtt b)}
 \frac{\delta}{\delta I_{\mathtt a}^{\alpha}(x_1)}
 \frac{\delta}{\delta I^\dagger_{\mathtt b\dot\beta}(x_2)}
 Z_{\psi,0}\right|_{I=I^\dagger=0}
 =\langle\widetilde\psi_{\mathtt a\alpha}(x_1)
 \widetilde\psi_{\mathtt b}^{\dagger\dot\beta}(x_2)\rangle_0,\\[5pt]
 D_{\mathtt{ab}}{}^{\dot\alpha}{}_{\beta}(x_1,x_2)
 &=\left.\frac{1}{(\ii\mathtt a)(\ii\mathtt b)}
 \frac{\delta}{\delta I^\dagger_{\mathtt a\dot\alpha}(x_1)}
 \frac{\delta}{\delta I_{\mathtt b}^{\beta}(x_2)}
 Z_{\psi,0}\right|_{I=I^\dagger=0}
 =\langle\widetilde\psi_{\mathtt a}^{\dagger\dot\alpha}(x_1)
 \widetilde\psi_{\mathtt b\beta}(x_2)\rangle_0,\\[5pt]
 D_{\mathtt{ab}}{}^{\dot\alpha\dot\beta}(x_1,x_2)
 &=\left.\frac{1}{(\ii\mathtt a)(\ii\mathtt b)}
 \frac{\delta}{\delta I^\dagger_{\mathtt a\dot\alpha}(x_1)}
 \frac{\delta}{\delta I^\dagger_{\mathtt b\dot\beta}(x_2)}
 Z_{\psi,0}\right|_{I=I^\dagger=0}
 =\langle\widetilde\psi_{\mathtt a}^{\dagger\dot\alpha}(x_1)
 \widetilde\psi_{\mathtt b}^{\dagger\dot\beta}(x_2)\rangle_0.
\end{aligned}
 \label{appeq:sk_fermion_source}
\end{equation}
Here brackets with SK-labeled fields denote the free correlation
functions generated by $Z_{\psi,0}$, with contour ordering
implicit in the branch labels.
Following the spinor conventions of Section~\ref{sec:intro}, the
propagator indices are raised and lowered with epsilon tensors:
\begin{align}
 D_{\mathtt{ab}}{}^{\alpha\beta}
 &=\epsilon^{\alpha\gamma}\epsilon^{\beta\delta}
 D_{\mathtt{ab}\gamma\delta},\nonumber\\
 D_{\mathtt{ab}}{}^\alpha{}_{\dot\beta}
 &=\epsilon^{\alpha\gamma}\epsilon_{\dot\beta\dot\delta}
 D_{\mathtt{ab}\gamma}{}^{\dot\delta}.
 \label{appeq:sk_raised}
\end{align}

The momentum-space propagators are obtained by a spatial Fourier
transform. For the first spinor type,
\begin{equation}
 D_{\mathtt{ab}\alpha\beta}(\mathbf p;\tau_1,\tau_2)
 =\int\dd^3\mathbf x\,e^{-\ii\mathbf p\cdot\mathbf x}
 D_{\mathtt{ab}\alpha\beta}(x_1,x_2),
 \qquad \mathbf x=\mathbf x_1-\mathbf x_2,
 \label{appeq:sk_fermion_fourier}
\end{equation}
and the same transform defines the other three types.

We first consider $D_{\mathtt{ab}\alpha\beta}$. The forward and
backward path integrals give time-ordered and anti-time-ordered
operator products, respectively~\cite{Chen:2017ryl,Chen:2018xck}.
Since the $-$ branch is later on the contour than the $+$ branch,
an operator on the $-$ branch is placed to the left of an operator
on the $+$ branch. The four components are therefore
\begin{equation}
\begin{aligned}
 D_{++\alpha\beta}(x_1,x_2)
 &=\langle T\{\widetilde\psi_\alpha(x_1)
 \widetilde\psi_\beta(x_2)\}\rangle_0,\\
 D_{--\alpha\beta}(x_1,x_2)
 &=\langle\overline T\{\widetilde\psi_\alpha(x_1)
 \widetilde\psi_\beta(x_2)\}\rangle_0,\\
 D_{-+\alpha\beta}(x_1,x_2)
 &=\langle\widetilde\psi_\alpha(x_1)
 \widetilde\psi_\beta(x_2)\rangle_0,\\
 D_{+-\alpha\beta}(x_1,x_2)
 &=-\langle\widetilde\psi_\beta(x_2)
 \widetilde\psi_\alpha(x_1)\rangle_0.
\end{aligned}
 \label{appeq:sk_offdiagonal}
\end{equation}
Here $T$ and $\overline T$ denote fermionic time ordering and
anti-time ordering, and the brackets denote free-vacuum operator
expectation values. In $D_{+-}$, placing the $-$ branch operator
to the left exchanges the two fermion fields and produces a minus
sign.

Substituting the mode expansion in Eq.~\eqref{planewave} into
$D_{-+\alpha\beta}$, the nonzero vacuum contraction is
\[
 \langle b^s_{\mathbf p}b^{r\dagger}_{\mathbf q}\rangle_0
 =(2\pi)^3\delta_{sr}\delta^{(3)}(\mathbf p-\mathbf q).
\]
Performing the spatial Fourier transform then gives
\begin{equation}
 D_{-+\alpha\beta}(\mathbf p;\tau_1,\tau_2)
 =\sum_s u_s(p,\tau_1)v_s^*(p,\tau_2)
 h^s_\alpha h^{s\dagger}_{\dot\gamma}
 \bar\sigma^{0\dot\gamma\delta}\epsilon_{\beta\delta}.
 \label{appeq:sk_dpsipsi}
\end{equation}
Here $p=|\mathbf p|$, $s=\pm1$, and all helicity spinors have argument
$\hat{\mathbf p}$. The factor $u_s(p,\tau_1)v_s^*(p,\tau_2)$
comes from the annihilation coefficient of the first field and the
creation coefficient of the second.
The last two lines of Eq.~\eqref{appeq:sk_offdiagonal} give
$D_{+-\alpha\beta}(x_1,x_2)=-D_{-+\beta\alpha}(x_2,x_1)$.
In the Fourier transform, exchanging the two positions changes
$\mathbf x_1-\mathbf x_2$ to its negative, so that
\begin{equation}
 D_{+-\alpha\beta}(\mathbf p;\tau_1,\tau_2)
 =-D_{-+\beta\alpha}(-\mathbf p;\tau_2,\tau_1).
 \label{appeq:sk_lessergeneral}
\end{equation}
For $D_{++\alpha\beta}$, time ordering places the field at the later time on
the left; while for $D_{--\alpha\beta}$, anti-time ordering places the earlier field
on the left. They are given by
\begin{equation}
\begin{aligned}
 D_{++\alpha\beta}(\mathbf p;\tau_1,\tau_2)
 &=\theta_{12}D_{-+\alpha\beta}(\mathbf p;\tau_1,\tau_2)
 +\theta_{21}D_{+-\alpha\beta}(\mathbf p;\tau_1,\tau_2),\\
 D_{--\alpha\beta}(\mathbf p;\tau_1,\tau_2)
 &=\theta_{21}D_{-+\alpha\beta}(\mathbf p;\tau_1,\tau_2)
 +\theta_{12}D_{+-\alpha\beta}(\mathbf p;\tau_1,\tau_2).
\end{aligned}
 \label{appeq:sk_skmatrix}
\end{equation}
At unequal times, $\theta_{12}+\theta_{21}=1$, so adding these two
equations gives $D_{++\alpha\beta}+D_{--\alpha\beta}
=D_{-+\alpha\beta}+D_{+-\alpha\beta}$.
Exchanging both field positions and spinor indices also gives
\begin{equation}
 D_{\mathtt{ab}\alpha\beta}(\mathbf p;\tau_1,\tau_2)
 =-D_{\mathtt{ba}\beta\alpha}(-\mathbf p;\tau_2,\tau_1),
 \qquad \tau_1\ne\tau_2,
 \label{appeq:sk_exchange}
\end{equation}
since $D_{++}$ and $D_{--}$ are fermionically ordered, while $D_{-+}$ and $D_{+-}$ obey Eq.~\eqref{appeq:sk_lessergeneral}.

The same derivation applies to the other three spinor-index
combinations.
Using the mode expansion and its Hermitian conjugate, their $-+$
components are
\begin{align}
 D_{-+\alpha}{}^{\dot\beta}(\mathbf p;\tau_1,\tau_2)
 &=\sum_s u_s(p,\tau_1)u_s^*(p,\tau_2)
 h^s_\alpha\epsilon^{\dot\beta\dot\gamma}h^{s\dagger}_{\dot\gamma},
 \label{appeq:sk_dpsidagger}\\
 D_{-+}{}^{\dot\alpha}{}_{\beta}(\mathbf p;\tau_1,\tau_2)
 &=\sum_s v_s(p,\tau_1)v_s^*(p,\tau_2)
 \bar\sigma^{0\dot\alpha\gamma}h^s_\gamma
 h^{s\dagger}_{\dot\gamma}
 \bar\sigma^{0\dot\gamma\delta}\epsilon_{\beta\delta},
 \label{appeq:sk_ddaggerpsi}\\
 D_{-+}{}^{\dot\alpha\dot\beta}(\mathbf p;\tau_1,\tau_2)
 &=\sum_s v_s(p,\tau_1)u_s^*(p,\tau_2)
 \bar\sigma^{0\dot\alpha\gamma}h^s_\gamma
 \epsilon^{\dot\beta\dot\gamma}h^{s\dagger}_{\dot\gamma}.
 \label{appeq:sk_ddaggerdagger}
\end{align}
Together with the corresponding ordering and exchange relations,
these expressions determine all sixteen SK components and reproduce
Eqs.~\eqref{psipropagator1}--\eqref{psipropagator4}.

Different types of propagators can be related by Hermitian conjugation. For example,
\[
 [D_{-+\alpha\beta}(x_1,x_2)]^*
 =\langle\widetilde\psi^\dagger_{\dot\beta}(x_2)
 \widetilde\psi^\dagger_{\dot\alpha}(x_1)\rangle_0
 =-D_{+-\dot\alpha\dot\beta}(x_1,x_2).
\]
The minus sign follows from the definition of $D_{+-}$. For the same-branch components $D_{++}$ and $D_{--}$, complex conjugation interchanges $T$ and $\overline{T}$, giving $[D_{\pm\pm\alpha\beta}]^*=-D_{\mp\mp\dot\alpha\dot\beta}$.
Finally, conjugating the Fourier phase sends $\mathbf p$ to
$-\mathbf p$, so
\begin{align}
 [D_{\mathtt{ab}\alpha\beta}(\mathbf p;\tau_1,\tau_2)]^*
 &=-D_{-\mathtt a,-\mathtt b\dot\alpha\dot\beta}
 (-\mathbf p;\tau_1,\tau_2),\nonumber\\
 [D_{\mathtt{ab}\alpha}{}^{\dot\beta}(\mathbf p;\tau_1,\tau_2)]^*
 &=-D_{-\mathtt a,-\mathtt b\dot\alpha}{}^\beta
 (-\mathbf p;\tau_1,\tau_2).
 \label{appeq:sk_componentconjugation}
\end{align}

\subsection{Fermion bubble function}
\label{app:sk_bubblehelicity}

We consider the fermion bubble diagram shown in Fig.~\ref{fig:1},
which contributes to the inflaton four-point function through
two insertions of the quartic interaction in Eq.~\eqref{modelint}.
For the $s$ channel, contracting external fields $(1,2)$ with
the scalar fields at $x_1$ and $(3,4)$ with those at $x_2$
gives four bulk-to-boundary propagators. The fermion fields
at the two vertices contract into a pair of internal
propagators, forming the closed loop. We first compute the fermion bubble loop,
leaving the external propagators and vertex coefficients to the next subsection.

The interactions in Eq.~\eqref{modelint} contain
$\psi\psi+\psi^\dagger\psi^\dagger$ at each vertex.
In terms of the canonical field,
\begin{equation}
 \psi\psi+\psi^\dagger\psi^\dagger
 =a^{-3}\bigl(
 \widetilde\psi^\alpha\widetilde\psi_\alpha
 +\widetilde\psi^\dagger_{\dot\alpha}
  \widetilde\psi^{\dagger\dot\alpha}\bigr).
 \label{appeq:sk_bilinear_rescaling}
\end{equation}
Contracting the fermion bilinears at the two vertices forms
the fermion loop. The first, undaggered term at each vertex gives
\begin{align}
 &\langle
 \widetilde\psi_{\mathtt a}^{\alpha}(x_1)
 \widetilde\psi_{\mathtt a\alpha}(x_1)
 \widetilde\psi_{\mathtt b}^{\beta}(x_2)
 \widetilde\psi_{\mathtt b\beta}(x_2)\rangle_{0,c}\nonumber\\
 &\qquad=-D_{\mathtt{ab}}{}^{\alpha\beta}
 D_{\mathtt{ab}\alpha\beta}
 +D_{\mathtt{ab}}{}^\alpha{}_\beta
 D_{\mathtt{ab}\alpha}{}^\beta
 =-2D_{\mathtt{ab}}{}^{\alpha\beta}
 D_{\mathtt{ab}\alpha\beta}.
 \label{appeq:sk_wickpsipsi}
\end{align}
Here the subscript $c$ denotes the connected part, with the
product of the two single-vertex contractions subtracted.
All propagators on the right have arguments $(x_1,x_2)$.
The relative sign between the two propagator products follows
from the epsilon contractions:
\begin{align}
 D_{\mathtt{ab}}{}^\alpha{}_\beta
 D_{\mathtt{ab}\alpha}{}^\beta
 =\epsilon^{\alpha\gamma}\epsilon^{\beta\delta}
 D_{\mathtt{ab}\gamma\beta}D_{\mathtt{ab}\alpha\delta}
 =-\epsilon^{\alpha\gamma}\epsilon^{\beta\delta}
 D_{\mathtt{ab}\gamma\delta}D_{\mathtt{ab}\alpha\beta}
 =-D_{\mathtt{ab}}{}^{\alpha\beta}D_{\mathtt{ab}\alpha\beta}.
 \label{appeq:sk_wick_index_sign}
\end{align}
The second equality follows from interchanging $\beta$ and $\delta$ and using the antisymmetry of the epsilon tensor. This compensates the relative Wick sign, so both pairings give the same contribution and their sum produces the factor $-2$.

Including the daggered bilinear at each vertex, the total
fermion contraction is
\begin{align}
 -2\Big[
 D_{\mathtt{ab}}{}^{\alpha\beta}D_{\mathtt{ab}\alpha\beta}
 +D_{\mathtt{ab}}{}^\alpha{}_{\dot\beta}
  D_{\mathtt{ab}\alpha}{}^{\dot\beta}+D_{\mathtt{ab}\dot\alpha}{}^\beta
  D_{\mathtt{ab}}{}^{\dot\alpha}{}_\beta
 +D_{\mathtt{ab}\dot\alpha\dot\beta}
  D_{\mathtt{ab}}{}^{\dot\alpha\dot\beta}\Big].
 \label{appeq:sk_position_bubble}
\end{align}
Spinor indices are summed, while SK labels are fixed.

At fixed $\tau_1,\tau_2$, spatial homogeneity makes this result
depend only on $\mathbf r=\mathbf x_1-\mathbf x_2$.
Fourier transforming $\mathbf r$ with momentum
$\mathbf k_s=\mathbf k_1+\mathbf k_2=-(\mathbf k_3+\mathbf k_4)$
converts each propagator product into a momentum convolution.
For the first term in Eq.~\eqref{appeq:sk_position_bubble},
the inverse Fourier transforms give, with the common time
arguments suppressed
\begin{align}
 &\int\dd^3\mathbf r\,e^{-\ii\mathbf k_s\cdot\mathbf r}
 D_{\mathtt{ab}}{}^{\alpha\beta}(\mathbf r)
 D_{\mathtt{ab}\alpha\beta}(\mathbf r)\nonumber\\
 &\quad=\int\!\frac{\dd^3\mathbf p_1\dd^3\mathbf p_2}{(2\pi)^6}
 (2\pi)^3\delta^{(3)}(\mathbf p_1+\mathbf p_2-\mathbf k_s)
 D_{\mathtt{ab}}{}^{\alpha\beta}(\mathbf p_1)
 D_{\mathtt{ab}\alpha\beta}(\mathbf p_2)\nonumber\\
 &\quad=\int\!\frac{\dd^3\mathbf p_1}{(2\pi)^3}
 D_{\mathtt{ab}}{}^{\alpha\beta}(\mathbf p_1)
 D_{\mathtt{ab}\alpha\beta}(\mathbf k_s-\mathbf p_1).
 \label{appeq:sk_convolution}
\end{align}
The momentum delta function fixes
$\mathbf p_2=\mathbf k_s-\mathbf p_1$, leaving one loop
integral. Transforming the other three terms in the same way
gives the fermion bubble function
\begin{align}
 \mathcal Q_{\mathtt{ab}}(k_s;\tau_1,\tau_2)
 =-2\int\!\frac{\dd^3\mathbf p_1}{(2\pi)^3}\Big\{&
 D_{\mathtt{ab}}{}^{\alpha\beta}(\mathbf p_1)
 D_{\mathtt{ab}\alpha\beta}(\mathbf p_2)+D_{\mathtt{ab}}{}^\alpha{}_{\dot\beta}(\mathbf p_1)
 D_{\mathtt{ab}\alpha}{}^{\dot\beta}(\mathbf p_2)\nonumber\\
 &+D_{\mathtt{ab}\dot\alpha}{}^\beta(\mathbf p_1)
 D_{\mathtt{ab}}{}^{\dot\alpha}{}_{\beta}(\mathbf p_2)+D_{\mathtt{ab}\dot\alpha\dot\beta}(\mathbf p_1)
 D_{\mathtt{ab}}{}^{\dot\alpha\dot\beta}(\mathbf p_2)\Big\}.
 \label{appeq:sk_qindex}
\end{align}
All propagators on the right have the same time arguments
$(\tau_1,\tau_2)$, which are suppressed.
The bubble function of Eq.~\eqref{bubblefunction} thus includes
the fermion contractions and the integral over the loop momentum at fixed
vertex times and SK labels.

The factor two can also be understood from momentum-space
Feynman rules. Each bilinear in Eq.~\eqref{modelint} contains
two identical fermion fields without a compensating $1/2!$,
so its two possible attachments give a factor two in the
vertex rule. Exchanging the two identical internal lines
leaves the bubble diagram unchanged and gives a symmetry
factor $1/2$. For fixed bilinears and vertex assignment,
the resulting multiplicity is therefore $2\times2/2=2$,
with the fermionic sign fixed by the contractions above.
This multiplicity is already included in $\mathcal Q$.

Under vertex exchange, the two propagators in each term of
Eq.~\eqref{appeq:sk_qindex} transform according to
Eq.~\eqref{appeq:sk_exchange}. Their minus signs cancel, and
relabeling the spinor indices interchanges the mixed products
while leaving the sum unchanged.
After $\mathbf p_{1,2}\to-\mathbf p_{1,2}$, the external momentum is
$-\mathbf k_s$. Isotropy makes the bubble depend only on
$k_s=|\mathbf k_s|$, giving
\begin{equation}
 \mathcal Q_{\mathtt{ab}}(k_s;\tau_1,\tau_2)
 =\mathcal Q_{\mathtt{ba}}(k_s;\tau_2,\tau_1).
 \label{appeq:sk_q_exchange}
\end{equation}

Under complex conjugation, Eq.~\eqref{appeq:sk_componentconjugation}
flips both SK labels and exchanges undaggered and daggered
fields in each propagator, while keeping $\tau_1,\tau_2$ fixed.
The two minus signs again cancel. Relabeling the spinor indices
and loop momenta restores the four-term sum, and the same
isotropy argument gives
\begin{equation}
 [\mathcal Q_{\mathtt{ab}}(k_s;\tau_1,\tau_2)]^*
 =\mathcal Q_{-\mathtt{a},-\mathtt{b}}(k_s;\tau_1,\tau_2).
 \label{appeq:sk_q_general_conjugation}
\end{equation}
Together, these relations give
\begin{equation}
 \mathcal Q_{+-}(k_s;\tau_1,\tau_2)
 =[\mathcal Q_{-+}(k_s;\tau_1,\tau_2)]^*
 =\mathcal Q_{-+}(k_s;\tau_2,\tau_1).
 \label{appeq:sk_qconjugate}
\end{equation}
For equal SK branches, substituting the two propagator
ordering relations into each product gives
$\theta_{12}^2=\theta_{12}$,
$\theta_{21}^2=\theta_{21}$, and
$\theta_{12}\theta_{21}=0$ at unequal times. Thus
\begin{align}
 \mathcal Q_{++}&=\theta_{12}\mathcal Q_{-+}
 +\theta_{21}\mathcal Q_{+-},\nonumber\\
 \mathcal Q_{--}&=\theta_{21}\mathcal Q_{-+}
 +\theta_{12}\mathcal Q_{+-},\nonumber\\
 \mathcal Q_{++}+\mathcal Q_{--}
 &=\mathcal Q_{-+}+\mathcal Q_{+-}.
 \label{appeq:sk_qdiagonal}
\end{align}
It therefore suffices to evaluate $\mathcal Q_{-+}$.
Substituting the propagators in
Eqs.~\eqref{appeq:sk_dpsipsi}--\eqref{appeq:sk_ddaggerdagger}
separates each term into mode-function coefficients and a
spinor contraction. Consider the first term in
Eq.~\eqref{appeq:sk_qindex}.
For fixed momenta and helicities, the remaining spinor contraction is
\begin{align*}
 &\bigl(\epsilon^{\alpha\gamma}
 h^{s_1}_{\gamma}(\hat{\mathbf p}_1)
 h^{s_2}_{\alpha}(\hat{\mathbf p}_2)\bigr)
 \bigl(h^{s_1\dagger}_{\dot\mu}(\hat{\mathbf p}_1)
 \bar\sigma^{0\dot\mu\beta}
 h^{s_2\dagger}_{\dot\nu}(\hat{\mathbf p}_2)
 \bar\sigma^{0\dot\nu\rho}\epsilon_{\beta\rho}\bigr)\\
 &\quad=\bigl(\epsilon^{\alpha\gamma}
 h^{s_1}_{\gamma}(\hat{\mathbf p}_1)
 h^{s_2}_{\alpha}(\hat{\mathbf p}_2)\bigr)
 \bigl(\epsilon^{\dot\alpha\dot\gamma}
 h^{s_1\dagger}_{\dot\gamma}(\hat{\mathbf p}_1)
 h^{s_2\dagger}_{\dot\alpha}(\hat{\mathbf p}_2)\bigr)\\
 &\quad=\bigl|h^{s_1\alpha}(\hat{\mathbf p}_1)
 h^{s_2}_{\alpha}(\hat{\mathbf p}_2)\bigr|^2.
\end{align*}
The second line uses
$\bar\sigma^{0\dot\mu\beta}\bar\sigma^{0\dot\nu\rho}
\epsilon_{\beta\rho}=\epsilon^{\dot\nu\dot\mu}$,
which follows from $\bar\sigma^0=\mathbf 1$ and
$\epsilon_{\beta\rho}=\epsilon^{\rho\beta}$.
The two brackets are complex conjugates because epsilon is real.
We denote this squared modulus by $F_{s_1s_2}$.
Using the normalization and projector in
Eq.~\eqref{eigenspinoridentity} together with
\begin{equation}
 \epsilon^{\alpha\beta}\epsilon^{\dot\gamma\dot\delta}
 =\bar\sigma^{0\dot\gamma\alpha}\bar\sigma^{0\dot\delta\beta}
 -\bar\sigma^{0\dot\gamma\beta}\bar\sigma^{0\dot\delta\alpha},
 \label{appeq:sk_epsilon_overlap}
\end{equation}
we obtain
\begin{align}
 F_{s_1s_2}(\hat{\mathbf p}_1,\hat{\mathbf p}_2)
 &\equiv
 \bigl|h^{s_1\alpha}(\hat{\mathbf p}_1)
 h^{s_2}_{\alpha}(\hat{\mathbf p}_2)\bigr|^2\nonumber\\
 &=1-
 h^{s_1\dagger}_{\dot\alpha}(\hat{\mathbf p}_1)
 \bar\sigma^{0\dot\alpha\beta}
 h^{s_2}_{\beta}(\hat{\mathbf p}_2)
 h^{s_2\dagger}_{\dot\gamma}(\hat{\mathbf p}_2)
 \bar\sigma^{0\dot\gamma\delta}
 h^{s_1}_{\delta}(\hat{\mathbf p}_1)\nonumber\\
 &=\frac12(1-s_1s_2\hat{\mathbf p}_1\cdot\hat{\mathbf p}_2).
 \label{appeq:sk_f}
\end{align}
The epsilon identity gives the second line: the two unit norms
produce $1$, while the subtracted term is the squared inner
product. The helicity projectors evaluate this term to
$\frac12(1+s_1s_2\hat{\mathbf p}_1\cdot\hat{\mathbf p}_2)$,
yielding the last line.

The first and fourth terms in Eq.~\eqref{appeq:sk_qindex}
have spinor factor $+F_{s_1s_2}$. For the second term,
the spinor contraction is
\begin{align}
 &\bigl(\epsilon^{\alpha\gamma}
 h^{s_1}_{\gamma}(\hat{\mathbf p}_1)
 h^{s_2}_{\alpha}(\hat{\mathbf p}_2)\bigr)
 \bigl(h^{s_1\dagger}_{\dot\beta}(\hat{\mathbf p}_1)
 \epsilon^{\dot\beta\dot\delta}
 h^{s_2\dagger}_{\dot\delta}(\hat{\mathbf p}_2)\bigr)\nonumber\\
 &\quad=-\bigl|h^{s_1\alpha}(\hat{\mathbf p}_1)
 h^{s_2}_{\alpha}(\hat{\mathbf p}_2)\bigr|^2
 =-F_{s_1s_2}.
 \label{appeq:sk_mixed_index_contraction}
\end{align}
The minus sign follows from the antisymmetry of
$\epsilon^{\dot\beta\dot\delta}$. The third term likewise
gives $-F_{s_1s_2}$. The four terms in
Eq.~\eqref{appeq:sk_qindex} therefore have the following
factors, listed in their original order:
\begin{center}
\begin{tabular}{@{}cccc@{}}
\toprule
Term & Mode coefficient at $\tau_1$ & Mode coefficient at $\tau_2$
& Spinor factor\\
\midrule
1 & $u_{s_1}(p_1,\tau_1)u_{s_2}(p_2,\tau_1)$
  & $v_{s_1}^*(p_1,\tau_2)v_{s_2}^*(p_2,\tau_2)$
  & $+F_{s_1s_2}$\\[3pt]
2 & $u_{s_1}(p_1,\tau_1)u_{s_2}(p_2,\tau_1)$
  & $u_{s_1}^*(p_1,\tau_2)u_{s_2}^*(p_2,\tau_2)$
  & $-F_{s_1s_2}$\\[3pt]
3 & $v_{s_1}(p_1,\tau_1)v_{s_2}(p_2,\tau_1)$
  & $v_{s_1}^*(p_1,\tau_2)v_{s_2}^*(p_2,\tau_2)$
  & $-F_{s_1s_2}$\\[3pt]
4 & $v_{s_1}(p_1,\tau_1)v_{s_2}(p_2,\tau_1)$
  & $u_{s_1}^*(p_1,\tau_2)u_{s_2}^*(p_2,\tau_2)$
  & $+F_{s_1s_2}$\\
\bottomrule
\end{tabular}
\end{center}
Each row contributes the product of its three factors.
Grouping the first two and last two rows gives the product
of differences below, with an overall minus sign.
Combining this sign with the Wick factor $-2$ gives $+2$.
Restoring the helicity sum and loop integral, we obtain
\begin{align}
 \mathcal Q_{-+}(k_s;\tau_1,\tau_2)
 ={}&2\int\!\frac{\dd^3\mathbf p_1}{(2\pi)^3}
 \sum_{s_1,s_2}F_{s_1s_2}\bigl[
 u_{s_1}(p_1,\tau_1)u_{s_2}(p_2,\tau_1)
 -v_{s_1}(p_1,\tau_1)v_{s_2}(p_2,\tau_1)\bigr]\nonumber\\
 &\times\bigl[
 u_{s_1}^*(p_1,\tau_2)u_{s_2}^*(p_2,\tau_2)
 -v_{s_1}^*(p_1,\tau_2)v_{s_2}^*(p_2,\tau_2)\bigr].
 \label{appeq:sk_qhelicity}
\end{align}
This reproduces Eq.~\eqref{bubblefunctionexpress}.
The exact mode functions and their normalization are given in
Eq.~\eqref{appeq:unified_mode_eta} and
Table~\ref{tab:normalization_of_mode_function}.

\subsection{Correlation functions and seed integrals}
\label{app:sk_seed}
\label{app:sk_correlators}

The correlation functions follow by attaching the external
propagators of Eq.~\eqref{appeq:sk_bulkboundary} to the bubble
$\mathcal Q_{\mathtt{ab}}$. Using the field rescaling in
Eq.~\eqref{appeq:sk_bilinear_rescaling}, the interactions in
Eq.~\eqref{modelint} become
\begin{equation}
\begin{aligned}
 \mathcal L_{{\rm int},3}
 &=\frac{1}{\Lambda_1}\delta\phi'
 \bigl(\widetilde\psi^\alpha\widetilde\psi_\alpha
 +\widetilde\psi^\dagger_{\dot\alpha}
  \widetilde\psi^{\dagger\dot\alpha}\bigr),\\
 \mathcal L_{{\rm int},4}
 &=\frac{1}{a\Lambda_2^3}(\delta\phi')^2
 \bigl(\widetilde\psi^\alpha\widetilde\psi_\alpha
 +\widetilde\psi^\dagger_{\dot\alpha}
  \widetilde\psi^{\dagger\dot\alpha}\bigr).
\end{aligned}
 \label{appeq:sk_rescaledvertices}
\end{equation}
The original factors $a^3$ and $a^2$ combine with the $a^{-3}$
from each fermion bilinear to give $1$ and $a^{-1}$.
Here $S_{\rm int}=\int\dd\tau\dd^3\mathbf x\,\mathcal L_{\rm int}$.
A vertex on branch $\mathtt a$ contributes $\ii\mathtt a$
times its coupling, and each external $\delta\phi'$ contributes
$\partial_\tau G_{\mathtt a}$ at that vertex.
The two branch factors therefore give
$(\ii\mathtt a)(\ii\mathtt b)=-\mathtt{ab}$.

\subsubsection{Four-point function}
\label{app:sk_fourpoint}

The four-point function receives a contribution from two
quartic vertices connected by the fermion bubble.
In the $s$ channel, external legs $(1,2)$ and $(3,4)$ attach
to the vertices at $\tau_1$ and $\tau_2$, respectively, with
$\mathbf k_s=\mathbf k_1+\mathbf k_2=-(\mathbf k_3+\mathbf k_4)$.
The two assignments of these pairs to the vertices cancel
$1/2!$ from the interaction expansion. At each vertex, the
two scalar fields can be contracted with the assigned external
legs in two ways, giving $2\times2=4$.
The fermion contractions, including their sign and multiplicity,
are already contained in $\mathcal Q_{\mathtt{ab}}$.
The Feynman rules thus give
\begin{align}
 \langle\delta\phi_{\mathbf k_1}\delta\phi_{\mathbf k_2}
 \delta\phi_{\mathbf k_3}\delta\phi_{\mathbf k_4}\rangle'_{c,s}
 ={}&\frac{4}{\Lambda_2^6}
 \sum_{\mathtt a,\mathtt b}(-\mathtt{ab})
 \int_{-\infty}^{0}\frac{\dd\tau_1\dd\tau_2}
 {a(\tau_1)a(\tau_2)}\,
 \mathcal Q_{\mathtt{ab}}(k_s;\tau_1,\tau_2)\nonumber\\
 &\times\partial_{\tau_1}G_{\mathtt a}(k_1,\tau_1)
 \partial_{\tau_1}G_{\mathtt a}(k_2,\tau_1)\partial_{\tau_2}G_{\mathtt b}(k_3,\tau_2)
 \partial_{\tau_2}G_{\mathtt b}(k_4,\tau_2).
 \label{appeq:sk_four_feynman}
\end{align}
The prime removes $(2\pi)^3\delta^{(3)}(\sum_i\mathbf k_i)$,
and the subscripts denote the connected $s$-channel contribution.

Set $\eta_i=-\tau_i>0$, so that
$\int_{-\infty}^{0}\dd\tau_i=\int_0^\infty\dd\eta_i$.
The two derivatives at each vertex supply $\eta_i^2$ by
Eq.~\eqref{appeq:sk_derivative}, while
$a^{-1}(-\eta_i)=H\eta_i$ gives the weight $\eta_i^3$. 
Writing $k_{12}=k_1+k_2$, $k_{34}=k_3+k_4$, and using
$\mathcal Q_{\mathtt{ab}}(k_s;\eta_1,\eta_2)$ for its value
at $\tau_i=-\eta_i$, we obtain
\begin{align}
 \langle\delta\phi_{\mathbf k_1}\delta\phi_{\mathbf k_2}
 \delta\phi_{\mathbf k_3}\delta\phi_{\mathbf k_4}\rangle'_{c,s}
 ={}&\frac{H^{10}}{4\Lambda_2^6k_1k_2k_3k_4}
 \sum_{\mathtt{a},\mathtt{b}}(-\mathtt{ab})\int_0^\infty\dd\eta_1\dd\eta_2\,
 \eta_1^3\eta_2^3\nonumber\\
 &\times e^{-\ii\mathtt{a} k_{12}\eta_1-\ii\mathtt{b} k_{34}\eta_2}
 \mathcal Q_{\mathtt{ab}}(k_s;\eta_1,\eta_2).
 \label{appeq:sk_fourraw}
\end{align}

Comparing Eq.~\eqref{appeq:sk_fourraw} with the seed integral
defined in Eq.~\eqref{seedintdefine}, we identify
$(q_1,q_2)=(-1,-1)$ from the time weight $\eta_1^3\eta_2^3$.
The inverse normalization contributes $r_1^4r_2^4/k_s^5$, giving
\begin{equation}
 \langle\delta\phi_{\mathbf k_1}\delta\phi_{\mathbf k_2}
 \delta\phi_{\mathbf k_3}\delta\phi_{\mathbf k_4}\rangle'_{c,s}
 =\frac{H^{10}}{4\Lambda_2^6k_1k_2k_3k_4}
 \frac{r_1^4r_2^4}{k_s^5}
 \sum_{\mathtt{a},\mathtt{b}=\pm1}\mathcal I_{\mathtt{ab}}^{-1,-1}(r_1,r_2).
 \label{appeq:sk_fourfinal}
\end{equation}

The pair partitions $(1,3)|(2,4)$ and $(1,4)|(2,3)$ give
the $t$ and $u$ channels.

\subsubsection{Three-point function}
\label{app:sk_threepoint}

The bubble $\mathcal Q_{\mathtt{ab}}$ also contributes to the
three-point function through one quartic and one cubic vertex.
We take external legs $(1,2)$ to meet at the quartic vertex
at $\tau_1$, and leg $3$ at the cubic vertex at $\tau_2$.
The two cubic--quartic assignments cancel $1/2!$,
leaving two scalar pairings at the quartic vertex.
Including the two branch factors and the fermion bubble,
the Feynman rules give
\begin{align}
 \langle\delta\phi_{\mathbf k_1}\delta\phi_{\mathbf k_2}
 \delta\phi_{\mathbf k_3}\rangle'_{c,s}
 ={}&\frac{2}{\Lambda_1\Lambda_2^3}
 \sum_{\mathtt a,\mathtt b}(-\mathtt{ab})
 \int_{-\infty}^{0}\frac{\dd\tau_1\dd\tau_2}{a(\tau_1)}\,
 \mathcal Q_{\mathtt{ab}}(k_s;\tau_1,\tau_2)\nonumber\\
 &\times\partial_{\tau_1}G_{\mathtt a}(k_1,\tau_1)
 \partial_{\tau_1}G_{\mathtt a}(k_2,\tau_1)
 \partial_{\tau_2}G_{\mathtt b}(k_3,\tau_2).
 \label{appeq:sk_three_feynman}
\end{align}
From Eq.~\eqref{appeq:sk_derivative}, the two derivatives and $a^{-1}$ at the quartic vertex
give the weight $\eta_1^3$, while the single
derivative at the cubic vertex gives $-\eta_2$. Substitution yields
\begin{align}
 \langle\delta\phi_{\mathbf k_1}\delta\phi_{\mathbf k_2}
 \delta\phi_{\mathbf k_3}\rangle'_{c,s}
 ={}&-\frac{H^7}{4\Lambda_1\Lambda_2^3k_1k_2k_3}
 \sum_{\mathtt{a},\mathtt{b}}(-\mathtt{ab})\int_0^\infty\dd\eta_1\dd\eta_2\,
 \eta_1^3\eta_2\nonumber\\
 &\times e^{-\ii\mathtt{a} k_{12}\eta_1-\ii\mathtt{b} k_3\eta_2}
 \mathcal Q_{\mathtt{ab}}(k_s;\eta_1,\eta_2).
 \label{appeq:sk_threeraw}
\end{align}
Momentum conservation gives $k_s=k_3$ and
$r_1=k_3/k_{12}$. The time weights select
$(q_1,q_2)=(-1,-3)$, and the phase at the cubic vertex
matches the seed at $k_{34}=k_3$, hence $r_2=1$.
In the four-point seed notation this is the limit
$k_4\to0$; the three physical momenta form a general
closed triangle.
The inverse normalization becomes
\begin{equation}
 \left.\frac{r_1^4r_2^2}{k_s^3}\right|_{r_2=1}
 =\frac{(k_s/k_{12})^4}{k_s^3}
 =\frac{k_s}{k_{12}^4}.
 \label{appeq:sk_threenormalization}
\end{equation}
Since $k_s=k_3$, this cancels the explicit $1/k_3$
in Eq.~\eqref{appeq:sk_threeraw}, giving
\begin{equation}
 \langle\delta\phi_{\mathbf k_1}\delta\phi_{\mathbf k_2}
 \delta\phi_{\mathbf k_3}\rangle'_{c,s}
 =-\frac{H^7}{4\Lambda_1\Lambda_2^3k_1k_2k_{12}^4}
\sum_{\mathtt{a},\mathtt{b}=\pm1}\lim_{k_4\to0}
 \mathcal I_{\mathtt{ab}}^{-1,-3}(r_1,r_2).
 \label{appeq:sk_threefinal}
\end{equation}
The remaining two terms attach leg $1$ or leg $2$ to the
cubic vertex.

\section{Proof of the Cosmological Collider Cutting Rule}
\label{appendixC}

The calculation in Section~\ref{sec3} uses the CC cutting rule
to extract the fermion-bubble signals from factorized time
integrals. The seed integral is decomposed so that the
nonlocal and local signals reside in the factorized part,
while the nested remainder contributes to the background.
Here we use the partial Mellin--Barnes (PMB) method to prove
the cutting rule and identify the appropriate decomposition
for each ordering of the external energies $k_{12}$ and $k_{34}$.

The MB representation used in the main text has kernels
built entirely from Gamma functions, which facilitate signal
calculations using Barnes' first lemma. The exponential
factors retained in this representation generate mixed
energy sums such as \mbox{$k_{34}+p_1+p_2$} after time integration.

For the proof, we use an alternative MB representation in
which each internal mode is expressed through powers of
momentum and time, without an explicit exponential prefactor.
The time and loop-momentum integrations can therefore be
performed separately, yielding compact expressions that
simplify the proof. We also use this representation in
Appendix~\ref{appendixD} to cross-check the signal calculation
in the main text.

\subsection{Factorized and nested time integrals}
\label{app:C_decomposition}

To extract the CC signals from factorized time integrals,
we separate the seed into a factorized part and a nested
remainder. The cutting rule then requires that the nested
remainder contain neither nonlocal nor local clock signals
in the appropriate momentum region.

With $\eta_i=-\tau_i>0$ and the common arguments
$(k_s;\eta_1,\eta_2)$ of the bubble functions left implicit,
the opposite-branch seeds in Eq.~\eqref{seedintdefine} read
\begin{align}
 \mathcal I_{\pm\mp}^{q_1q_2}
 ={}\frac{k_s^{7+q_{12}}}{r_1^{5+q_1}r_2^{5+q_2}}
 \int_0^\infty\dd\eta_1\,\dd\eta_2\,
 \eta_1^{q_1+4}\eta_2^{q_2+4}e^{\mp\ii k_{12}\eta_1\pm\ii k_{34}\eta_2}
 \mathcal Q_{\pm\mp}.
 \label{appeq:C_opposite_branch_seed}
\end{align}
Each term in the Wightman bubble integrand separates into
a function of $\eta_1$ times a function of $\eta_2$, so the
two time integrations already factorize.

For the same-branch seeds, time ordering prevents direct
factorization. To separate their factorized and nested
contributions, we use the bubble relations in
Appendix~\ref{appendixB}:
\begin{align}
 \mathcal Q_{++}
 &=\theta(\eta_2-\eta_1)\mathcal Q_{-+}
   +\theta(\eta_1-\eta_2)\mathcal Q_{+-},\nonumber\\
 \mathcal Q_{--}
 &=\theta(\eta_2-\eta_1)\mathcal Q_{+-}
   +\theta(\eta_1-\eta_2)\mathcal Q_{-+}.
 \label{appeq:C_bubble_ordering}
\end{align}
Since $\eta_i=-\tau_i$, later conformal time corresponds to
smaller $\eta_i$, and
$\theta(\eta_1-\eta_2)=\theta(\tau_2-\tau_1)$.

For $r_1<r_2$, or equivalently $k_{12}>k_{34}$, using
$\theta(\eta_2-\eta_1)=1-\theta(\eta_1-\eta_2)$ gives
\begin{equation}
 \mathcal Q_{\pm\pm}
 =\mathcal Q_{\mp\pm}
 +\theta(\eta_1-\eta_2)
  (\mathcal Q_{\pm\mp}-\mathcal Q_{\mp\pm}).
 \label{appeq:C_decomposition_less}
\end{equation}
Substituting the Wightman term into Eq.~\eqref{seedintdefine}
yields the factorized contribution
\begin{equation}
 \mathcal I_{\pm\pm,\mathrm F}^{q_1q_2}
 ={}-\frac{k_s^{7+q_{12}}}{r_1^{5+q_1}r_2^{5+q_2}}
 \int_0^\infty\dd\eta_1\,\dd\eta_2\,
 \eta_1^{q_1+4}\eta_2^{q_2+4}
 e^{\mp\ii(k_{12}\eta_1+k_{34}\eta_2)}
 \mathcal Q_{\mp\pm}.
 \label{appeq:C_factorized_branch_seed}
\end{equation}
The step-function term instead restricts the integration
domain to $0<\eta_2<\eta_1$ and gives
\begin{equation}
 \mathcal I_{\pm\pm,\TO}^{q_1q_2}
 ={}\mp\frac{k_s^{7+q_{12}}}{r_1^{5+q_1}r_2^{5+q_2}}
 \int_0^\infty\dd\eta_1\int_0^{\eta_1}\dd\eta_2\,
 \eta_1^{q_1+4}\eta_2^{q_2+4}
  e^{\mp\ii(k_{12}\eta_1+k_{34}\eta_2)}
 (\mathcal Q_{+-}-\mathcal Q_{-+}).
 \label{appeq:C_nested_time_definition}
\end{equation}

For $r_1>r_2$, or equivalently $k_{12}<k_{34}$, we use
$\theta(\eta_1-\eta_2)=1-\theta(\eta_2-\eta_1)$:
\begin{equation}
 \mathcal Q_{\pm\pm}
 =\mathcal Q_{\pm\mp}
 +\theta(\eta_2-\eta_1)
  (\mathcal Q_{\mp\pm}-\mathcal Q_{\pm\mp}),
 \label{appeq:C_decomposition_greater}
\end{equation}
The factorized integral is obtained from
Eq.~\eqref{appeq:C_factorized_branch_seed} by replacing
$\mathcal Q_{\mp\pm}$ with $\mathcal Q_{\pm\mp}$.
The nested domain becomes $0<\eta_1<\eta_2$, and the
difference of Wightman functions changes sign.
The two decompositions are related by exchanging the two
vertices together with their time weights.

Both rearrangements are exact for arbitrary external energies.
The momentum region determines which one isolates the clock
signals in the factorized part: using the other decomposition
can leave local clock signals in the nested
remainder~\cite{Tong:2021wai,Qin:2022lva}.
We explain this dependence in Section~\ref{app:C_local}.

Adding the opposite-branch seeds to the factorized
same-branch contributions gives the complete decomposition
\begin{align}
 \mathcal I^{q_1q_2}
 &=\mathcal I_{\rm F}^{q_1q_2}
  +\mathcal I_{\TO}^{q_1q_2},\nonumber\\
 \mathcal I_{\rm F}^{q_1q_2}
 &=\mathcal I_{-+}^{q_1q_2}+\mathcal I_{+-}^{q_1q_2}
  +\mathcal I_{++,\mathrm F}^{q_1q_2}
  +\mathcal I_{--,\mathrm F}^{q_1q_2},\nonumber\\
 \mathcal I_{\TO}^{q_1q_2}
 &=\mathcal I_{++,\TO}^{q_1q_2}
  +\mathcal I_{--,\TO}^{q_1q_2}
  =2\operatorname{Re}\mathcal I_{++,\TO}^{q_1q_2}.
 \label{appeq:C_seed_decomposition}
\end{align}
The proof of the cutting rule therefore reduces to showing
that the nested terms contain neither nonlocal nor local
clock signals in the region assigned to each decomposition:
\begin{equation}
 \left.\mathcal I_{\pm\pm,\TO}^{q_1q_2}\right|_{\rm NS}=0,
 \qquad
 \left.\mathcal I_{\pm\pm,\TO}^{q_1q_2}\right|_{\rm LS}=0.
 \label{appeq:C_cutting_statement}
\end{equation}
Here NS and LS denote the two signal contributions defined
in the main text. We first consider $0<r_1<r_2<1$; the
other region follows by exchanging the vertices and $q_1,q_2$.

\subsection{Mellin representation of the bubble and seed integrals}
\label{app:C_MB}

To analyze the signal content of the nested remainder, we
adopt an MB representation of the internal fermion modes
that differs from the one used in the main text(DLMF~13.23.4~\cite{NIST:DLMF}):
\begin{align}
 W_{\kappa,\ii\tnu}(w)
 &=\int_{\mathcal C}\frac{\dd z}{2\pi\ii}\,
 \mathcal W_\kappa(z)w^{-z},\qquad |\arg w|<\frac{\pi}{2},\nonumber\\
 \mathcal W_\kappa(z)
 &=\Gamma\!\left[z+\tfrac12+\ii\tnu,z+\tfrac12-\ii\tnu\right]
 {}_2\mathbf F_1\!\left[
 \begin{matrix}z+\frac12-\ii\tnu,\ z+\frac12+\ii\tnu\\
 z-\kappa+1\end{matrix};\frac12\right].
 \label{appeq:C_Whittaker_Mellin}
\end{align}
The hypergeometric function is regularized as in
Eq.~\eqref{reghyperfun}. The contour $\mathcal C$ runs from
$-\ii\infty$ to $+\ii\infty$, keeping the poles of
$\Gamma(z+\frac12\pm\ii\tnu)$ to its left and ensuring
convergence of the MB integral.
Setting $w=-2\ii p\eta$ and using the mode functions in
Eq.~\eqref{appeq:unified_mode_eta} gives
\begin{equation}
 f_s(p,-\eta)=\frac{\mathfrak n_{f,s}}{\sqrt{2p\eta}}
 \int_{\mathcal C}\frac{\dd z}{2\pi\ii}\,
 e^{\ii\pi z/2}\mathcal W_{\kappa_{f,s}}(z)(2p\eta)^{-z},
 \qquad f\in\{u,v\}.
 \label{appeq:C_mode_Mellin}
\end{equation}
For the conjugated mode, $\mathfrak n_{f,s}$ and
$\kappa_{f,s}$ are replaced by their complex conjugates,
and the factor $e^{\ii\pi z/2}$ becomes
$e^{-\ii\pi z/2}$ because the Whittaker argument changes
from $-2\ii p\eta$ to $2\ii p\eta$.
The BD $\ii\epsilon$ prescription continues
$\eta\to\eta e^{\pm\ii\delta}$ for the mode and its conjugate,
respectively, with $0<\delta<\pi/2$.
This damps their early-time oscillations by
$e^{-p\eta\sin\delta}$ and places the Whittaker arguments
$2p\eta e^{\mp\ii(\pi/2-\delta)}$ inside the
absolute-convergence sector of
Eq.~\eqref{appeq:C_Whittaker_Mellin}.
The physical modes in Eq.~\eqref{appeq:C_mode_Mellin}
are recovered as $\delta\to0^+$, when the arguments
approach the imaginary axis from $\operatorname{Re}w>0$.

\paragraph{Bubble function.}
In the bubble of Eq.~\eqref{appeq:sk_qhelicity}, we assign
Mellin variables $z_1,z_2$ to the modes at the first vertex
and $z_3,z_4$ to those at the second; $s_1,s_2=\pm1$ label
the helicities of the two internal lines.
The combination $u_{s_1}u_{s_2}-v_{s_1}v_{s_2}$ at the first
vertex and its conjugate at the second then give the
Mellin coefficients
\begin{align}
 \mathcal V^L_{s_1s_2}(z_1,z_2)
 ={}&e^{\ii\pi z_{12}/2}\Bigl[
 \mathfrak n_{u,s_1}\mathfrak n_{u,s_2}
 \mathcal W_{\kappa_{u,s_1}}(z_1)
 \mathcal W_{\kappa_{u,s_2}}(z_2)
 -\mathfrak n_{v,s_1}\mathfrak n_{v,s_2}
 \mathcal W_{\kappa_{v,s_1}}(z_1)
 \mathcal W_{\kappa_{v,s_2}}(z_2)\Bigr],\nonumber\\
 \mathcal V^R_{s_1s_2}(z_3,z_4)
 ={}&e^{-\ii\pi z_{34}/2}\Bigl[
 \mathfrak n_{u,s_1}^*\mathfrak n_{u,s_2}^*
 \mathcal W_{\kappa_{u,s_1}^*}(z_3)
 \mathcal W_{\kappa_{u,s_2}^*}(z_4)
 -\mathfrak n_{v,s_1}^*\mathfrak n_{v,s_2}^*
 \mathcal W_{\kappa_{v,s_1}^*}(z_3)
 \mathcal W_{\kappa_{v,s_2}^*}(z_4)\Bigr].
 \label{appeq:C_vertex_Mellin}
\end{align}
Here $z_{ij}=z_i+z_j$ and $z_{1234}=\sum_{i=1}^4z_i$.
Substituting these coefficients into
Eq.~\eqref{appeq:sk_qhelicity} gives
\begin{align*}
 \mathcal Q_{-+}(k_s;\eta_1,\eta_2)
 ={}&\frac12\sum_{s_1,s_2}
 \int\prod_{i=1}^4\frac{\dd z_i}{2\pi\ii}\,
 2^{-z_{1234}}\eta_1^{-1-z_{12}}\eta_2^{-1-z_{34}}\\
 &\times\mathcal V^L_{s_1s_2}(z_1,z_2)
 \mathcal V^R_{s_1s_2}(z_3,z_4)
 \int\!\frac{\dd^3\mathbf p_1}{(2\pi)^3}\,
 F_{s_1s_2}\,p_1^{-1-z_{13}}p_2^{-1-z_{24}}.
\end{align*}
The prefactor $1/2$ combines the bubble factor $2$ with
$1/4$ from the four factors $1/\sqrt{2p\eta}$.
This MB representation separates the momentum and time
dependence into powers, allowing the loop-momentum integral
to be evaluated independently of the time integrals.
Writing $A=-1-z_{13}$, $B=-1-z_{24}$ and
$\mathbf p_2=\mathbf k_s-\mathbf p_1$, the helicity factor
$F_{s_1s_2}=(1-s_1s_2\hat{\mathbf p}_1\cdot\hat{\mathbf p}_2)/2$
splits the loop integral into two scalar terms:
\begin{equation}
 \int\!\frac{\dd^3\mathbf p_1}{(2\pi)^3}
 F_{s_1s_2}\,p_1^A p_2^B
 =\frac{k_s^{A+B+3}}{8\pi^{3/2}}
 \bigl[J_0(A,B)+s_1s_2J_1(A,B)\bigr],
 \label{appeq:C_loop_Mellin}
\end{equation}
where
\begin{align}
 J_0(A,B)
 &=\frac{\Gamma[-(A+B+3)/2]}{2\Gamma[(A+B+6)/2]}
 \frac{\Gamma[(A+3)/2]\Gamma[(B+3)/2]}
 {\Gamma(-A/2)\Gamma(-B/2)},\nonumber\\
 J_1(A,B)
 &=\frac{\Gamma[-(A+B+3)/2]}{2\Gamma[(A+B+6)/2]}
 \frac{\Gamma[(A+4)/2]\Gamma[(B+4)/2]}
 {\Gamma[(1-A)/2]\Gamma[(1-B)/2]}.
 \label{appeq:C_angular_kernels}
\end{align}
This is equivalent to the four-term expansion in
Eq.~\eqref{loopintexpansion}. The $\chi=0$ term gives $J_0$,
while the $\chi=1,2,3$ terms combine into $s_1s_2J_1$
by Gamma-function recurrence relations.

Using Eq.~\eqref{appeq:C_loop_Mellin} then gives
\begin{align}
 \mathcal Q_{-+}(k_s;\eta_1,\eta_2)
 ={}&\frac{1}{16\pi^{3/2}}
 \sum_{s_1,s_2}\int\prod_{i=1}^4\frac{\dd z_i}{2\pi\ii}\,
 2^{-z_{1234}}k_s^{1-z_{1234}}\nonumber\\
 &\times\mathcal V^L_{s_1s_2}(z_1,z_2)
 \mathcal V^R_{s_1s_2}(z_3,z_4)
 [J_0(A,B)+s_1s_2J_1(A,B)]\nonumber\\
 &\times\eta_1^{-1-z_{12}}\eta_2^{-1-z_{34}}.
 \label{appeq:C_bubble_Mellin}
\end{align}
The expression for $\mathcal Q_{+-}$ follows
by exchanging $\eta_1$ and $\eta_2$, leaving the Mellin
coefficients unchanged.

\paragraph{Factorized seed integral.}
Multiplying the seed weights in Eq.~\eqref{seedintdefine}
by the time factors in Eq.~\eqref{appeq:C_bubble_Mellin}
gives $\eta_1^{q_1+3-z_{12}}\eta_2^{q_2+3-z_{34}}$;
the $\mathcal Q_{+-}$ term has $z_{12}$ and $z_{34}$ exchanged.
Since the bubble depends on time only through powers of
$\eta_1$ and $\eta_2$, the two independent time integrals
follow from
\begin{equation}
 \int_0^\infty\dd\eta\,
 \eta^{q+3-z}e^{-\ii\mathtt a k\eta}
 =e^{-\ii\mathtt a\pi(q+4-z)/2}
  \Gamma(q+4-z)k^{-q-4+z},
 \qquad\mathtt a=\pm1,
 \label{appeq:C_vertex_time_integral}
\end{equation}
initially for $\operatorname{Re}(q+4-z)>0$.
The integral is evaluated with the BD vacuum $\ii\epsilon$
prescription, implemented by inserting $e^{-\epsilon\eta}$
with $\epsilon>0$. Taking $\epsilon\to0^+$ after integration
fixes the phase shown above. Applying this formula separately
at the two vertices evaluates the factorized time integrals.

Using $k_{12}=k_s/r_1$ and $k_{34}=k_s/r_2$, the momentum
powers from the seed normalization, the bubble and the
two time integrals combine as
\begin{align*}
 \frac{k_s^{7+q_{12}}}{r_1^{5+q_1}r_2^{5+q_2}}
 k_s^{1-z_{1234}}
 k_{12}^{-q_1-4+z_{12}}k_{34}^{-q_2-4+z_{34}}=r_1^{-1-z_{12}}r_2^{-1-z_{34}}.
\end{align*}
The relative signs of $\mathcal I_{-+}$ and
$\mathcal I_{++,\mathrm F}$ give the phase combination
\begin{align*}
 &e^{-\ii\pi(q_2+4-z_{34})/2}
 \left[e^{+\ii\pi(q_1+4-z_{12})/2}
       -e^{-\ii\pi(q_1+4-z_{12})/2}\right]\\
 &\qquad=2\ii\,e^{-\ii\pi(q_2+4-z_{34})/2}
 \sin\!\left[\frac{\pi}{2}(q_1+4-z_{12})\right].
\end{align*}
The phases of $\mathcal I_{+-}$ and
$\mathcal I_{--,\mathrm F}$ instead combine to
\begin{equation*}
 -2\ii\,e^{+\ii\pi(q_2+4-z_{12})/2}
 \sin\!\left[\frac{\pi}{2}(q_1+4-z_{34})\right].
\end{equation*}
The factor $2\ii$ combines with the bubble coefficient
$1/(16\pi^{3/2})$ to give $\ii/(8\pi^{3/2})$.
Substituting Eq.~\eqref{appeq:C_bubble_Mellin} into the
factorized contributions in Section~\ref{app:C_decomposition}
and summing the SK branches gives, for $r_1<r_2$,
\begin{align}
 \mathcal I_{\mathrm F}^{q_1q_2}
 ={}&\frac{\ii}{8\pi^{3/2}}
 \sum_{s_1,s_2}\int\prod_{i=1}^4\frac{\dd z_i}{2\pi\ii}\,
 2^{-z_{1234}}\nonumber\\
 &\times\mathcal V^L_{s_1s_2}(z_1,z_2)
 \mathcal V^R_{s_1s_2}(z_3,z_4)
 \bigl[J_0(A,B)+s_1s_2J_1(A,B)\bigr]\nonumber\\
 &\times\Bigl\{
 r_1^{-1-z_{12}}r_2^{-1-z_{34}}
 \Gamma(q_1+4-z_{12})\Gamma(q_2+4-z_{34})\nonumber\\
 &\qquad\quad\times
 e^{-\ii\pi(q_2+4-z_{34})/2}
 \sin\!\left[\frac{\pi}{2}(q_1+4-z_{12})\right]\nonumber\\
 &\qquad-
 r_1^{-1-z_{34}}r_2^{-1-z_{12}}
 \Gamma(q_1+4-z_{34})\Gamma(q_2+4-z_{12})\nonumber\\
 &\qquad\quad\times
 e^{+\ii\pi(q_2+4-z_{12})/2}
 \sin\!\left[\frac{\pi}{2}(q_1+4-z_{34})\right]
 \Bigr\}.
 \label{appeq:C_factorized_seed_Mellin}
\end{align}
The signal residues of this expression are
evaluated in Appendix~\ref{appendixD}.

\paragraph{Nested seed integral.}
For the nested terms, define $\Delta=q_{12}+8-z_{1234}$
and write the power of $\eta_2$ as $\eta_2^{\beta-1}$.
The substitution $\eta_2=t\eta_1$ maps the ordered domain
to $\eta_1>0$, $0<t<1$, with measure and time powers
$\dd\eta_1\,\dd t\,\eta_1^{\Delta-1}t^{\beta-1}$.
The phase becomes $e^{\mp\ii\eta_1(k_{12}+k_{34}t)}$,
where $k_{12}+k_{34}t=k_{12}[1+(r_1/r_2)t]$.
Performing the $\eta_1$ integral gives
\begin{align}
 \int_0^\infty\dd\eta_1\int_0^{\eta_1}\dd\eta_2\,
 \eta_1^{\Delta-\beta-1}\eta_2^{\beta-1}
 e^{\mp\ii(k_{12}\eta_1+k_{34}\eta_2)}=e^{\mp\ii\pi\Delta/2}k_{12}^{-\Delta}
 \mathcal G\!\left(\Delta,\beta;\frac{r_1}{r_2}\right),
 \label{appeq:C_nested_time_evaluation}
\end{align}
where
\begin{align}
 \mathcal G\!\left(\Delta,\beta;\frac{r_1}{r_2}\right)
 \equiv\Gamma(\Delta)\int_0^1\dd t\,
 t^{\beta-1}\left(1+\frac{r_1}{r_2}t\right)^{-\Delta}
 =\frac{\Gamma(\Delta)}{\beta}
 {}_2F_1\!\left[\begin{matrix}\Delta,\beta\\\beta+1\end{matrix};
 -\frac{r_1}{r_2}\right].
 \label{appeq:C_nested_time_kernel}
\end{align}
The initial convergence conditions are
$\operatorname{Re}\Delta>0$ for the $\eta_1$ integral and
$\operatorname{Re}\beta>0$ at $t=0$; other values follow
by analytic continuation of the regulated integral.
The hypergeometric form follows from Euler's integral,
with prefactor $\Gamma(\beta)/\Gamma(\beta+1)=1/\beta$.
In Eq.~\eqref{appeq:C_nested_time_definition}, the
$\mathcal Q_{+-}$ and $\mathcal Q_{-+}$ terms have
$\beta=q_2+4-z_{12}$ and $\beta=q_2+4-z_{34}$, respectively.
Their difference therefore gives
\begin{align}
 \mathcal I_{\pm\pm,\TO}^{q_1q_2}
 ={}&\frac{\mp1}{16\pi^{3/2}r_1^2}\left(\frac{r_1}{r_2}\right)^{5+q_2}
 \sum_{s_1,s_2}\int\prod_{i=1}^4\frac{\dd z_i}{2\pi\ii}\,
 2^{-z_{1234}}r_1^{-z_{1234}}\nonumber\\
 &\times e^{\mp\ii\pi\Delta/2}\mathcal V^L_{s_1s_2}(z_1,z_2)
 \mathcal V^R_{s_1s_2}(z_3,z_4)
 \bigl[J_0(A,B)+s_1s_2J_1(A,B)\bigr]\nonumber\\
 &\times\bigl[
 \mathcal G\!\left(\Delta,q_2+4-z_{12};\frac{r_1}{r_2}\right)
 -\mathcal G\!\left(\Delta,q_2+4-z_{34};\frac{r_1}{r_2}\right)\bigr].
 \label{appeq:C_nested_branch_Mellin}
\end{align}
The prefactor combines the bubble coefficient in
Eq.~\eqref{appeq:C_bubble_Mellin} with the seed normalization
in Eq.~\eqref{seedintdefine}.
In particular, $\Delta=q_{12}+8-z_{1234}$ gives
\begin{align*}
 \frac{k_s^{7+q_{12}}k_s^{1-z_{1234}}k_{12}^{-\Delta}}
 {r_1^{5+q_1}r_2^{5+q_2}}
 =\frac{r_1^\Delta}{r_1^{5+q_1}r_2^{5+q_2}}=\frac{1}{r_1^2}\left(\frac{r_1}{r_2}\right)^{5+q_2}
 r_1^{-z_{1234}}.
\end{align*}
This yields the momentum factors in
Eq.~\eqref{appeq:C_nested_branch_Mellin}. The Mellin contours
keep the mode and UV poles to the left and the right pole
families from the loop and time integrals to the right.

\subsection{Absence of clock signals in the nested seed}
\label{app:C_nested_signals}
\label{app:C_poles}

We examine the clock-signal content of the nested seed in
Eq.~\eqref{appeq:C_nested_branch_Mellin}, beginning with
$0<r_1<r_2<1$. In this region, the time-kernel expansion
excludes local clock phases, while the spectral residues
that could produce nonlocal oscillations cancel in pairs.
We establish these two results in turn below.
The region $r_1>r_2$ is treated in
Section~\ref{app:C_local}, while the explicit calculation of
signals from the factorized seed is deferred to
Appendix~\ref{appendixD}.

\subsubsection{Local clock signals}
\label{app:C_local_signals}
The time kernel
$\mathcal G\!\left(\Delta,\beta;\frac{r_1}{r_2}\right)$
depends on momenta through $r_1/r_2=k_{34}/k_{12}$.
For $r_1/r_2<1$, it admits the expansion
\begin{equation}
 \mathcal G\!\left(\Delta,\beta;\frac{r_1}{r_2}\right)
 =\sum_{j=0}^{\infty}
 \frac{\Gamma(\Delta+j)}{j!(\beta+j)}\left(-\frac{r_1}{r_2}\right)^j.
 \label{appeq:C_Gauss_series}
\end{equation}
Together with the prefactor
$r_1^{-2}(r_1/r_2)^{5+q_2}$, this series gives only integer
powers of $r_1$ and $r_1/r_2$ for the time weights used here.
Any mass-dependent oscillation must therefore come from
$r_1^{-z_{1234}}$ at a pole with
$\operatorname{Im}z_{1234}\ne0$. Since $r_1=k_s/k_{12}$,
such an oscillation is nonanalytic in $k_s$ and belongs
to the nonlocal signal.
The local part of the nested residue expansion therefore
vanishes in this region:
\begin{equation}
 \left.\mathcal I_{\pm\pm,\TO}^{q_1q_2}\right|_{\rm LS}=0,
 \qquad 0<r_1<r_2<1.
 \label{appeq:C_nested_local_zero}
\end{equation}
The remaining candidate nonlocal contributions are
examined next.

\subsubsection{Nonlocal clock signals}
\label{app:C_nonlocal}
\label{app:C_residues}

\paragraph{Selection of nonlocal poles.}
The poles in the MB representation of the nested seed
arise from three sources.

\noindent\textit{(1) Mode-function poles.}
Since the regularized hypergeometric factor in the mode kernel
is entire in its parameters~\cite{NIST:DLMF}, the two
Gamma functions in Eq.~\eqref{appeq:C_Whittaker_Mellin}
give the candidate spectral poles
\begin{equation}
 z_j^{(\rho_j,n_j)}=-\frac12-\ii\rho_j\tnu-n_j,
 \qquad \rho_j=\pm1,\quad n_j=0,1,\ldots,
 \quad j=1,2,3,4,
 \label{appeq:C_mode_poles}
\end{equation}
forming two left pole families for each Mellin variable.
The corresponding spectral residues can contribute to
the nonlocal signal through $r_1^{-z_{1234}}$ only if
$\operatorname{Im}z_{1234}\ne0$.
The four mode poles then give
$\operatorname{Im}z_{1234}=\pm2\tnu$ or $\pm4\tnu$.

\noindent\textit{(2) Loop-integral poles.}
The loop kernels $J_0$ and $J_1$ in
Eq.~\eqref{appeq:C_angular_kernels} give the candidate pole families
\begin{align}
 z_{1234}&=1-2n
 &&\text{(UV)},\nonumber\\
 z_{13}\ \text{or}\ z_{24}&=2+2n
 &&\text{(from $J_0$)},\nonumber\\
 z_{13}\ \text{or}\ z_{24}&=3+2n
 &&\text{(from $J_1$)},
 \qquad n=0,1,\ldots.
 \label{appeq:C_loop_pole_families}
\end{align}
At a UV pole $z_{1234}=1-2n$, the factor
$r_1^{-z_{1234}}$ becomes $r_1^{2n-1}$ and hence
cannot generate a mass-dependent oscillation in $r_1$. The UV residues therefore do not contribute to the nonlocal clock signal.
The other two loop-pole families, in $z_{13}$ and $z_{24}$,
lie to the right of the original Mellin contours and are
not picked up by the leftward contour used here.

\noindent\textit{(3) Time-integral poles.}
The time kernels produce poles through
$\Gamma(\Delta+j)$ and $(\beta+j)^{-1}$ in
Eq.~\eqref{appeq:C_Gauss_series}.
With $\Delta=q_{12}+8-z_{1234}$ and
$\beta=q_2+4-z_{12}$ or $q_2+4-z_{34}$, the candidate
locations are
\begin{equation}
 z_{1234}=q_{12}+8+j+m,\qquad
 z_{12}\ \text{or}\ z_{34}=q_2+4+j,\qquad j,m\geq0.
 \label{appeq:C_time_parameter_poles}
\end{equation}
These poles also lie to the right of the original
Mellin contours.

The nonlocal signal is therefore determined by the
four families of spectral poles with
$\operatorname{Im}z_{1234}\ne0$.
We next show that the spectral contributions related by
$(z_1,z_2)\leftrightarrow(z_3,z_4)$ cancel in pairs,
so that the nested seed contains no nonlocal signal.

\paragraph{Pairing the spectral residues.}
We observe that the time-kernel difference in
Eq.~\eqref{appeq:C_nested_branch_Mellin} changes sign under
$(z_1,z_2)\leftrightarrow(z_3,z_4)$, which interchanges $z_{12}$ and $z_{34}$ without changing $\Delta$. Since the spectral residue sum includes all pole combinations of the four Mellin variables, exchanging the pole
labels at the two vertices merely relabels the summed configurations.
Therefore this antisymmetry suggests that the nonlocal spectral
contributions may cancel in pairs, provided the remaining
residue factors are invariant under the same exchange.

In the following we implement this pairing explicitly. First, we label a choice of the four
spectral poles $z_j=z_j^{(\rho_j,n_j)}$ by $P$,
\begin{equation}
 P=\bigl\{(\rho_1,n_1),(\rho_2,n_2),
 (\rho_3,n_3),(\rho_4,n_4)\bigr\}.
\end{equation}
With $n_{1234}=\sum_{j=1}^4n_j$ and
$\rho_{1234}=\sum_{j=1}^4\rho_j$, we have
\begin{align}
 z_{1234}=-2-n_{1234}-\ii\rho_{1234}\tnu,
 \qquad \operatorname{Im}z_{1234}=-\rho_{1234}\tnu.
 \label{appeq:C_spectral_frequency}
\end{align}
For the nonlocal configurations,
$\rho_{1234}=\pm2,\pm4$. The vertex exchange acts on
$P$ through the involution
\begin{equation}
 \mathfrak I P=\bigl\{(\rho_3,n_3),(\rho_4,n_4),
 (\rho_1,n_1),(\rho_2,n_2)\bigr\},
 \qquad \mathfrak I^2P=P.
 \label{appeq:C_nonlocal_pairing}
\end{equation}
At fixed external momenta and time weights,
$\mathfrak I$ preserves $n_{1234},\rho_{1234}$ and
$z_{13},z_{24},z_{1234}$. Then we denote the time-kernel difference at $P$ by
\begin{align}
 \mathcal K_{\TO}(P)\equiv{}
 \mathcal G\!\left(\Delta,q_2+4-z_{12};\frac{r_1}{r_2}\right)-\mathcal G\!\left(\Delta,q_2+4-z_{34};\frac{r_1}{r_2}\right),
 \label{appeq:C_time_kernel_involution}
\end{align}
satisfying $\mathcal K_{\TO}(\mathfrak I P)={}-\mathcal K_{\TO}(P)$.

Next we define the helicity-summed residue coefficient as
\begin{align}
 \mathcal C(P)\equiv\sum_{s_1,s_2}\operatorname*{Res}_{P}
 \Bigl\{&\mathcal V^L_{s_1s_2}(z_1,z_2)
 \mathcal V^R_{s_1s_2}(z_3,z_4)\bigl[J_0(A,B)+s_1s_2J_1(A,B)\bigr]\Bigr\},
 \label{appeq:C_helicity_sum}
\end{align}
where $\operatorname*{Res}_{P}$ takes the residues at
$z_j=z_j^{(\rho_j,n_j)}$.
Using the Residue Theorem and substituting into Eq.~\eqref{appeq:C_nested_branch_Mellin},
we obtain the nested seed integral from nonlocal spectral poles,
\begin{align}
 \left.\mathcal I_{\pm\pm,\TO}^{q_1q_2}\right|_{\rm NS,spec}
 ={}&\frac{\mp1}{16\pi^{3/2}r_1^2}
 \left(\frac{r_1}{r_2}\right)^{5+q_2}
 \sum_{P:\,\rho_{1234}\ne0}
 2^{-z_{1234}}r_1^{-z_{1234}}
 \nonumber\\
 &\times e^{\mp\ii\pi\Delta/2}
 \mathcal K_{\TO}(P)\mathcal C(P).
 \label{appeq:C_nonlocal_spectral_sum}
\end{align}
Pairing $P$ with $\mathfrak I P$ and using
$\mathcal K_{\TO}(\mathfrak I P)=-\mathcal K_{\TO}(P)$ gives
\begin{align}
 \left.\mathcal I_{\pm\pm,\TO}^{q_1q_2}\right|_{\rm NS,spec}
 ={}&\frac{\mp1}{32\pi^{3/2}r_1^2}
 \left(\frac{r_1}{r_2}\right)^{5+q_2}
 \sum_{P:\,\rho_{1234}\ne0}
 2^{-z_{1234}}r_1^{-z_{1234}}
 \nonumber\\
 &\times e^{\mp\ii\pi\Delta/2}\mathcal K_{\TO}(P)
 \bigl[\mathcal C(P)-\mathcal C(\mathfrak I P)\bigr].
 \label{appeq:C_involution_averaged_sum}
\end{align}
The factor $1/2$ compensates for counting each pair twice
in the sum over all $P$.
Equation~\eqref{appeq:C_involution_averaged_sum} shows that
the nonlocal spectral sum vanishes if
$\mathcal C(\mathfrak I P)=\mathcal C(P)$.
We will establish this symmetry below using relations among
the mode residues.

\paragraph{Relations among mode residues.}
To evaluate $\mathcal C(P)$, we need the spectral-pole
residues of the individual mode functions in their MB
representation.
The relations below express conjugated and negative-helicity
residues in terms of unconjugated, positive-helicity
coefficients. These relations allow us to extract from the product of
the two vertex residues a common mode-function factor
that is symmetric under $\mathfrak I$.
We then combine the remaining prefactors with the loop
kernels $J_0$ and $J_1$ in the helicity sum to establish
$\mathcal C(\mathfrak I P)=\mathcal C(P)$.
For a mode $f_s$, with $f\in\{u,v\}$, let
$C_{f_s,n_j}^{(\rho_j)}$ denote the residue of its Mellin
kernel at $z_j=z_j^{(\rho_j,n_j)}$:
\begin{align}
 C_{f_s,n_j}^{(\rho_j)}
 &\equiv\operatorname*{Res}_{z_j=z_j^{(\rho_j,n_j)}}
 \bigl[\mathfrak n_{f,s}e^{\ii\pi z_j/2}
 \mathcal W_{\kappa_{f,s}}(z_j)\bigr]\nonumber\\
 &=\mathfrak n_{f,s}\frac{(-1)^{n_j}}{n_j!}
 e^{-\ii\pi(1/2+\ii\rho_j\tnu+n_j)/2}
 \Gamma(-n_j-2\ii\rho_j\tnu)\,
 {}_2\mathbf F_1\!\left[
 \begin{matrix}
 -n_j,\,-n_j-2\ii\rho_j\tnu\\
 \frac12-\kappa_{f,s}-\ii\rho_j\tnu-n_j
 \end{matrix};\frac12\right].
 \label{appeq:C_residue_coefficients}
\end{align}
Then the mode functions can be expressed as a sum over
spectral-pole residues, equivalently as their late-time
expansion:
\begin{equation}
 f_s(p,-\eta)
 \sim\sum_{\rho=\pm1}\sum_{n\geq0}
 C_{f_s,n}^{(\rho)}(2p\eta)^{n+\ii\rho\tnu}.
 \label{appeq:C_mode_residue_expansion}
\end{equation}
The leading residues reproduce the late-time coefficients
used in Eq.~\eqref{latetimeexp},
$C_{f_s,0}^{(\rho)}=C_{f_s}^{(\rho)}$.
At $n=0$, one upper hypergeometric parameter vanishes,
so Eq.~\eqref{appeq:C_residue_coefficients} gives
\begin{equation}
 C_{f_s,0}^{(\rho)}
 =\mathfrak n_{f,s}e^{-\ii\pi(1/2+\ii\rho\tnu)/2}
 \frac{\Gamma(-2\ii\rho\tnu)}
      {\Gamma(1/2-\kappa_{f,s}-\ii\rho\tnu)}.
 \label{appeq:C_leading_residue}
\end{equation}
Conjugating the mode expansion reverses the oscillatory
exponent, so the coefficients of the conjugated modes are
\begin{equation}
 \overline C_{f_s,n}^{(\rho)}
 =\bigl(C_{f_s,n}^{(-\rho)}\bigr)^*.
 \label{appeq:C_conjugate_residues}
\end{equation}
To relate the mode residues at successive orders,
we substitute the expansion
\eqref{appeq:C_mode_residue_expansion} into the coupled
first-order mode equations \eqref{appeq:coupled_first_order}
and match powers of $2p\eta$.
Solving the resulting two equations for the order-$n$
coefficients gives, for $n\geq1$,
\begin{align}
 C_{u_s,n}^{(\rho)}
 &=-\frac{\ii s}{2n(n+2\ii\rho\tnu)}
 \Bigl[
 (n+\ii\rho\tnu+\ii\tlam)C_{u_s,n-1}^{(\rho)}
 -\ii\mR C_{v_s,n-1}^{(\rho)}
 \Bigr],\nonumber\\
 C_{v_s,n}^{(\rho)}
 &=\frac{\ii s}{2n(n+2\ii\rho\tnu)}
 \Bigl[
 (n+\ii\rho\tnu-\ii\tlam)C_{v_s,n-1}^{(\rho)}
 -\ii\mR C_{u_s,n-1}^{(\rho)}
 \Bigr].
 \label{appeq:C_mode_recurrence}
\end{align}
For $\mR>0$, Eq.~\eqref{appeq:C_leading_residue} and
the definition of $r_\rho$ in Eq.~\eqref{appeq:r_rho_def}
give the leading-order relations
\begin{align*}
 \frac{C_{f_-,0}^{(\rho)}}{C_{f_+,0}^{(\rho)}}
 &=r_\rho,\qquad f\in\{u,v\},\\
 \frac{\overline C_{u_+,0}^{(\rho)}}{C_{v_+,0}^{(\rho)}}
 &=-\frac{\overline C_{v_+,0}^{(\rho)}}{C_{u_+,0}^{(\rho)}}
 =\ii e^{\pi(\tlam-\rho\tnu)}r_\rho,
\end{align*}
where 
$r_\rho=e^{-\pi\tlam}\frac{\rho\tnu-\tlam}{\mR}
\frac{\Gamma[-\ii(\rho\tnu-\tlam)]}
{\Gamma[-\ii(\rho\tnu+\tlam)]}$.
The two helicity ratios are equal at $n=0$.
If they are equal at order $n-1$, their common value
factors out of both brackets in
Eq.~\eqref{appeq:C_mode_recurrence}.
Changing $s=+1$ to $s=-1$ also reverses the overall sign,
so
\[
 \frac{C_{f_-,n}^{(\rho)}}{C_{f_+,n}^{(\rho)}}
 =-\frac{C_{f_-,n-1}^{(\rho)}}{C_{f_+,n-1}^{(\rho)}},
 \qquad f\in\{u,v\}.
\]
Thus the two ratios remain equal and change sign at each
order. Their initial value $r_\rho$ gives
$C_{f_-,n}^{(\rho)}/C_{f_+,n}^{(\rho)}=(-1)^n r_\rho$.

For the conjugation relations, conjugating
Eq.~\eqref{appeq:C_mode_recurrence} at $-\rho$ and $s=+1$,
and using Eq.~\eqref{appeq:C_conjugate_residues}, gives
\begin{align*}
 \overline C_{u_+,n}^{(\rho)}
 &=\frac{\ii}{2n(n+2\ii\rho\tnu)}
 \Bigl[
 (n+\ii\rho\tnu-\ii\tlam)\overline C_{u_+,n-1}^{(\rho)}
 +\ii\mR\overline C_{v_+,n-1}^{(\rho)}
 \Bigr],\\
 \overline C_{v_+,n}^{(\rho)}
 &=-\frac{\ii}{2n(n+2\ii\rho\tnu)}
 \Bigl[
 (n+\ii\rho\tnu+\ii\tlam)\overline C_{v_+,n-1}^{(\rho)}
 +\ii\mR\overline C_{u_+,n-1}^{(\rho)}
 \Bigr].
\end{align*}
The leading-order conjugation relations give
$\overline C_{u_+,0}^{(\rho)}/C_{v_+,0}^{(\rho)}
=-\overline C_{v_+,0}^{(\rho)}/C_{u_+,0}^{(\rho)}$.
If these signed ratios are equal at order $n-1$,
their common value factors out of the first bracket,
leaving the bracket for $C_{v_+,n}^{(\rho)}$ in
Eq.~\eqref{appeq:C_mode_recurrence}.
Factoring it out of the second bracket instead leaves
the negative of the bracket for $C_{u_+,n}^{(\rho)}$.
Hence
\[
 \frac{\overline C_{u_+,n}^{(\rho)}}{C_{v_+,n}^{(\rho)}}
 =-\frac{\overline C_{v_+,n}^{(\rho)}}{C_{u_+,n}^{(\rho)}}
 =\frac{\overline C_{u_+,n-1}^{(\rho)}}{C_{v_+,n-1}^{(\rho)}}
 =-\frac{\overline C_{v_+,n-1}^{(\rho)}}{C_{u_+,n-1}^{(\rho)}}.
\]
Combining the helicity and conjugation results yields,
for all $n\geq0$,
\begin{align}
 C_{f_-,n}^{(\rho)}
 &=(-1)^n r_\rho C_{f_+,n}^{(\rho)},
 \qquad f\in\{u,v\},\nonumber\\
 \overline C_{u_+,n}^{(\rho)}
 &=\ii e^{\pi(\tlam-\rho\tnu)}r_\rho C_{v_+,n}^{(\rho)},\nonumber\\
 \overline C_{v_+,n}^{(\rho)}
 &=-\ii e^{\pi(\tlam-\rho\tnu)}r_\rho C_{u_+,n}^{(\rho)}.
 \label{appeq:C_helicity_residue_relation}
\end{align}

\begin{samepage}
\paragraph{Vertex residues and helicity sum.}
We now use Eq.~\eqref{appeq:C_helicity_residue_relation}
to express the two vertex residues in terms of
unconjugated, positive-helicity coefficients.
For $s_1=s_2=+1$, denote the left-vertex residue by
$\mathcal A(P)$,
\begin{equation}
  \mathcal A(P)\equiv
  C_{u_+,n_1}^{(\rho_1)}C_{u_+,n_2}^{(\rho_2)}
  -C_{v_+,n_1}^{(\rho_1)}C_{v_+,n_2}^{(\rho_2)}.
\end{equation}
Applying the conjugation relations
to the right vertex gives
\begin{align}
 \overline C_{u_+,n_3}^{(\rho_3)}
  \overline C_{u_+,n_4}^{(\rho_4)}
 -\overline C_{v_+,n_3}^{(\rho_3)}
  \overline C_{v_+,n_4}^{(\rho_4)}=
 e^{\pi[2\tlam-(\rho_3+\rho_4)\tnu]}
 r_{\rho_3}r_{\rho_4}\mathcal A(\mathfrak I P).
 \label{appeq:C_vertex_residue_factors}
\end{align}
\end{samepage}
To include negative helicities, we can conjugate the first
relation in Eq.~\eqref{appeq:C_helicity_residue_relation}
at $-\rho$ and obtain
$\overline C_{f_-,n}^{(\rho)}
=(-1)^n r_{-\rho}^*\overline C_{f_+,n}^{(\rho)}$.
The mode residues on the first internal line are
labelled $1,3$, and those on the second are labelled $2,4$.
Changing a line's helicity from $+1$ to $-1$ therefore
multiplies its two mode residues by
\begin{equation}
 w_{13}=(-1)^{n_1+n_3}r_{\rho_1}r_{-\rho_3}^*,
 \qquad
 w_{24}=(-1)^{n_2+n_4}r_{\rho_2}r_{-\rho_4}^*.
 \label{appeq:C_line_helicity_factors}
\end{equation}
Since the bubble includes contributions from both helicities
of each internal fermion line, we now perform the helicity sum to obtain $\mathcal C(P)$ in Eq.~\eqref{appeq:C_helicity_sum}.
The four choices $(s_1,s_2)=(+,+),(-,+),(+,-),(-,-)$
have weights $1,w_{13},w_{24},w_{13}w_{24}$, respectively.
Summing these weights with
$J_0(A,B)+s_1s_2J_1(A,B)$ gives
\begin{align}
 \mathcal C(P)={}&
 e^{\pi[2\tlam-(\rho_3+\rho_4)\tnu]}
 r_{\rho_3}r_{\rho_4}
 \mathcal A(P)\mathcal A(\mathfrak I P)
 \nonumber\\
 &\times\Bigl\{
 J_0(A,B)(1+w_{13})(1+w_{24})+
 J_1(A,B)(1-w_{13})(1-w_{24})
 \Bigr\}.
 \label{appeq:C_helicity_sum_factored}
\end{align}
Under $\mathfrak I$, the product
$\mathcal A(P)\mathcal A(\mathfrak I P)$ is unchanged.
The loop kernels are also unchanged because
$A=-1-z_{13}$ and $B=-1-z_{24}$ are preserved.
It remains to check the exponential and $r_\rho$ factors
together with the $J_0,J_1$-weighted helicity sum.
We do this separately for the two possible nonlocal
frequencies.

\paragraph{Frequency $4\tnu$.}
For $\rho_{1234}=\pm4$, all four spectral signs $\rho_i$ are equal.
Exchanging the vertices therefore leaves the prefactor
$e^{\pi[2\tlam-(\rho_3+\rho_4)\tnu]}r_{\rho_3}r_{\rho_4}$
unchanged.
The factors $w_{13}$ and $w_{24}$ are unchanged as well:
they depend only on the common spectral sign and the
preserved sums $n_1+n_3$ and $n_2+n_4$.
Equation~\eqref{appeq:C_helicity_sum_factored}
is therefore invariant, proving
$\mathcal C(\mathfrak I P)=\mathcal C(P)$ in this sector.

\paragraph{Frequency $2\tnu$.}
For $\rho_{1234}=\pm2$, one spectral sign is opposite
to the other three. The two vertices therefore carry
opposite signs on one internal line and equal signs
on the other.
By symmetry between the lines, we may take
$\rho_1=\rho$, $\rho_3=-\rho$, and $\rho_2=\rho_4$.
At these spectral poles, the loop exponents become
\[
 A=-1-z_{13}=n_1+n_3,\qquad
 B=-1-z_{24}=n_2+n_4+2\ii\rho_2\tnu.
\]
Here $A$ is a nonnegative integer, while $B$ has a
nonzero imaginary part.
Observing that the factors $1/\Gamma(-A/2)$ and
$1/\Gamma[(1-A)/2]$ in Eq.~\eqref{appeq:C_angular_kernels}
vanish for even and odd $A$, respectively, we have
\begin{equation}
 J_0(A,B)=0\quad(A\ \text{even}),\qquad
 J_1(A,B)=0\quad(A\ \text{odd}).
 \label{appeq:C_parity_zeros}
\end{equation}

Using $\rho_3=-\rho_1$ in
Eq.~\eqref{appeq:C_line_helicity_factors} gives
$w_{13}=(-1)^A|r_\rho|^2$.
Together with the zeros in Eq.~\eqref{appeq:C_parity_zeros},
this reduces the helicity sum of the loop kernels to
\begin{align*}
 &J_0(A,B)(1+w_{13})(1+w_{24})
  +J_1(A,B)(1-w_{13})(1-w_{24})\\
 &\qquad=(1-|r_\rho|^2)
 \begin{cases}
  J_1(A,B)(1-w_{24}),& A\ \text{even},\\
  J_0(A,B)(1+w_{24}),& A\ \text{odd}.
 \end{cases}
\end{align*}
Thus the loop kernels select different terms for even
and odd $A$, but the helicity weight from the first
line is $1-|r_\rho|^2$ in both cases.
The factors $J_0(A,B)$, $J_1(A,B)$ and $w_{24}$ are
invariant under $\mathfrak I$, so the helicity sum
transforms only through $1-|r_\rho|^2$.

Under $\mathfrak I$, $\rho_3=-\rho$ is replaced by
$\rho$, and the first-line helicity weight becomes
$1-|r_{-\rho}|^2$.
The conjugation factor from the second line is unchanged
because $\rho_2=\rho_4$.
Since $\mathcal A(P)\mathcal A(\mathfrak I P)$ is also
invariant, it suffices to verify
\begin{equation}
 e^{\pi(\tlam-\rho\tnu)}r_\rho(1-|r_{-\rho}|^2)
 =e^{\pi(\tlam+\rho\tnu)}r_{-\rho}(1-|r_\rho|^2).
 \label{appeq:C_connection_identity}
\end{equation}
To verify this equality, we can evaluate $|r_\rho|^2$ and
$r_\rho/r_{-\rho}$ from Eq.~\eqref{appeq:r_rho_def}.
Using $|\Gamma(\ii y)|^2=\pi/[y\sinh(\pi y)]$,
the Gamma reflection formula, and
$\tnu^2-\tlam^2=\mR^2$ then gives
\begin{equation}
 |r_\rho|^2=e^{-2\pi\tlam}
 \frac{\sinh\pi(\tnu+\rho\tlam)}{\sinh\pi(\tnu-\rho\tlam)},
 \qquad
 \frac{r_\rho}{r_{-\rho}}
 =-\frac{\sinh\pi(\tnu+\rho\tlam)}{\sinh\pi(\tnu-\rho\tlam)}.
 \label{appeq:C_connection_moduli}
\end{equation}
Substitution into Eq.~\eqref{appeq:C_connection_identity}
verifies the equality and hence establishes
$\mathcal C(\mathfrak I P)=\mathcal C(P)$ in the
$2\tnu$ sector.

\paragraph{Cancellation of the spectral residues.}
Since $\mathcal C(\mathfrak I P)=\mathcal C(P)$ for all
$\rho_{1234}\ne0$, the paired sum in
Eq.~\eqref{appeq:C_involution_averaged_sum} vanishes.
The nested seed integral therefore has no nonlocal
clock signal from the spectral poles,
\begin{equation}
 \left.\mathcal I_{\pm\pm,\TO}^{q_1q_2}\right|_{\rm NS}=0,
 \qquad 0<r_1<r_2<1.
 \label{appeq:C_nested_nonlocal_zero}
\end{equation}
Together with Eq.~\eqref{appeq:C_nested_local_zero},
this excludes both types of CC clock signal from the
nested term obtained from
Eq.~\eqref{appeq:C_decomposition_less} for $0<r_1<r_2<1$.
For $r_1>r_2$, the seed integral is instead split using
Eq.~\eqref{appeq:C_decomposition_greater}. In Section~\ref{app:C_local} we will explain why this change of
nested time ordering is needed to isolate the signals
in the factorized part.

\subsection{Dependence on the momentum region}
\label{app:C_local}

Recall that, for $r_1<r_2$, the same-branch seed is
split using Eq.~\eqref{appeq:C_decomposition_less}:
the Wightman term $\mathcal Q_{\mp\pm}$ gives the
factorized contribution, while the nested term integrates
over $0<\eta_2<\eta_1$.
For $r_1>r_2$, Eq.~\eqref{appeq:C_decomposition_greater}
instead uses $\mathcal Q_{\pm\mp}$ and the nested domain
$0<\eta_1<\eta_2$.
Both identities are exact for arbitrary external energies,
but continuing the first seed decomposition to $r_1>r_2$
can leave local clock contributions in its nested term.
To identify these contributions, we retain
Eq.~\eqref{appeq:C_decomposition_less} and continue the
time kernels of its nested term from $r_1<r_2$ to $r_1>r_2$.
The pole structure of the time kernel is explicit in the Barnes
representation
\begin{equation}
 \mathcal G\!\left(\Delta,\beta;\frac{r_1}{r_2}\right)
 =\int_{\mathcal C_\xi}\frac{\dd\xi}{2\pi\ii}\,
 \frac{\Gamma(-\xi)\Gamma(\Delta+\xi)}{\beta+\xi}\left(\frac{r_1}{r_2}\right)^\xi.
 \label{appeq:C_time_Barnes}
\end{equation}
This follows by inserting the Barnes representation
of $\left(1+\frac{r_1}{r_2}t\right)^{-\Delta}$
into Eq.~\eqref{appeq:C_nested_time_kernel} and using
$\int_0^1\dd t\,t^{\beta+\xi-1}=1/(\beta+\xi)$
\cite{NIST:DLMF}.
Convergence at $t=0$ requires
$\operatorname{Re}(\beta+\xi)>0$.
The integrand has poles
\begin{equation}
 \begin{aligned}
  \xi&=j &&\text{from }\Gamma(-\xi),\\
  \xi&=-\Delta-j &&\text{from }\Gamma(\Delta+\xi),\\
  \xi&=-\beta &&\text{from }(\beta+\xi)^{-1},
 \end{aligned}
 \qquad j=0,1,\ldots.
 \label{appeq:C_time_poles}
\end{equation}
For $\operatorname{Re}\Delta,\operatorname{Re}\beta>0$,
we take $\mathcal C_\xi$ upward, with $\xi=-\Delta-j$
to its left and $\xi=j$ to its right, so the contour can
initially be chosen in the strip
\[
 -\min\{\operatorname{Re}\Delta,\operatorname{Re}\beta\}
 <\operatorname{Re}\xi<0.
\]

For $r_1<r_2$, closing to the right picks up the poles $\xi=j$ only and reproduces Eq.~\eqref{appeq:C_Gauss_series}.
For $r_1>r_2$, closing to the left instead includes
$\xi=-\beta$ and $\xi=-\Delta-j$, leading to
\begin{align}
 \mathcal G\!\left(\Delta,\beta;\frac{r_1}{r_2}\right)
 ={}\Gamma(\beta)\Gamma(\Delta-\beta)
 \left(\frac{r_1}{r_2}\right)^{-\beta}-\left(\frac{r_1}{r_2}\right)^{-\Delta}
 \mathcal G\!\left(\Delta,\Delta-\beta;\frac{r_2}{r_1}\right).
 \label{appeq:C_time_connection}
\end{align}
Here the first term comes from the residue of the isolated-pole $\xi=-\beta$ and the second term comes from the residue sum at $\xi=-\Delta-j$.

Now the critical observation is that the first term may contribute to the local clock signal. For the two time kernels in
Eq.~\eqref{appeq:C_nested_branch_Mellin}, the isolated
poles are
\begin{equation}
 \xi=z_{12}-q_2-4,\qquad
 \xi=z_{34}-q_2-4,
 \label{appeq:C_wrong_region_poles}
\end{equation}
Hence their combined contribution to the time-kernel difference is
\begin{align}
 &\Gamma(q_2+4-z_{12})\Gamma(q_1+4-z_{34})
   \left(\frac{r_1}{r_2}\right)^{z_{12}-q_2-4}\nonumber\\
 &\qquad-
 \Gamma(q_2+4-z_{34})\Gamma(q_1+4-z_{12})
   \left(\frac{r_1}{r_2}\right)^{z_{34}-q_2-4}.
 \label{appeq:C_wrong_region_residues}
\end{align}
Here $\beta=q_2+4-z_{12}$ for the first term and
$\beta=q_2+4-z_{34}$ for the second.
At the UV pole $z_{1234}=1$, the bubble momentum
factor becomes $k_s^{1-z_{1234}}=1$, while the mode-pole
residues can give $\beta$ a nonzero imaginary part.
The factor $(r_1/r_2)^{-\beta}$ then carries an oscillatory
phase in the external-momentum ratio, allowing a local clock
signal in the nested seed for $r_1>r_2$.

Another key observation is that the second term in
Eq.~\eqref{appeq:C_time_connection} yields the nested seed
integral of the second decomposition,
Eq.~\eqref{appeq:C_decomposition_greater}.
The second decomposition therefore corresponds to adding the isolated-pole
contribution in Eq.~\eqref{appeq:C_time_connection}
to the factorized seed of the first decomposition.

From another perspective, the second decomposition at
$(r_1,r_2;q_1,q_2)$ is the first decomposition evaluated at
$(r_2,r_1;q_2,q_1)$, after relabeling the vertex times.
For $r_1>r_2$, the mapped ratio is $r_2/r_1<1$.
The result established in Section~\ref{app:C_nested_signals}
for the first decomposition therefore gives the cutting
rule for the second decomposition in $r_1>r_2$.
The parameter exchange yields the same second decomposition
as continuing the first through
Eq.~\eqref{appeq:C_time_connection} and reassigning the
isolated-pole contribution to the factorized part.
We therefore use Eq.~\eqref{appeq:C_decomposition_greater}
for $r_1>r_2$ and Eq.~\eqref{appeq:C_decomposition_less}
for $r_1<r_2$.

The cutting rule for the fermion bubble therefore reduces
the calculation of local and nonlocal CC signals to the
factorized seed $\mathcal I_{\rm F}^{q_1q_2}$ in
Eq.~\eqref{appeq:C_seed_decomposition}.
The decomposition is chosen as
Eq.~\eqref{appeq:C_decomposition_less} for $r_1<r_2$
and Eq.~\eqref{appeq:C_decomposition_greater} for $r_1>r_2$.
The corresponding nested same-branch seeds contain neither
type of clock signal and can therefore be omitted when
extracting the CC signals.
The time integrations at the two vertices can then be
performed independently.

\section{Signal Extraction from the alternative MB Representation}
\label{appendixD}

In this appendix, we compute the CC signals using the alternative
MB representation employed in the proof of the cutting rule in Appendix~\ref{appendixC}, providing a cross-check of the main-text
calculations in Sections~\ref{4} and~\ref{sec5}.
Starting from the factorized seed in
Eq.~\eqref{appeq:C_factorized_seed_Mellin},
Section~\ref{app:D_factorized_poles} analyzes its pole structure
and classifies the signal contributions from different pole families.
Section~\ref{app:D_squeezed_late_time} then justifies the
left-vertex late-time expansion used to extract the leading
single-squeezed CC signal. In Section~\ref{app:D_signal_comparison} we take the leading
spectral residues at the left vertex of the factorized seed and obtain the leading late-time expansion. Then we calculate the total signal through evaluating the remaining double MB integral numerically to cross-check the results obtained in the main text.
Finally, Section~\ref{app:D_component_comparison} evaluates
the nonlocal and local CC signals separately.
Together, these comparisons provide a numerical cross-check
between the two MB representations for both the total CC
signal and its nonlocal and local components.

\subsection{Pole structure of the factorized seed}
\label{app:D_factorized_poles}

We begin with the factorized seed integral obtained from the alternative MB representation in
Eq.~\eqref{appeq:C_factorized_seed_Mellin}, which for
$0<r_1<r_2<1$ reads
\begin{align}
 \mathcal I_{\mathrm F}^{q_1q_2}
 ={}&\frac{\ii}{8\pi^{3/2}}\sum_{s_1,s_2}
 \int\prod_{j=1}^4\frac{\dd z_j}{2\pi\ii}\,
 2^{-z_{1234}}
 \mathcal V^L_{s_1s_2}(z_1,z_2)
 \mathcal V^R_{s_1s_2}(z_3,z_4)\nonumber\\
 &\times[J_0(A,B)+s_1s_2J_1(A,B)]\nonumber\\
 &\times\Bigg\{
 r_1^{-1-z_{12}}r_2^{-1-z_{34}}
 \Gamma(q_1+4-z_{12})\Gamma(q_2+4-z_{34})\nonumber\\
 &\hspace{2.2em}\times e^{-\ii\pi(q_2+4-z_{34})/2}
 \sin\!\left[\frac\pi2(q_1+4-z_{12})\right]\nonumber\\
 &\hspace{1em}-r_1^{-1-z_{34}}r_2^{-1-z_{12}}
 \Gamma(q_1+4-z_{34})\Gamma(q_2+4-z_{12})\nonumber\\
 &\hspace{2.2em}\times e^{+\ii\pi(q_2+4-z_{12})/2}
 \sin\!\left[\frac\pi2(q_1+4-z_{34})\right]\Bigg\}.
 \label{appeq:D_factorized_seed}
\end{align}
Here $A=-1-z_{13}$, $B=-1-z_{24}$. The vertex
coefficients $\mathcal V^{L,R}$ are given by
Eq.~\eqref{appeq:C_vertex_Mellin}, and the loop kernels are
given in Eq.~\eqref{appeq:C_angular_kernels}. As in Section~\ref{app:C_poles}, the poles arise from the
mode functions, the loop integral, and the time integrals.

\paragraph{Mode-function poles.}
The two Gamma functions in the mode kernel give the spectral poles on the left of the contour,
\begin{equation}
 z_j=z_j^{(\rho_j,n_j)}
 =-\tfrac12-\ii\rho_j\tnu-n_j,
 \qquad \rho_j=\pm1,\quad n_j=0,1,\ldots,
 \quad j=1,2,3,4,
 \label{appeq:D_spectral_poles}
\end{equation}
with mode residues given in
Eqs.~\eqref{appeq:C_residue_coefficients} and
\eqref{appeq:C_conjugate_residues}. 

For the term proportional to
$r_1^{-1-z_{12}}r_2^{-1-z_{34}}$, four spectral residues give
\begin{equation}
 r_1^{-1-z_{12}}r_2^{-1-z_{34}}
 =r_1^{n_{12}}r_2^{n_{34}}
 r_1^{\ii\rho_{12}\tnu}r_2^{\ii\rho_{34}\tnu},
 \label{appeq:D_spectral_momentum}
\end{equation}
where $n_{12}=n_1+n_2$, $\rho_{12}=\rho_1+\rho_2$,
and similarly for the other summed subscripts.
For $\rho_{1234}\ne0$, the factor
$k_s^{\ii\rho_{1234}\tnu}$ gives a nonlocal clock phase, while for $\rho_{1234}=0$, any remaining
oscillation in the external-momentum ratio
$r_1/r_2=k_{34}/k_{12}$ corresponds to a local signal.
Table~\ref{tab:D_spectral_signals} summarizes these possibilities.
The second term in Eq.~\eqref{appeq:D_factorized_seed}
follows by interchanging $z_{12}$ and $z_{34}$ and has
the same classification.

\begin{table}[htbp]
 \centering
 \renewcommand{\arraystretch}{1.2}
 \begin{tabular}{cccc l}
 \toprule
 $\rho_{12}$ & $\rho_{34}$ & $\rho_{1234}$
 & Clock phase & Signal type \\
 \midrule
 $2\rho$ & $2\rho$ & $4\rho$
 & $(r_1r_2)^{2\ii\rho\tnu}$ & Nonlocal \\
 $2\rho$ & $0$ & $2\rho$
 & $r_1^{2\ii\rho\tnu}$ & Nonlocal \\
 $0$ & $2\rho$ & $2\rho$
 & $r_2^{2\ii\rho\tnu}$ & Nonlocal \\
 $2\rho$ & $-2\rho$ & $0$
 & $(r_1/r_2)^{2\ii\rho\tnu}$ & Local \\
 $0$ & $0$ & $0$
 & $1$ & nonoscillatory \\
 \bottomrule
 \end{tabular}
 \caption{Clock phases and signal types of the spectral-pole
 branches. Here $\rho=\pm1$ in each row, and the common
 integer-power factor $r_1^{n_{12}}r_2^{n_{34}}$ is omitted.}
 \label{tab:D_spectral_signals}
\end{table}

\paragraph{Loop-integral poles.}
The kernels $J_0,J_1$ give the candidate families
\begin{align}
 z_{1234}&=1-2n
 &&\text{(UV)},\nonumber\\
 z_{13}\ \text{or}\ z_{24}&=2+2m
 &&\text{(from $J_0$)},\nonumber\\
 z_{13}\ \text{or}\ z_{24}&=3+2m
 &&\text{(from $J_1$)},
 \qquad n,m=0,1,\ldots.
 \label{appeq:D_loop_poles}
\end{align}
The UV family lies to the left of the original contours.
The other two families arise from the soft endpoints of the
loop integral and lie to the right, so they are not picked up
when the contours are closed to the left.
After taking a UV residue at $z_{1234}=1-2n$, however,
the substitution $z_{24}=1-2n-z_{13}$ maps the right
$z_{24}$ families to the apparent left families
\begin{equation}
 z_{13}=-1-2n-2m\quad(J_0),\qquad
 z_{13}=-2-2n-2m\quad(J_1).
 \label{appeq:D_spurious_IR_poles}
\end{equation}
To determine whether these candidates survive, we evaluate
the UV residues of the full loop kernels. Using $B=2n-3-A$,
$\operatorname{Res}\Gamma[(z_{1234}-1)/2]=2(-1)^n/n!$,
and $\Gamma(a+n)/\Gamma(a)=(a)_n$ gives
\begin{align}
 \underset{z_{1234}=1-2n}{\operatorname{Res}}J_0
 &=\frac{(-1)^n}{n!\Gamma(n+3/2)}
 \left(\frac{A+3}{2}-n\right)_n
 \left(-\frac A2\right)_n,\nonumber\\
 \underset{z_{1234}=1-2n}{\operatorname{Res}}J_1
 &=\frac{(-1)^n}{n!\Gamma(n+3/2)}
 \left(\frac{A+4}{2}-n\right)_n
 \left(\frac{1-A}{2}\right)_n.
 \label{appeq:D_UV_loop_polynomials}
\end{align}
Both residues are polynomials in $A$. The reciprocal Gamma
factors therefore cancel the apparent left poles in
Eq.~\eqref{appeq:D_spurious_IR_poles}, as well as the original
right $z_{13}$ poles. None of these endpoint poles survives
in the UV loop residues.

At a UV pole, the bubble momentum factor becomes
$k_s^{1-z_{1234}}=k_s^{2n}$, so the UV residues carry
no nonlocal clock phase. To determine their dependence on
the external-momentum ratio, we take the left-vertex spectral
poles, which give $z_{12}=-1-n_{12}-\ii\rho_{12}\tnu$.
The UV condition then fixes
$z_{34}=2-2n+n_{12}+\ii\rho_{12}\tnu$, and hence
\begin{equation}
 \left.r_1^{-1-z_{12}}r_2^{-1-z_{34}}
 \right|_{\mathrm{UV},\,z_1,z_2\ \mathrm{spectral}}
 =r_2^{2n-3}
 \left(\frac{r_1}{r_2}\right)^{n_{12}+\ii\rho_{12}\tnu}.
 \label{appeq:D_UV_local_phase}
\end{equation}
Thus the UV residues can contribute to the local clock
signal for $\rho_{12}=\pm2$, whereas the
$\rho_{12}=0$ branch is nonoscillatory.

\paragraph{Time-integral poles.}
In the first term of Eq.~\eqref{appeq:D_factorized_seed},
the Gamma functions from the time integral give the candidate families
\begin{equation}
 z_{12}=q_1+4+j,\qquad
 z_{34}=q_2+4+j,\qquad j=0,1,\ldots.
 \label{appeq:D_original_time_poles}
\end{equation}
Both families lie to the right of the original contours
and therefore do not contribute when the contours are
closed to the left.

After taking a UV residue at $z_{1234}=1-2n$, however,
$z_{34}$ is fixed by $z_{12}$. Substituting
$z_{34}=1-2n-z_{12}$ into $\Gamma(q_2+4-z_{34})$ gives
\begin{align}
 \Gamma(q_2+4-z_{34})
 &=\Gamma(q_2+3+2n+z_{12}),\nonumber\\
 z_{12}&=-q_2-3-2n-j,\qquad j=0,1,\ldots.
 \label{appeq:D_mapped_time_poles}
\end{align}
The original right family in $z_{34}$ has thus become
a left family in $z_{12}$.
Unlike the loop-integral candidates in
Eq.~\eqref{appeq:D_spurious_IR_poles}, these poles
are not generally cancelled by the UV loop residues.
At a surviving pole, the momentum factor is
\begin{equation}
 \left.r_1^{-1-z_{12}}r_2^{-1-z_{34}}
 \right|_{\mathrm{UV},\,\mathrm{time}}
 =r_1^{q_2+2+2n+j}\,r_2^{-q_2-5-j}.
 \label{appeq:D_UV_time_background}
\end{equation}
For the physical time weights, both exponents are integers
and carry no mass-dependent clock phase. These time-pole
contributions therefore belong to the nonoscillatory
background. Exchanging $z_{12}$ and $z_{34}$ gives the
same conclusion for the second term of
Eq.~\eqref{appeq:D_factorized_seed}.

\subsection{Single-squeezed limit and the left-vertex late-time expansion}
\label{app:D_squeezed_late_time}

In the single-squeezed limit $r_1\to0$ at fixed
$0<r_2<1$, we can keep $k_s$ and $k_{34}$ fixed and
take $k_{12}\to\infty$. The external phase
$e^{\pm\ii k_{12}\eta_1}$ then oscillates rapidly,
so contributions from times much earlier than
$k_{12}^{-1}$ tend to cancel. The characteristic time
scale is therefore $\eta_1\sim k_{12}^{-1}$, at which
the left-vertex mode arguments satisfy, for fixed
internal momenta,
\begin{equation*}
 p_j\eta_1\sim\frac{p_j}{k_{12}}
 =\frac{p_j}{k_s}r_1\longrightarrow0,
 \qquad j=1,2.
\end{equation*}
This motivates the leading late-time approximation at the
left vertex. We show below that the leading oscillatory
contribution in the single-squeezed limit is reproduced by
the leading late-time expansion.

The MB representation in Eq.~\eqref{appeq:D_factorized_seed}
allows us to establish this correspondence after the loop
integration has been performed. Since the leading spectral
residues already reproduce the leading late-time coefficients,
as shown in Eqs.~\eqref{appeq:C_mode_residue_expansion}
and~\eqref{appeq:C_leading_residue},
it remains to show that they determine the leading oscillatory
powers of $r_1$, including the local signal from the UV poles.
We first consider the term proportional to
$r_1^{-1-z_{12}}r_2^{-1-z_{34}}$; the other term follows
by exchanging the two Mellin blocks.

\paragraph{Spectral poles and squeezed powers.}
At the left-vertex spectral poles in
Eq.~\eqref{appeq:D_spectral_poles}, the $r_1$ factor becomes
\begin{equation}
 r_1^{-1-z_{12}}
 =r_1^{n_{12}+\ii\rho_{12}\tnu},
 \qquad n_{12}=n_1+n_2,\quad \rho_{12}=\rho_1+\rho_2,
 \label{appeq:D_squeezed_spectral_order}
\end{equation}
and the remaining factors are independent of $r_1$.
For the same-sign branches $\rho_1=\rho_2=\rho$, the
clock factor is $r_1^{2\ii\rho\tnu}$ multiplied by
$r_1^{n_{12}}$. Its leading coefficient is therefore
selected by $n_{12}=0$, which requires $n_1=n_2=0$.
Every higher left-vertex spectral order carries at least
one additional power of $r_1$. The opposite-sign branches
have $\rho_{12}=0$ and give no oscillation in $r_1$.

\paragraph{UV contributions.}
We should also check that the loop UV condition preserves this
ordering. At $z_{1234}=1-2n$, the left spectral poles
fix $z_{34}=2-2n+n_{12}+\ii\rho_{12}\tnu$.
For $\rho_{12}=2\rho$, Eq.~\eqref{appeq:D_UV_local_phase}
can consequently be written as
\begin{align}
 \left.r_1^{-1-z_{12}}r_2^{-1-z_{34}}\right|_{\mathrm{UV}}
 &=r_2^{2n-3}
 \left(\frac{r_1}{r_2}\right)^{n_{12}+2\ii\rho\tnu}
 \nonumber\\
 &=r_1^{n_{12}+2\ii\rho\tnu}
 r_2^{2n-3-n_{12}-2\ii\rho\tnu}.
 \label{appeq:D_UV_squeezed_order}
\end{align}
The UV level $n$ changes the $r_2$ dependence but leaves
the $r_1$ order controlled by $n_{12}$. Thus, in the
single-squeezed limit $r_1\to0$ at fixed $r_2$, the
leading local clock contribution from the UV residues
is selected by $n_1=n_2=0$. Higher left spectral orders
remain suppressed by additional powers of $r_1$.

The remaining pole families do not add a leading clock
term. As shown in Section~\ref{app:D_factorized_poles},
the loop endpoint poles are right poles, and their
apparent left counterparts cancel in the UV residues
\eqref{appeq:D_UV_loop_polynomials}. The time poles
mapped to the left by the UV condition instead satisfy
\begin{equation}
 z_{12}=-q_2-3-2n-j
 \quad\Longrightarrow\quad
 r_1^{-1-z_{12}}=r_1^{q_2+2+2n+j},
 \qquad j=0,1,\ldots.
 \label{appeq:D_squeezed_background_order}
\end{equation}
For $q_2=-1,-3$, these are integer powers and belong
to the background.

The leading left-vertex late-time approximation therefore
reproduces the leading single-squeezed clock contribution.
Exchanging the Mellin blocks gives the same result
for the second term of Eq.~\eqref{appeq:D_factorized_seed}
and hence for the complete factorized seed used in
Eq.~\eqref{JF}, including its nonlocal and local
signals. In the double-squeezed regime $r_1,r_2\ll1$,
an analogous argument justifies the leading late-time
approximation at the right vertex as well.

\subsection{Total signal and the three-point limit}
\label{app:D_signal_comparison}

We now use the alternative MB representation to compute
the leading single-squeezed signal in Eq.~\eqref{JF}.
Following the main text, we retain the same-sign
left-vertex branches $\rho_1=\rho_2=\rho$.
Taking their leading spectral residues, $n_1=n_2=0$,
in Eq.~\eqref{appeq:D_factorized_seed} gives
\begin{equation}
 \mathcal J_{\mathrm F}^{q_1q_2}(r_1,r_2)
 =2\operatorname{Re}\sum_{\rho=\pm1}
 r_1^{2\ii\rho\tnu}
 \widetilde{\mathcal T}_\rho^{q_1q_2}(r_2),
 \label{appeq:D_projected_total_signal}
\end{equation}
where the coefficients $\widetilde{\mathcal T}_\rho^{q_1q_2}(r_2)$
include the helicity sum and the SK phases in
Eq.~\eqref{JF}. Keeping the right modes exact gives
the remaining double MB integral,
\begin{align}
 \widetilde{\mathcal T}_\rho^{q_1q_2}(r_2)
 ={}&\frac{\ii\,2^{2\ii\rho\tnu}}{4\pi^{3/2}}
 \Gamma(q_1+5+2\ii\rho\tnu)
 \sin\!\left[\frac\pi2(q_1+5+2\ii\rho\tnu)\right]
 \nonumber\\
 &\times\sum_{s_1,s_2}\mathcal C_{s_1s_2}^{\rho}
 \int_{\mathcal C_3}\frac{\dd z_3}{2\pi\ii}
 \int_{\mathcal C_4}\frac{\dd z_4}{2\pi\ii}\,
 \frac{(2r_2)^{-z_{34}}}{r_2}
 \Gamma(q_2+4-z_{34})\nonumber\\
 &\times e^{-\ii\pi(q_2+4-z_{34})/2}
 \mathcal V^R_{s_1s_2}(z_3,z_4)
 [J_0(A,B)+s_1s_2J_1(A,B)].
 \label{appeq:D_reduced_MB_coefficient}
\end{align}
Here $A=-\tfrac12+\ii\rho\tnu-z_3$ and
$B=-\tfrac12+\ii\rho\tnu-z_4$.
The left-vertex combination is the one used in
Eq.~\eqref{partialJ},
\begin{equation*}
 \mathcal C_{s_1s_2}^{\rho}
 =C_{u_{s_1}}^{(\rho)}C_{u_{s_2}}^{(\rho)}
 -C_{v_{s_1}}^{(\rho)}C_{v_{s_2}}^{(\rho)}.
\end{equation*}
The right-vertex coefficient $\mathcal V^R$ and loop
kernels $J_0,J_1$ are defined in
Eqs.~\eqref{appeq:C_vertex_Mellin} and
\eqref{appeq:C_angular_kernels}.
Equation~\eqref{appeq:D_reduced_MB_coefficient} combines
the nonlocal and local contributions for each left branch.
We evaluate this complete coefficient numerically here;
Section~\ref{app:D_component_comparison} then separates
its pole contributions and compares them with the
main-text signals individually.

\paragraph{Contours and numerical evaluation.}
We choose Mellin contours that separate the left
spectral and UV pole families from the right time-integral
poles, allowing bends where necessary.
For numerical evaluation, we deform these contours to
the vertical lines
$\operatorname{Re}z_3=\operatorname{Re}z_4=c=1/4$.
This deformation crosses the UV pole
\begin{equation}
 z_{34}=2+2\ii\rho\tnu.
 \label{appeq:D_numerical_UV_pole}
\end{equation}
Therefore the result is obtained by the sum of the double MB integral along these
lines and the correction from the crossed UV pole,
which we denote by
$\Delta\widetilde{\mathcal T}_\rho^{q_1q_2}(r_2)$.
At this pole, $A+B=-3$ and the loop-kernel residue is
\begin{equation}
 \operatorname*{Res}_{z_4=2+2\ii\rho\tnu-z_3}
 [J_0(A,B)+s_1s_2J_1(A,B)]
 =\frac{1+s_1s_2}{\Gamma(3/2)}.
 \label{appeq:D_numerical_UV_residue}
\end{equation}
Substituting this residue into
Eq.~\eqref{appeq:D_reduced_MB_coefficient} gives
\begin{align}
 \Delta\widetilde{\mathcal T}_\rho^{q_1q_2}(r_2)
 ={}&\frac{\ii\,2^{2\ii\rho\tnu}}
 {4\pi^{3/2}\Gamma(3/2)}
 \Gamma(q_1+5+2\ii\rho\tnu)
 \sin\!\left[\frac\pi2(q_1+5+2\ii\rho\tnu)\right]
 \nonumber\\
 &\times\frac{(2r_2)^{-2-2\ii\rho\tnu}}{r_2}
 \Gamma(q_2+2-2\ii\rho\tnu)
 e^{-\ii\pi(q_2+2-2\ii\rho\tnu)/2}
 \nonumber\\
 &\times\sum_{s_1,s_2}(1+s_1s_2)\mathcal C_{s_1s_2}^{\rho}
 \int_{c-\ii\infty}^{c+\ii\infty}\frac{\dd z_3}{2\pi\ii}\,
 \mathcal V^R_{s_1s_2}
 (z_3,2+2\ii\rho\tnu-z_3).
 \label{appeq:D_numerical_UV_correction}
\end{align}

\paragraph{Three-point comparison.}
At $q_1=-1$, $q_2=-3$, and $r_2=1$, the coefficient
$\widetilde{\mathcal T}_\rho^{q_1q_2}(1)$ reproduces the
three-point coefficient
$\widetilde{\mathcal T}_{\rho,\mathrm{3pt}}^{q_1q_2}$
defined in Eq.~\eqref{JF3pt}. The oscillation amplitude and phase are obtained
from $\widetilde{\mathcal T}_{+,\mathrm{3pt}}^{q_1q_2}
+(\widetilde{\mathcal T}_{-,\mathrm{3pt}}^{q_1q_2})^*$
using Eq.~\eqref{ampandphase}.

\begin{figure}[!htbp]
 \centering
 \includegraphics[width=\linewidth]{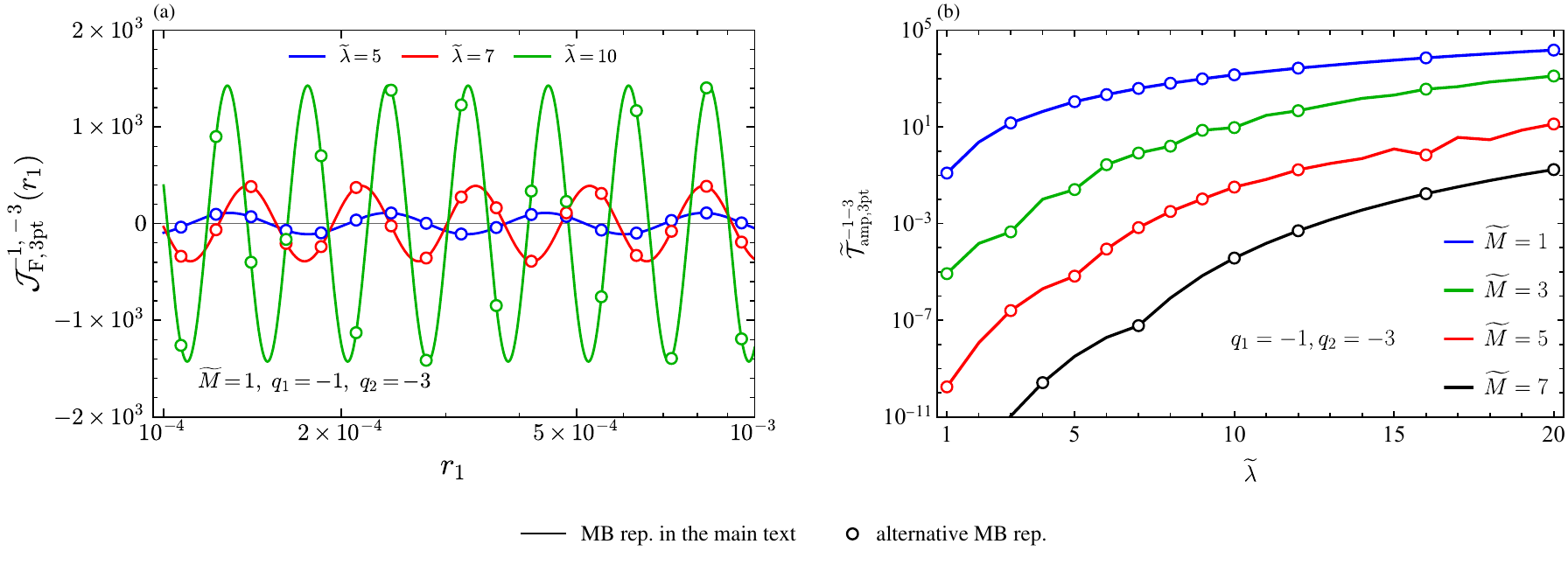}
 \caption{Three-point cross-check of the two MB representations,
 with $q_1=-1$, $q_2=-3$, and $r_2=1$.
 Left: $\mathcal J_{\mathrm{F,3pt}}^{-1,-3}(r_1)$
 for $\mR=1$ and $\tlam=5,7,10$ in blue, red, and green.
 Right: the oscillation amplitude as a function of $\tlam$
 for $\mR=1,3,5,7$ in blue, green, red, and black.
 The solid curves are obtained from the MB representation in the main text, while the open circles are evaluated from
 Eq.~\eqref{appeq:D_reduced_MB_coefficient}.}
 \label{fig:D_figure8_crosscheck}
\end{figure}

We numerically evaluate the alternative MB representation
to obtain the three-point signal
$\mathcal J_{\mathrm{F,3pt}}^{-1,-3}(r_1)$ and its
oscillation amplitude.
Figure~\ref{fig:D_figure8_crosscheck} compares the
oscillations in $r_1$ and the chemical-potential
dependence of the amplitude with the main-text
results in Fig.~\ref{fig3pt}. The alternative-MB points agree closely with the
main-text curves in both panels.

\subsection{Nonlocal and local contributions}
\label{app:D_component_comparison}

Section~\ref{app:D_signal_comparison} evaluates the
nonlocal and local signals together. Here we separate
these contributions to compare them individually with
the main-text results.
Closing the $z_3$ contour in
Eq.~\eqref{appeq:D_reduced_MB_coefficient} to the left
gives two types of contributions: taking spectral
residues in both $z_3,z_4$ yields NS and LSSP, whereas
taking the UV residues of $z_3$ with a one-dimensional $z_4$ integral leaves LSUV.
This classification agrees with
Eqs.~\eqref{partialJNSsec5}--\eqref{partialJNLSUVSsec5}.
As in Section~\ref{app:D_signal_comparison}, the
coefficients $\widetilde{\mathcal T}_{\rho,X}$ include
the helicity and SK sums, with the oscillatory factor
$r_1^{2\ii\rho\tnu}$ written separately.

\paragraph{Spectral contributions.}
Before taking residues, we simplify the phases in
Eq.~\eqref{appeq:D_reduced_MB_coefficient}.
The two conjugated modes in $\mathcal V^R$ supply
$e^{-\ii\pi z_{34}/2}$, as shown in
Eq.~\eqref{appeq:C_vertex_Mellin}. Multiplying this
factor by the phase from the right time integral gives
\begin{equation*}
 e^{-\ii\pi z_{34}/2}
 e^{-\ii\pi(q_2+4-z_{34})/2}
 =e^{-\ii\pi(q_2+4)/2},
\end{equation*}
which is independent of $z_3$ and $z_4$.
We can therefore extract this overall phase and take
residues of the remaining mode kernels. For each
conjugated mode, define
\begin{align}
 \mathcal R_{f_s}(z)
 &=\mathfrak n_{f,s}^*\mathcal W_{\kappa_{f,s}^*}(z),
 \qquad f\in\{u,v\},\nonumber\\
 \mathcal R_{f_s,n}^{(\sigma)}
 &\equiv\underset{z=-1/2-\ii\sigma\tnu-n}{\operatorname{Res}}
 \mathcal R_{f_s}(z)
 =e^{-\ii\pi(1/2+\ii\sigma\tnu+n)/2}
 \overline C_{f_s,n}^{(\sigma)}.
 \label{appeq:D_right_mode_residues}
\end{align}
Here $\sigma=\pm1$ labels the two spectral families.
In this notation, the right-vertex coefficient is
\begin{equation*}
 \mathcal V^R_{s_1s_2}(z_3,z_4)
 =e^{-\ii\pi z_{34}/2}
 \bigl[\mathcal R_{u_{s_1}}(z_3)\mathcal R_{u_{s_2}}(z_4)
 -\mathcal R_{v_{s_1}}(z_3)\mathcal R_{v_{s_2}}(z_4)\bigr].
\end{equation*}
After combining the phases, we collect the factors
independent of $z_3$ and $z_4$ into
\begin{equation}
 K_\rho^{q_1q_2}
 =\frac{\ii\,2^{2\ii\rho\tnu}}{4\pi^{3/2}}
 \Gamma(q_1+5+2\ii\rho\tnu)
 \sin\!\left[\frac\pi2(q_1+5+2\ii\rho\tnu)\right]
 e^{-\ii\pi(q_2+4)/2}.
 \label{appeq:D_component_common_factor}
\end{equation}
Taking the spectral residues at
$z_j=-1/2-\ii\sigma_j\tnu-n_j$, $j=3,4$, now gives
\begin{align}
 \widetilde{\mathcal T}_{\rho,\sigma_3\sigma_4}^{q_1q_2}(r_2)
 ={}&2K_\rho^{q_1q_2}
 \sum_{n_3,n_4\geq0}(2r_2)^{n_{34}+\ii\sigma_{34}\tnu}
 \Gamma(q_2+5+n_{34}+\ii\sigma_{34}\tnu)\nonumber\\
 &\times\sum_{s_1,s_2}\mathcal C_{s_1s_2}^{\rho}
 \Bigl[
 \mathcal R_{u_{s_1},n_3}^{(\sigma_3)}
 \mathcal R_{u_{s_2},n_4}^{(\sigma_4)}
 -\mathcal R_{v_{s_1},n_3}^{(\sigma_3)}
 \mathcal R_{v_{s_2},n_4}^{(\sigma_4)}\Bigr]\nonumber\\
 &\times[J_0(A,B)+s_1s_2J_1(A,B)],
 \label{appeq:D_component_spectral_sum}
\end{align}
where $n_{34}=n_3+n_4$, $\sigma_{34}=\sigma_3+\sigma_4$,
$A=n_3+\ii(\rho+\sigma_3)\tnu$, and
$B=n_4+\ii(\rho+\sigma_4)\tnu$.
Restoring the left-vertex factor
$r_1^{2\ii\rho\tnu}$, the choice
$\sigma_3=\sigma_4=-\rho$ gives the local oscillation
$(r_1/r_2)^{2\ii\rho\tnu}$.
The other sign choices belong to NS, according to
Table~\ref{tab:D_spectral_signals}. Thus
\begin{align}
 \widetilde{\mathcal T}_{\rho,\mathrm{LSSP}}
 &=\widetilde{\mathcal T}_{\rho,-\rho,-\rho},\nonumber\\
 \widetilde{\mathcal T}_{\rho,\mathrm{NS}}
 &=\sum_{(\sigma_3,\sigma_4)\ne(-\rho,-\rho)}
 \widetilde{\mathcal T}_{\rho,\sigma_3\sigma_4}.
 \label{appeq:D_component_spectral_classification}
\end{align}
For LSSP, $A=n_3$ and $B=n_4$.
At odd $n_{34}$, a divergent UV Gamma factor multiplies
a vanishing reciprocal Gamma factor in the loop kernel.
We evaluate the finite limit of the complete kernel
with the common shift $A\to A+\epsilon$,
$B\to B+\epsilon$, followed by $\epsilon\to0$.
This is the prescription used in
Eqs.~\eqref{ACgamma1} and~\eqref{ACgamma2}, and preserves
the symmetry under $A\leftrightarrow B$.

\paragraph{UV contributions.}
The loop UV poles fix the sum of the two Mellin
variables, reducing the double integral to a single
integral. Writing their locations as
$z_{34}=\alpha_{\rho n}\equiv2-2n+2\ii\rho\tnu$,
we take the $z_3$ residue at
$z_3=\alpha_{\rho n}-z_4$.
With $B=-\tfrac12+\ii\rho\tnu-z_4$,
Eq.~\eqref{appeq:D_UV_loop_polynomials} gives
\begin{align}
 \underset{z_3=\alpha_{\rho n}-z_4}{\operatorname{Res}}
 [J_0(A,B)+s_1s_2J_1(A,B)]
 &=\frac{(-1)^n}{n!\Gamma(n+3/2)}
 [P_{0,n}(B)+s_1s_2P_{1,n}(B)],\nonumber\\
 P_{0,n}(B)&=\left(-\frac B2\right)_n
 \left(\frac{B+3}{2}-n\right)_n,\nonumber\\
 P_{1,n}(B)&=\left(\frac{1-B}{2}\right)_n
 \left(\frac{B+4}{2}-n\right)_n.
 \label{appeq:D_component_UV_residue}
\end{align}
Substituting this loop-kernel residue into
Eq.~\eqref{appeq:D_reduced_MB_coefficient} and summing
over the UV levels gives
\begin{align}
 \widetilde{\mathcal T}_{\rho,\mathrm{LSUV}}^{q_1q_2}(r_2)
 ={}&K_\rho^{q_1q_2}
 \sum_{n\geq0}\frac{(-1)^n2^{-\alpha_{\rho n}}
 r_2^{-1-\alpha_{\rho n}}\Gamma(q_2+4-\alpha_{\rho n})}
 {n!\Gamma(n+3/2)}\nonumber\\
 &\times\int_{\mathcal C_{4,n}}\frac{\dd z_4}{2\pi\ii}
 \sum_{s_1,s_2}\mathcal C_{s_1s_2}^{\rho}
 [P_{0,n}(B)+s_1s_2P_{1,n}(B)]\nonumber\\
 &\times\Bigl[
 \mathcal R_{u_{s_1}}(\alpha_{\rho n}-z_4)
 \mathcal R_{u_{s_2}}(z_4)
 -\mathcal R_{v_{s_1}}(\alpha_{\rho n}-z_4)
 \mathcal R_{v_{s_2}}(z_4)\Bigr].
 \label{appeq:D_component_UV_integral}
\end{align}
Each UV residue, including the left-vertex factor, carries
$r_1^{2\ii\rho\tnu}r_2^{-1-\alpha_{\rho n}}
=r_2^{2n-3}(r_1/r_2)^{2\ii\rho\tnu}$.
The oscillation is therefore local at every level. Together
with LSSP, these residues make up the local signal.

The contour $\mathcal C_{4,n}$ and its continuation
are inherited from the original double MB integral.
We evaluate the remaining one-dimensional integrals
numerically, including any residues crossed when
deforming the contours to a common vertical line.

\paragraph{Numerical comparison.}
We now compare NS, LSSP, and LSUV, as well as their sum,
with the corresponding main-text results.
For $X\in\{\mathrm{NS},\mathrm{LSSP},\mathrm{LSUV}\}$,
summing over the two branches $\rho=\pm1$ gives
\begin{equation}
 \mathcal J_X^{q_1q_2}
 =2\operatorname{Re}\sum_{\rho=\pm1}
 r_1^{2\ii\rho\tnu}\widetilde{\mathcal T}_{\rho,X}^{q_1q_2}(r_2),
 \qquad
 \mathcal J_{\mathrm{LS}}=\mathcal J_{\mathrm{LSSP}}
 +\mathcal J_{\mathrm{LSUV}}.
 \label{appeq:D_component_reconstruction}
\end{equation}
Adding NS, LSSP, and LSUV gives $\mathcal J_{\mathrm F}$ in
Eq.~\eqref{appeq:D_projected_total_signal}.

\begin{figure}[!htbp]
 \centering
 \includegraphics[width=\linewidth]{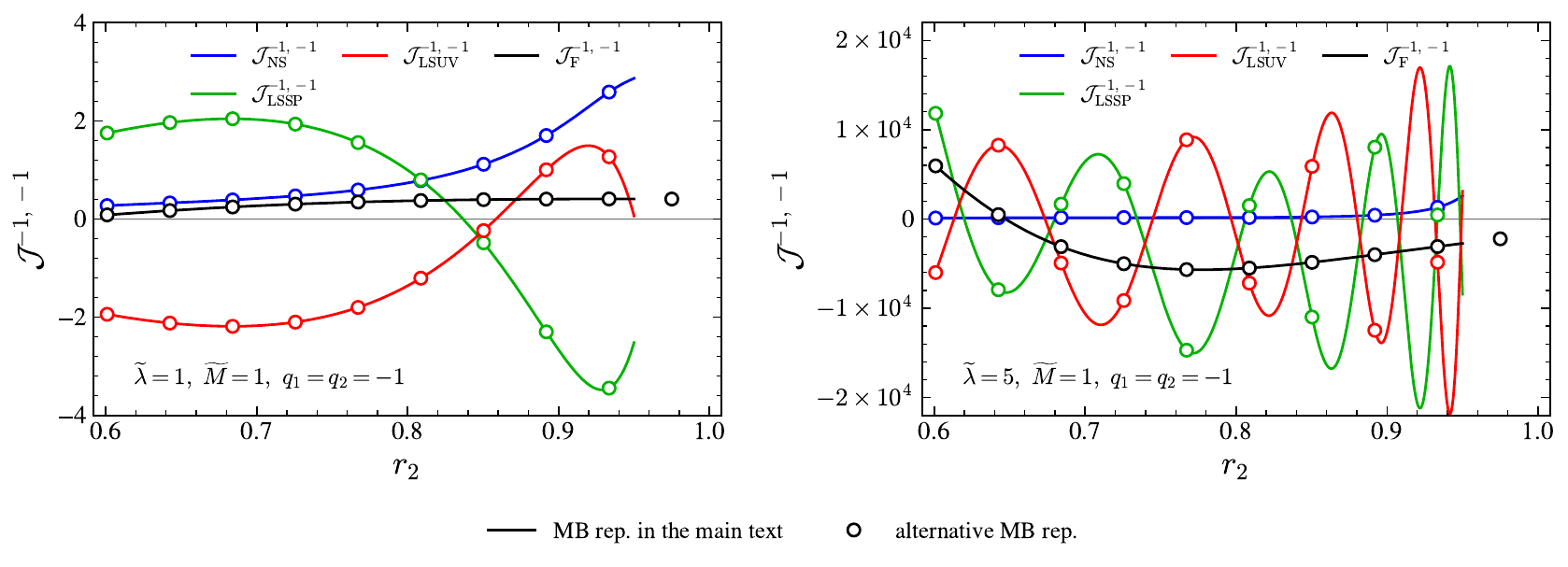}
 \caption{Four-point cross-check of NS, LSSP, LSUV,
 and the total signal (blue, green, red, and black),
 corresponding to Fig.~\ref{figcrosscheck}.
 Both panels use $q_1=q_2=-1$, $\mR=1$, and $r_1=10^{-3}$,
 with $\tlam=1$ on the left and $\tlam=5$ on the right.
 Solid curves are calculated from the MB representation
 in the main text; open circles are obtained from the
 alternative-MB spectral sums and UV contour integrals.}
 \label{fig:D_factorized_components_crosscheck}
\end{figure}

Figure~\ref{fig:D_factorized_components_crosscheck}
shows the four-point signals for $\tlam=1$ and $5$,
with the same parameters as Fig.~\ref{figcrosscheck}.
The alternative-MB points agree with the main-text
curves for NS, LSSP, LSUV, and their sum.
At $\tlam=5$, LSSP and LSUV nearly cancel, leaving
a much smaller total signal. Agreement for this small
total signal also checks the relative normalization
and phase of the two local contributions.

To test both amplitude and phase, we compare the
complex coefficients at fixed parameter and momentum, defining the relative difference
\begin{equation}
 \epsilon_{\mathrm{MB}}
 =\frac{\displaystyle\max_{\rho=\pm1}
 |\widetilde{\mathcal T}_{\rho,\mathrm{main}}
 -\widetilde{\mathcal T}_{\rho,\mathrm{alt}}|}
 {\displaystyle\max_{\rho=\pm1}
 |\widetilde{\mathcal T}_{\rho,\mathrm{alt}}|}.
 \label{appeq:D_numerical_relative_difference}
\end{equation}
\begin{samepage}
The subscripts $\mathrm{main}$ and $\mathrm{alt}$ refer
to the main-text and alternative MB representations.
Both maxima in Eq.~\eqref{appeq:D_numerical_relative_difference}
run over $\rho=\pm1$ at the same parameter--momentum point;
$\epsilon_{\mathrm{MB}}$ is therefore defined pointwise.
For the three-point comparison in
Section~\ref{app:D_signal_comparison}, the maximum
$\epsilon_{\mathrm{MB}}$ is $6.72\times10^{-6}$ over
the 18 parameter pairs $\mR=1,3,5$ and
$\tlam=5,6,\ldots,10$.
For the four-point comparison,
Table~\ref{tab:D_component_accuracy} reports the largest
value of this pointwise relative difference over all
140 tested parameter--momentum pairs, separately for
each component and for the total coefficient.
\par
\end{samepage}

\begin{samepage}
The sum of the NS, LSSP, and LSUV coefficients agrees
with the total coefficient computed in
Section~\ref{app:D_signal_comparison}, with a maximum
relative difference of $2.23\times10^{-10}$ over
the common four-point test set.
\par
\end{samepage}

\begin{table}[htbp]
 \centering
 \begin{tabular}{lc}
 \toprule
 Component & Maximum $\epsilon_{\mathrm{MB}}$ over test points\\
 \midrule
 NS & $1.62\times10^{-19}$\\
 LSSP & $5.10\times10^{-21}$\\
 LSUV & $1.01\times10^{-21}$\\
 LS $=$ LSSP $+$ LSUV & $3.22\times10^{-21}$\\
 Total & $1.30\times10^{-17}$\\
 \bottomrule
 \end{tabular}
 \caption{Largest pointwise relative differences between
 the complex coefficients from the two MB representations.
 In each row, Eq.~\eqref{appeq:D_numerical_relative_difference}
 is evaluated at every test point, and its maximum over
 all 140 parameter--momentum pairs is reported.
 The test set has
 $\mR=1$, $q_1=q_2=-1$, $\tlam=1,4,5,6$, and
 $10^{-3}\leq r_2\leq0.98$.}
 \label{tab:D_component_accuracy}
\end{table}

The two MB representations thus agree well for
both the total clock signal and its nonlocal and local
contributions across the tested configurations.

\section{Evaluation of the $\xi$ Integral for the UV Local Signal}
\label{appendixE}
In this appendix, we present a detailed evaluation of the $\xi$ integral appearing in Eq.~\eqref{JLSUVmedi},
\begin{equation}
\label{xiint}
    \begin{aligned}
        \int \frac{\dd \xi}{2\pi\ii}
    \,
    (2+\ii(\rho-1)\tnu+\xi+\alpha_{\chi}-t)_t
    \,
    \Gamma\!\left[
    \begin{matrix}
    -\xi,a_{s_1}+\xi,-\xi+2\ii\tnu,D_{\rho\chi}-t+\xi,\bar{D}_{\rho\chi}-t+\xi\\
    a_{s_1},\bar{a}_{s_1},
    a_{s_2}+\bar{D}_{\rho\chi}-t+\xi
    \end{matrix}
    \right].
    \end{aligned}
\end{equation}
The $\xi$ integration contour is chosen to separate the left and right pole families, while the definitions of $D_{\rho\chi}$ and $\bar{D}_{\rho\chi}$ are given in Eq.~\eqref{DandDbar}.
\par 
We employ different methods to evaluate the $\xi$ integral for $\rho=+1$ and $\rho=-1$. We first consider the case $\rho=+1$. To eliminate the $\xi$ dependence from the Pochhammer factor, we use the identity
\begin{equation}
\label{localUVidentity1}
    \begin{aligned}
    &(2+\xi+\alpha_{\chi}-t)_t
    \,\Gamma[\delta_\chi+3-t+\xi]
    =
    \sum_{j=0}^{t}
    \binom{t}{j}
    (-1-\beta_\chi)_j\, 
    \Gamma[\delta_\chi+3-j+\xi].
    \end{aligned}
\end{equation}
After applying the above identity to the integrand of the $\xi$ integral, we can close the integration contour to the right and evaluate the integral using the residue theorem by summing over the two right pole families arising from $\Gamma[-\xi,-\xi+2\ii\tnu]$. This yields two infinite series. Using the series definition of the generalized hypergeometric function in Eq.~\eqref{pFqseries}, we can further resum these series into two ${}_3\text F_2$ functions multiplied by the corresponding Gamma-function factors. The explicit result is
\begin{equation}
\label{localUVBdata}
    \begin{aligned}
&\int\frac{\dd \xi}{2\pi \ii}\, 
        \Gamma\!\left[
\begin{matrix}
-\xi,a_{s_1}+\xi,-\xi+2\ii\tnu,\delta_\chi+3-j+\xi,2\ii\tnu+\delta_\chi+3-t+\xi\\
a_{s_1},\bar{a}_{s_1},
\bar{a}_{s_2}+\delta_\chi+3-t+\xi
\end{matrix}
\right]
\\&=
\Gamma\!\left[
\begin{matrix}
2\ii\tnu,\delta_\chi+3-j,2\ii\tnu+\delta_\chi+3-t\\
\bar{a}_{s_1},a_{s_2}+2\ii\tnu+\delta_\chi+3-t
\end{matrix}
\right]
{}_3\text F_2\!\left[
\begin{matrix}
a_{s_1},\delta_\chi+3-j,2\ii\tnu+\delta_\chi+3-t\\
a_{s_2}+2\ii\tnu+\delta_\chi+3-t,1-2\ii\tnu
\end{matrix};1
\right]
\\&+
\Gamma\!\left[
\begin{matrix}
-2\ii\tnu,2\ii\tnu+\delta_\chi+3-j,4\ii\tnu+\delta_\chi+3-t\\
a_{s_1},a_{s_2}+4\ii\tnu+\delta_\chi+3-t
\end{matrix}
\right]
{}_3\text F_2\!\left[
\begin{matrix}
\bar{a}_{s_1},2\ii\tnu+\delta_\chi+3-j,4\ii\tnu+\delta_\chi+3-t\\
a_{s_2}+4\ii\tnu+\delta_\chi+3-t,1+2\ii\tnu
\end{matrix};1
\right].
    \end{aligned}
\end{equation}
\par 
Before analyzing the ${}_3\text F_2$ functions appearing in the integral result in detail, we first address another issue. It should be noted that a similar $0\times\infty$ ambiguity associated with analytic continuation arises here. When $j\geq\delta_\chi+3$, the factor $\Gamma[\delta_\chi+3-j]$ in Eq.~\eqref{localUVBdata} diverges, whereas the factor $(-1-\beta_\chi)_j$ in Eq.~\eqref{localUVidentity1} vanishes. Therefore, their product is expected to remain finite, and it is necessary to properly regularize this finite value. Similar to the regularization prescription adopted in Eqs.~\eqref{ACgamma1} and \eqref{ACgamma2}, in order to preserve the exchange symmetry between $\alpha_\chi$ and $\beta_\chi$, we also introduce a common infinitesimal shift $\epsilon$ in both $\alpha_\chi$ and $\beta_\chi$
\begin{equation}
    \alpha_\chi\rightarrow \alpha_\chi + \epsilon,\quad \beta_\chi\rightarrow\beta_\chi + \epsilon.
\end{equation}
We then define the analytically continued value of $(-1-\beta_\chi)_j\times\Gamma(\delta_\chi+3-j)$ as
\begin{equation}
\begin{aligned}
    \mathcal{C}_{\chi j}&\equiv \text{AC}\left[
    (-1-\beta_\chi)_j\, \Gamma(\delta_\chi+3-j)
    \right]
    \equiv \lim_{\epsilon\to 0}(-1-\beta_\chi-\epsilon)_j\, \Gamma(\delta_\chi+3-j+2\epsilon)
    \\&=
\begin{cases}
(-1)^j\
\Gamma\!\left[\begin{matrix}
    \beta_\chi+2,\delta_\chi+3-j\\
    \beta_\chi+2-j
\end{matrix}\right]
,
& 0\leq j\leq \beta_\chi+1,
\\[4mm]
\dfrac{(-1)^{j-\alpha_\chi-1}}{2}\
\Gamma\!\left[\begin{matrix}
    \beta_\chi+2,j-\beta_\chi-1\\
    j-\delta_\chi-2
\end{matrix}\right],
& j\geq \delta_\chi+3,
\\[4mm]
0,&\text{other}.
\end{cases}
\end{aligned}
\end{equation}
For any value of $j$, $\mathcal{C}_{\chi j}$ remains finite. For compactness, we combine $(-1-\beta_\chi)_j$ with the result of the $\xi$ integral in Eq.~\eqref{localUVBdata}, and define
\begin{equation}
    \begin{aligned}
    \mathcal{A}_{tj,+\chi}^{s_1s_2}&\equiv 
    \mathcal{C}_{\chi j}\, 
\Gamma\!\left[
\begin{matrix}
2\ii\tnu,2\ii\tnu+\delta_\chi+3-t\\
\bar{a}_{s_1},a_{s_2}+2\ii\tnu+\delta_\chi+3-t
\end{matrix}
\right]
{}_3\text F_2\!\left[
\begin{matrix}
a_{s_1},\delta_\chi+3-j,2\ii\tnu+\delta_\chi+3-t\\
a_{s_2}+2\ii\tnu+\delta_\chi+3-t,1-2\ii\tnu
\end{matrix};1
\right]
\\&+
(-1-\beta_\chi)_j\, 
\Gamma\!\left[
\begin{matrix}
-2\ii\tnu,2\ii\tnu+\delta_\chi+3-j,4\ii\tnu+\delta_\chi+3-t\\
a_{s_1},a_{s_2}+4\ii\tnu+\delta_\chi+3-t
\end{matrix}
\right]
\\&\times 
{}_3\text F_2\!\left[
\begin{matrix}
\bar{a}_{s_1},2\ii\tnu+\delta_\chi+3-j,4\ii\tnu+\delta_\chi+3-t\\
a_{s_2}+4\ii\tnu+\delta_\chi+3-t,1+2\ii\tnu
\end{matrix};1
\right].
    \end{aligned}
\end{equation}
Here, the subscript $+$ indicates that we are considering the $\rho=+1$ branch.
\par 
Next, we analyze the generalized hypergeometric functions ${}_3\text F_2$ appearing in $\mathcal{A}_{tj,+\chi}^{s_1s_2}$, all of which have unit argument. Their nonterminating series representations converge only when the parametric excess has a positive real part and diverge otherwise. For a detailed discussion, see Ref.~\cite{NIST:DLMF}. Both ${}_3\text F_2$ functions in Eq.~\eqref{localUVBdata} have the same parametric excess, which we denote by
\begin{equation}
    \mathfrak s_+=a_{s_2}-a_{s_1}+j-\delta_\chi-2-2\ii\tnu,
\end{equation}
its real part is given by
\begin{equation}
    \operatorname{Re}\mathfrak s_+=\frac{s_1-s_2}{2}+j-\delta_\chi-2.
\end{equation}
Here, the subscript $+$ indicates that we are considering the $\rho=+1$ branch. When $j\leq \delta_\chi +2+(s_2-s_1)/2$, we have $\text{Re}\, \mathfrak s_+\leq0$, and the ${}_3\text F_2[1]$ functions appearing in Eq.~\eqref{localUVBdata} are divergent. Strictly speaking, in this case, the infinite series arising from the evaluation of the $\xi$ integral in Eq.~\eqref{localUVBdata} using the residue theorem cannot be summed in the usual sense, since they are divergent. Our procedure above therefore amounts to formally expressing these divergent infinite series in terms of ${}_3\text F_2[1]$ functions whose defining series are also divergent. To avoid the divergences of ${}_3\text F_2[1]$, we can perform an analytic continuation using the Thomae transformation~\cite{NIST:DLMF}. For the helicity branch $s_1=+$, we apply the following Thomae transformation
\begin{equation}
\label{Thomae1}
{}_3\text F_2\!\left[
\begin{matrix}
a,b,c\\
e,f
\end{matrix}
;1
\right]
=
\Gamma\!\left[
\begin{matrix}
f,\mathfrak s\\
f-a,\mathfrak s+a
\end{matrix}
\right]\, 
{}_3\text F_2\!\left[
\begin{matrix}
a,e-b,e-c\\
e,\mathfrak s+a
\end{matrix}
;1
\right],
\end{equation}
where $\mathfrak{s}=e+f-a-b-c$ denotes the parametric excess of the ${}_3\text F_2[1]$ function on the left-hand side of the above equation. Under this transformation, $\mathcal{A}_{tj,+\chi}^{+s_2}$ can be written as
\begin{equation}
    \begin{aligned}
    \label{B++}
    \mathcal{A}_{tj,+\chi}^{+s_2}&=
\Bigg\{\mathcal{C}_{\chi j}
\sin(\pi\bar{a}_{+})
\Gamma[2\ii\tnu+\delta_\chi+3-t]\, 
{}_3\mathbf F_2\!\left[
\begin{matrix}
a_{+},a_{s_2},a_{s_2}+2\ii\tnu+j-t\\
a_{s_2}+2\ii\tnu+\delta_\chi+3-t,\mathfrak s_++a_+
\end{matrix};1
\right]
\\&-
(-1-\beta_\chi)_j
\sin(\pi a_{+})
\Gamma[2\ii\tnu+\delta_\chi+3-j,4\ii\tnu+\delta_\chi+3 -t]\, 
\\&\times 
{}_3\mathbf F_2\!
\left[
\begin{matrix}
\bar{a}_{+},a_{s_2},a_{s_2}+2\ii\tnu+j-t\\
a_{s_2}+4\ii\tnu+\delta_\chi+3 -t,\mathfrak s_++\bar{a}_+
\end{matrix};1
\right]
\Bigg\}\times \frac{\Gamma[\mathfrak s_+]}{\sin(2\pi\ii\tnu)},
    \end{aligned}
\end{equation}
where ${}_3\mathbf{F}_2$ denotes the regularized generalized hypergeometric function, defined in Eq.~\eqref{reghyperfun}. It can be readily checked that both ${}_3\text F_2[1]$ functions appearing in $\mathcal{A}_{tj,+\chi}^{+s_2}$ have parametric excesses with real part equal to $+1$. For the helicity branch $s_1=-$, we apply the following Thomae transformation
\begin{equation}
\label{Thomae2}
{}_3\text F_2\!\left[
\begin{matrix}
a,b,c\\
e,f
\end{matrix}
;1
\right]
=
\Gamma\!\left[
\begin{matrix}
e,f,\mathfrak s\\
a,\mathfrak s+b,\mathfrak s+c
\end{matrix}
\right]
\, 
{}_3\text F_2\!\left[
\begin{matrix}
e-a,f-a,\mathfrak s\\
\mathfrak s+b,\mathfrak s+c
\end{matrix}
;1
\right].
\end{equation}
Under this transformation, $\mathcal{A}_{tj,+\chi}^{-s_2}$ can be written as
\begin{equation}
    \begin{aligned}
    \label{B+-}
    \mathcal{A}_{tj,+\chi}^{-s_2}&=
\Bigg\{
\mathcal{C}_{\chi j}\, \Gamma[2\ii\tnu+\delta_\chi+3-t]\, 
{}_3\mathbf F_2\!\left[
\begin{matrix}
a_{s_2}+2\ii\tnu+\delta_\chi+3-t-a_-,a_+,\mathfrak s_+\\
\mathfrak s_++\delta_\chi+3-j,\mathfrak s_++2\ii\tnu+\delta_\chi+3-t
\end{matrix};1
\right]
\\&-
(-1-\beta_\chi)_j\, 
\Gamma[2\ii\tnu+\delta_\chi+3-j,4\ii\tnu+\delta_\chi+3 -t]\\&\times 
{}_3\mathbf F_2\!
\left[
\begin{matrix}
a_{s_2}+4\ii\tnu+\delta_\chi+3 -t-\bar{a}_-,\bar{a}_+,\mathfrak s_+\\
\mathfrak s_++2\ii\tnu+\delta_\chi+3-j,\mathfrak s_++4\ii\tnu+\delta_\chi+3 -t
\end{matrix};1
\right]
\Bigg\}\frac{\pi\, \Gamma[\mathfrak s_+]}{\sin(2\pi\ii\tnu )\, \Gamma[a_-,\bar{a}_-]}.
    \end{aligned}
\end{equation}
It can be verified that the real parts of the balance parameters of both ${}_3\text F_2[1]$ functions appearing in $\mathcal{A}_{tj,+\chi}^{-s_2}$ are also equal to $+1$.
\par 
For the $\rho=-1$ branch, the calculation proceeds similarly, with the only difference being that the identity~\eqref{localUVidentity1} is replaced by
\begin{equation}
    \begin{aligned}
    &(2-2\ii\tnu+\xi+\alpha_{\chi}-t)_t\,
    \Gamma[-2\ii\tnu+\delta_\chi+3-t+\xi]
    =
    \sum_{j=0}^{t}
    \binom{t}{j}
    (-1-\beta_\chi)_j
    \Gamma[-2\ii\tnu+\delta_\chi+3-j+\xi].
    \end{aligned}
\end{equation}
Using the same procedure, we obtain the result of the $\xi$ integral
\begin{equation}
    \begin{aligned}
        &\int\frac{\dd \xi}{2\pi \ii}\, 
        \Gamma\!\left[
\begin{matrix}
-\xi,
a_{s_1}+\xi,
-\xi+2\ii\tnu,
-4\ii\tnu+\delta_\chi+3-t+\xi,
-2\ii\tnu+\delta_\chi+3-j+\xi
\\
a_{s_1},
\bar{a}_{s_1},
a_{s_2}-2\ii\tnu+\delta_\chi+3-t+\xi
\end{matrix}
\right]
\\
&=
\Gamma\!\left[
\begin{matrix}
2\ii\tnu,
-4\ii\tnu+\delta_\chi+3-t,
-2\ii\tnu+\delta_\chi+3-j
\\
\bar{a}_{s_1},
a_{s_2}-2\ii\tnu+\delta_\chi+3-t
\end{matrix}
\right]
{}_3\text F_2\!\left[
\begin{matrix}
a_{s_1},
-4\ii\tnu+\delta_\chi+3-t,
-2\ii\tnu+\delta_\chi+3-j
\\
a_{s_2}-2\ii\tnu+\delta_\chi+3-t,
1-2\ii\tnu
\end{matrix};1
\right]
\\
&+
\Gamma\!\left[
\begin{matrix}
-2\ii\tnu,
-2\ii\tnu+\delta_\chi+3-t,
\delta_\chi+3-j
\\
a_{s_1},
a_{s_2}+\delta_\chi+3-t
\end{matrix}
\right]
{}_3\text F_2\!\left[
\begin{matrix}
\bar{a}_{s_1},
-2\ii\tnu+\delta_\chi+3-t,
\delta_\chi+3-j
\\
a_{s_2}+\delta_\chi+3-t,
1+2\ii\tnu
\end{matrix};1
\right].
    \end{aligned}
\end{equation}
Similarly, we can define
\begin{equation}
    \begin{aligned}
        \mathcal{A}_{tj,-\chi}^{s_1s_2}
&=
(-1-\beta_\chi)_j\, 
\Gamma\!\left[
\begin{matrix}
2\ii\tnu,
-4\ii\tnu+\delta_\chi+3-t,
-2\ii\tnu+\delta_\chi+3-j
\\
\bar{a}_{s_1},
a_{s_2}-2\ii\tnu+\delta_\chi+3-t
\end{matrix}
\right]
\\&\times
{}_3\text F_2\!\left[
\begin{matrix}
a_{s_1},
-4\ii\tnu+\delta_\chi+3-t,
-2\ii\tnu+\delta_\chi+3-j
\\
a_{s_2}-2\ii\tnu+\delta_\chi+3-t,
1-2\ii\tnu
\end{matrix};1
\right]
\\
&+
\mathcal{C}_{\chi j}\, 
\Gamma\!\left[
\begin{matrix}
-2\ii\tnu,
-2\ii\tnu+\delta_\chi+3-t
\\
a_{s_1},
a_{s_2}+\delta_\chi+3-t
\end{matrix}
\right]
{}_3\text F_2\!\left[
\begin{matrix}
\bar{a}_{s_1},
-2\ii\tnu+\delta_\chi+3-t,
\delta_\chi+3-j
\\
a_{s_2}+\delta_\chi+3-t,
1+2\ii\tnu
\end{matrix};1
\right].
    \end{aligned}
\end{equation}
For the two ${}_3\text F_2[1]$ functions appearing in
$\mathcal{A}_{tj,-\chi}^{s_1s_2}$, they have the same balance parameter
\begin{equation}
    \mathfrak s_-
    =
    a_{s_2}-a_{s_1}
    +j-\delta_\chi-2+2\ii\tnu,
\end{equation}
\begin{equation}
    \text{Re}\, \mathfrak s_-
    =\text{Re}\, \mathfrak s_+
    =
    \frac{s_1-s_2}{2}
    +j-\delta_\chi-2.
\end{equation}
When $j\leq \delta_\chi+2+(s_2-s_1)/2$, we have $\operatorname{Re}\mathfrak{s}_-\leq 0$, and the ${}_3\text F_2[1]$ functions appearing in $\mathcal{A}_{tj,-\chi}^{s_1s_2}$ are divergent. Nevertheless, the divergent ${}_3\text F_2[1]$ functions can still be analytically continued using the Thomae transformation. For the helicity branch $s_1=+$, applying the Thomae transformation in Eq.~\eqref{Thomae1}, we obtain
\begin{equation}
\label{B-+}
    \begin{aligned}
    \mathcal{A}_{tj,-\chi}^{+s_2}
    &=
\Bigg\{(-1-\beta_\chi)_j\, \sin(\pi\bar{a}_{+})\, 
\Gamma[
-4\ii\tnu+\delta_\chi+3-t,\,
-2\ii\tnu+\delta_\chi+3-j
]
\\&\times
{}_3\mathbf F_2\!\left[
\begin{matrix}
a_{+},\,
\bar a_{s_2},\,
a_{s_2}+j-t
\\
a_{s_2}-2\ii\tnu+\delta_\chi+3-t,\,
\mathfrak s_-+a_+
\end{matrix};1
\right]
\\&-
\mathcal{C}_{\chi j}\, \sin(\pi a_{+})\, 
\Gamma[-2\ii\tnu+\delta_\chi+3-t]
\\&\times 
{}_3\mathbf F_2\!
\left[
\begin{matrix}
\bar{a}_{+},\,
\bar a_{s_2},\,
a_{s_2}+j-t
\\
a_{s_2}+\delta_\chi+3-t,\,
\mathfrak s_-+\bar{a}_+
\end{matrix};1
\right]
\Bigg\}\times \frac{\Gamma[\mathfrak s_-]}{\sin(2\pi\ii\tnu)}.
    \end{aligned}
\end{equation}
For the helicity branch $s_1=-$, applying the Thomae transformation in Eq.~\eqref{Thomae2}, we obtain
\begin{equation}
    \begin{aligned}
    \label{B--}
    \mathcal{A}_{tj,-\chi}^{-s_2}
    &=
\Bigg\{
(-1-\beta_\chi)_j\, 
\Gamma[
-4\ii\tnu+\delta_\chi+3-t,\,
-2\ii\tnu+\delta_\chi+3-j
]
\\&\times
{}_3\mathbf F_2\!\left[
\begin{matrix}
a_{s_2}-2\ii\tnu+\delta_\chi+3-t-a_-,
\,
a_+,\,
\mathfrak s_-
\\
\mathfrak s_--4\ii\tnu+\delta_\chi+3-t,\,
\mathfrak s_--2\ii\tnu+\delta_\chi+3-j
\end{matrix};1
\right]
\\&-
\mathcal{C}_{\chi j}\, \Gamma[-2\ii\tnu+\delta_\chi+3-t]
\\&\times
{}_3\mathbf F_2\!
\left[
\begin{matrix}
a_{s_2}+\delta_\chi+3-t-\bar{a}_-,
\,
\bar{a}_+,\,
\mathfrak s_-
\\
\mathfrak s_--2\ii\tnu+\delta_\chi+3-t,\,
\mathfrak s_-+\delta_\chi+3-j
\end{matrix};1
\right]
\Bigg\}\frac{\pi\, \Gamma[\mathfrak s_-]}
    {\sin(2\pi\ii\tnu)\, \Gamma[a_-,\bar{a}_-]}.
    \end{aligned}
\end{equation}
It can be verified that the ${}_3\text F_2[1]$ functions appearing in $\mathcal{A}_{tj,-\chi}^{\pm s_2}$ all have parametric excesses with real part equal to $+1$.
\par 
Combining the results above, we can write the result of the $\xi$ integral in Eq.~\eqref{xiint} as
\begin{equation}
\label{xiintresult}
    \begin{aligned}
        &\int \frac{\dd \xi}{2\pi\ii}
    \, 
    (2+\ii(\rho-1)\tnu +\xi+\alpha_{\chi}-t)_t
    \\&\times \Gamma\!\left[
\begin{matrix}
-\xi,a_{s_1}+\xi,-\xi+2\ii\tnu,D_{\rho\chi}-t+\xi,\bar{D}_{\rho\chi}-t+\xi\\
a_{s_1},\bar{a}_{s_1},
a_{s_2}+\bar{D}_{\rho\chi}-t+\xi
\end{matrix}
\right]
\\&=
\sum_{j=0}^{t} \binom{t}{j}\, \mathcal{A}_{tj,\rho \chi}^{s_1 s_2},
    \end{aligned}
\end{equation}
where the definitions of $\mathcal{A}_{tj,\pm\chi}^{\pm s_2}$ are given in Eqs.~\eqref{B++}, \eqref{B+-}, \eqref{B-+}, and \eqref{B--}.

\section{Cancellation of Folded Singularities}
\label{appendixF}
In this appendix, we prove that $\mathcal B^{q_2,\chi}_\rho(X,\zeta;r_2)$, defined in Eq.~\eqref{eq:complete-bracket}, remains regular in the folded limit $r_2\to1$. This kernel is constructed from $\widehat{\mathcal O}_{1,\rho}^{q_2,\chi}$ and $\widehat{\mathcal O}_{2,\rho}^{q_2,\chi}$, given in Eqs.~\eqref{O1} and \eqref{O2}, respectively. We show that the potentially singular terms in the regularized hypergeometric functions ${}_4\mathbf F_3$ appearing in these expressions cancel as $r_2\to1$. These terms can arise when the parametric excess has a nonpositive real part, but their cancellation leaves the complete kernel $\mathcal B^{q_2,\chi}_\rho(X,\zeta;r_2)$ regular in this limit. We also derive an analytic expression for the regular kernel that remains after this cancellation.
\par 
We note that the regularized hypergeometric functions ${}_4\mathbf F_3$ appearing in $\widehat{\mathcal O}_{1,\rho}^{q_2,\chi}$ and $\widehat{\mathcal O}_{2,\rho}^{q_2,\chi}$ can be expressed as combinations of two ${}_2\mathbf F_1$ functions
\begin{equation}
\label{4F3expan1}
    {}_4\mathbf F_3\!\left[
\begin{matrix}
\frac q2,\frac{q+1}{2},\frac b2,\frac{b+1}{2}
\\
\frac12,\frac c2,\frac{c+1}{2}
\end{matrix}
;r_2^2
\right]
=
\frac{2^{c-2}}{\pi}
\left\{
{}_2\mathbf F_1\!\left[
\begin{matrix}
q,b
\\
c
\end{matrix}
;r_2
\right]
+
{}_2\mathbf F_1\!\left[
\begin{matrix}
q,b
\\
c
\end{matrix}
;-r_2
\right]
\right\},
\end{equation}

\begin{equation}
\begin{aligned}
\label{4F3expan2}
    {}_4\mathbf F_3\!\left[
\begin{matrix}
\frac q2,\frac{q+1}{2},\frac b2,\frac{b+1}{2}
\\
\frac32,\frac c2,\frac{c+1}{2}
\end{matrix}
;r_2^2
\right]
=
\frac{2^{c-1}}{\pi (q-1)(b-1)r_2}
\left\{
{}_2\mathbf F_1\!\left[
\begin{matrix}
q-1,b-1
\\
c-1
\end{matrix}
;r_2
\right]
-
{}_2\mathbf F_1\!\left[
\begin{matrix}
q-1,b-1
\\
c-1
\end{matrix}
;-r_2
\right]
\right\}.
\end{aligned}
\end{equation}
Here, for convenience, we have denoted the $X$- and $\zeta$-dependent parameters by the generic symbols $q$, $b$, and $c$. The above expressions can be derived using the series definition of the generalized hypergeometric function in Eq.~\eqref{pFqseries} and are valid only for $r_2\in[0,1)$. The potentially singular behavior of these two classes of ${}_4\mathbf F_3$ functions as $r_2\to1$ is therefore entirely encoded in the Gauss hypergeometric functions ${}_2\mathbf F_1(q,b;c;r_2)$ and ${}_2\mathbf F_1(q-1,b-1;c-1;r_2)$. To separate their regular and potentially singular parts in this limit, we use the connection formula~\cite{NIST:DLMF}
\begin{equation}
\label{2F1 connection formula}
\begin{aligned}
    {}_2\mathbf F_1\!\left[
        \begin{matrix}
            q,b\\
            c
        \end{matrix}
        ;r_2
    \right]
    &=
    \Gamma\!\left[
        \begin{matrix}
            \mathfrak s\\
            q+\mathfrak s,b+\mathfrak s
        \end{matrix}
    \right]
    {}_2\text F_1\!\left[
        \begin{matrix}
            q,b\\
            1-\mathfrak s
        \end{matrix}
        ;1-r_2
    \right]
    \\
    &+
    \Gamma\!\left[
        \begin{matrix}
            -\mathfrak s\\
            q,b
        \end{matrix}
    \right]
    (1-r_2)^{\mathfrak s}\,
    {}_2\text F_1\!\left[
        \begin{matrix}
            q+\mathfrak s,b+\mathfrak s\\
            1+\mathfrak s
        \end{matrix}
        ;1-r_2
    \right],
    \ \ \ r_2\in(0,1),\ \mathfrak s\in\mathbb C\setminus\mathbb Z,
\end{aligned}
\end{equation}
where
\begin{equation}
    \mathfrak s\equiv c-q-b
\end{equation}
is the parametric excess of the Gauss hypergeometric function. For $\mathfrak s\notin\mathbb Z$, the first term on the right-hand side remains regular as $r_2\to1$, whereas the factor $(1-r_2)^{\mathfrak s}$ in the second term diverges in magnitude for $\text{Re}\, \mathfrak s<0$ and oscillates without a limit for $\text{Re}\, \mathfrak s=0$. The connection formula thus separates the regular and potentially singular contributions. We use this decomposition to prove that the singular contributions cancel in the complete kernel $\mathcal B^{q_2,\chi}_\rho(X,\zeta;r_2)$ as $r_2\to1$, and to derive an analytic expression for the regular remainder. Equation~\eqref{2F1 connection formula} cannot be applied directly when $\mathfrak s\in\mathbb Z$. In such cases, we first work with generic noninteger $\mathfrak s$ and take the integer limit only after combining the relevant terms. The final expression remains valid irrespective of whether $\mathfrak s$ is an integer.
\par 
We now apply the above identities to our case. We first define
\begin{equation}
    \mathcal{H}_\pm(X,\zeta;r_2)\equiv
\tF{q_2+5-2\ii\tnu+X+\zeta}{\ii(\rho-1)\tnu+\zeta+\beta_\chi+2}{2\ii(\rho-1)\tnu+X+\zeta+\delta_\chi+5}{\pm r_2},
\end{equation}
\begin{equation}
    \mathcal{G}_\pm(X,\zeta;r_2)\equiv
\tF{q_2+1-2\ii\rho\tnu -\delta_\chi}{-\ii(\rho-1)\tnu-X-\alpha_\chi-2}{-2\ii(\rho-1)\tnu-X-\zeta-\delta_\chi-3}{\pm r_2},
\end{equation}
where for convenience, we have suppressed the indices $q_2$, $\rho$, and $\chi$ on $\mathcal{H}_\pm(X,\zeta;r_2)$ and $\mathcal{G}_\pm(X,\zeta;r_2)$. We adopt the same convention in the expressions below. Using the ${}_4\mathbf F_3$ expansions in Eqs.~\eqref{4F3expan1} and \eqref{4F3expan2}, we can expand $\widehat{\mathcal O}_{1,\rho}^{q_2,\chi}$ and $\widehat{\mathcal O}_{2,\rho}^{q_2,\chi}$, defined in Eqs.~\eqref{O1} and \eqref{O2}, as
\begin{equation}
    \widehat{\mathcal{O}}_{1,\rho}^{q_2,\chi}(X,\zeta;r_2)
=\mathcal K_{1+}(X,\zeta)\mathcal{H}_+(X,\zeta;r_2)+\mathcal K_{1-}(X,\zeta)\mathcal{H}_-(X,\zeta;r_2),
\label{eq:I1-H}
\end{equation}
\begin{equation}
    \widehat{\mathcal{O}}_{2,\rho}^{q_2,\chi}(X,\zeta;r_2)
=-\frac{\mathcal K_2(X,\zeta)}{r_2}\left[\mathcal{G}_+(X,\zeta;r_2)
-\mathcal{G}_-(X,\zeta;r_2)\right],
\label{eq:I2-G}
\end{equation}
where
\begin{equation}
\begin{aligned}
    \mathcal K_{1+}(X,\zeta)
&=-2\sqrt{\pi}\, \Gamma\!\left[
\begin{matrix}
q_2+5-2\ii\tnu+X+\zeta,\ii(\rho-1)\tnu+\zeta+\beta_\chi+2
\\
-\ii(\rho-1)\tnu-X-\alpha_\chi-1
\end{matrix}
\right]
\\&\times
\csc\left[
\pi(2\ii(\rho-1)\tnu+X+\zeta+\delta_\chi+3)
\right],
\end{aligned}
\label{eq:K1plus}
\end{equation}
\begin{equation}
\begin{aligned}
    \mathcal K_{1-}(X,\zeta)&=\frac{4}{\sqrt{\pi}}
    \frac{\Gamma\left[q_2+5-2\ii\tnu+X+\zeta,\ii(\rho-1)\tnu+\zeta+\beta_\chi+2,\ii(\rho-1)\tnu+X+\alpha_\chi+2\right]}{\sin\left[\pi(2\ii(\rho-1)\tnu+X+\zeta+\delta_\chi+3)\right]}
    \\&\times 
    \sin\left[\frac{\pi}{2}(\ii(\rho-1)\tnu +X+\alpha_\chi)\right]
    \cos\left[\pi\left(\frac{3}{2}\ii(\rho-1)\tnu +\frac{X}{2}+\zeta+\frac{\alpha_\chi}{2}+\beta_\chi\right)\right],
\end{aligned}
\label{eq:K1minus}
\end{equation}
and
\begin{equation}
\begin{aligned}
    \mathcal K_{2}(X,\zeta)
&=
\frac{
2\sqrt{\pi}\,
\Gamma\!\left[
q_2+1-2\ii\rho\tnu-\delta_\chi
\right]
}{
\ii(\rho-1)\tnu+X+\alpha_\chi+2
}\, 
\csc\left[
\pi(2\ii(\rho-1)\tnu+X+\zeta+\delta_\chi+3)
\right].
\end{aligned}
\label{eq:K2}
\end{equation}
The functions $\mathcal{H}_-(X,\zeta;r_2)$ and $\mathcal{G}_-(X,\zeta;r_2)$ remain regular as $r_2\to1$, since their arguments approach $-1$. In contrast, $\mathcal{H}_+(X,\zeta;r_2)$ and $\mathcal{G}_+(X,\zeta;r_2)$ can develop singular behavior in this limit, as their arguments approach $+1$.
\par 
Using Eq.~\eqref{2F1 connection formula}, we decompose $\mathcal{H}_+(X,\zeta;r_2)$ and $\mathcal{G}_+(X,\zeta;r_2)$ into regular and potentially singular parts as $r_2\to1$. For $\mathcal{H}_+(X,\zeta;r_2)$, we obtain
\begin{equation}
\begin{aligned}
\mathcal{H}_+(X,\zeta;r_2)\big|_{\rm reg}
&=
\Gamma\!\left[
\begin{matrix}
\ii(\rho+1)\tnu+\alpha_\chi-\zeta-q_2-2
\\
2\ii\rho\tnu+\delta_\chi-q_2,
\ii(\rho-1)\tnu+X+\alpha_\chi+3
\end{matrix}
\right]
\\
&\quad\times
{}_2\text F_1\!\left[
\begin{matrix}
q_2+5-2\ii\tnu+X+\zeta,
\ii(\rho-1)\tnu+\zeta+\beta_\chi+2
\\
-\ii(\rho+1)\tnu-\alpha_\chi+\zeta+q_2+3
\end{matrix}
;1-r_2
\right],
\end{aligned}
\label{H+reg}
\end{equation}
and
\begin{equation}
\begin{aligned}
\mathcal{H}_+(X,\zeta;r_2)\big|_{\rm sin}
&=
\Gamma\!\left[
\begin{matrix}
-\ii(\rho+1)\tnu-\alpha_\chi+\zeta+q_2+2
\\
q_2+5-2\ii\tnu+X+\zeta,
\ii(\rho-1)\tnu+\zeta+\beta_\chi+2
\end{matrix}
\right]
(1-r_2)^{
\ii(\rho+1)\tnu+\alpha_\chi-\zeta-q_2-2
}
\\
&\quad\times
{}_2\text F_1\!\left[
\begin{matrix}
2\ii\rho\tnu+\delta_\chi-q_2,
\ii(\rho-1)\tnu+X+\alpha_\chi+3
\\
\ii(\rho+1)\tnu+\alpha_\chi-\zeta-q_2-1
\end{matrix}
;1-r_2
\right].
\end{aligned}
\label{H+sin}
\end{equation}
Here, the label ``sin'' denotes the potentially singular part, whose behavior as $r_2\to1$ depends on the corresponding parametric excess; it does not imply that this part necessarily singular. We adopt the same convention below. Similarly, for $\mathcal{G}_+(X,\zeta;r_2)$, we find
\begin{equation}
\begin{aligned}
\mathcal{G}_+(X,\zeta;r_2)\big|_{\rm reg}
&=
\Gamma\!\left[
\begin{matrix}
\ii(\rho+1)\tnu+\alpha_\chi-\zeta-q_2-2
\\
2\ii\tnu-\zeta-X-q_2-4,
-\ii(\rho-1)\tnu-\zeta-\beta_\chi-1
\end{matrix}
\right]
\\
&\quad\times
{}_2\text F_1\!\left[
\begin{matrix}
-\ii(\rho-1)\tnu-X-\alpha_\chi-2,
q_2+1-2\ii\rho\tnu-\delta_\chi
\\
-\ii(\rho+1)\tnu-\alpha_\chi+\zeta+q_2+3
\end{matrix}
;1-r_2
\right],
\end{aligned}
\label{eq:Gplusreg_expand}
\end{equation}
and
\begin{equation}
\begin{aligned}
\mathcal{G}_+(X,\zeta;r_2)\big|_{\rm sin}
&=
\Gamma\!\left[
\begin{matrix}
-\ii(\rho+1)\tnu-\alpha_\chi+\zeta+q_2+2
\\
q_2+1-2\ii\rho\tnu-\delta_\chi,
-\ii(\rho-1)\tnu-X-\alpha_\chi-2
\end{matrix}
\right]
(1-r_2)^{
\ii(\rho+1)\tnu+\alpha_\chi-\zeta-q_2-2
}
\\
&\quad\times
{}_2\text F_1\!\left[
\begin{matrix}
2\ii\tnu-\zeta-X-q_2-4,
-\ii(\rho-1)\tnu-\zeta-\beta_\chi-1
\\
\ii(\rho+1)\tnu+\alpha_\chi-\zeta-q_2-1
\end{matrix}
;1-r_2
\right].
\end{aligned}
\label{eq:Gplussin_expand}
\end{equation}
\par 
The Gauss hypergeometric functions appearing in $\mathcal{H}_+(X,\zeta;r_2)\big|_{\rm reg,sin}$ and $\mathcal{G}_+(X,\zeta;r_2)\big|_{\rm reg,sin}$ have different parameter structures, which complicates the subsequent proof. To bring them into a common form, we apply the following Euler transformation~\cite{NIST:DLMF} separately to the two ${}_2\text F_1$ functions in Eqs.~\eqref{H+reg} and \eqref{H+sin}
\begin{equation}
\label{Eulertran1}
\begin{aligned}
&{}_2\text F_1
\left[
\begin{matrix}
q_2+5-2\ii\tnu+X+\zeta,
\ii(\rho-1)\tnu+\zeta+\beta_\chi+2
\\
q_2+3+\zeta-\alpha_\chi-\ii(\rho+1)\tnu
\end{matrix}
;
1-r_2
\right]
\\
&=
r_2^{-2\ii(\rho-1)\tnu-X-\zeta-\delta_\chi-4}\, 
{}_2\text F_1
\left[
\begin{matrix}
-\ii(\rho-1)\tnu-X-\alpha_\chi-2,
q_2+1-2\ii\rho\tnu-\delta_\chi
\\
q_2+3+\zeta-\alpha_\chi-\ii(\rho+1)\tnu
\end{matrix}
;
1-r_2
\right],
\end{aligned}
\end{equation}
\begin{equation}
\label{Eulertran2}
\begin{aligned}
&{}_2\text F_1
\left[
\begin{matrix}
2\ii\rho\tnu+\delta_\chi-q_2,
\ii(\rho-1)\tnu+X+\alpha_\chi+3
\\
\ii(\rho+1)\tnu+\alpha_\chi-\zeta-q_2-1
\end{matrix}
;
1-r_2
\right]
\\
&=
r_2^{-2\ii(\rho-1)\tnu-X-\zeta-\delta_\chi-4}
\, 
{}_2\text F_1
\left[
\begin{matrix}
2\ii\tnu-\zeta-X-q_2-4,
-\ii(\rho-1)\tnu-\zeta-\beta_\chi-1
\\
\ii(\rho+1)\tnu+\alpha_\chi-\zeta-q_2-1
\end{matrix}
;
1-r_2
\right].
\end{aligned}
\end{equation}
We can get
\begin{equation}
\begin{aligned}
\mathcal{H}_+(X,\zeta;r_2)\big|_{\rm reg}
&=
\Gamma\!\left[
\begin{matrix}
\ii(\rho+1)\tnu+\alpha_\chi-\zeta-q_2-2
\\
2\ii\rho\tnu+\delta_\chi-q_2,
\ii(\rho-1)\tnu+X+\alpha_\chi+3
\end{matrix}
\right]
r_2^{-2\ii(\rho-1)\tnu-X-\zeta-\delta_\chi-4}
\\&
\times
{}_2\text F_1
\left[
\begin{matrix}
-\ii(\rho-1)\tnu-X-\alpha_\chi-2,
q_2+1-2\ii\rho\tnu-\delta_\chi
\\
q_2+3+\zeta-\alpha_\chi-\ii(\rho+1)\tnu
\end{matrix}
;
1-r_2
\right],
\end{aligned}
\label{H+regnew}
\end{equation}
\begin{equation}
\begin{aligned}
\mathcal{H}_+(X,\zeta;r_2)\big|_{\rm sin}
&=
\Gamma\!\left[
\begin{matrix}
-\ii(\rho+1)\tnu-\alpha_\chi+\zeta+q_2+2
\\
q_2+5-2\ii\tnu+X+\zeta,
\ii(\rho-1)\tnu+\zeta+\beta_\chi+2
\end{matrix}
\right]
\\&\times 
(1-r_2)^{
\ii(\rho+1)\tnu+\alpha_\chi-\zeta-q_2-2
}\, 
r_2^{-2\ii(\rho-1)\tnu-X-\zeta-\delta_\chi-4}
\\&\times
{}_2\text F_1
\left[
\begin{matrix}
2\ii\tnu-\zeta-X-q_2-4,
-\ii(\rho-1)\tnu-\zeta-\beta_\chi-1
\\
\ii(\rho+1)\tnu+\alpha_\chi-\zeta-q_2-1
\end{matrix}
;
1-r_2
\right].
\end{aligned}
\label{H+divnew}
\end{equation}
As a result, the ${}_2\text F_1$ functions appearing in $\mathcal{H}_+(X,\zeta;r_2)$ and $\mathcal{G}_+(X,\zeta;r_2)$ are identical.
\par 
After decomposing $\mathcal{H}_+(X,\zeta;r_2)$ and $\mathcal{G}_+(X,\zeta;r_2)$ into their regular and singular parts, we can perform the same decomposition for $\widehat{\mathcal{O}}_{1,\rho}^{q_2,\chi}(X,\zeta;r_2)$ and $\widehat{\mathcal{O}}_{2,\rho}^{q_2,\chi}(X,\zeta;r_2)$
\begin{equation}
\label{222}
   \left.\widehat{\mathcal{O}}_{1,\rho}^{q_2,\chi}(X,\zeta;r_2)\right|_{\mathrm{reg}}= \mathcal K_{1+}(X,\zeta)\mathcal{H}_+(X,\zeta;r_2)|_{\mathrm{reg}}+\mathcal K_{1-}(X,\zeta)\mathcal{H}_-(X,\zeta;r_2),
\end{equation}
\begin{equation}
   \left.\widehat{\mathcal{O}}_{1,\rho}^{q_2,\chi}(X,\zeta;r_2)\right|_{\mathrm{sin}}= \mathcal K_{1+}(X,\zeta)\mathcal{H}_+(X,\zeta;r_2)|_{\mathrm{sin}},
\end{equation}
\begin{equation}
    \widehat{\mathcal{O}}_{1,\rho}^{q_2,\chi}(X,\zeta;r_2)=\left.\widehat{\mathcal{O}}_{1,\rho}^{q_2,\chi}(X,\zeta;r_2)\right|_{\mathrm{reg}}
    +
    \left.\widehat{\mathcal{O}}_{1,\rho}^{q_2,\chi}(X,\zeta;r_2)\right|_{\mathrm{sin}},
\end{equation}
\begin{equation}
   \left.\widehat{\mathcal{O}}_{2,\rho}^{q_2,\chi}(X,\zeta;r_2)\right|_{\mathrm{reg}}= -\frac{\mathcal K_2(X,\zeta)}{r_2}\left[\mathcal{G}_+(X,\zeta;r_2)|_{\mathrm{reg}}
-\mathcal{G}_-(X,\zeta;r_2)\right],
\end{equation}
\begin{equation}
   \left.\widehat{\mathcal{O}}_{2,\rho}^{q_2,\chi}(X,\zeta;r_2)\right|_{\mathrm{sin}}= -\frac{\mathcal K_2(X,\zeta)}{r_2}\mathcal{G}_+(X,\zeta;r_2)|_{\mathrm{sin}},
\end{equation}
\begin{equation}
    \widehat{\mathcal{O}}_{2,\rho}^{q_2,\chi}(X,\zeta;r_2)=\left.\widehat{\mathcal{O}}_{2,\rho}^{q_2,\chi}(X,\zeta;r_2)\right|_{\mathrm{reg}}
    +
    \left.\widehat{\mathcal{O}}_{2,\rho}^{q_2,\chi}(X,\zeta;r_2)\right|_{\mathrm{sin}}.
\end{equation}
In this way, we can also directly decompose $\mathcal{B}_{\rho}^{q_2,\chi}$ as
\begin{equation}
\label{Breg}
    \begin{aligned}
       \left.\mathcal B^{q_2,\chi}_\rho(X,\zeta;r_2)\right|_{\mathrm{reg}}
       &=r_2^{-2\ii\tnu +X+\zeta}\left[\mathcal K_{1+}(X,\zeta)\mathcal{H}_+(X,\zeta;r_2)|_{\mathrm{reg}}+\mathcal K_{1-}(X,\zeta)\mathcal{H}_-(X,\zeta;r_2)\right]
       \\&
       -r_2^{-\delta_\chi-4-2\ii\rho\widetilde\nu}\mathcal K_2(X,\zeta)\left[\mathcal{G}_+(X,\zeta;r_2)|_{\mathrm{reg}}
-\mathcal{G}_-(X,\zeta;r_2)\right],
    \end{aligned}
\end{equation}
\begin{equation}
\label{Bdiv}
    \begin{aligned}
       \left.\mathcal B^{q_2,\chi}_\rho(X,\zeta;r_2)\right|_{\mathrm{sin}}
       &=r_2^{-2\ii\widetilde\nu+X+\zeta} \mathcal K_{1+}(X,\zeta)\mathcal{H}_+(X,\zeta;r_2)|_{\mathrm{sin}}
       \\&-
       r_2^{-\delta_\chi-4-2\ii\rho\widetilde\nu} \mathcal K_2(X,\zeta)\mathcal{G}_+(X,\zeta;r_2)|_{\mathrm{sin}}.
    \end{aligned}
\end{equation}
\begin{equation}
\label{228}
    \mathcal B^{q_2,\chi}_\rho(X,\zeta;r_2)=
    \left.\mathcal B^{q_2,\chi}_\rho(X,\zeta;r_2)\right|_{\mathrm{reg}}+\left.\mathcal B^{q_2,\chi}_\rho(X,\zeta;r_2)\right|_{\mathrm{sin}}.
\end{equation}
It should be noted that the decomposition in Eqs.~\eqref{222}--\eqref{228} is valid throughout the entire interval $0<r_2<1$, rather than only in the limit $r_2\to1$.
\par 
We now show that the potentially singular contributions to $\mathcal{B}_\rho^{q_2,\chi}$ cancel identically. Substituting Eqs.~\eqref{eq:K1plus}, \eqref{H+divnew}, \eqref{eq:K2}, and \eqref{eq:Gplussin_expand} into Eq.~\eqref{Bdiv}, we obtain
\begin{equation}
\begin{aligned}
\left.\mathcal B^{q_2,\chi}_\rho(X,\zeta;r_2)\right|_{\mathrm{sin}}
&=
\frac{
2\sqrt{\pi}\,
\csc\!\left[
\pi\bigl(2\ii(\rho-1)\tnu+X+\zeta+\delta_\chi+3\bigr)
\right]
}{
\Gamma\!\left[
-\ii(\rho-1)\tnu-X-\alpha_\chi-1
\right]
}
\\
&\quad\times
\Gamma\!\left[
-\ii(\rho+1)\tnu-\alpha_\chi+\zeta+q_2+2
\right]
(1-r_2)^{
\ii(\rho+1)\tnu+\alpha_\chi-\zeta-q_2-2
}
\\
&\quad\times
{}_2F_1\!\left[
\begin{matrix}
2\ii\tnu-\zeta-X-q_2-4,\,
-\ii(\rho-1)\tnu-\zeta-\beta_\chi-1
\\
\ii(\rho+1)\tnu+\alpha_\chi-\zeta-q_2-1
\end{matrix}
;1-r_2
\right]
\\
&\quad\times
\left(
-r_2^{-2\ii\rho\tnu-\delta_\chi-4}
+r_2^{-\delta_\chi-4-2\ii\rho\tnu}
\right)
\\
&=0.
\end{aligned}
\end{equation}
Thus, the potentially singular contributions from the individual hypergeometric functions cancel exactly in the complete kernel. For $0<r_2< 1$, we therefore have
\begin{equation}
    \mathcal{B}_\rho^{q_2,\chi}(X,\zeta;r_2)
    =
    \left.\mathcal B^{q_2,\chi}_\rho(X,\zeta;r_2)\right|_{\mathrm{reg}},
\end{equation}
whose right-hand side remains regular as $r_2\to1$. This provides a representation of the complete kernel that extends regularly to the folded point $r_2=1$.
\par 
Under this decomposition and cancellation, Eq.~\eqref{seedintLambda} can be rewritten as
\begin{equation}
\begin{aligned}
    \widehat{\mathcal{J}}_{-+,\rho,\lambda}^{q_1q_2,s_1s_2}
	&=\mathcal{N}_{\rho}^{q_1q_2,s_1s_2}\, r_1^{2\ii\rho\tnu}
    \int \frac{\dd X \dd \xi\dd \zeta}{(2\pi\ii)^3}\, 
    2^{\xi+\zeta}\, \Gamma(\xi-X)\, 
    \widetilde{\Gamma}^{s_1s_2}(\xi,\zeta)\, 
    \sum_{\chi=0}^3d_\chi\left.\mathcal B^{q_2,\chi}_\rho(X,\zeta;r_2)\right|_{\mathrm{reg}}.
\end{aligned}
\end{equation}
For the convenience of the subsequent calculations, we can further simplify $\left.\mathcal B^{q_2,\chi}_\rho(X,\zeta;r_2)\right|_{\mathrm{reg}}$. Notice that the first and the third terms in Eq.~\eqref{Breg} can be combined and simplified
\begin{equation}
\begin{aligned}
&r_2^{-2\ii\tnu +X+\zeta}\mathcal K_{1+}(X,\zeta)\mathcal{H}_+(X,\zeta;r_2)|_{\mathrm{reg}}-r_2^{-\alpha_\chi-\beta_\chi-4-2\ii\rho\widetilde\nu}\mathcal K_2(X,\zeta)\mathcal{G}_+(X,\zeta;r_2)|_{\mathrm{reg}}
\\&=
\frac{2}{\sqrt{\pi}}\,
r_2^{-2\ii\tnu+X+\zeta}
\frac{
\Gamma\!\left[
q_2+5-2\ii\tnu+X+\zeta
,
\ii(\rho-1)\tnu+\zeta+\beta_\chi+2
,
q_2+1-2\ii\rho\tnu-\delta_\chi
\right]
}
{
\ii(\rho-1)\tnu+X+\alpha_\chi+2
}
\\
&\times
{}_2\mathbf F_1\!
\left[
\begin{matrix}
q_2+5-2\ii\tnu+X+\zeta,\,
\ii(\rho-1)\tnu+\zeta+\beta_\chi+2
\\
q_2+3+\zeta-\alpha_\chi-\ii(\rho+1)\tnu
\end{matrix}
;
1-r_2
\right].
\end{aligned}
\label{eq:T1T3_expand}
\end{equation}
We summarize the complete expression of the simplified $\left.\mathcal B^\chi_\rho(X,\zeta;r_2)\right|_{\mathrm{reg}}$ here
\begin{equation}
    \begin{aligned}
       \left.\mathcal B^{q_2,\chi}_\rho(X,\zeta;r_2)\right|_{\mathrm{reg}}
       &=
\frac{2}{\sqrt{\pi}}\frac{
\Gamma\!\left[
q_2+5-2\ii\tnu+X+\zeta
,
\ii(\rho-1)\tnu+\zeta+\beta_\chi+2
,
q_2+1-2\ii\rho\tnu-\delta_\chi
\right]
}
{
\ii(\rho-1)\tnu+X+\alpha_\chi+2
}
\\
&\times
r_2^{-2\ii\tnu+X+\zeta}\, 
{}_2\mathbf F_1\!
\left[
\begin{matrix}
q_2+5-2\ii\tnu+X+\zeta,\,
\ii(\rho-1)\tnu+\zeta+\beta_\chi+2
\\
q_2+3+\zeta-\alpha_\chi-\ii(\rho+1)\tnu
\end{matrix}
;
1-r_2
\right]
\\&+\frac{4}{\sqrt{\pi}}
    \frac{\Gamma\left[q_2+5-2\ii\tnu+X+\zeta,\ii(\rho-1)\tnu+\zeta+\beta_\chi+2,\ii(\rho-1)\tnu+X+\alpha_\chi+2\right]}{\sin\left[\pi(2\ii(\rho-1)\tnu+X+\zeta+\delta_\chi+3)\right]}
    \\&\times 
    \sin\left[\frac{\pi}{2}(\ii(\rho-1)\tnu +X+\alpha_\chi)\right]
    \cos\left[\pi\left(\frac{3}{2}\ii(\rho-1)\tnu +\frac{X}{2}+\zeta+\frac{\alpha_\chi}{2}+\beta_\chi\right)\right]
    \\&\times r_2^{-2\ii\tnu +X+\zeta}
    \tF{q_2+5-2\ii\tnu+X+\zeta}{\ii(\rho-1)\tnu+\zeta+\beta_\chi+2}{2\ii(\rho-1)\tnu+X+\zeta+\delta_\chi+5}{-r_2}
    \\&+
    \frac{
2\sqrt{\pi}\,
\Gamma\!\left[
q_2+1-2\ii\rho\tnu-\delta_\chi
\right]
}{
\ii(\rho-1)\tnu+X+\alpha_\chi+2
}\, 
\csc\left[
\pi(2\ii(\rho-1)\tnu+X+\zeta+\delta_\chi+3)
\right]
\\&\times 
r_2^{-\delta_\chi-4-2\ii\rho\widetilde\nu}\, 
\tF{q_2+1-2\ii\rho\tnu -\delta_\chi}{-\ii(\rho-1)\tnu-X-\alpha_\chi-2}{-2\ii(\rho-1)\tnu-X-\zeta-\delta_\chi-3}{- r_2}.
    \end{aligned}
\end{equation}

\section{Convergence of the Double Mellin--Barnes Integral}
\label{appendixG}
In this appendix, we establish the absolute convergence of the double Mellin--Barnes integral in Eq.~\eqref{totalint} using the explicit kernel in Eq.~\eqref{Bregfinalexp}. We restrict the analysis to $q_2>-5$ and parameterize the integration contours as
\begin{equation}
\label{appG:eq1}
    X=c_X+\ii x,\qquad
    \zeta=c_\zeta+\ii y,\qquad
    x,y\in\mathbb R,
\end{equation}
where $c_X$ and $c_\zeta$ satisfy
\begin{equation}
\label{appG:eq2}
    -1<c_X<0,\qquad
    -1<c_\zeta<0,\qquad
    c_X+c_\zeta>-(q_2+5).
\end{equation}
This contour choice is justified in Section~\ref{5.2}. We additionally impose $c_X+c_\zeta\notin\mathbb Z$ so that the contours also avoid the removable singularities arising from the explicit cosecant factors. The integrand is therefore holomorphic in a neighborhood of the integration contours.
\par 
To estimate the integrand, we introduce the following factors
\begin{equation}
\label{UXzeta}
    \mathcal U(X,\zeta)
    \equiv 2^{X+\zeta}\, \Gamma\!\left[
		\begin{matrix}
			-X,2\ii\tnu - X,-\zeta,a_{s_2}+\zeta,-\zeta+2\ii\tnu \\
			 a_{s_2},\bar{a}_{s_2}
    		\end{matrix}
		\right]\, {}_2\mathbf F_1\!
\left[
\begin{matrix}
-X,2\ii\tnu - X\\
\bar{a}_{s_1} - X
\end{matrix}
;
\frac{1}{2}
\right],
\end{equation}
\begin{equation}
\label{M1}
\begin{aligned}
    \mathcal{M}_1(X,\zeta;r_2)&\equiv
    \frac{2}{\sqrt{\pi}}\frac{
\Gamma\!\left[
q_2+5-2\ii\tnu+X+\zeta
,
\ii(\rho-1)\tnu+\zeta+\beta_\chi+2
,
q_2+1-2\ii\rho\tnu-\delta_\chi
\right]
}
{
\ii(\rho-1)\tnu+X+\alpha_\chi+2
}
\\
&\times
r_2^{-2\ii\tnu+X+\zeta}\, 
{}_2\mathbf F_1\!
\left[
\begin{matrix}
q_2+5-2\ii\tnu+X+\zeta,\,
\ii(\rho-1)\tnu+\zeta+\beta_\chi+2
\\
q_2+3+\zeta-\alpha_\chi-\ii(\rho+1)\tnu
\end{matrix}
;
1-r_2
\right],
\end{aligned}
\end{equation}
\begin{equation}
\label{M2}
\begin{aligned}
    \mathcal{M}_2(X,\zeta;r_2)&\equiv
    \frac{4}{\sqrt{\pi}}
    \frac{\Gamma\left[q_2+5-2\ii\tnu+X+\zeta,\ii(\rho-1)\tnu+\zeta+\beta_\chi+2,\ii(\rho-1)\tnu+X+\alpha_\chi+2\right]}{\sin\left[\pi(2\ii(\rho-1)\tnu+X+\zeta+\delta_\chi+3)\right]}
    \\&\times 
    \sin\left[\frac{\pi}{2}(\ii(\rho-1)\tnu +X+\alpha_\chi)\right]
    \cos\left[\pi\left(\frac{3}{2}\ii(\rho-1)\tnu +\frac{X}{2}+\zeta+\frac{\alpha_\chi}{2}+\beta_\chi\right)\right]
    \\&\times r_2^{-2\ii\tnu +X+\zeta}
    \tF{q_2+5-2\ii\tnu+X+\zeta}{\ii(\rho-1)\tnu+\zeta+\beta_\chi+2}{2\ii(\rho-1)\tnu+X+\zeta+\delta_\chi+5}{-r_2},
\end{aligned}
\end{equation}
\begin{equation}
\label{M3}
\begin{aligned}
    \mathcal{M}_3(X,\zeta;r_2)&\equiv
    \frac{
2\sqrt{\pi}\,
\Gamma\!\left[
q_2+1-2\ii\rho\tnu-\delta_\chi
\right]
}{
\ii(\rho-1)\tnu+X+\alpha_\chi+2
}\, 
\csc\left[
\pi(2\ii(\rho-1)\tnu+X+\zeta+\delta_\chi+3)
\right]
\\&\times 
r_2^{-\delta_\chi-4-2\ii\rho\widetilde\nu}\, 
\tF{q_2+1-2\ii\rho\tnu -\delta_\chi}{-\ii(\rho-1)\tnu-X-\alpha_\chi-2}{-2\ii(\rho-1)\tnu-X-\zeta-\delta_\chi-3}{- r_2}.
\end{aligned}
\end{equation}
The full integrand in Eq.~\eqref{totalint} then takes the form
\begin{equation}
\label{Yint}
    \mathcal{Y}(X,\zeta;r_2) = \mathcal U(X,\zeta)\sum_\chi d_\chi \, \left[\mathcal{M}_1(X,\zeta;r_2)+\mathcal{M}_2(X,\zeta;r_2)+\mathcal{M}_3(X,\zeta;r_2)\right].
\end{equation}
Here, we have suppressed all indices for notational simplicity. Along the contours, we write these functions in terms of $x$ and $y$ and omit their dependence on the fixed value of $r_2$ when no confusion can arise.
\par 
For each fixed $r_2\in(0,1]$, absolute convergence of the double Mellin--Barnes integral requires
\begin{equation}
\label{appG:eq8}
\int_{\mathbb R^2}
\left|
\mathcal Y(x,y)
\right|\,\dd x\,\dd y
<\infty.
\end{equation}
For any fixed $R>0$, we divide the integration domain into the compact region $|x|+|y|\leq R$ and its complement. Since the integrand is continuous on the chosen contours, it is bounded on the compact region, and hence
\begin{equation}
\label{appG:eq9}
\int_{|x|+|y|\leq R}
\left|
\mathcal Y(x,y)
\right|\,\dd x\,\dd y
<\infty.
\end{equation}
It remains to prove that, for some $R>0$,
\begin{equation}
\label{appG:eq10}
\int_{|x|+|y|>R}
\left|
\mathcal Y(x,y)
\right|\,\dd x\,\dd y
<\infty.
\end{equation}
The required estimates must control every direction in which $|x|+|y|\to\infty$, including regions where one variable remains bounded while the other becomes large.
\par 
Throughout the proof, $C$ denotes a positive constant whose value may change from one estimate to another. It is independent of $x$ and $y$, but may depend on the fixed parameters, the contours, and the chosen value of $r_2\in(0,1]$.

\subsection{Asymptotic Expansions and Bounds for Special Functions}
We first collect the asymptotic formulas and inequalities needed below. For fixed real $a$, Stirling's formula gives
\begin{equation}
\label{stirling}
    \left|\Gamma[a+\ii z]\right| \sim \sqrt{2\pi}\, |z|^{a-\frac{1}{2}}e^{-\frac{\pi}{2}|z|},\quad a,z\in\mathbb R,\quad|z|\to\infty.
\end{equation}
The exponential suppression in this formula is central to the convergence proof. For fixed $a\in\mathbb R\setminus\{0,-1,-2,\ldots\}$, it also yields the following Stirling-type bounds, valid uniformly for all real $z$
\begin{equation}
\label{GammaBound}
    \left|\Gamma[a+\ii z]\right|
    \leq C(1+|z|)^{a-\frac12}
    e^{-\frac{\pi}{2}|z|},
    \quad 
    \frac{1}{|\Gamma[a+\ii z]|}
    \leq C(1+|z|)^{\frac12-a}
    e^{\frac{\pi}{2}|z|},
    \quad z\in\mathbb R.
\end{equation}
Here, $C$ may depend on $a$ but is independent of $z$. Stirling's formula establishes both bounds for sufficiently large $|z|$. On the remaining compact interval, the ratio
\begin{equation*}
    \frac{|\Gamma[a+\ii z]|}
    {(1+|z|)^{a-\frac12}e^{-\frac{\pi}{2}|z|}}
\end{equation*}
is continuous and strictly positive, and thus has a finite maximum and a positive minimum. The finite maximum bounds the ratio from above, while the positive
minimum ensures that its reciprocal is also bounded. Thus, both quantities are bounded on the compact interval, as well as for sufficiently large $|z|$ by Stirling's formula. Choosing $C$ large enough to cover both regions establishes both inequalities in Eq.~\eqref{GammaBound} for all $z\in\mathbb R$.

For the regularized Gauss hypergeometric function ${}_2\mathbf F_1$, the large-$|c|$ expansion with fixed upper parameters $a$ and $b$ is
\begin{equation}
\label{2F1asymbeha}
    \tF{a}{b}{c}{\frac{1}{2}} = \frac{1}{\Gamma[c]}\left[1+\mathcal{O}\left(|c|^{-1}\right)\right],\quad |c|\to\infty,\quad |\text{Arg}\, c|<\pi-\varepsilon ,
\end{equation}
where $\varepsilon$ denotes an arbitrarily small positive constant. This expansion follows from Ref.~\cite[Eq.~(15.12.2)]{NIST:DLMF} and will be applied to the
hypergeometric function in Eq.~\eqref{UXzeta} after an Euler transformation.
\par 
For the hypergeometric functions in Eqs.~\eqref{M1}--\eqref{M3}, both upper and lower parameters depend on the integration variables. We control these functions
using bounds obtained from Euler's integral representation. For real $z<1$, $\operatorname{Re}b>0$, and $\operatorname{Re}(c-b)>0$, this representation reads~\cite[Eq.~(15.6.1)]{NIST:DLMF}
\begin{equation}
\label{appG:eq14}
    \Gamma[b,c-b]\,
    {}_2\mathbf F_1\!\left[
        \begin{matrix}a,b\\c\end{matrix};z
    \right]
    =
    \int_0^1 \dd t\,
    t^{b-1}(1-t)^{c-b-1}(1-zt)^{-a}.
\end{equation}
Since $1-zt>0$ for $0\leq t\leq1$, taking the absolute value and applying the triangle inequality gives
\begin{equation}
\label{appG:eq15}
    \begin{aligned}
        \left|\int_0^1 \dd t\,
        t^{b-1}(1-t)^{c-b-1}(1-zt)^{-a}\right|
        &\leq \int_0^1 \dd t\,
        \left|t^{b-1}(1-t)^{c-b-1}(1-zt)^{-a}\right|
        \\
        &=\int_0^1 \dd t\,
        t^{\operatorname{Re}b-1}
        (1-t)^{\operatorname{Re}(c-b)-1}
        (1-zt)^{-\operatorname{Re}a}.
    \end{aligned}
\end{equation}
Substituting this estimate into Eq.~\eqref{appG:eq14}, we obtain
\begin{equation}
\label{appG:eq16}
    \left|\Gamma[b,c-b]\,
    {}_2\mathbf F_1\!\left[
        \begin{matrix}a,b\\c\end{matrix};z
    \right]\right|
    \leq
    \int_0^1 \dd t\,
    t^{\operatorname{Re}b-1}
    (1-t)^{\operatorname{Re}(c-b)-1}
    (1-zt)^{-\operatorname{Re}a}.
\end{equation}
The factor $1-zt$ lies between the positive numbers $1$ and $1-z$, so
\begin{equation}
\label{appG:eq17}
    (1-zt)^{-\operatorname{Re}a}
    \leq
    \max\left\{1,(1-z)^{-\operatorname{Re}a}\right\}.
\end{equation}
Using this bound and evaluating the remaining beta integral, we find
\begin{equation}
\label{2F1parabound}
    \begin{aligned}
        \left|
    \Gamma[b,c-b]\,
    {}_2\mathbf F_1\!\left[
        \begin{matrix}a,b\\c\end{matrix};z
    \right]
    \right|
    &\leq \max\left\{1,(1-z)^{-\operatorname{Re}a}\right\}
    \int_0^1 dt\,
    t^{\operatorname{Re}b-1}
    (1-t)^{\operatorname{Re}(c-b)-1}
    \\&=
    \Gamma\!\left[
        \begin{matrix}
            \operatorname{Re}b,\operatorname{Re}(c-b)\\
            \operatorname{Re}c
        \end{matrix}
    \right]\, 
    \max\left\{1,(1-z)^{-\operatorname{Re}a}\right\}.
    \end{aligned}
\end{equation}
For fixed $z$ and fixed real parts of the parameters, the right-hand side is a finite constant independent of their imaginary parts. This is the property needed to control the dependence on $x$ and $y$. When $\operatorname{Re}b$ or $\operatorname{Re}(c-b)$ is nonpositive, the bound cannot be applied directly. We instead use the following contiguous relations to shift the parameters into the required range~\cite[Eqs.~(15.5.15) and (15.5.16\_5)]{NIST:DLMF}
\begin{equation}
\label{contiguousrelations1}
    \begin{aligned}
        {}_2\mathbf F_1\!\left[
        \begin{matrix}a,b\\c\end{matrix};z
    \right]=
    c\, {}_2\mathbf F_1\!\left[
        \begin{matrix}a,b\\c+1\end{matrix};z
    \right]
    +abz\,{}_2\mathbf F_1\!\left[
        \begin{matrix}a+1,b+1\\c+2\end{matrix};z
    \right],
    \end{aligned}
\end{equation}
\begin{equation}
\label{contiguousrelations2}
    \begin{aligned}
        {}_2\mathbf F_1\!\left[
        \begin{matrix}a,b\\c\end{matrix};z
    \right]
    =
    c\,{}_2\mathbf F_1\!\left[
        \begin{matrix}a,b+1\\c+1\end{matrix};z
    \right]-a(c-b)z\,{}_2\mathbf F_1\!\left[
        \begin{matrix}a+1,b+1\\c+2\end{matrix};z
    \right].
    \end{aligned}
\end{equation}
\par 
We also need bounds for the trigonometric factors appearing in the $\mathcal M_i$. For real $a$ and $z$,
\begin{equation}
\label{appG:eq21}
    \begin{aligned}
        |\sin(a+\ii z)|^2
        &=\sin^2a+\sinh^2z
        \leq1+\sinh^2z=\cosh^2z,\\
        |\cos(a+\ii z)|^2
        &=\cos^2a+\sinh^2z
        \leq1+\sinh^2z=\cosh^2z.
    \end{aligned}
\end{equation}
Together with $\cosh z\leq e^{|z|}$, these relations imply
\begin{equation}
\label{sincosbound}
    |\sin(a+\ii z)|\leq e^{|z|},
    \qquad
    |\cos(a+\ii z)|\leq e^{|z|}\qquad z\in\mathbb R.
\end{equation}
For the cosecant function, we fix $a\notin\pi\mathbb Z$ and use
\begin{equation}
\label{appG:eq23}
    \begin{aligned}
        |\sin(a+\ii z)|^2
        =\sin^2a+\sinh^2z\geq\sin^2a\,(1+\sinh^2z)
        =\sin^2a\,\cosh^2z.
    \end{aligned}
\end{equation}
Since $\cosh z\geq\frac12e^{|z|}$, it follows that
\begin{equation}
\label{cscbound}
    |\csc(a+\ii z)|
    \leq\frac{2}{|\sin a|}e^{-|z|}
    =C_a e^{-|z|},
    \qquad z\in\mathbb R,
\end{equation}
where $C_a$ depends only on the fixed parameter $a$.

\subsection{Absolute Convergence of the Integral}
We now apply these estimates to $\mathcal U$ and the three kernel terms $\mathcal M_i$. The resulting bounds hold for all $x,y\in\mathbb R$ and therefore control every direction at infinity. Combining them will give
an integrable upper bound for the full integrand $\mathcal Y$.

\paragraph{Upper bound for $\mathcal M_1$.}
We introduce the parameters
\begin{equation}
\label{appG:eq29}
    \begin{aligned}
        a&=q_2+5-2\ii\tnu+X+\zeta,\\
        b&=\ii(\rho-1)\tnu+\zeta+\beta_\chi+2,\\
        c&=q_2+3+\zeta-\alpha_\chi-\ii(\rho+1)\tnu,
    \end{aligned}
\end{equation}
and rewrite Eq.~\eqref{M1} as
\begin{equation}
\label{g30}
    \mathcal M_1(X,\zeta;r_2)
    =
    \frac{2}{\sqrt{\pi}}\,
    \frac{\Gamma[a]\,r_2^{-2\ii\tnu+X+\zeta}}
    {\ii(\rho-1)\tnu+X+\alpha_\chi+2}\,
    \Gamma[b,c-b]\,
    {}_2\mathbf F_1\!\left[
        \begin{matrix}
            a,b\\c
        \end{matrix};1-r_2
    \right].
\end{equation}
Here, $\operatorname{Re}b=c_\zeta+\beta_\chi+2>0$, but
$\operatorname{Re}(c-b)=q_2+1-\delta_\chi$ need not be positive. Since $\delta_\chi\in\{0,-2\}$ and $q_2>-5$, four successive applications of Eq.~\eqref{contiguousrelations1} give
\begin{equation}
\label{M12F1iden}
    {}_2\mathbf F_1\!\left[
        \begin{matrix}
            a,b\\c
        \end{matrix};1-r_2
    \right]
    =
    \sum_{j=0}^{4}
    \binom{4}{j}
    (c+j)_{4-j}(a)_j(b)_j (1-r_2)^j\,
    {}_2\mathbf F_1\!\left[
        \begin{matrix}
            a+j,b+j\\
            c+4+j
        \end{matrix};1-r_2
    \right].
\end{equation}
The parameters of every shifted hypergeometric function satisfy
\begin{equation}
\label{appG:eq32}
    \operatorname{Re}(b+j)>0,\qquad
    \operatorname{Re}\!\left[(c+4+j)-(b+j)\right]
    =q_2+5-\delta_\chi>0.
\end{equation}
We can therefore apply Eq.~\eqref{2F1parabound} to each term
\begin{equation}
\label{appG:eq33}
    \left|
    \Gamma[b+j,c-b+4]\,
    {}_2\mathbf F_1\!\left[
        \begin{matrix}
            a+j,b+j\\
            c+4+j
        \end{matrix};1-r_2
    \right]
    \right|
    \leq C.
\end{equation}
For fixed $r_2$, the right-hand side of Eq.~\eqref{2F1parabound} depends only on the
fixed real parts of the shifted parameters, so the bound is independent of $x$ and $y$. Since only finitely many shifts occur, the same $C$ can be used for all $j=0,1,2,3,4$. Multiplying Eq.~\eqref{M12F1iden} by $\Gamma[b,c-b]$, using $(b)_j\Gamma[b]=\Gamma[b+j]$, and absorbing the fixed ratio $\Gamma[c-b]/\Gamma[c-b+4]$ into $C$, we obtain
\begin{equation}
\label{G25}
    \begin{aligned}
        \left|
        \Gamma[b,c-b]\,
        {}_2\mathbf F_1\!\left[
            \begin{matrix}
                a,b\\c
            \end{matrix};1-r_2
        \right]
        \right|
        &\leq C\sum_{j=0}^{4}
        \left|(c+j)_{4-j}(a)_j\right|\leq C(1+|x|+|y|)^4.
    \end{aligned}
\end{equation}
The second inequality follows by bounding each factor in $(c+j)_{4-j}$ and $(a)_j$ by $C(1+|x|+|y|)$ using the triangle inequality. These Pochhammer symbols contain $4-j$ and $j$ factors, respectively, giving four factors in total.
\par 
Substituting Eq.~\eqref{G25} into Eq.~\eqref{g30} yields
\begin{equation}
\label{appG:eq35}
    |\mathcal{M}_1(x,y)|\leq C\, \frac{(1+|x|+|y|)^4}{1+|x|}\left|\Gamma[q_2+5+c_X+c_\zeta+\ii(x+y-2\tnu)]\right|,
\end{equation}
where we have used
\begin{equation}
\label{G36}
    \frac{1}{\left|\ii(\rho-1)\tnu+X+\alpha_\chi+2\right|}
    \leq \frac{C}{1+|x|}.
\end{equation}
Applying the Stirling-type bound~\eqref{GammaBound} to the remaining Gamma function, we find
\begin{equation}
\label{appG:eq37}
    |\mathcal{M}_1(x,y)|\leq C\, \frac{(1+|x|+|y|)^4}{1+|x|}(1+|x+y-2\tnu|)^{q_2+\frac{9}{2}+c_X+c_\zeta}e^{-\frac{\pi}{2}|x+y-2\tnu|}.
\end{equation}
To remove the fixed shift from the algebraic and exponential factors, we use the following consequences of the triangle inequality
\begin{equation}
\label{appG:eq38}
    \begin{aligned}
        (1+|t-2\tnu|)^p
        &\leq (1+2\tnu)^{|p|}(1+|t|)^p,\\
        e^{-\frac{\pi}{2}|t-2\tnu|}
        &\leq e^{\pi\tnu}e^{-\frac{\pi}{2}|t|}.
    \end{aligned}
\end{equation}
Taking $t=x+y$ and $p=q_2+\frac92+c_X+c_\zeta$, and absorbing the fixed prefactors into $C$, gives
\begin{equation}
\label{M1upperbound}
    |\mathcal M_1(x,y)|
    \leq C\,
    \frac{(1+|x|+|y|)^4}{1+|x|}
    (1+|x+y|)^{q_2+\frac92+c_X+c_\zeta}
    e^{-\frac{\pi}{2}|x+y|}
    ,\quad x,y\in\mathbb R.
\end{equation}
The same treatment of fixed imaginary shifts will be used below.

\paragraph{Upper bound for $\mathcal M_2$.}
For this term, we choose
\begin{equation}
\label{appG:eq40}
    \begin{aligned}
        a&=q_2+5-2\ii\tnu+X+\zeta,\\
        b&=\ii(\rho-1)\tnu+\zeta+\beta_\chi+2,\\
        c&=2\ii(\rho-1)\tnu+X+\zeta+\delta_\chi+5,
    \end{aligned}
\end{equation}
so that Eq.~\eqref{M2} becomes
\begin{equation}
\label{appG:eq41}
    \begin{aligned}
        \mathcal M_2(X,\zeta;r_2)
        &=\frac{4}{\sqrt{\pi}}\,
        \frac{\Gamma[a,b,c-b-1]}{\sin[\pi(c-2)]}
        \sin\!\left[\frac{\pi}{2}(c-b-3)\right]
        \cos\!\left[\frac{\pi}{2}(c+b-7)\right]
        \\&\times 
        r_2^{a-q_2-5}\,
        {}_2\mathbf F_1\!\left[
            \begin{matrix}
                a,b\\c
            \end{matrix};-r_2
        \right].
    \end{aligned}
\end{equation}
Since
\begin{equation}
\label{appG:eq42}
    \begin{aligned}
        \operatorname{Re}b&=c_\zeta+\beta_\chi+2>0,\\
        \operatorname{Re}(c-b)&=c_X+\alpha_\chi+3>0,
    \end{aligned}
\end{equation}
the hypergeometric bound~\eqref{2F1parabound} applies without parameter shifts. To use it, we factor the Gamma functions as
\begin{equation}
\label{appG:eq43}
    \begin{aligned}
        &\Gamma[a,b,c-b-1]\,
        {}_2\mathbf F_1\!\left[
            \begin{matrix}
                a,b\\c
            \end{matrix};-r_2
        \right]
=
        \frac{\Gamma[a]}{c-b-1}
        \left\{
        \Gamma[b,c-b]\,
        {}_2\mathbf F_1\!\left[
            \begin{matrix}
                a,b\\c
            \end{matrix};-r_2
        \right]
        \right\},
    \end{aligned}
\end{equation}
Applying Eq.~\eqref{2F1parabound} then gives
\begin{equation}
\label{appG:eq44}
    |\mathcal M_2|
    \leq
    C\,\frac{|\Gamma[a]|}{|c-b-1|}
    \frac{
        \left|\sin\!\left[\frac{\pi}{2}(c-b-3)\right]\right|
        \left|\cos\!\left[\frac{\pi}{2}(c+b-7)\right]\right|
    }{
        |\sin[\pi(c-2)]|
    }.
\end{equation}
Combining the Stirling-type bound~\eqref{GammaBound}, the denominator estimate~\eqref{G36}, and the trigonometric bounds~\eqref{sincosbound} and \eqref{cscbound}, and using the triangle inequality to absorb fixed imaginary shifts, we obtain
\begin{equation}
\label{M2upperbound}
    |\mathcal M_2(x,y)|
    \leq
    C\,\frac{(1+|x+y|)^{q_2+\frac92+c_X+c_\zeta}}{1+|x|}
    e^{\frac{\pi}{2}\left(|x|+|y|-2|x+y|\right)},\quad x,y\in\mathbb R.
\end{equation}

\paragraph{Upper bound for $\mathcal M_3$.}
For $\mathcal M_3$, we introduce
\begin{equation}
\label{appG:eq46}
    \begin{aligned}
        a&=q_2+1-2\ii\rho\tnu-\delta_\chi,\\
        b&=-\ii(\rho-1)\tnu-X-\alpha_\chi-2,\\
        c&=-2\ii(\rho-1)\tnu-X-\zeta-\delta_\chi-3,
    \end{aligned}
\end{equation}
and rewrite Eq.~\eqref{M3} as
\begin{equation}
\label{M3abc}
    \mathcal M_3(X,\zeta;r_2)
    =
    \frac{2\sqrt{\pi}\,\Gamma[a]}{b\sin(\pi c)}
    r_2^{a-q_2-5}
    {}_2\mathbf F_1\!\left[
    \begin{matrix}
        a,b\\c
    \end{matrix};-r_2
    \right].
\end{equation}
In this case, $a$ is fixed, whereas
\begin{equation}
\label{appG:eq48}
    \begin{aligned}
        \operatorname{Re}b
        &=-c_X-\alpha_\chi-2>-3,\\
        \operatorname{Re}(c-b)
        &=-c_\zeta-\beta_\chi-1>-2.
    \end{aligned}
\end{equation}
Neither real part is necessarily positive. We therefore shift both $b$ and $c-b$ before applying Eq.~\eqref{2F1parabound}. First, we apply Eq.~\eqref{contiguousrelations2} three times, applying the relation to every term generated at each step.
Each application raises the second upper parameter by one, while the difference between the lower parameter and the second upper parameter either stays fixed or increases by one. The result is
\begin{equation}
\label{appG:eq49}
    {}_2\mathbf F_1\!\left[
        \begin{matrix}
            a,b\\c
        \end{matrix};-r_2
    \right]
    =
    \sum_{k=0}^{3}
    \binom{3}{k}
    (c+k)_{3-k}(a)_k(c-b)_k r_2^k\,
    {}_2\mathbf F_1\!\left[
        \begin{matrix}
            a+k,b+3\\
            c+3+k
        \end{matrix};-r_2
    \right].
\end{equation}
Next, for each $k=0,1,2,3$, we apply Eq.~\eqref{contiguousrelations1} three times to the shifted hypergeometric function in Eq.~\eqref{appG:eq49}. Combining the two finite
sums gives
\begin{equation}
    \begin{aligned}
        {}_2\mathbf F_1\!\left[
        \begin{matrix}
            a,b\\c
        \end{matrix};-r_2
        \right]
        =
        \sum_{j,k=0}^{3}P_{jk}(x,y)\,
        {}_2\mathbf F_1\!\left[
        \begin{matrix}
            a+j+k,b+3+j\\
            c+6+j+k
        \end{matrix};-r_2
        \right],
    \end{aligned}
    \label{M3shifted2F1}
\end{equation}
where
\begin{equation}
\label{appG:eq51}
    \begin{aligned}
        P_{jk}(x,y)
        ={}&
        \binom{3}{j}\binom{3}{k}
        (c+k)_{3-k}(c+3+j+k)_{3-j}
        (a)_{j+k}(c-b)_k(b+3)_j
        r_2^k(-r_2)^j.
    \end{aligned}
\end{equation}
Since $a$ is fixed, the dependence of $P_{jk}$ on $x$ and $y$ comes from the four Pochhammer symbols $(c+k)_{3-k}$, $(c+3+j+k)_{3-j}$, $(c-b)_k$, and $(b+3)_j$. They contain $3-k$, $3-j$, $k$, and $j$ factors, respectively, giving six factors in total. The triangle
inequality bounds each factor by $C(1+|x|+|y|)$, and hence
\begin{equation}
\label{appG:eq52}
    |P_{jk}(x,y)|
    \leq C(1+|x|+|y|)^6.
\end{equation}
\par 
The shifted parameters in every term of Eq.~\eqref{M3shifted2F1} satisfy
\begin{equation}
\label{appG:eq53}
    \begin{aligned}
        \operatorname{Re}(b+3+j)
        &=-c_X-\alpha_\chi+1+j>0,\\
        \operatorname{Re}\bigl[
        (c+6+j+k)-(b+3+j)
        \bigr]
        &=-c_\zeta-\beta_\chi+2+k>0.
    \end{aligned}
\end{equation}
Equation~\eqref{2F1parabound} therefore gives
\begin{equation}
    \left|
    \Gamma[b+3+j,c-b+3+k]\,
    {}_2\mathbf F_1\!\left[
    \begin{matrix}
        a+j+k,b+3+j\\
        c+6+j+k
    \end{matrix};-r_2
    \right]
    \right|
    \leq C.
    \label{M3shiftedbound}
\end{equation}
As in the estimate for $\mathcal M_1$, $C$ is independent of $x$ and $y$ and can be chosen for all $0\leq j,k\leq3$. To bound the hypergeometric function itself, we divide by the accompanying Gamma factors. Applying the reciprocal Gamma bound in Eq.~\eqref{GammaBound}, absorbing fixed imaginary
shifts into $C$, and dropping the factors $(1+|x|)^{-j}\leq1$ and $(1+|y|)^{-k}\leq1$, we obtain
\begin{equation}
\label{appG:eq55}
    \left|\frac{1}{\Gamma[b+3+j,c-b+3+k]}\right|\leq 
    C(1+|x|)^{c_X+\alpha_\chi-\frac12}
(1+|y|)^{c_\zeta+\beta_\chi-\frac32}e^{\frac{\pi}{2}(|x|+|y|)}.
\end{equation}
Combining Eqs.~\eqref{M3shifted2F1}, \eqref{appG:eq52}, \eqref{M3shiftedbound}, and \eqref{appG:eq55}, we find
\begin{equation}
\label{G51}
    \begin{aligned}
        \left|
        {}_2\mathbf F_1\!\left[
            \begin{matrix}
                a,b\\c
            \end{matrix};-r_2
        \right]
        \right|
        &\leq 
        \sum_{j,k=0}^{3}|P_{jk}(x,y)|
        \left|
        {}_2\mathbf F_1\!\left[
            \begin{matrix}
                a+j+k,b+3+j\\
                c+6+j+k
            \end{matrix};-r_2
        \right]
        \right|
        \\
        &\leq
        C(1+|x|+|y|)^6
        (1+|x|)^{c_X+\alpha_\chi-\frac12}
        (1+|y|)^{c_\zeta+\beta_\chi-\frac32}
        e^{\frac{\pi}{2}(|x|+|y|)}.
    \end{aligned}
\end{equation}
Substituting this bound into Eq.~\eqref{M3abc}, and using the cosecant bound~\eqref{cscbound} together with the denominator estimate~\eqref{G36}, yields
\begin{equation}
\label{M3upperbound}
    \begin{aligned}
        |\mathcal M_3(x,y)|
        \leq
        C\,(1+|x|+|y|)^6
        (1+|x|)^{c_X+\alpha_\chi-\frac32}
        (1+|y|)^{c_\zeta+\beta_\chi-\frac32}
        e^{\frac{\pi}{2}\left(|x|+|y|-2|x+y|\right)},
        \quad x,y\in\mathbb R.
    \end{aligned}
\end{equation}

\paragraph{Upper bound for $\mathcal U$.}
Applying the Euler transformation to the hypergeometric function in Eq.~\eqref{UXzeta}, we obtain
\begin{equation}
\label{appG:eq25}
    2^X\,{}_2\mathbf F_1\!\left[
    \begin{matrix}
        -X,2\ii\tnu-X\\
        \bar a_{s_1}-X
    \end{matrix};
    \frac12
    \right]
    =
    2^{-a_{s_1}}\,{}_2\mathbf F_1\!\left[
    \begin{matrix}
        \bar a_{s_1},a_{s_1}\\
        \bar a_{s_1}-X
    \end{matrix};
    \frac12
    \right].
\end{equation}
The upper parameters of the transformed function are fixed, whereas the modulus of its lower parameter grows with $|x|$. Equation~\eqref{2F1asymbeha} therefore gives
\begin{equation}
\label{appG:eq26}
    \Gamma[\bar a_{s_1}-X]\,
    {}_2\mathbf F_1\!\left[
        \begin{matrix}
            \bar a_{s_1},a_{s_1}\\
            \bar a_{s_1}-X
        \end{matrix};\frac12
    \right]
    \sim 
    1,
    \qquad |x|\to\infty.
\end{equation}
The left-hand side is continuous on every compact interval in $x$ and tends to unity as $|x|\to\infty$. It is therefore bounded on the entire real axis. Combining this observation with Eq.~\eqref{appG:eq25}, and absorbing the fixed factor $2^{-a_{s_1}}$ into $C$, gives
\begin{equation}
\label{appG:eq27}
    \left|
    2^X\,{}_2\mathbf F_1\!\left[
        \begin{matrix}
            -X,2\ii\tnu-X\\
            \bar a_{s_1}-X
        \end{matrix};\frac12
    \right]
    \right|
    \leq
    \frac{C}{|\Gamma[\bar a_{s_1}-X]|},
    \qquad x\in\mathbb R.
\end{equation}
Applying the Stirling-type bounds~\eqref{GammaBound} to the Gamma factors in $\mathcal U$, and absorbing fixed prefactors and imaginary shifts into $C$, we obtain
\begin{equation}
\label{Uupperbound}
    \left|\mathcal U(x,y)\right|
    \leq 
    C\,(1+|x|)^{\frac{s_1}{2}-c_X-1}
    (1+|y|)^{-\frac{s_2}{2}-c_\zeta-1}
    e^{-\frac{\pi}{2}|x|-\frac{3\pi}{2}|y|},
    \qquad x,y\in\mathbb R.
\end{equation}

\paragraph{Upper bound for $\mathcal Y$.}
Equations~\eqref{M1upperbound}, \eqref{M2upperbound}, \eqref{M3upperbound}, and \eqref{Uupperbound} give global upper bounds for $\mathcal U$ and $\mathcal M_i$. Taking their products, we obtain
\begin{equation}
\label{appG:eq58}
\begin{aligned}
|\mathcal U(x,y)\mathcal M_1(x,y)|
&\leq
C(1+|x|+|y|)^4
(1+|x|)^{\frac{s_1}{2}-c_X-2}
(1+|y|)^{-\frac{s_2}{2}-c_\zeta-1}
\\
&\times
(1+|x+y|)^{q_2+\frac92+c_X+c_\zeta}
e^{-\frac{\pi}{2}\left(|x|+3|y|+|x+y|\right)},
\qquad x,y\in\mathbb R,
\end{aligned}
\end{equation}
\begin{equation}
\label{appG:eq59}
\begin{aligned}
|\mathcal U(x,y)\mathcal M_2(x,y)|
&\leq
C(1+|x|)^{\frac{s_1}{2}-c_X-2}
(1+|y|)^{-\frac{s_2}{2}-c_\zeta-1}
\\
&\times
(1+|x+y|)^{q_2+\frac92+c_X+c_\zeta}
e^{-\pi\left(|y|+|x+y|\right)},
\qquad x,y\in\mathbb R,
\end{aligned}
\end{equation}
\begin{equation}
\label{appG:eq60}
\begin{aligned}
|\mathcal U(x,y)\mathcal M_3(x,y)|
&\leq
C(1+|x|+|y|)^6
(1+|x|)^{\frac{s_1}{2}+\alpha_\chi-\frac52}
\\
&\times
(1+|y|)^{-\frac{s_2}{2}+\beta_\chi-\frac52}
e^{-\pi\left(|y|+|x+y|\right)},
\qquad x,y\in\mathbb R.
\end{aligned}
\end{equation}

To obtain a common bound for these three terms, we use several consequences of the triangle inequality, in particular
\begin{equation}
\label{appG:eq61}
    |x|+|y|\leq 2\left(|y|+|x+y|\right).
\end{equation}
This controls the exponential factors in Eqs.~\eqref{appG:eq59} and \eqref{appG:eq60} by $e^{-\frac{\pi}{2}(|x|+|y|)}$; the exponential factor in Eq.~\eqref{appG:eq58} decays at least as fast. Bounding the remaining algebraic factors by a common power of $1+|x|+|y|$, and absorbing the finite $\chi$ sum and
the coefficients $d_\chi$ in Eq.~\eqref{Yint} into $C$, we obtain
\begin{equation}
\label{appG:eq62}
    |\mathcal Y(x,y)|
    \leq
    C(1+|x|+|y|)^N
    e^{-\frac{\pi}{2}(|x|+|y|)},
    \qquad x,y\in\mathbb R,
\end{equation}
where $N$ is a finite nonnegative exponent independent of $x$ and $y$. The exponential decay dominates the polynomial prefactor as $|x|+|y|\to\infty$, so, for any $R>0$,
\begin{equation}
\label{appG:eq63}
    \int_{|x|+|y|>R}
    |\mathcal Y(x,y)|\,\dd x\,\dd y
    <\infty.
\end{equation}
Together with Eq.~\eqref{appG:eq9}, this establishes the absolute convergence of the double Mellin--Barnes integral in Eq.~\eqref{totalint} for each fixed $r_2\in(0,1]$.
\par 
For $s_2=+1$, the contour deformation discussed in Section~\ref{5.2} also produces the one-dimensional residue integral in Eq.~\eqref{seedintNint}. After taking the residue at $\zeta=-a_+$, the common $X$-dependent factor and each kernel term contribute exponential factors $e^{-\pi|x|/2}$, up to algebraic prefactors. Applying the same estimation procedure as above to the residue integrand, denoted by $\mathcal R(x)$, we obtain
\begin{equation}
    |\mathcal R(x)|
    \leq C(1+|x|)^{N_{\mathrm{res}}}e^{-\pi|x|},
    \qquad x\in\mathbb R,
\end{equation}
where $N_{\mathrm{res}}$ is a finite nonnegative exponent independent of $x$. This bound establishes the absolute convergence of the residue integral.

\newpage


\bibliographystyle{JHEP}
\bibliography{biblio.bib}



\end{document}